\documentclass{ptephy_v1_for_Arxiv}
\usepackage{epstopdf}

\preprintnumber{XXXX-XXXX} 
\usepackage{hyperref}

\usepackage{amsmath} 
\usepackage{amsthm} 
\usepackage{hyperref} 
\usepackage{graphics} 
\usepackage{algorithmic} 
\usepackage{subfig} 
\usepackage{url} 
\usepackage{newunicodechar}
\newunicodechar{ }{\thinspace}
\usepackage{array}
\usepackage{comment}
\usepackage{rotating}
\usepackage{makecell}
\usepackage{url}
\usepackage{adjustbox}
\usepackage{multirow}
\usepackage{orcidlink}
\usepackage{ulem}

\newcolumntype{P}[1]{>{\centering\arraybackslash}p{#1}}

\usepackage{lscape}

\begin{document}

\title{Identification of Noise-Associated Glitches in KAGRA O3GK with Hveto}

\author{
T.~Akutsu\,\orcidlink{0000-0003-0733-7530}$^{1,2}$, 
M.~Ando$^{3,4}$, 
M.~Aoumi$^{5}$, 
A.~Araya\,\orcidlink{0000-0002-6884-2875}$^{6}$, 
Y.~Aso\,\orcidlink{0000-0002-1902-6695}$^{1,7}$, 
L.~Baiotti\,\orcidlink{0000-0003-0458-4288}$^{8}$, 
R.~Bajpai\,\orcidlink{0000-0003-0495-5720}$^{9}$, 
K.~Cannon\,\orcidlink{0000-0003-4068-6572}$^{4}$, 
A.~H.-Y.~Chen$^{10}$, 
D.~Chen\,\orcidlink{0000-0003-1433-0716}$^{11}$
% ...
\
(The KAGRA Collaboration)
and Y.-M~Kim\,\orcidlink{0000-0001-8720-6113}$^{40}$, 
}

\affil{
$^{1}$Gravitational Wave Science Project, National Astronomical Observatory of Japan, 2-21-1 Osawa, Mitaka City, Tokyo 181-8588, Japan\\
$^{2}$Advanced Technology Center, National Astronomical Observatory of Japan, 2-21-1 Osawa, Mitaka City, Tokyo 181-8588, Japan\\
$^{3}$Department of Physics, The University of Tokyo, 7-3-1 Hongo, Bunkyo-ku, Tokyo 113-0033, Japan\\
$^{4}$Research Center for the Early Universe (RESCEU), The University of Tokyo, 7-3-1 Hongo, Bunkyo-ku, Tokyo 113-0033, Japan\\
$^{5}$Institute for Cosmic Ray Research, KAGRA Observatory, The University of Tokyo, 238 Higashi-Mozumi, Kamioka-cho, Hida City, Gifu 506-1205, Japan\\
$^{6}$Earthquake Research Institute, The University of Tokyo, 1-1-1 Yayoi, Bunkyo-ku, Tokyo 113-0032, Japan\\
$^{7}$The Graduate University for Advanced Studies (SOKENDAI), 2-21-1 Osawa, Mitaka City, Tokyo 181-8588, Japan\\
$^{8}$International College, Osaka University, 1-1 Machikaneyama-cho, Toyonaka City, Osaka 560-0043, Japan\\
$^{9}$Accelerator Laboratory, High Energy Accelerator Research Organization (KEK), 1-1 Oho, Tsukuba City, Ibaraki 305-0801, Japan\\
$^{10}$Institute of Physics, National Yang Ming Chiao Tung University, 101 Univ. Street, Hsinchu, Taiwan\\
$^{11}$Kamioka Branch, National Astronomical Observatory of Japan, 238 Higashi-Mozumi, Kamioka-cho, Hida City, Gifu 506-1205, Japan\\
$^{12}$Department of Physics, National Tsing Hua University, No. 101 Section 2, Kuang-Fu Road, Hsinchu 30013, Taiwan\\
$^{13}$Faculty of Science, University of Toyama, 3190 Gofuku, Toyama City, Toyama 930-8555, Japan\\
$^{14}$Department of Electrophysics, National Yang Ming Chiao Tung University, 101 Univ. Street, Hsinchu, Taiwan\\
$^{15}$Department of Physics, Graduate School of Science, Osaka Metropolitan University, 3-3-138 Sugimoto-cho, Sumiyoshi-ku, Osaka City, Osaka 558-8585, Japan\\
$^{16}$Graduate School of Science, Institute of Science Tokyo, 2-12-1 Ookayama, Meguro-ku, Tokyo 152-8551, Japan\\
$^{17}$Institute of Physics, Academia Sinica, 128 Sec. 2, Academia Rd., Nankang, Taipei 11529, Taiwan\\
$^{18}$Department of Applied Physics, Fukuoka University, 8-19-1 Nanakuma, Jonan, Fukuoka City, Fukuoka 814-0180, Japan\\
$^{19}$Institute for Cosmic Ray Research, KAGRA Observatory, The University of Tokyo, 5-1-5 Kashiwa-no-Ha, Kashiwa City, Chiba 277-8582, Japan\\
$^{20}$College of Industrial Technology, Nihon University, 1-2-1 Izumi, Narashino City, Chiba 275-8575, Japan\\
$^{21}$Faculty of Engineering, Niigata University, 8050 Ikarashi-2-no-cho, Nishi-ku, Niigata City, Niigata 950-2181, Japan\\
$^{22}$Institute of Astronomy, National Tsing Hua University, No. 101 Section 2, Kuang-Fu Road, Hsinchu 30013, Taiwan\\
$^{23}$Institute of Photonics Technologies, National Tsing Hua University, No. 101 Section 2, Kuang-Fu Road, Hsinchu 30013, Taiwan\\
$^{24}$Department of Physics, Tamkang University, No. 151, Yingzhuan Rd., Danshui Dist., New Taipei City 25137, Taiwan\\
$^{25}$Department of Physical Sciences, Aoyama Gakuin University, 5-10-1 Fuchinobe, Sagamihara City, Kanagawa 252-5258, Japan\\
$^{26}$Nambu Yoichiro Institute of Theoretical and Experimental Physics (NITEP), Osaka Metropolitan University, 3-3-138 Sugimoto-cho, Sumiyoshi-ku, Osaka City, Osaka 558-8585, Japan\\
$^{27}$School of Astronomy and Space Science, University of Chinese Academy of Sciences, 20A Datun Road, Chaoyang District, Beijing, China\\
$^{28}$National Astronomical Observatories, Chinese Academic of Sciences, 20A Datun Road, Chaoyang District, Beijing, China\\
$^{29}$Department of Physics, Ulsan National Institute of Science and Technology (UNIST), 50 UNIST-gil, Ulju-gun, Ulsan 44919, Republic of Korea\\
$^{30}$Institute for Cosmic Ray Research, The University of Tokyo, 5-1-5 Kashiwa-no-Ha, Kashiwa City, Chiba 277-8582, Japan\\
$^{31}$Department of Astronomy and Space Science, Chungnam National University, 9 Daehak-ro, Yuseong-gu, Daejeon 34134, Republic of Korea\\
$^{32}$Institute of Particle and Nuclear Studies (IPNS), High Energy Accelerator Research Organization (KEK), 1-1 Oho, Tsukuba City, Ibaraki 305-0801, Japan\\
$^{33}$Division of Science, National Astronomical Observatory of Japan, 2-21-1 Osawa, Mitaka City, Tokyo 181-8588, Japan\\
$^{34}$School of Physics and Astronomy, Cardiff University, The Parade, Cardiff, CF24 3AA, UK\\
$^{35}$Department of Physics and Astronomy, University of Padova, Via Marzolo, 8-35151 Padova, Italy\\
$^{36}$Sezione di Padova, Istituto Nazionale di Fisica Nucleare (INFN), Via Marzolo, 8-35131 Padova, Italy\\
$^{37}$Instituto de Fisica Teorica UAM-CSIC, Universidad Autonoma de Madrid, 28049 Madrid, Spain\\
$^{38}$Department of Physics, Nagoya University, ES building, Furocho, Chikusa-ku, Nagoya, Aichi 464-8602, Japan\\
$^{39}$Department of Computer Simulation, Inje University, 197 Inje-ro, Gimhae, Gyeongsangnam-do 50834, Republic of Korea\\
$^{40}$Technology Center for Astronomy and Space Science, Korea Astronomy and Space Science Institute (KASI), 776 Daedeokdae-ro, Yuseong-gu, Daejeon 34055, Republic of Korea\\
$^{41}$Department of Physics, National Cheng Kung University, No.1, University Road, Tainan City 701, Taiwan\\
$^{42}$Research Center for Space Science, Advanced Research Laboratories, Tokyo City University, 3-3-1 Ushikubo-Nishi, Tsuzuki-Ku, Yokohama, Kanagawa 224-8551, Japan\\
$^{43}$Faculty of Law, Ryukoku University, 67 Fukakusa Tsukamoto-cho, Fushimi-ku, Kyoto City, Kyoto 612-8577, Japan\\
$^{44}$Istituto Nazionale di Fisica Nucleare (INFN), Universita di Roma "La Sapienza", P.le A. Moro 2, 00185 Roma, Italy\\
$^{45}$Phenikaa Institute for Advanced Study (PIAS), Phenikaa University, Yen Nghia, Ha Dong, Hanoi, Vietnam\\
$^{46}$Department of Astronomy, The University of Tokyo, 7-3-1 Hongo, Bunkyo-ku, Tokyo 113-0033, Japan\\
$^{47}$Physics Program, Graduate School of Advanced Science and Engineering, Hiroshima University, 1-3-1 Kagamiyama, Higashihiroshima City, Hiroshima 739-8526, Japan\\
$^{48}$Graduate School of Science and Technology, Niigata University, 8050 Ikarashi-2-no-cho, Nishi-ku, Niigata City, Niigata 950-2181, Japan\\
$^{49}$Niigata Study Center, The Open University of Japan, 754 Ichibancho, Asahimachi-dori, Chuo-ku, Niigata City, Niigata 951-8122, Japan\\
$^{50}$Department of Astronomy, Yonsei University, 50 Yonsei-Ro, Seodaemun-Gu, Seoul 03722, Republic of Korea\\
$^{51}$Department of Physics, University of Guadalajara, Av. Revolucion 1500, Colonia Olimpica C.P. 44430, Guadalajara, Jalisco, Mexico\\
$^{52}$Department of Electronic Control Engineering, National Institute of Technology, Nagaoka College, 888 Nishikatakai, Nagaoka City, Niigata 940-8532, Japan\\
$^{53}$Faculty of Science, Toho University, 2-2-1 Miyama, Funabashi City, Chiba 274-8510, Japan\\
$^{54}$Kavli Institute for Astronomy and Astrophysics, Peking University, Yiheyuan Road 5, Haidian District, Beijing 100871, China\\
$^{55}$Faculty of Information Science and Technology, Osaka Institute of Technology, 1-79-1 Kitayama, Hirakata City, Osaka 573-0196, Japan\\
$^{56}$Astronomical course, The Graduate University for Advanced Studies (SOKENDAI), 2-21-1 Osawa, Mitaka City, Tokyo 181-8588, Japan\\
$^{57}$Faculty of Science and Technology, Kochi University, 2-5-1 Akebono-cho, Kochi-shi, Kochi 780-8520, Japan\\
$^{58}$Laser Interferometry and Gravitational Wave Astronomy, Max Planck Institute for Gravitational Physics, Callinstrasse 38, 30167 Hannover, Germany\\
$^{59}$Department of Physics, Kyoto University, Kita-Shirakawa Oiwake-cho, Sakyou-ku, Kyoto City, Kyoto 606-8502, Japan\\
$^{60}$The Hakubi Center for Advanced Research, Kyoto University, Yoshida-honmachi, Sakyou-ku, Kyoto City, Kyoto 606-8501, Japan\\
$^{61}$Yukawa Institute for Theoretical Physics (YITP), Kyoto University, Kita-Shirakawa Oiwake-cho, Sakyou-ku, Kyoto City, Kyoto 606-8502, Japan\\
$^{62}$National Institute of Technology, Fukui College, Geshi-cho, Sabae-shi, Fukui 916-8507, Japan\\
$^{63}$Department of Communications Engineering, National Defense Academy of Japan, 1-10-20 Hashirimizu, Yokosuka City, Kanagawa 239-8686, Japan\\
$^{64}$Kavli Institute for the Physics and Mathematics of the Universe (Kavli IPMU), WPI, The University of Tokyo, 5-1-5 Kashiwa-no-Ha, Kashiwa City, Chiba 277-8583, Japan\\
$^{65}$Department of Astronomy, Beijing Normal University, Xinjiekouwai Street 19, Haidian District, Beijing 100875, China\\
$^{66}$School of Physics and Technology, Wuhan University, Bayi Road 299, Wuchang District, Wuhan, Hubei, 430072, China\\
}

\begin{abstract} 

Transient noise ("glitches") in gravitational wave detectors can mimic or obscure true signals, significantly reducing detection sensitivity. Identifying and excluding glitch-contaminated data segments is therefore crucial for enhancing the performance of gravitational-wave searches.
We perform a noise analysis of the KAGRA data obtained during the O3GK observation. Our analysis is performed with hierarchical veto (Hveto) which identifies noises based on the statistical time correlation between the main channel and the auxiliary channels. A total of 2,531 noises were vetoed by 28 auxiliary channels with the configuration (i.e., signal-to-noise threshold set to 8) that we chose for Hveto.
We identify vetoed events as glitches on the spectrogram via visual examination after plotting them with Q-transformation. 
By referring to the Gravity Spy project, we categorize 2,354 glitches into six types: blip, helix, scratchy, and scattered light, which correspond to those listed in Gravity Spy, and dot and line, which are not found in the Gravity Spy classification and are thus named based on their spectrogram morphology in KAGRA data. The remaining 177 glitches are determined not to belong to any of these six types. We show how the KAGRA glitch types are related to each subsystem of KAGRA. To investigate the possible correlation between the main channel and the round winner — an auxiliary channel statistically associated with the main channel for vetoing purposes — we visually examine the similarity or difference in the glitch pattern on the spectrogram.
We compare the qualitative correlation found through visual examination with coherence, which is known to provide quantitative measurement for the correlation between the main channel and each auxiliary channel. 
Our comprehensive noise analysis will help improve the data quality of KAGRA by applying it to future KAGRA observation data.

\end{abstract}

\maketitle

%
%
%
%
%

\section{Introduction}\label{sec:introduction}

KAGRA is a gravitational wave (GW) detector located in the Kamioka mine, Japan. KAGRA has a similar design to LIGO~\cite{Abbott_2009} and Virgo~\cite{Accadia_2011}, that is, a laser interferometer composed of two 3~km arms perpendicular to each other. However, what sets it apart from other detectors is its underground operation and its design to maintain the mirrors at cryogenic temperatures. These two features are expected to reduce environmental noises, for example, random vibration caused by heat. However, despite these noise-reducing features, transient noise artifacts known as glitches still occur and can significantly degrade the sensitivity of gravitational wave searches by mimicking or obscuring real signals \cite{PhysRevD.110.122002, Kwok_2022, Powell_2018, Mozzon_2022, Macas_2022, Hourihane_2022}. Therefore, identifying and characterizing these glitches is essential to improve the detectability of gravitational waves. Due to these two features, KAGRA is often called a 2.5-generation GW detector compared to current operating second-generation detectors such as advanced LIGO\cite{2015} and advanced Virgo~\cite{Acernese_2015}, and future third-generation detectors such as Einstein Telescope~\cite{Punturo_2010}, and Cosmic Explorer~\cite{galaxies10040090}. More details about KAGRA can be found in a series of articles~\cite{10.1093/ptep/ptaa120, 10.1093/ptep/ptaa125, 10.1093/ptep/ptab018, 10.1093/ptep/ptad112}.

Although KAGRA did not fully participate in the third observing run (O3) of LIGO and Virgo, it conducted its first observing run, called O3GK, in joint operation with GEO600—a German-British gravitational-wave detector located near Hannover, Germany—for approximately two weeks, from April 7, 2020, 08:00 UTC (GPS time: 1270281618) to April 21, 2020, 00:00 UTC (GPS time: 1271462418) \cite{Dooley_2016, 10.1093/ptep/ptac073}. This marked the beginning of KAGRA's data collection activities and coincided with the formation of the LIGO-Virgo-KAGRA (LVK) collaboration, initiating the integration of KAGRA into the global gravitational wave detection network \cite{10.1093/ptep/ptac093,Abbott_2023}. The performance of the KAGRA detector during O3GK and subsequent noise subtraction using independent component analysis were reported in Refs.~\cite{10.1093/ptep/ptac093, H_Abe_2023}.

Although the duration of O3GK was relatively short, the collected data provide a valuable opportunity to investigate the noise characteristics of the detector in its early operation phase. These data contain not only gravitational wave strain signals, but also signatures of various instrumental and environmental noise sources. Studying these noise sources allows us to identify and mitigate them. This process, often referred to as detector characterization (DetChar), ultimately aims to improve the data quality (DQ) and enhance the sensitivity of gravitational-wave detectors, including KAGRA. DetChar activities during the third observing run (O3) for the LIGO and Virgo detectors, as well as for KAGRA during its first observing run (O3GK), are described in Refs.~\cite{Davis_2021, Acernese_2023, 10.1093/ptep/ptab018}.

Auxiliary channels in gravitational wave detectors are used to monitor various instrumental or environmental noise sources that may affect the main gravitational wave signal. These channels track signals from subsystems, such as vibration sensors, temperature monitors, or power lines, which can contribute to noise. Noises identified in the main gravitational wave strain channel that originate from such sources often appear as glitches, which are short-duration transient signals. These auxiliary channels play a critical role in identifying and mitigating these glitches, improving the overall data quality of the detector.

To identify and analyze such glitches, a series of standard DetChar tools have been developed. One such tool is Omicron~\cite{ROBINET2020100620}, which detects excess power in the time-frequency representation of both the main and auxiliary channels using a Q-transform, assigning a signal-to-noise ratio (SNR) to each instance. These instances are referred to as triggers. Among them, short-duration transients in the main channel—typically caused by instrumental or environmental noise—are termed glitches. Another tool, hierarchical veto (Hveto)~\cite{Smith_2011}, builds on the Omicron results by statistically evaluating the time coincidence between triggers in the main and auxiliary channels. Triggers that exceed an additional SNR threshold defined by Hveto are referred to as events in this paper. Hveto compares these events between the main channel and the auxiliary channels to identify auxiliary channels that are time-correlated with the glitches, enabling noise-associated data to be vetoed, that is, excluded from further astrophysical analysis. In this context, a veto refers to the exclusion of data segments that are likely contaminated by noise, thereby improving the performance of gravitational wave searches by preventing the use of poor-quality data in the analysis.

In this study, we apply Hveto to the KAGRA O3GK data to identify statistically significant correlations between the Omicron triggers of the main channel and those of the auxiliary channels. Accordingly, this paper focuses not on all glitches found in the KAGRA data, but specifically on those that can be associated with known noise sources. Spectrograms, which are time-frequency representations of the glitches, are also used to investigate their morphology. These spectrograms are constructed using Q-transforms, which provide a detailed view of the glitches' frequency content over time.\footnote{We use the Q-transform module in GWpy~\cite{MACLEOD2021100657}, a Python package for gravitational wave astrophysics. An online example is available at \url{https://gwpy.github.io/docs/latest/examples/timeseries/qscan/}.}

According to their shape on the spectrogram, these glitches are categorized into six different types, some of which were previously identified in the LIGO and Virgo data. In particular, we adopt the naming convention used in the Gravity Spy project~\cite{Zevin_2017} for glitch types such as 'blip', 'helix', 'scratchy' and 'scattered light'. For glitch types such as 'dot' and 'line', which are not part of the Gravity Spy classification, we introduce these names based on their morphology observed in the KAGRA data. We also present which types of glitch appear in which subsystems of KAGRA. In addition to our approach based on auxiliary channel correlation, a recent study~\cite{Oshino2025} applied unsupervised machine learning to the same KAGRA O3GK dataset, identifying eight distinct glitch morphologies solely from the main channel spectrograms. Together, these approaches offer a more comprehensive understanding of transient noise artifacts in KAGRA.

One of the aims of DetChar is to identify noise sources and their coupling paths to the gravitational-wave strain channel, in addition to finding statistical correlations between the main and auxiliary channels. Although a strong correlation may suggest the possibility of subtracting the noise from the strain data, subtraction is not always feasible and often requires further analysis, such as verifying the linearity and stationarity of the coupling. Since Hveto identifies correlated auxiliary channels (round winners) based solely on the statistical time coincidence of triggers, additional investigations are required to confirm whether the correlation is physically meaningful and whether subtraction or mitigation is possible.
In this work, we search for additional evidence of correlation by visually comparing the spectrograms of the main and auxiliary channels. A similar glitch shape between the two spectrograms may suggest frequency-domain correlation, supplementing the time-domain correlation identified by Hveto. However, the absence of shape similarity does not necessarily imply a lack of correlation, as noise can couple through complex mechanisms; for example, broadband noise in an auxiliary channel may manifest at specific frequencies in the strain due to mechanical or optical resonances.

This paper is organized as follows. In the next section, we describe how we apply Hveto to the KAGRA O3GK data. Section \ref{sec:results} presents our results: the glitches found in the auxiliary channels, their classification, and their correlation with the KAGRA subsystem. In Section \ref{sec:discussion}, we discuss more about the correlation between the main channel and the auxiliary channels by comparing the results of our qualitative visual inspection with a quantitative measurement, coherence. Section \ref{sec:summary} summarizes our work. 

\section{Method}\label{sec:method}

During the O3GK observation run, the KAGRA interferometer was composed of several key subsystems, including Arm Length Stabilization (ALS; controls the differential arm length during lock acquisition), Auxiliary Optic System (AOS; monitors and controls auxiliary optics), Alignment Sensing and Control (ASC; maintains the alignment of optical components), Calibration (CAL; injects signals for calibration), Input Mode Cleaner (IMC; filters laser spatial modes before injection), Laser (LAS; provides the main laser source), Length Sensing and Control (LSC; maintains interferometric sensing and control), Output Mode Cleaner (OMC; filters output spatial modes before detection), Physical Environment Monitoring (PEM; monitors environmental disturbances), Pre-Stabilized Laser (PSL; stabilizes the frequency and intensity of the laser), and the Vibration Isolation System (VIS; suppresses seismic and mechanical vibrations). These subsystems collectively contribute to the stable operation and high sensitivity of the detector. 
\begin{figure*}[h]
\centering
\includegraphics[width=1\textwidth]{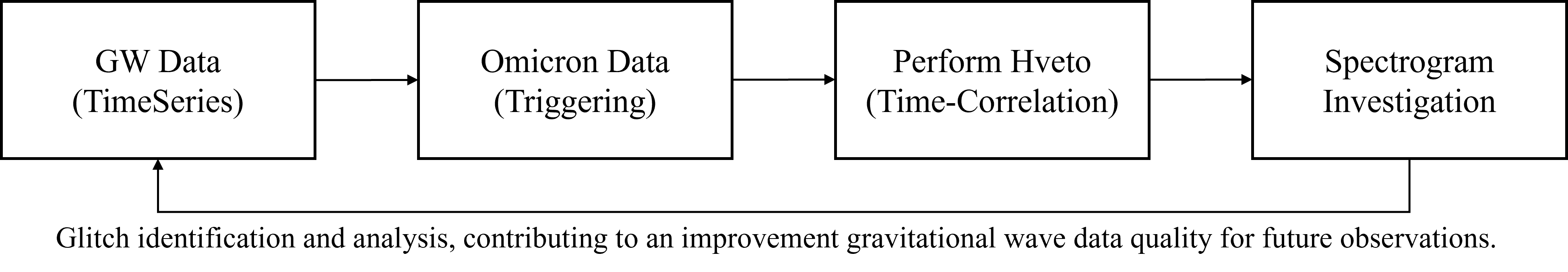}
\caption{Workflow of the noise-hunting: (1) Raw time-series data acquisition, (2) Multi-resolution time-frequency analysis using Omicron, (3) Hveto application to identify glitches based on time correlation, and (4) Glitch identification to analyze and categorize noise patterns, ultimately contributing to an improvement gravitational wave date quality for future observations.}
\label{fig:Workflow}
\end{figure*}
For our analysis, we used a total of 320 auxiliary channels distributed among the subsystems as follows: AOS (10), CAL (6), IMC (31), LAS (1), LSC (2), PEM (66), PSL (4) and VIS (200). The main gravitational wave strain channel used in this study was K1:DAC-STRAIN\_C20, which contains high-latency strain data calibrated after the observation run.

To identify and study instrumental glitches in the strain data, we employed a three-step pipeline composed of Omicron, Hveto, and visual inspection using spectrograms.
The workflow of this analysis is illustrated in Figure~\ref{fig:Workflow}, and is designed to identify noise transients associated with instrumental or environmental sources and exclude them from further astrophysical analysis. 

In the first step, the Omicron~\cite{ROBINET2020100620} identifies transient noise features -called triggers - by detecting excess power in time-frequency representations using a Q-transform. These triggers are generated for both the main strain channel and each of the auxiliary channels. Next, Hveto~\cite{Smith_2011} statistically evaluates the temporal coincidence between Omicron triggers in the main channel and those in the auxiliary channels. It performs this comparison over multiple time windows and thresholds, selecting the most statistically significant auxiliary channels, called "round winners," that are temporally correlated with glitches in the main channel. In the final step, we performed visual inspection of the spectrograms produced for both the main channel and each round-winning auxiliary channel. These spectrograms, generated using the Q-transform module in GWpy \cite{MACLEOD2021100657}, provide a time-frequency view of each glitch event. By comparing the morphology of glitches in both channels, we assessed whether there is additional evidence of correlation beyond time coincidence, e.g., similar frequency structure or duration. This inspection also allowed the classification of glitches into different morphological types, which are discussed in Section \ref{sec:results}.

In this way, the combined use of statistical and visual analysis enhances our ability to identify specific noise sources and their coupling mechanisms to the main gravitational-wave strain channel. The important list of parameter configurations used for Omicron and Hveto in our analysis is provided in the Appendix.

\section{Results}\label{sec:results} 
In this section, we present the KAGRA glitches identified during the O3GK observation.
Although the total number of glitches does not change with the fixed duration of the O3GK observation, we choose to analyze the O3GK KAGRA data on a daily basis.

Table \ref{table2:HvetoSummary} summarizes the Hveto results, including
the livetime of the KAGRA detector, the number of round winner channels, the number of total events and the number of vetoed events per day. A vetoed event refers to a trigger in the main channel that, after comparing with triggers in the auxiliary channels using the Hveto statistical analysis, is found to be statistically correlated with the auxiliary channels. In this paper, we investigate the spectrograms of the vetoed events to classify the glitches according to their morphology. During O3GK, Hveto identified a total of 2,531 vetoed events, which corresponds to the sum of the number of vetoed events per day for the entire O3GK observation period. We produced spectrogram images of these vetoed events (or glitches) for both the main channel and the roundwinner channel by using the Q-transform module in GWpy.
\begin{table*}[h!]
\caption{Daily summary of KAGRA glitches during O3GK} \label{table2:HvetoSummary}
\centering
\resizebox{0.8\textwidth}{!}
{%
\begin{tabular}{|@{}c|c|c|c|c|}
\hline
Date & Livetime (Ratio) & \begin{tabular}[c]{@{}c@{}}\# of \\ round winner\end{tabular} & \# of events & \# of vetoed events  \\ \hline
April 7th & 42,224s (73.31\%) & 6 & 1,101 & 164  \\
8th & 52,138s (60.34\%) & 4 & 1,200 & 118 \\
9th & 50,626s (58.59\%) & 5 & 1,465 & 194 \\
10th & 47,048s (54.45\%) & 3 & 1,232 & 83 \\
11th & 76,980s (89.10\%) & 7 & 1,065 & 153 \\ 
12th & 47,722s (55.23\%) & 3 & 958 & 169 \\
13th & 0s (0.00\%) & no data & no data & no data \\
14th & 8,074s (9.34\%) & not vetoed & not vetoed & not vetoed \\
15th & 28,284s (32.74\%) & 4 & 1,103 & 143 \\
16th & 55,718s (64.49\%) & 7 & 2,272 & 354 \\
17th & 52,998s (61.34\%) & 5 & 2,311 & 211 \\
18th & 55,718s (64.49\%) & 6 & 3,481 & 331 \\
19th & 67,692s (78.35\%) & 4 & 7,223 & 433 \\
20th & 63,000s (72.02\%) & 2 & 5,096 & 178 \\ \hline
\end{tabular} }
\end{table*}

In the following subsection, we describe the O3GK KAGRA glitches, including the search or identifying process, in more detail. 
The specific patterns that the glitches show on the spectrogram are often classified into certain types because the classification of the glitch types may play a role in the identification of the noise sources. In Section \ref{sec:GlitchType}, we present the classification of O3GK KAGRA glitches, together with their correlation with the KAGRA subsystems. 

\begin{figure*}[h!]
\begin{center}
\includegraphics[width=0.8\linewidth]{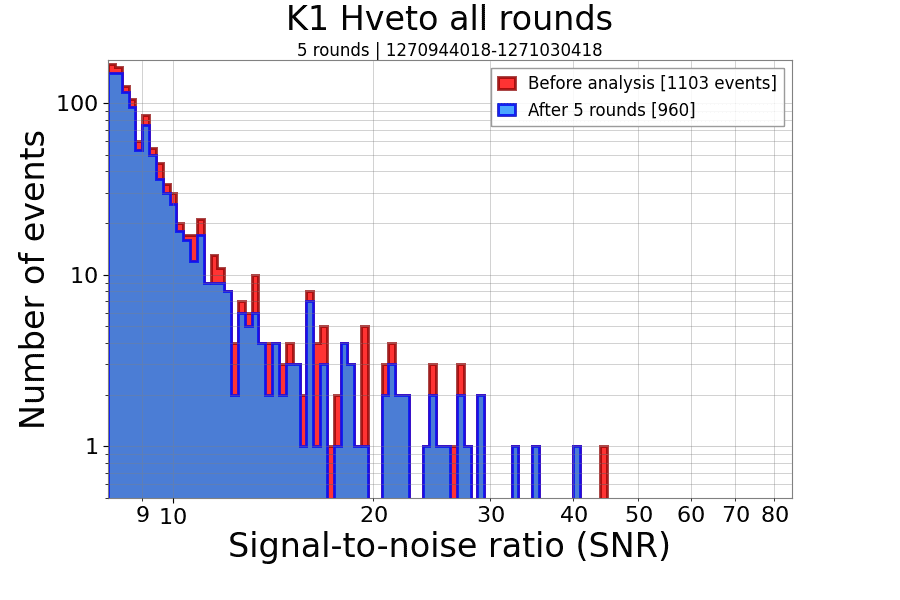} 
\caption{Hveto result on April 15, 2020. The GPS time of this day is shown in the second line of the figure title, from 00:00 UTC (GPS time: 1270944018) to 24:00 UTC (GPS time: 1271030418).
The red histogram displays the number of events generated by Omicron. On this day, the total number of events before applying Hveto was 1,103. The blue histogram shows the number of events (960) remaining after five rounds of veto. The difference between the red and blue histograms represents a total of 143 vetoed events on this day.}
\label{fig:ExampleofHveto}
\end{center}   
\end{figure*}

\subsection{O3GK KAGRA glitches}

Figure \ref{fig:ExampleofHveto} shows one of the results obtained with Hveto applied to the O3GK KAGRA data on April 15, 2020. On this day, four auxiliary channels were used to veto a total of 143 events of the main channel, with five rounds (also see Table \ref{table2:HvetoSummary}). On the other days of the O3GK observation, similar results (or plots) of Hveto are obtained. In fact, the results in Table \ref{table2:HvetoSummary} are collected from the daily Hveto analysis, and, as mentioned earlier, a total of 2,531 events are vetoed during O3GK.

\begin{table*}[h]
\centering
\caption{Round winner channels in the O3GK KAGRA data}\label{table3:Vetoedchannels}
\small
\resizebox{0.98\textwidth}{!}
{%
\begin{tabular}{|P{0.07\textwidth}|P{0.55\textwidth}|P{0.38\textwidth}|}
\hline
\begin{tabular}[c]{@{}c@{}}Sub-\\ System\end{tabular} & \begin{tabular}[c]{@{}c@{}}Round Winner \\ Auxiliary Channel\end{tabular} & \begin{tabular}[c]{@{}c@{}}Vetoed Date in April\\ (\# of Vetoed Events; Significance)\end{tabular} \\ \hline

AOS 
& K1:AOS-TMSX\_IR\_PDA1\_OUT\_DQ 
& \begin{tabular}[c]{@{}c@{}}10th (24; 7.37), 11th (3; 8.28), \\ 16th (102; 5.90), 17th (124; 6.99), \\ 18th (204; 7.60)\end{tabular} \\ \hline

\multirow{2}{*}{IMC}
& \begin{tabular}[c]{@{}c@{}}K1:IMC-IMMT1 \\ \_TRANS\_QPDA1\_DC\_PIT\_OUT\_DQ\end{tabular}  & \begin{tabular}[c]{@{}c@{}}8th (30; 7.75), 9th (100; 17.59),\\ 10th (55; 14.65), 11th (14; 8.36)\end{tabular}     \\ \cline{2-3} 
& \begin{tabular}[c]{@{}c@{}}K1:IMC-IMMT1 \\ \_TRANS\_QPDA1\_DC\_YAW\_OUT\_DQ\end{tabular} 
& \begin{tabular}[c]{@{}c@{}} 9th (14; 6.34), 11th (45; 25.59), \\ 19th (317; 10.15) \end{tabular} \\ \hline

LAS 
& K1:LAS-POW\_FIB\_OUT\_DQ
& 7th (81; 5.33) \\ \hline

\multirow{2}{*}{LSC}
& \begin{tabular}[c]{@{}c@{}}K1:LSC-ALS\_CARM\_OUT\_DQ\end{tabular}
& 16th (44; 6.97) 
\\ \cline{2-3}
& \begin{tabular}[c]{@{}c@{}}K1:LSC-ALS\_DARM\_OUT\_DQ\end{tabular}
& \begin{tabular}[c]{@{}c@{}} 12th (132; 6.36), 15th (91; 5.41), \\ 16th (67; 26.80) \end{tabular}
\\ \hline

\multirow{9}{*}{PEM}
& K1:PEM-ACC\_OMC\_TABLE\_AS\_Z\_OUT\_DQ
& 17th (4; 9.04) 
\\ \cline{2-3}
& K1:PEM-MAG\_BS\_BOOTH\_BS\_X\_OUT\_DQ
& 16th (8; 7.70) 
\\ \cline{2-3}
& K1:PEM-MAG\_BS\_BOOTH\_BS\_Y\_OUT\_DQ
& 8th (6; 5.07) 
\\ \cline{2-3}
& K1:PEM-MAG\_BS\_BOOTH\_BS\_Z\_OUT\_DQ
& 12th (18; 46.29), 18th (5; 7.30) 
\\ \cline{2-3}
& K1:PEM-MAG\_EXC\_BOOTH\_EXC\_Y\_OUT\_DQ
& 15th (5; 7.33) 
\\ \cline{2-3}
& K1:PEM-MIC\_MCF\_TABLE\_REFL\_Z\_OUT\_DQ
& 15th (5; 7.73) 
\\ \cline{2-3}
& K1:PEM-MIC\_SR\_BOOTH\_SR\_Z\_OUT\_DQ
& \begin{tabular}[c]{@{}c@{}} 16th (16; 10.23), 17th (14; 11.60), \\ 18th (13; 8.23) \end{tabular}
\\ \cline{2-3}
& K1:PEM-SEIS\_IXV\_GND\_EW\_IN1\_DQ
& 7th (14; 5.35) 
\\ \cline{2-3}
& K1:PEM-VOLT\_REFL\_TABLE\_GND\_OUT\_DQ
& 11th (7; 5.94) 
\\ \hline

\multirow{13}{*}{VIS}
& K1:VIS-ETMX\_MN\_PSDAMP\_R\_IN1\_DQ
& 7th (7; 5.36)
\\ \cline{2-3}
& K1:VIS-ETMX\_MN\_PSDAMP\_Y\_IN1\_DQ
& 7th (26; 6.57)
\\ \cline{2-3}
& K1:VIS-ETMY\_MN\_PSDAMP\_Y\_IN1\_DQ
& 19th (21; 5.27)
\\ \cline{2-3}
& K1:VIS-ITMY\_IM\_PSDAMP\_R\_IN1\_DQ
& \begin{tabular}[c]{@{}c@{}}7th (14; 42.60), 8th (50; 154.76), \\ 9th (37; 98.84), 11th (29; 92.86), \\ 18th (63; 146.92)\end{tabular}
\\ \cline{2-3}
& K1:VIS-ITMY\_MN\_OPLEV\_TILT\_YAW\_OUT\_DQ
& 9th(9; 5.58)
\\ \cline{2-3}
& K1:VIS-ITMY\_MN\_PSDAMP\_L\_IN1\_DQ
& \begin{tabular}[c]{@{}c@{}}15th (21; 57.66), 16th (97; 262.28),\\ 17th (49; 126.68), 19th (43; 87.83)\end{tabular} 
\\ \cline{2-3}
& K1:VIS-ITMY\_MN\_PSDAMP\_Y\_IN1\_DQ
& 10th (4; 10.65)
\\ \cline{2-3}
& K1:VIS-OMMT1\_TM\_OPLEV\_PIT\_OUT\_DQ
& \begin{tabular}[c]{@{}c@{}}8th (32; 34.21), 9th (34; 27.13), \\ 12th (19; 12.68), 16th (20; 9.45), \\ 18th (31; 5.00)\end{tabular} 
\\ \cline{2-3}
& K1:VIS-OMMT1\_TM\_OPLEV\_YAW\_OUT\_DQ
& \begin{tabular}[c]{@{}c@{}}11th (49; 65.56), 17th (20; 14.06),\\ 18th (15; 29.53), 19th (52; 30.93)\end{tabular}
\\ \cline{2-3}
& K1:VIS-OSTM\_TM\_OPLEV\_PIT\_OUT\_DQ
& 15th (21; 21.32) 
\\ \cline{2-3}
& K1:VIS-OSTM\_TM\_OPLEV\_YAW\_OUT\_DQ 
& 7th (22; 22.58), 20th (17; 11.15)
\\ \cline{2-3}
& K1:VIS-SR3\_TM\_OPLEV\_TILT\_YAW\_OUT\_DQ
& 11th (6; 7.44)
\\ \cline{2-3}
& K1:VIS-TMSY\_DAMP\_R\_IN1\_DQ 
& 20th (161; 5.07)
\\ \hline

\end{tabular} }
\end{table*}

Because our Hveto calculations were performed on a daily basis, some round winner channels appear on multiple days in Table~\ref{table2:HvetoSummary}. Identifying these repeated channels, we found that a total of 28 auxiliary channels were used to veto 2,531 events of the main channel throughout the O3GK period. Since identifying which vetoed auxiliary channel belongs to which subsystem of KAGRA helps to study the noise sources, we sorted the vetoed auxiliary channels into each KAGRA subsystem, as shown in Table~\ref{table3:Vetoedchannels}. This table also presents the number of events that were vetoed along with their Hveto significance values on each date that was vetoed, in the format \textit{(number of vetoed events; significance)}. For example, in the AOS subsystem, a single auxiliary channel (K1:AOS-TMSX\_IR\_PDA1\_OUT\_DQ) had 457 veto occurrences over five days of high significance (10 April (24; 7.37), 11 April (3; 8.28), 16 April (102; 5.90), 17 April (124; 6.99) and 18 April (204; 7.60)). This information helps to distinguish statistically robust correlations from possible chance coincidences.
\begin{table*}[!h]
\centering
\caption{O3GK KAGRA glitches with different types found in each subsystem}\label{table:Glitchtypes}
\begin{tabular}{ccccccccc}
\hline
      & Blip   & Dot & Helix & Line & Scratchy & scattered light & Total  & Not Classified\\ \hline\hline
AOS   & 227    & 4   & 92    & 82   & 31       & 2       & 438    & 19  \\ 
IMC   & 255    & 16  & 96    & 122  & 44       & 0       & 533    & 42  \\
LAS   & 58     & 0   & 7     & 0    & 12       & 0       & 77     & 4   \\
LSC   & 99     & 14  & 39    & 65   & 31       & 53      & 301    & 33  \\
PEM   & 83     & 3   & 9     & 10   & 5        & 0       & 110    & 5   \\
VIS   & 572    & 13  & 77    & 199  & 34       & 0       & 895    & 74  \\
Total & 1,294  & 50  & 320   & 478  & 157      & 55      & 2,354  & 177 \\
\hline
\end{tabular}
\end{table*}

Using the Q-transform module in GWpy~\cite{MACLEOD2021100657}, we plot spectrograms of the entire 2,531 vetoed events for both the main channel and the auxiliary channels (round winners). We visually examine the spectrogram images of the main channel and identify a total of 2,354 glitches that are apparent on the spectrogram. The remaining 177 spectrogram images are either difficult to identify or do not show any glitch-like features. We visually inspect all of these glitch spectrograms and classify them into six types: blip (1,294), dot (50), helix (320), line (478), scratchy (157), and scattered light (55) (the numbers in parentheses are the number of identified glitches) referring to the Gravity Spy project~\cite{Zevin_2017}.

Table \ref{table:Glitchtypes} summarizes the O3GK KAGRA glitches, classified into six types, and the KAGRA subsystems where they were found.
In the following subsection, we will show some examples of each glitch type and investigate the correlation between the glitch type and the KAGRA subsystem in more detail.

\begin{table*}[t]
\centering
\caption{\label{table:BlipGlitches} O3GK KAGRA blip glitches.}
\small
\resizebox{0.98\textwidth}{!}
{%
\begin{tabular}{|P{0.07\textwidth}|P{0.58\textwidth}|P{0.35\textwidth}|}
\hline
\begin{tabular}[c]{@{}c@{}}Sub-\\ System\end{tabular} & \begin{tabular}[c]{@{}c@{}}Round Winner\\ Auxiliary Channel\end{tabular} & \begin{tabular}[c]{@{}c@{}}Vetoed Date in April\\ (\# of Vetoed Events)\end{tabular} \\ \hline

\begin{tabular}[c]{@{}c@{}}AOS \\ (227) \end{tabular}
& \textit{K1:AOS-TMSX\_IR\_PDA1\_OUT\_DQ}
& \begin{tabular}[c]{@{}c@{}}10th (11), 16th (38), \\ 17th (78), 18th (100)\end{tabular} \\ \hline

\multirow{2}{*}{\begin{tabular}[c]{@{}c@{}} IMC \\ (255) \end{tabular}}
& \begin{tabular}[c]{@{}c@{}}\textit{K1:IMC-IMMT1}\\ \textit{\_TRANS\_QPDA1\_DC\_PIT\_OUT\_DQ}\end{tabular}  
& \begin{tabular}[c]{@{}c@{}}8th (9), 9th (31),\\ 10th (11), 11th (1)\end{tabular}\\ \cline{2-3} 
& \begin{tabular}[c]{@{}c@{}}\textit{K1:IMC-IMMT1}\\ \textit{\_TRANS\_QPDA1\_DC\_YAW\_OUT\_DQ}\end{tabular} 
& 9th (3), 11th (9), 19th (191)                                                          \\ \hline

\begin{tabular}[c]{@{}c@{}}LAS \\ (58) \end{tabular}
& \textbf{K1:LAS-POW\_FIB\_OUT\_DQ}
& 7th (58) \\ \hline

\multirow{2}{*}{\begin{tabular}[c]{@{}c@{}}LSC \\ (99) \end{tabular}}
& \begin{tabular}[c]{@{}c@{}}\textit{K1:LSC-ALS\_CARM\_OUT\_DQ}\end{tabular}
& 16th (13) 
\\ \cline{2-3}
& \begin{tabular}[c]{@{}c@{}}\textbf{K1:LSC-ALS\_DARM\_OUT\_DQ}\end{tabular}
& 12th (42), 15th (36), 16th (8) 
\\ \hline

\multirow{8}{*}{\begin{tabular}[c]{@{}c@{}}PEM \\ (83) \end{tabular}}
& \textbf{K1:PEM-MAG\_BS\_BOOTH\_BS\_X\_OUT\_DQ}
& 16th (6) 
\\ \cline{2-3}
& \textbf{K1:PEM-MAG\_BS\_BOOTH\_BS\_Y\_OUT\_DQ}
& 8th (6) 
\\ \cline{2-3}
& \textbf{K1:PEM-MAG\_BS\_BOOTH\_BS\_Z\_OUT\_DQ}
& 12th (16), 18th (4) 
\\ \cline{2-3}
& \begin{tabular}[c]{@{}c@{}}\textbf{K1:PEM-MAG}\\ \textbf{\_EXC\_BOOTH\_EXC\_Y\_OUT\_DQ}\end{tabular}
& 15th (3) 
\\ \cline{2-3}
& \textit{K1:PEM-MIC\_SR\_BOOTH\_SR\_Z\_OUT\_DQ}
& 16th (13), 17th (12), 18th (10)
\\ \cline{2-3}
& \textit{K1:PEM-SEIS\_IXV\_GND\_EW\_IN1\_DQ}
& 7th (9) 
\\ \cline{2-3}
& \textit{K1:PEM-VOLT\_REFL\_TABLE\_GND\_OUT\_DQ}
& 11th (4) 
\\ \hline

\multirow{12}{*}{\begin{tabular}[c]{@{}c@{}}VIS \\ (572) \end{tabular}}
& \textit{K1:VIS-ETMX\_MN\_PSDAMP\_R\_IN1\_DQ}
& 7th (2)
\\ \cline{2-3}
& \textit{K1:VIS-ETMX\_MN\_PSDAMP\_Y\_IN1\_DQ}
& 7th (21)
\\ \cline{2-3}
& \textit{K1:VIS-ETMY\_MN\_PSDAMP\_Y\_IN1\_DQ}
& 19th (15)
\\ \cline{2-3}
& \textbf{K1:VIS-ITMY\_IM\_PSDAMP\_R\_IN1\_DQ}
& \begin{tabular}[c]{@{}c@{}}7th (13), 8th (46), 9th (30),\\ 11th (25), 18th (47)\end{tabular}
\\ \cline{2-3}
& \textit{K1:VIS-ITMY\_MN\_OPLEV\_TILT\_YAW\_OUT\_DQ}
& 9th(5)
\\ \cline{2-3}
& \textbf{K1:VIS-ITMY\_MN\_PSDAMP\_L\_IN1\_DQ}
& \begin{tabular}[c]{@{}c@{}}15th (21), 16th (84),\\ 17th (45), 19th (34)\end{tabular} 
\\ \cline{2-3}
& \textbf{K1:VIS-ITMY\_MN\_PSDAMP\_Y\_IN1\_DQ}
& 10th (2)
\\ \cline{2-3}
& \textit{K1:VIS-OMMT1\_TM\_OPLEV\_PIT\_OUT\_DQ}
& \begin{tabular}[c]{@{}c@{}}8th (3), 9th (5), 12th (5),\\ 16th (3), 18th (14)\end{tabular} 
\\ \cline{2-3}
& \textit{K1:VIS-OMMT1\_TM\_OPLEV\_YAW\_OUT\_DQ}
& \begin{tabular}[c]{@{}c@{}}11th (2), 17th (2),\\ 18th (7), 19th (12)\end{tabular}
\\ \cline{2-3}
& \textit{K1:VIS-OSTM\_TM\_OPLEV\_YAW\_OUT\_DQ}
& 7th (5), 20th (5)
\\ \cline{2-3}
& \textit{K1:VIS-SR3\_TM\_OPLEV\_TILT\_YAW\_OUT\_DQ}
& 11th (2)
\\ \cline{2-3}
& \textit{K1:VIS-TMSY\_DAMP\_R\_IN1\_DQ}
& 20th (117)
\\ \hline

\end{tabular} }
\end{table*}

\subsection{Classification of O3GK KAGRA Glitches}\label{sec:GlitchType}
\subsubsection{Blip Glitch}
On a spectrogram, a blip glitch appears as a vertically elongated line. Understanding the blip glitch, including its source, is important, especially compared to other types of glitches, because it can often be confused with the chirp signal produced by a gravitational wave from the merger of high-mass binaries \cite{Abbott_2016}.
For the same reason, recognizing and distinguishing blip glitches is crucial to enhancing data quality, which improves GW signal detection, particularly in detecting GW emitted from CBC. 
We refer to Ref. \cite{Cabero_2019} for more details on the blip glitches in the LIGO detectors found during O1 and O2.

Blip glitches dominate the O3GK KAGRA glitches, accounting for about $54.97 \%$ (1,294 out of 2,354). We also find that the blip glitches are present in all subsystems of KAGRA. Table \ref{table:BlipGlitches} summarizes the O3GK KAGRA blip glitches, grouped by each KAGRA subsystem. In the first column, the number of blip glitches is presented in parentheses below the name of each subsystem. The second and third columns show the name of the round winner auxiliary channel and the number of blip glitches found in each auxiliary channel on a given date, respectively.

\begin{figure}[h]
\centering
\resizebox{!}{0.132\textheight}
{%
\begin{minipage}[t]{0.45\textwidth}
    \centering
    \includegraphics[width=0.48\textwidth]{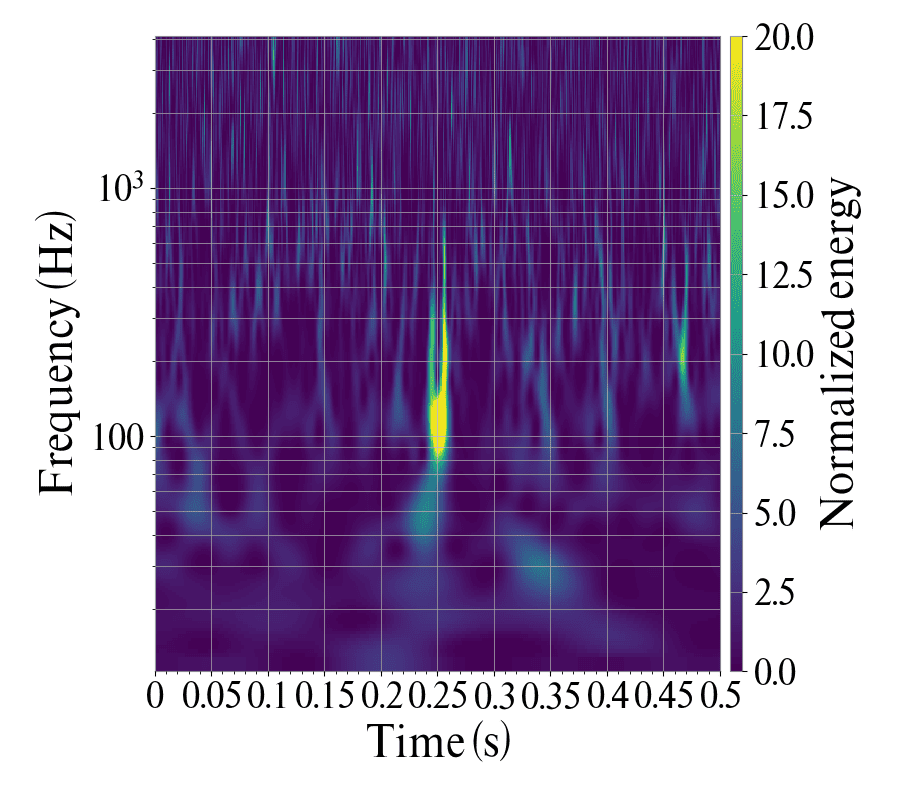}
    \includegraphics[width=0.48\textwidth]{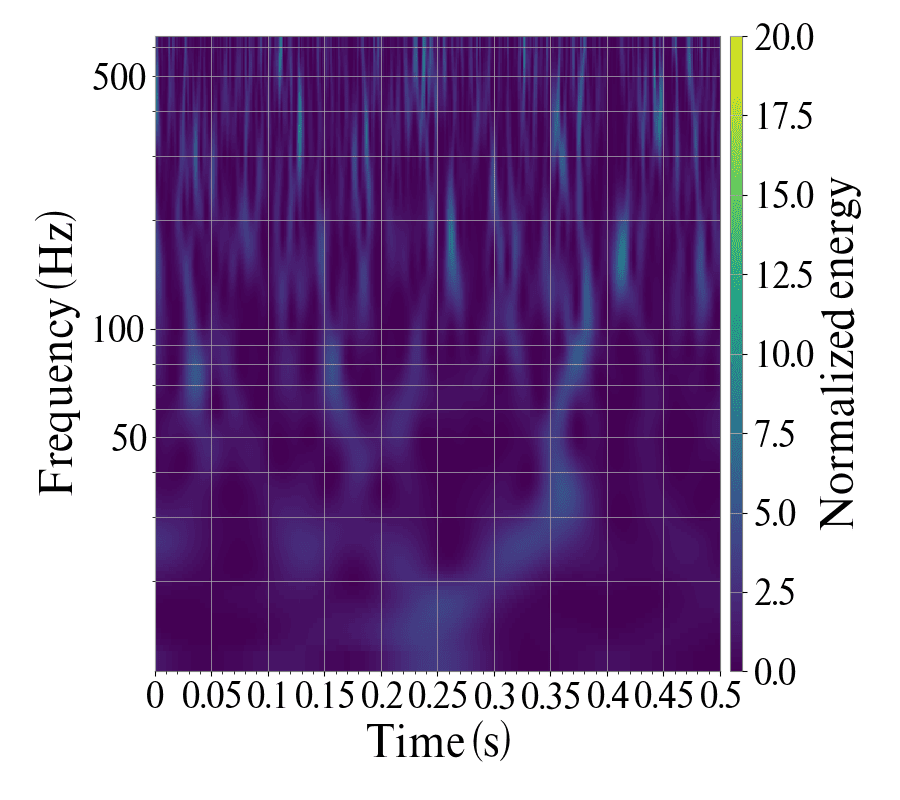}
    \par\smallskip
    {\scriptsize \scalebox{0.8}[1.0]{\textit{K1:AOS-TMSX\_IR\_PDA1\_OUT\_DQ}} \\ Apr 17, 2020 13:02:19 UTC (GPS: 1271163757)}
    \end{minipage}
\hfill
\vrule width 0.5pt
\hfill
  \begin{minipage}[t]{0.45\textwidth}
    \centering
    \includegraphics[width=0.48\textwidth]{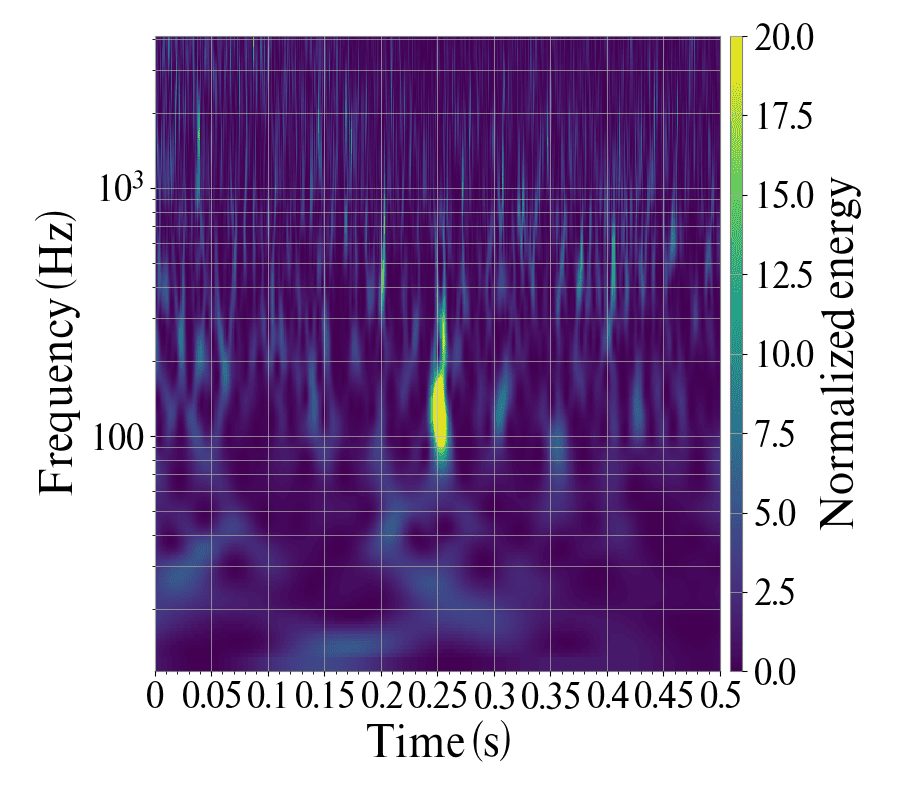}
    \includegraphics[width=0.48\textwidth]{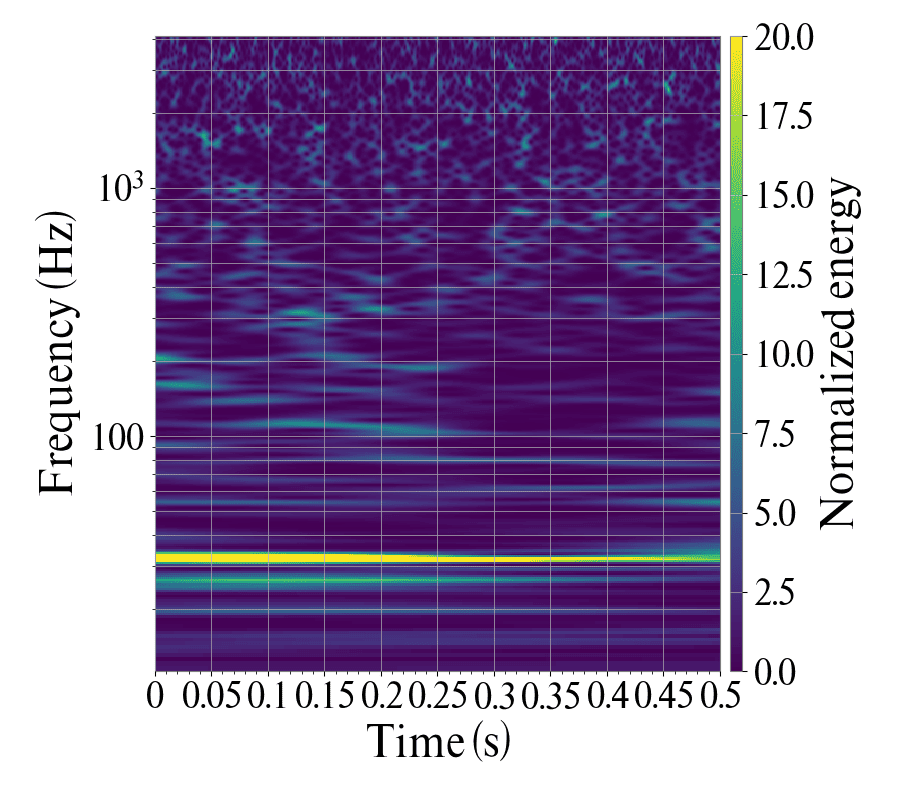}
    \par\smallskip
    {\scriptsize \scalebox{0.8}[1.0]{\textit{K1:IMC-IMMT1\_TRANS\_QPDA1\_DC\_PIT\_OUT\_DQ}} \\ Apr 09, 2020	01:14:08	UTC (GPS: 1270430066)}
    \end{minipage}
}
\end{figure} 
\clearpage

\begin{figure}[p]
\centering
\resizebox{!}{0.132\textheight}
{%
\begin{minipage}[t]{0.45\textwidth}
    \centering
    \includegraphics[width=0.48\textwidth]{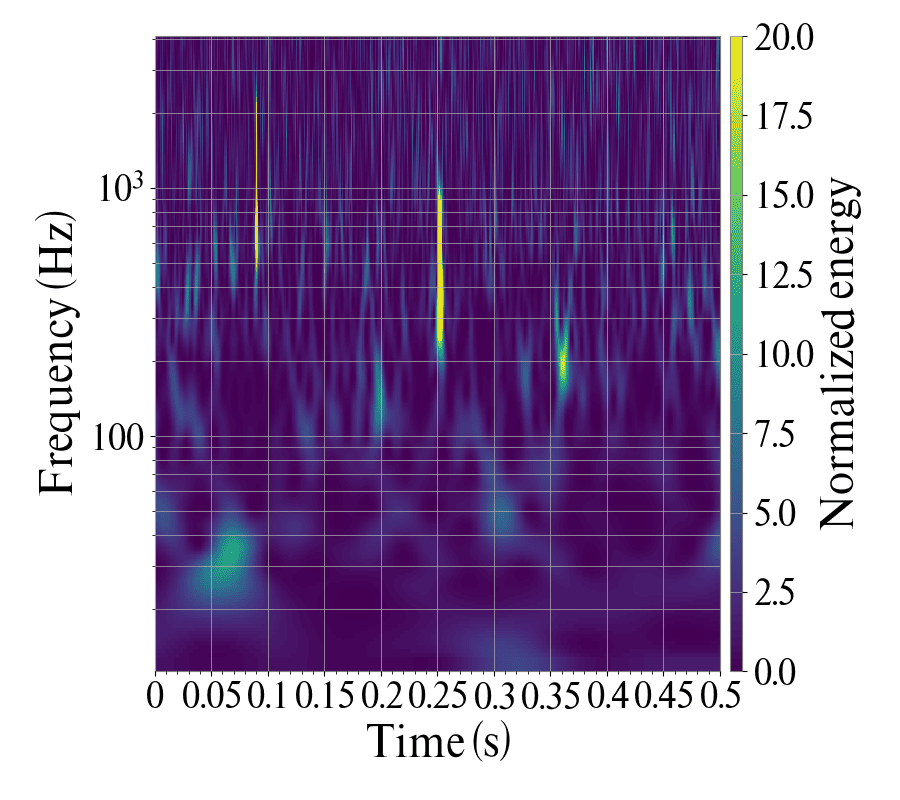}
    \includegraphics[width=0.48\textwidth]{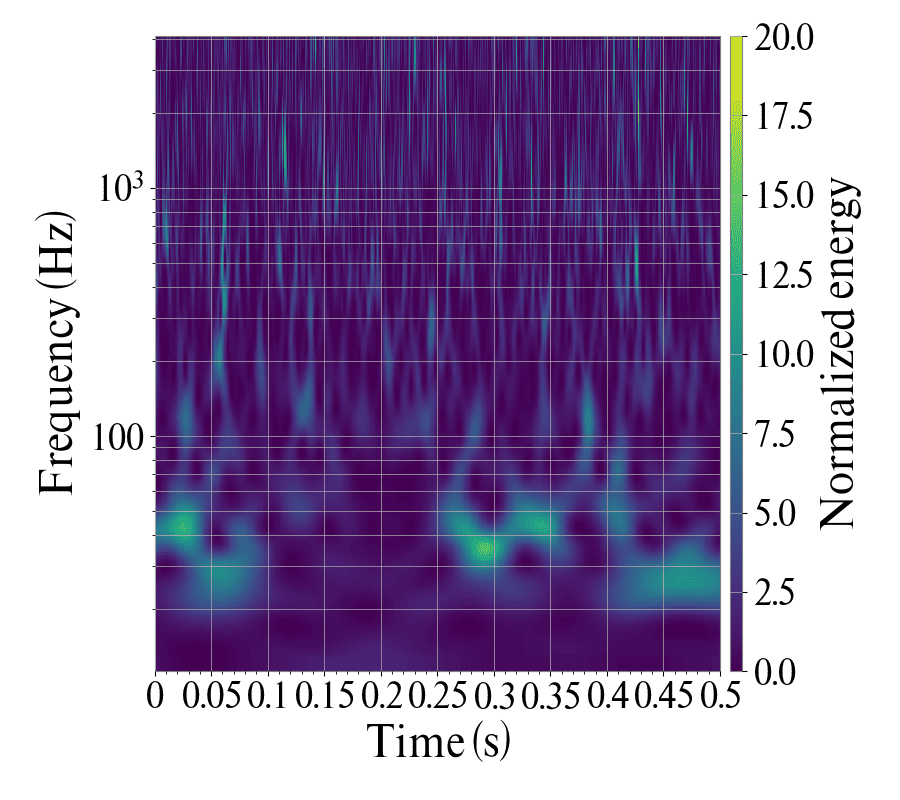}
    \par\smallskip
    {\scriptsize \scalebox{0.8}[1.0]{\textit{K1:IMC-IMMT1\_TRANS\_QPDA1\_DC\_YAW\_OUT\_DQ}} \\ Apr 11, 2020	00:19:49	UTC (GPS: 1270599607)}
    \end{minipage}
\hfill
\vrule width 0.5pt
\hfill
\begin{minipage}[t]{0.45\textwidth}   
    \centering
    \includegraphics[width=0.48\textwidth]{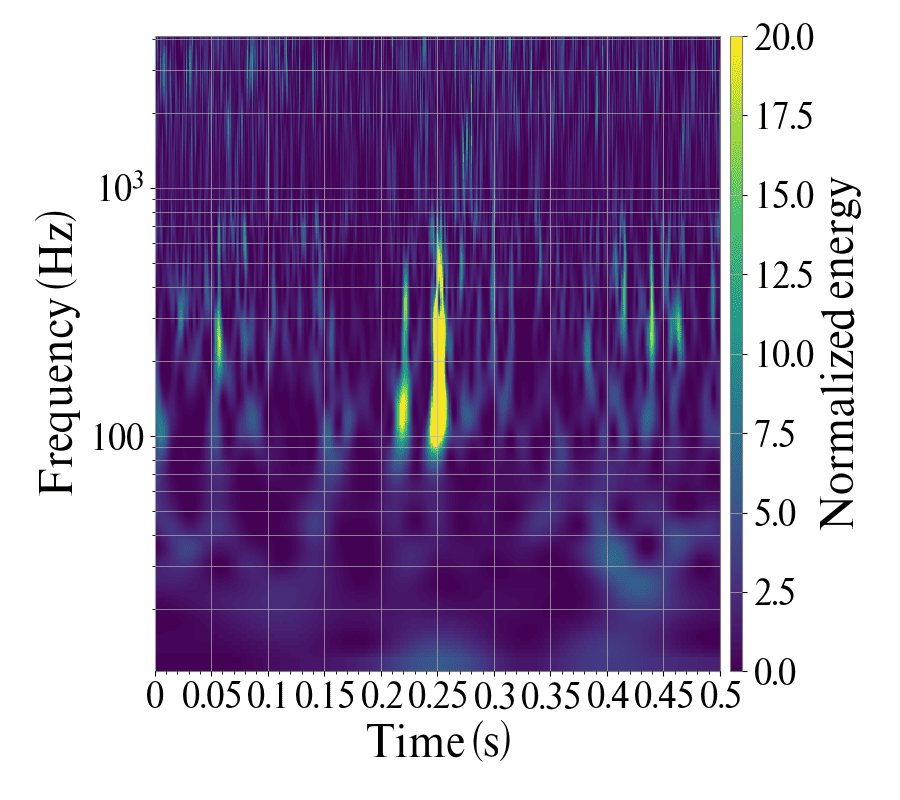}
    \includegraphics[width=0.48\textwidth]{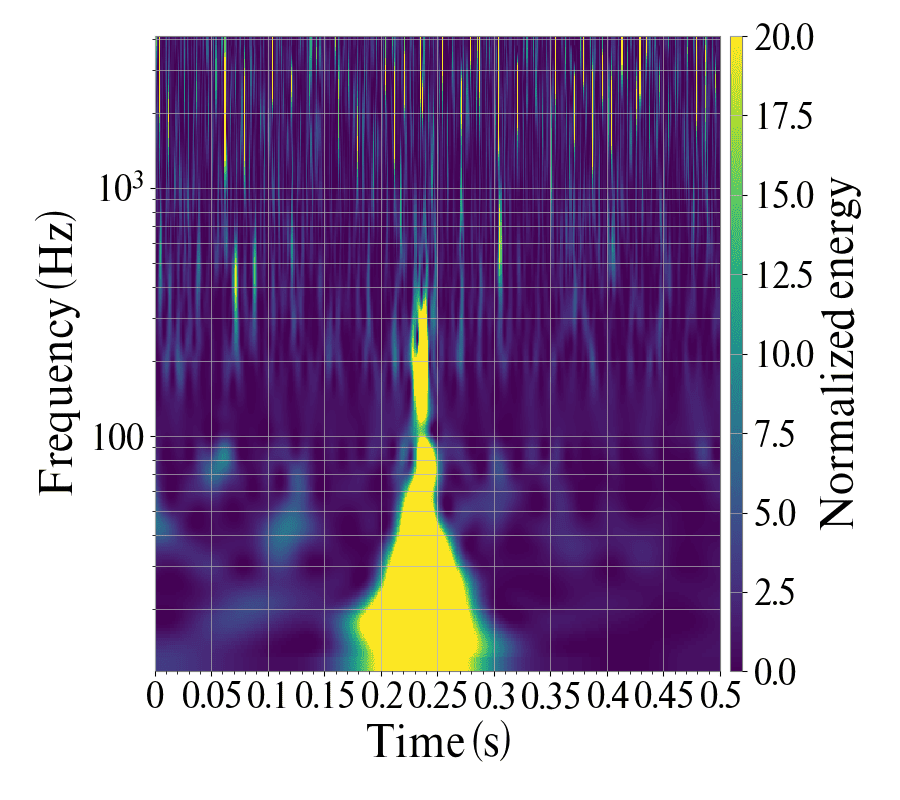}
    \par\smallskip
    {\scriptsize \scalebox{0.8}[1.0]{\textbf{K1:LAS-POW\_FIB\_OUT\_DQ}} \\ Apr 07, 2020	21:38:39	UTC (GPS: 1270330737)}
    \end{minipage}
}

\resizebox{!}{0.132\textheight}
{%
\begin{minipage}[t]{0.45\textwidth}
    \centering
    \includegraphics[width=0.48\textwidth]{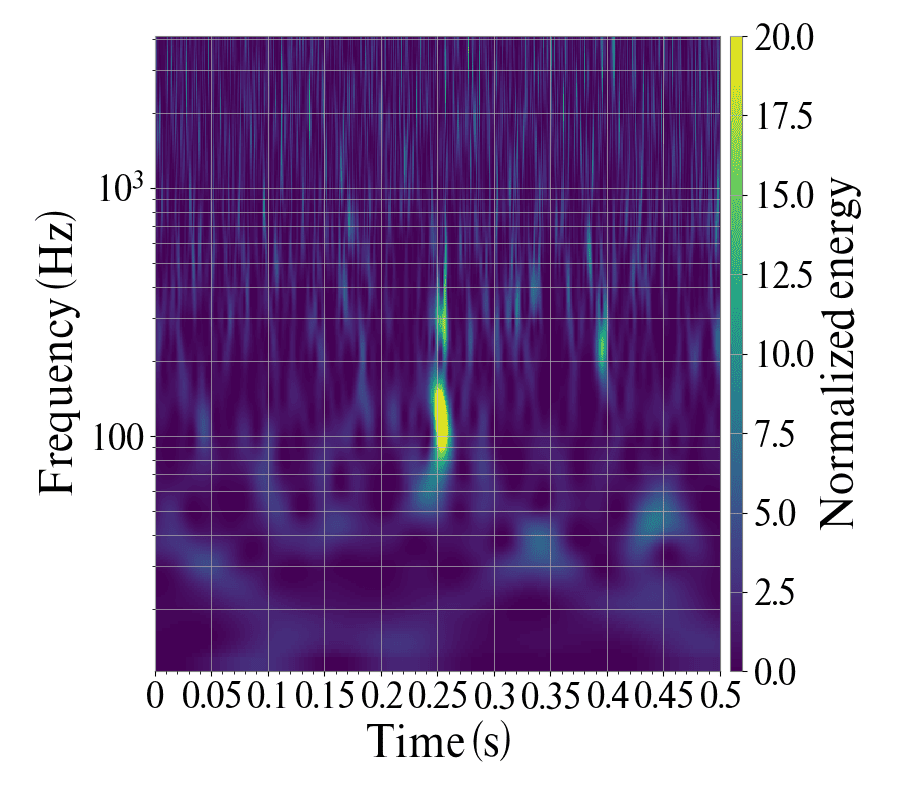}
    \includegraphics[width=0.48\textwidth]{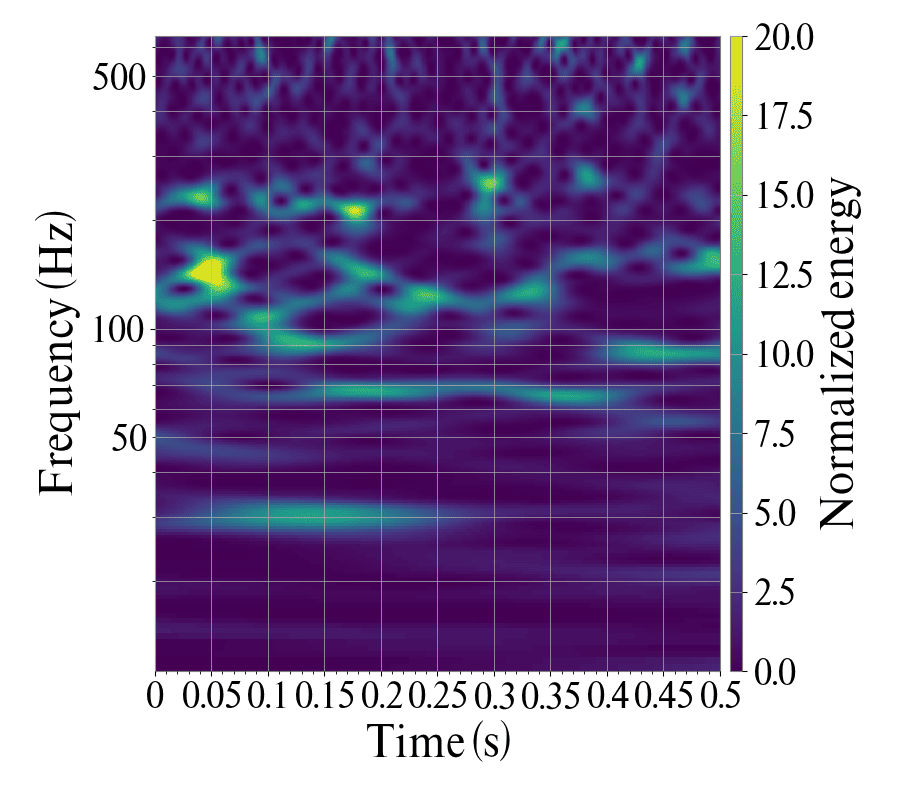}
    \par\smallskip
    {\scriptsize \scalebox{0.8}[1.0]{\textit{K1:LSC-ALS\_CARM\_OUT\_DQ}} \\ Apr 16, 2020	11:41:29	UTC (GPS: 1271072507)}
    \end{minipage}
\hfill
\vrule width 0.5pt
\hfill
\begin{minipage}[t]{0.45\textwidth}
    \centering
    \includegraphics[width=0.48\textwidth]{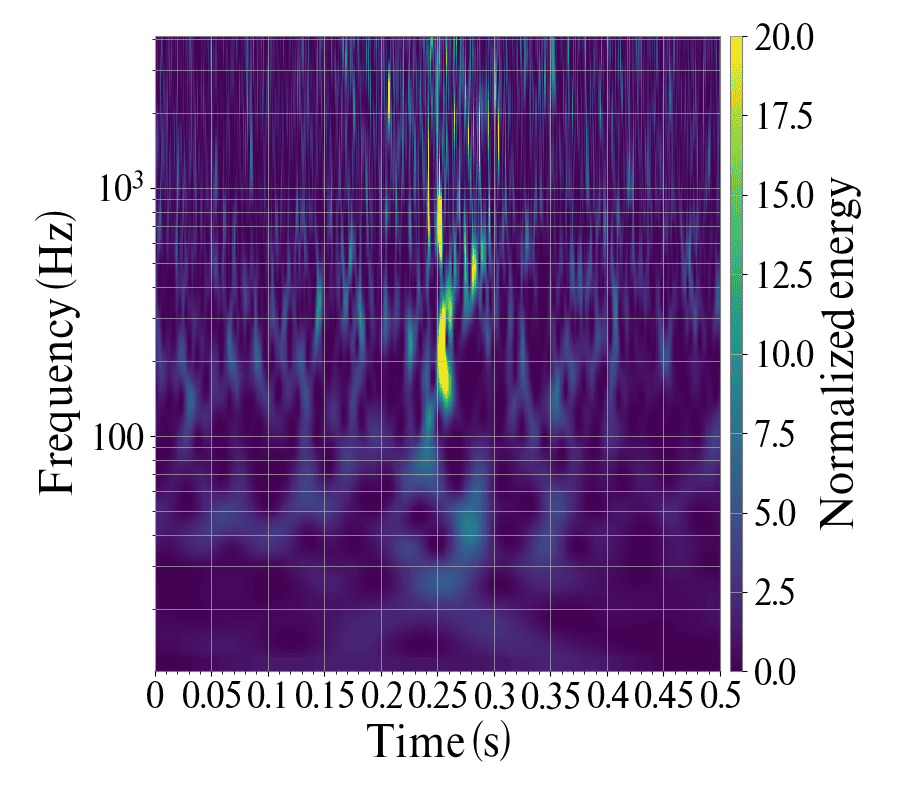}
    \includegraphics[width=0.48\textwidth]{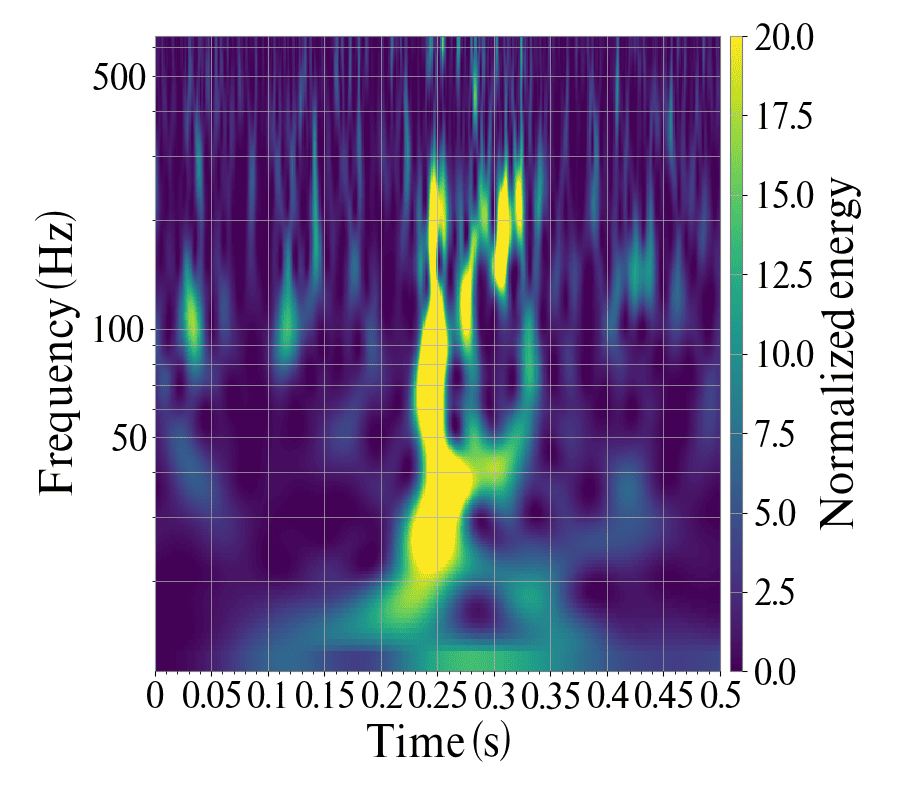}
    \par\smallskip
    {\scriptsize \scalebox{0.8}[1.0]{\textbf{K1:LSC-ALS\_DARM\_OUT\_DQ}} \\ Apr 12, 2020	16:33:38	UTC (GPS: 1270744436)}
    \end{minipage}
}

\resizebox{!}{0.132\textheight}
{%
\begin{minipage}[t]{0.45\textwidth}
    \centering
    \includegraphics[width=0.48\textwidth]{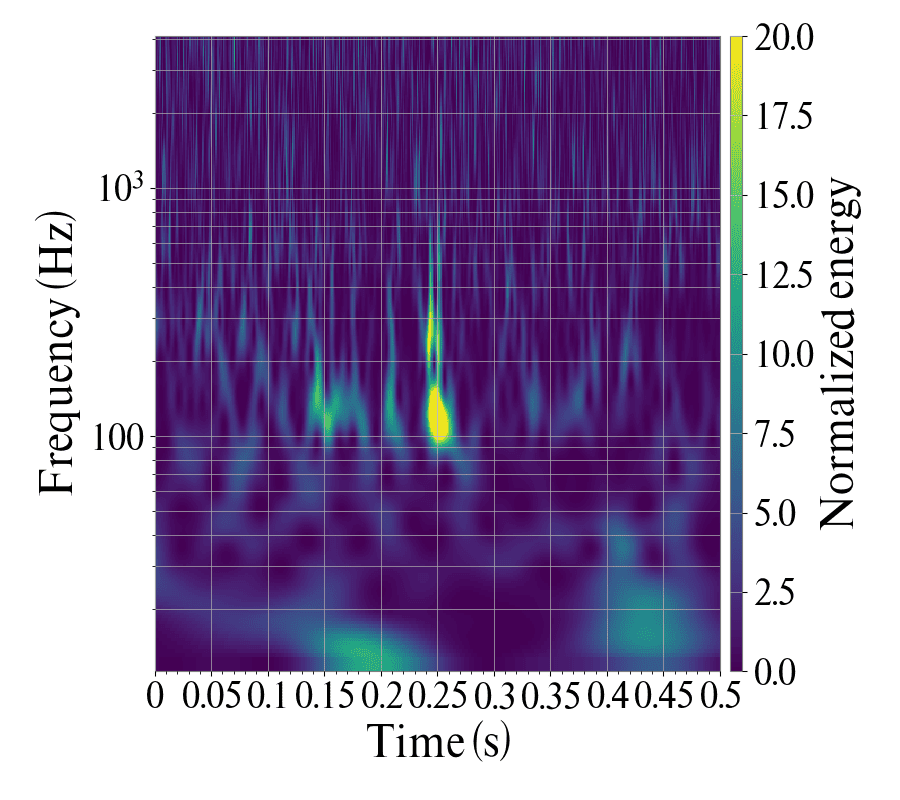}
    \includegraphics[width=0.48\textwidth]{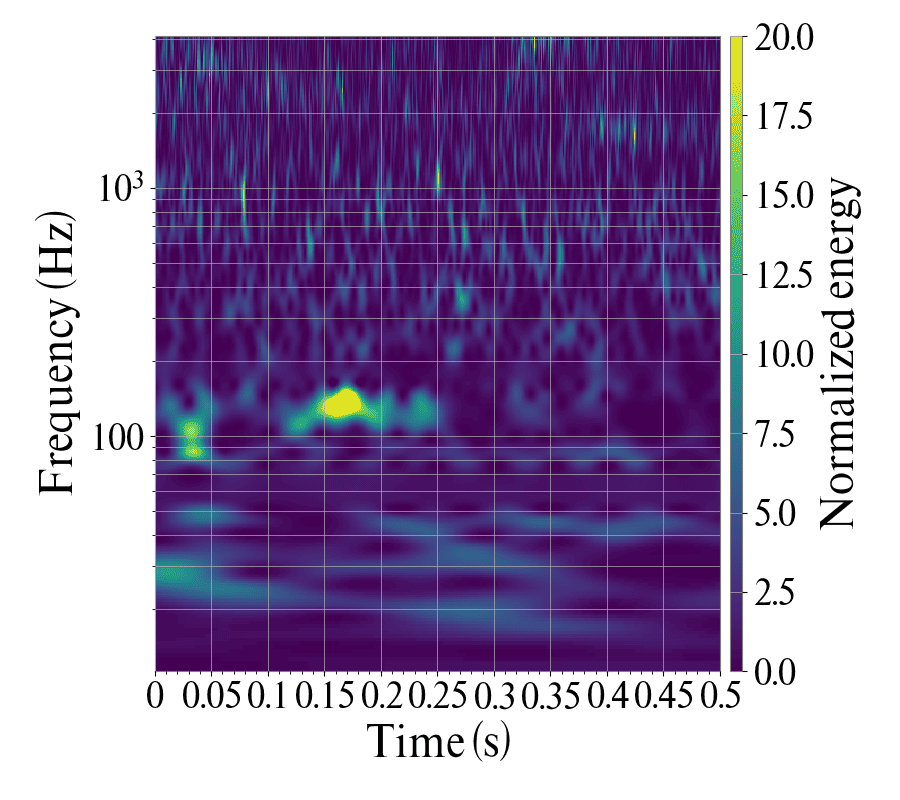}
    \par\smallskip
    {\scriptsize \scalebox{0.8}[1.0]{\textbf{K1:PEM-MAG\_BS\_BOOTH\_BS\_X\_OUT\_DQ}} \\ Apr 16, 2020	13:57:58	UTC	(GPS: 1271080696)}
    \end{minipage}
\hfill
\vrule width 0.5pt
\hfill    
\begin{minipage}[t]{0.45\textwidth}
    \centering
    \includegraphics[width=0.48\textwidth]{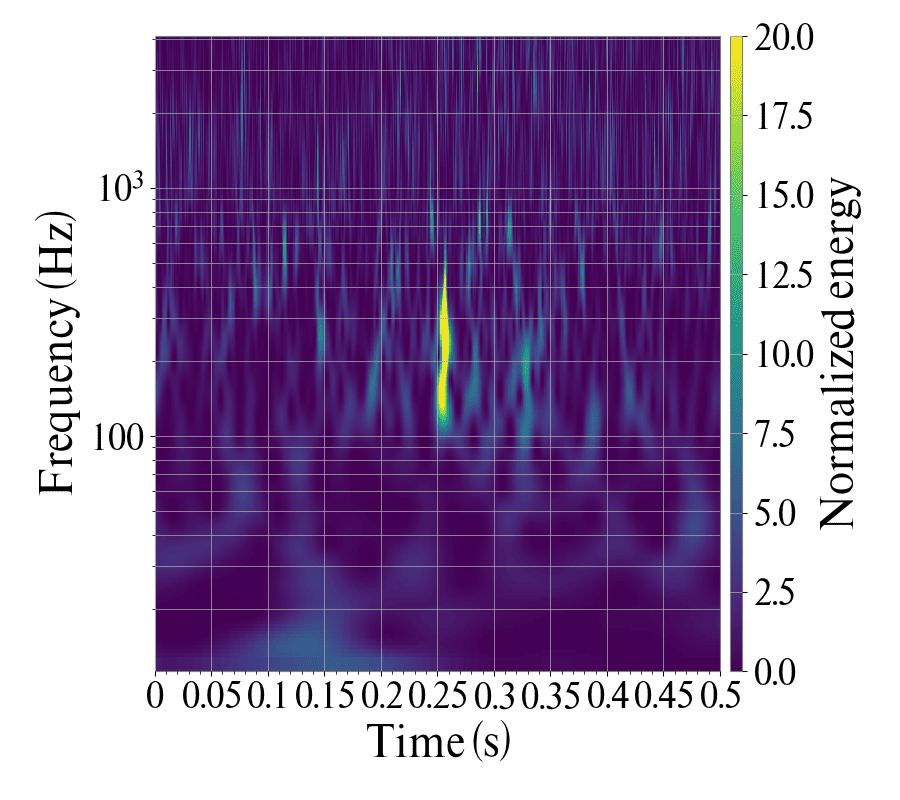}
    \includegraphics[width=0.48\textwidth]{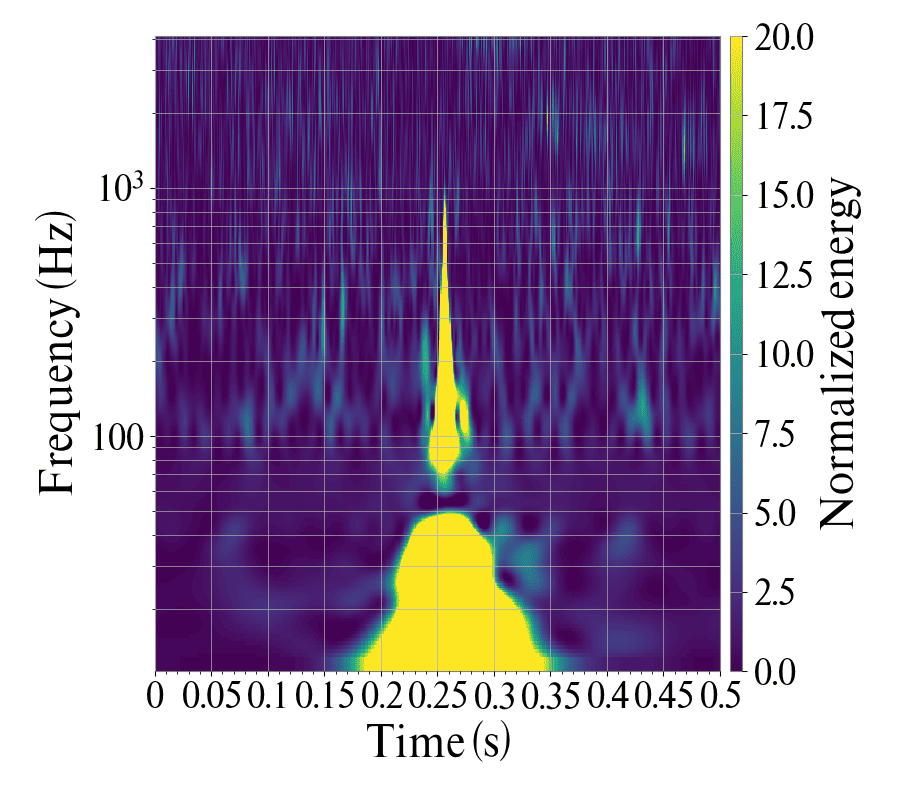}
    \par\smallskip
    {\scriptsize \scalebox{0.8}[1.0]{\textbf{K1:PEM-MAG\_BS\_BOOTH\_BS\_Y\_OUT\_DQ}} \\ Apr 08, 2020	21:48:45	UTC	(GPS: 1270417743)}
    \end{minipage}
}

\resizebox{!}{0.132\textheight}
{%
\begin{minipage}[t]{0.45\textwidth}
    \centering
    \includegraphics[width=0.48\textwidth]{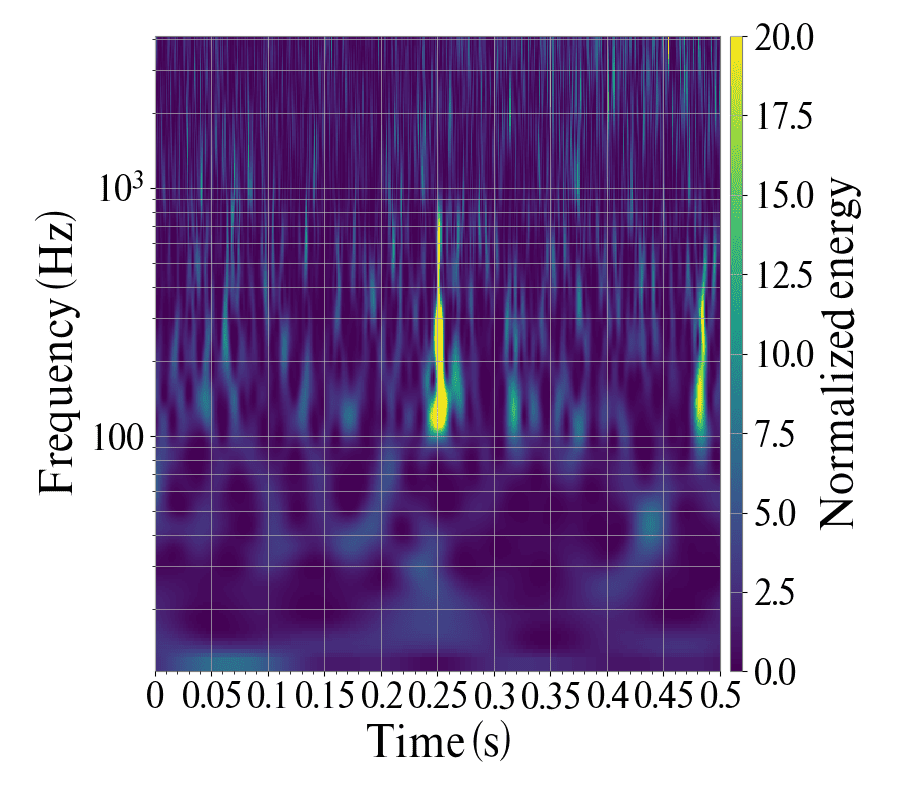}
    \includegraphics[width=0.48\textwidth]{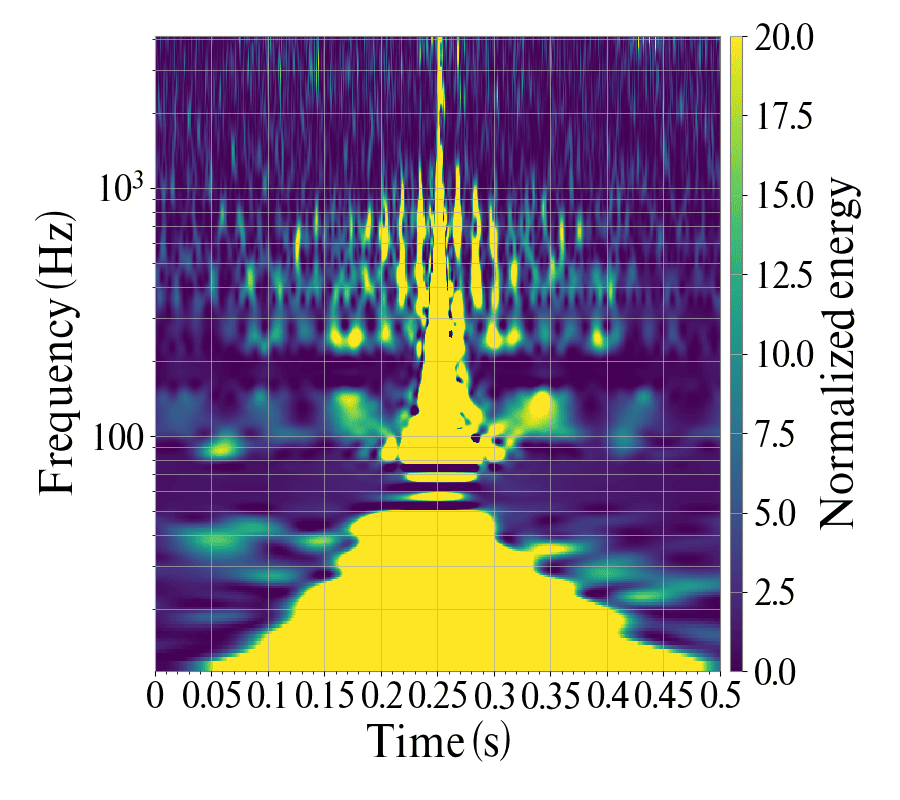}
    \par\smallskip
    {\scriptsize \scalebox{0.8}[1.0]{\textbf{K1:PEM-MAG\_BS\_BOOTH\_BS\_Z\_OUT\_DQ}} \\ Apr 12, 2020	01:29:11	UTC	(GPS: 1270417743)}
    \end{minipage}
\hfill
\vrule width 0.5pt
\hfill
\begin{minipage}[t]{0.45\textwidth}
    \centering
    \includegraphics[width=0.48\textwidth]{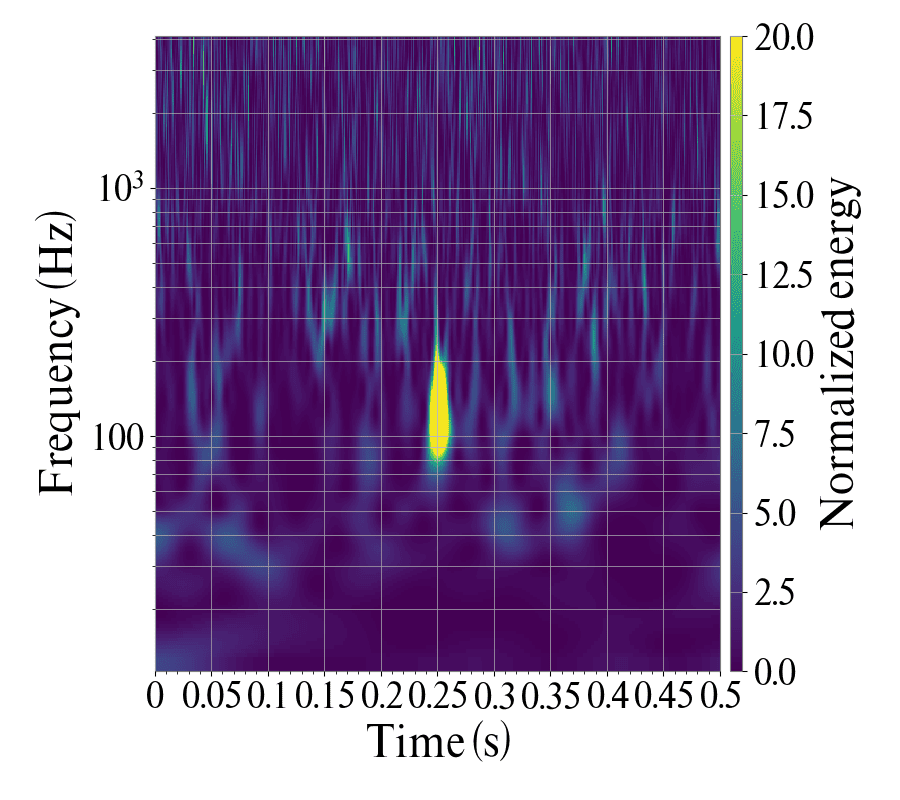}
    \includegraphics[width=0.48\textwidth]{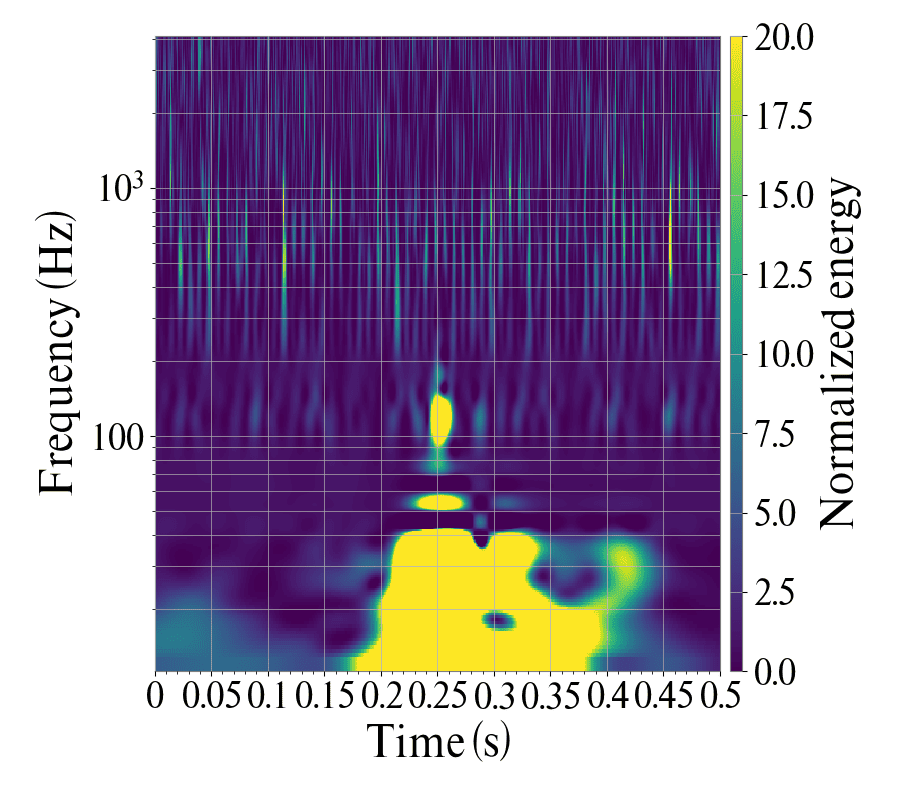}
    \par\smallskip
    {\scriptsize \scalebox{0.8}[1.0]{\textbf{K1:PEM-MAG\_EXC\_BOOTH\_EXC\_Y\_OUT\_DQ}} \\ Apr 15, 2020	18:52:49	UTC	(GPS: 1271011987)}
    \end{minipage}
}

\resizebox{!}{0.136\textheight}
{%
\begin{minipage}[t]{0.45\textwidth}
    \centering
    \includegraphics[width=0.48\textwidth]{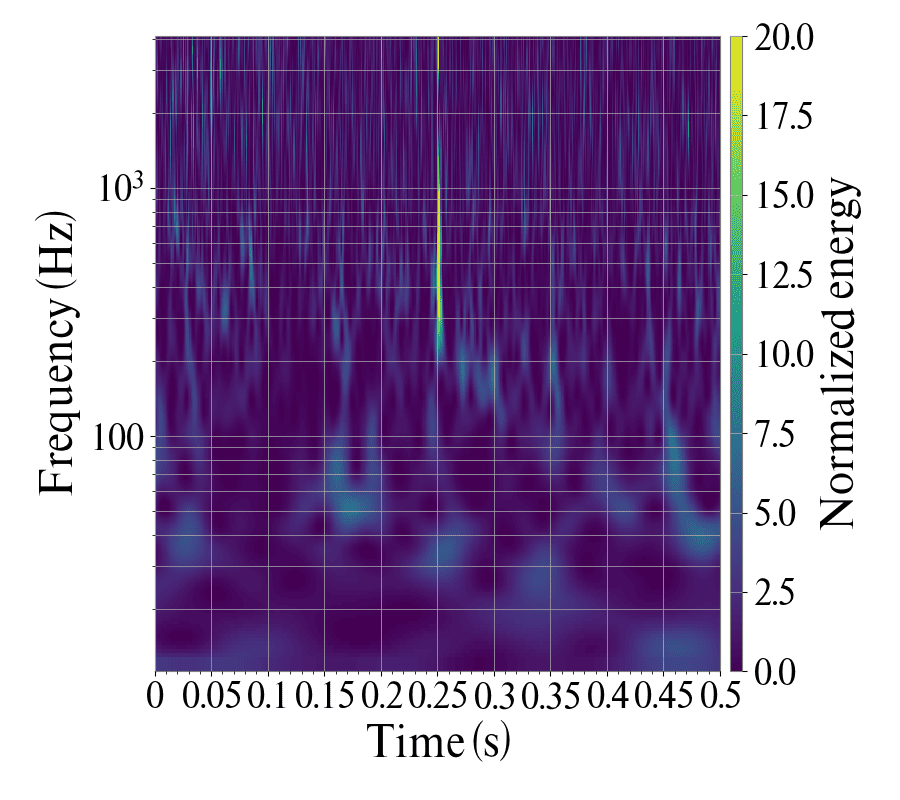}
    \includegraphics[width=0.48\textwidth]{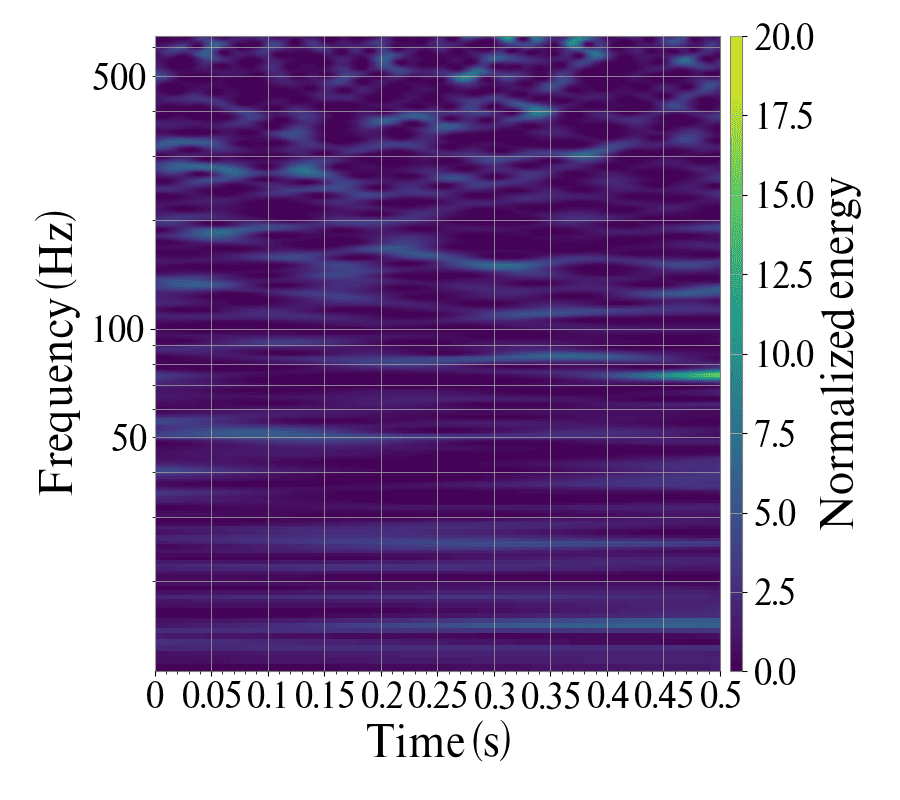}
    \par\smallskip
    {\scriptsize \scalebox{0.8}[1.0]{\textit{K1:PEM-MIC\_SR\_BOOTH\_SR\_Z\_OUT\_DQ}}\\ Apr 16, 2020	04:44:25	UTC	(GPS: 1271047483)}
    \end{minipage}
\hfill
\vrule width 0.5pt
\hfill    
\begin{minipage}[t]{0.45\textwidth}
    \centering
    \includegraphics[width=0.48\textwidth]{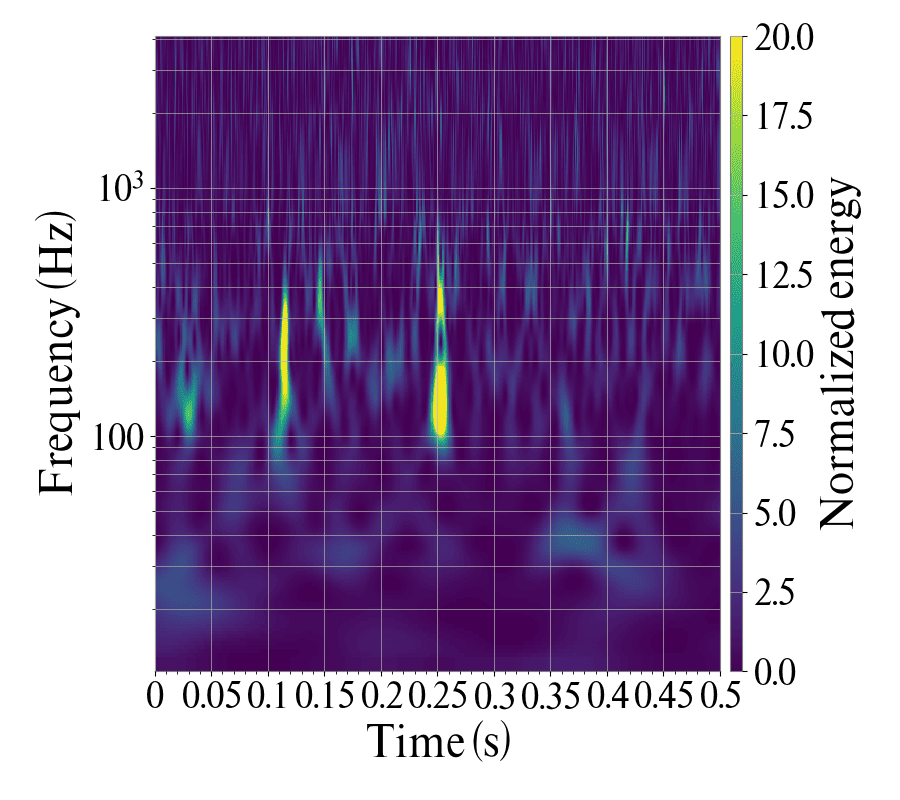}
    \includegraphics[width=0.48\textwidth]{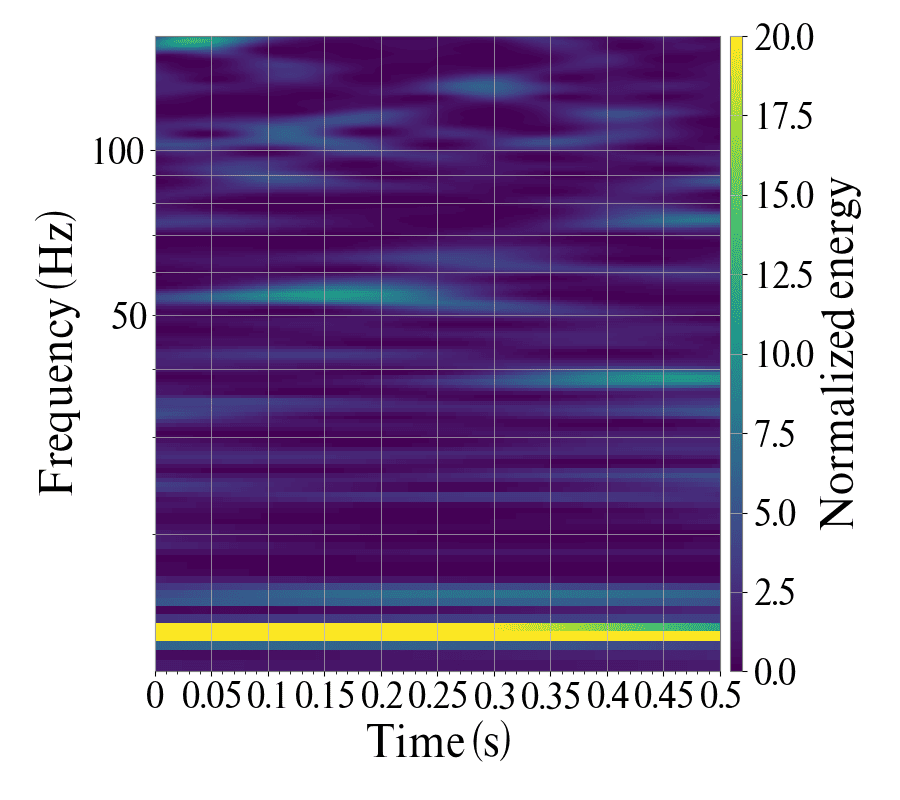}
    \par\smallskip
    {\scriptsize \scalebox{0.8}[1.0]{\textit{K1:PEM-SEIS\_IXV\_GND\_EW\_IN1\_DQ}}\\ Apr 07, 2020	20:45:57	UTC	(GPS: 1270327575)}
    \end{minipage}
}
\end{figure} 
\clearpage

\begin{figure}[p]
\centering
\resizebox{!}{0.136\textheight}
{%
\begin{minipage}[t]{0.45\textwidth}
    \centering
    \includegraphics[width=0.48\textwidth]{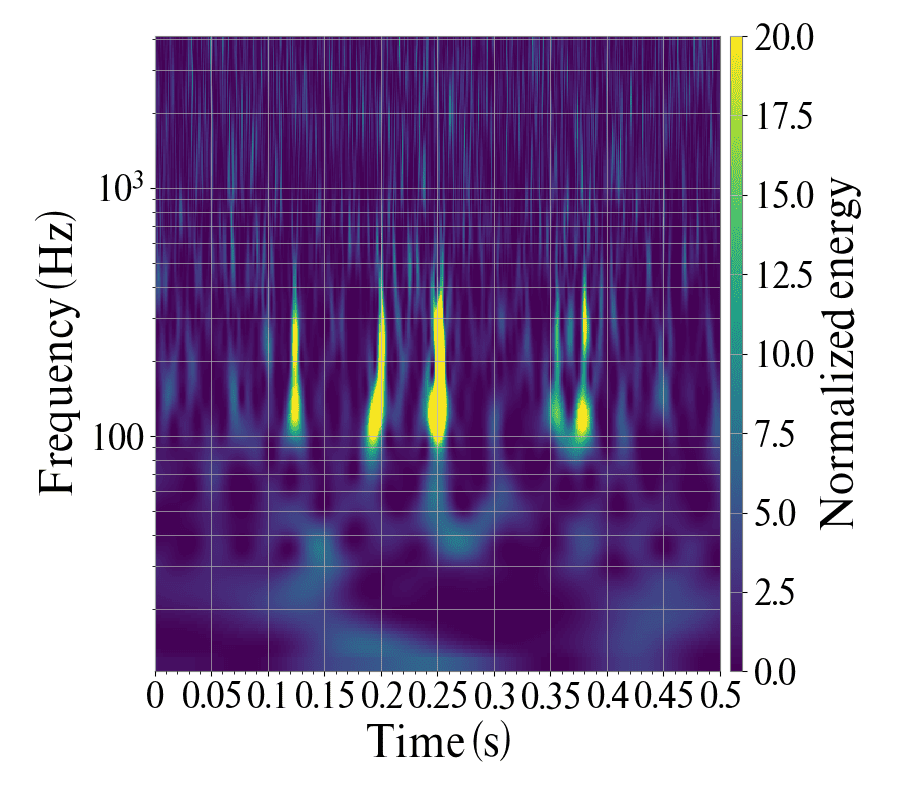}
    \includegraphics[width=0.48\textwidth]{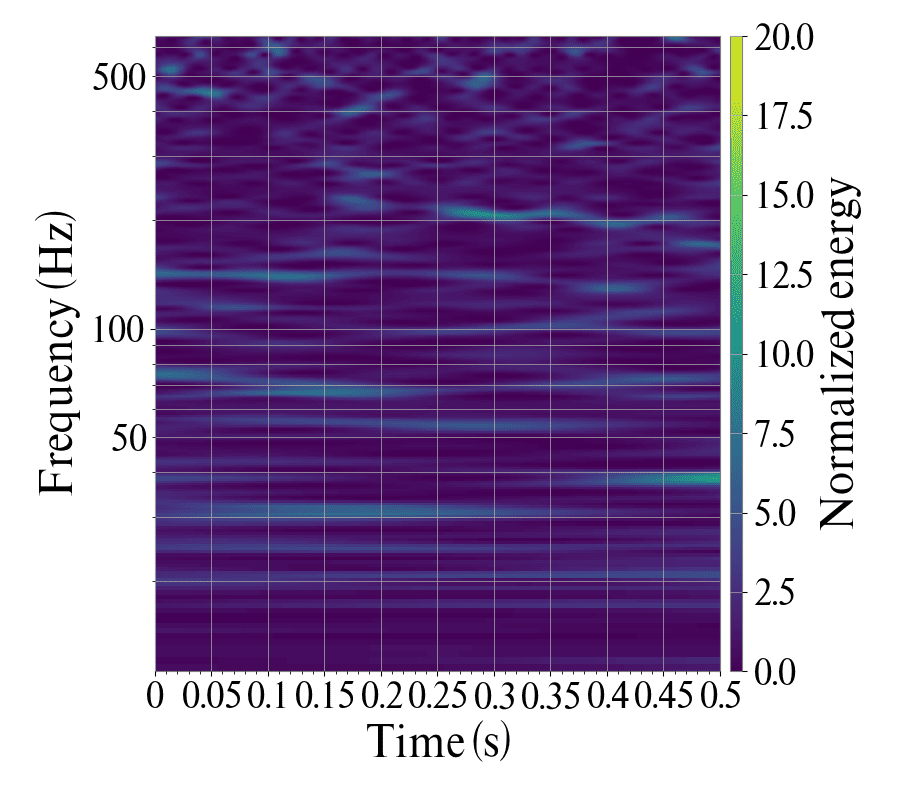}
    \par\smallskip
    {\scriptsize \scalebox{0.8}[1.0]{\textit{K1:PEM-VOLT\_REFL\_TABLE\_GND\_OUT\_DQ}}\\ Apr 11, 2020	01:47:59	UTC	(GPS: 1270604897)}
    \end{minipage}
\hfill
\vrule width 0.5pt
\hfill    
\begin{minipage}[t]{0.45\textwidth}
    \centering
    \includegraphics[width=0.48\textwidth]{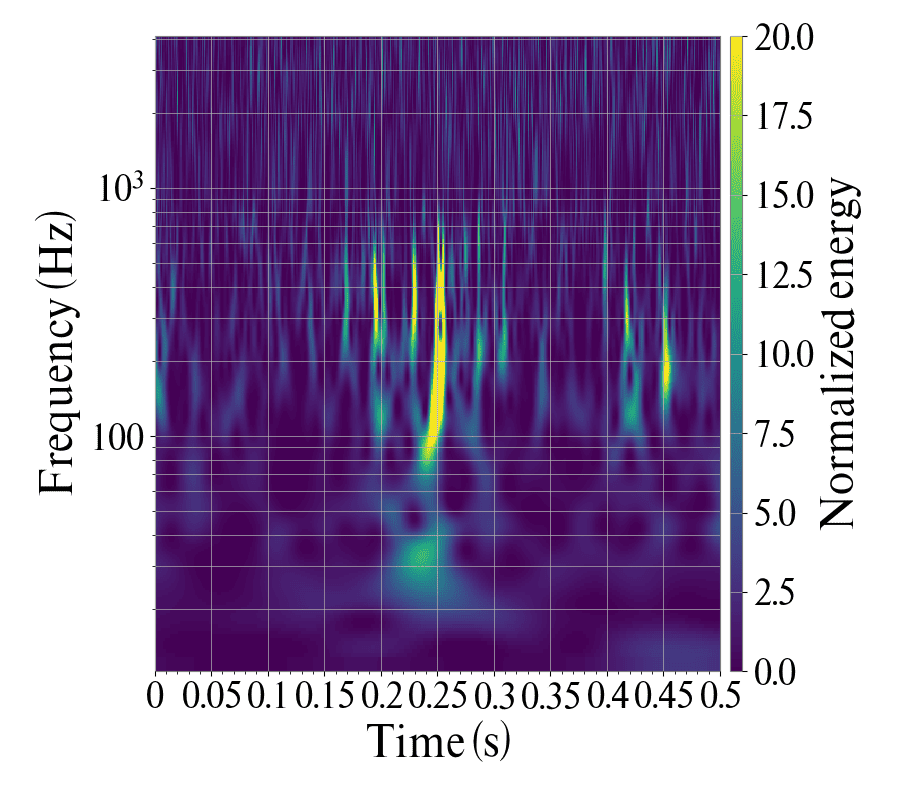}
    \includegraphics[width=0.48\textwidth]{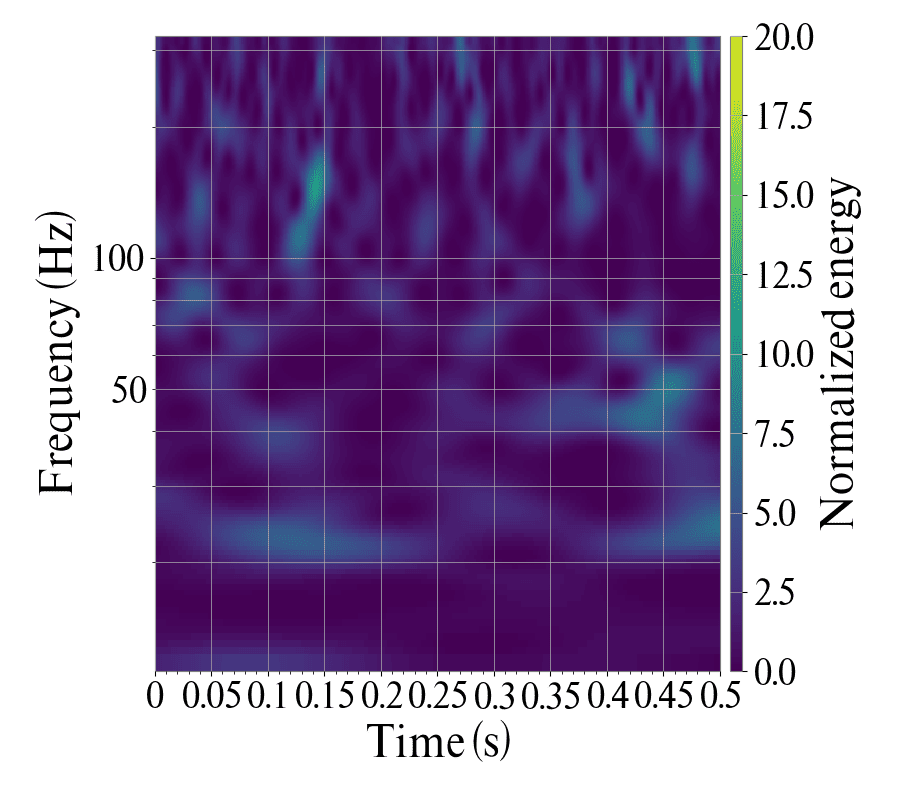}
    \par\smallskip
    {\scriptsize \scalebox{0.8}[1.0]{\textit{K1:VIS-ETMX\_MN\_PSDAMP\_R\_IN1\_DQ}}\\ Apr 07, 2020	23:17:12	UTC	(GPS: 1270336650)}
    \end{minipage}
}

\resizebox{!}{0.136\textheight}
{%
\begin{minipage}[t]{0.45\textwidth}
    \centering
    \includegraphics[width=0.48\textwidth]{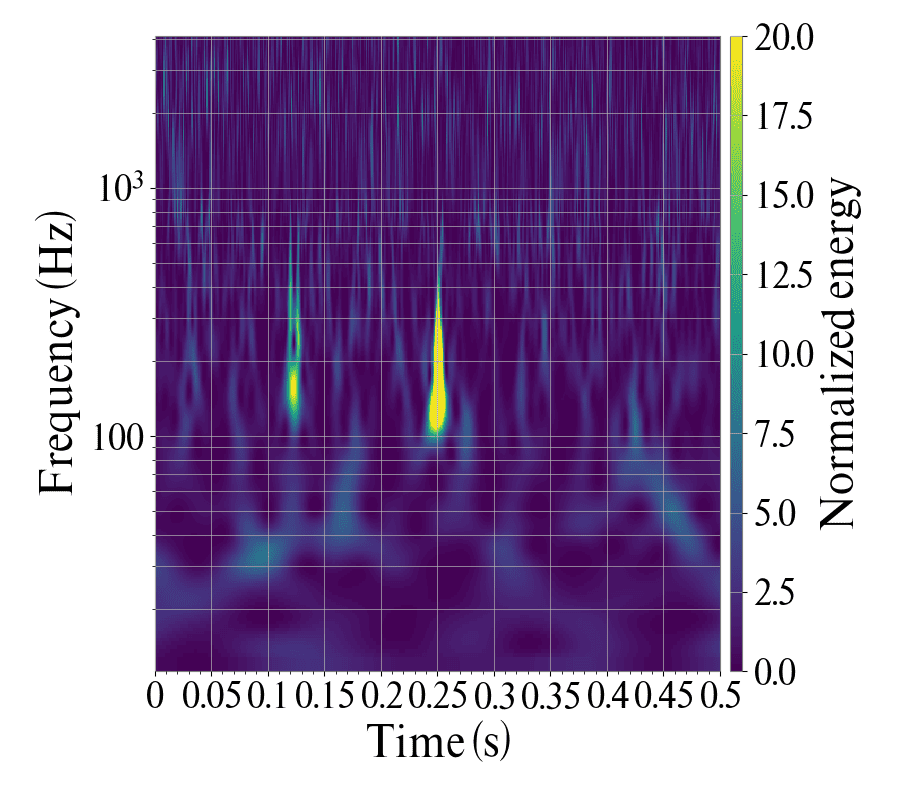}
    \includegraphics[width=0.48\textwidth]{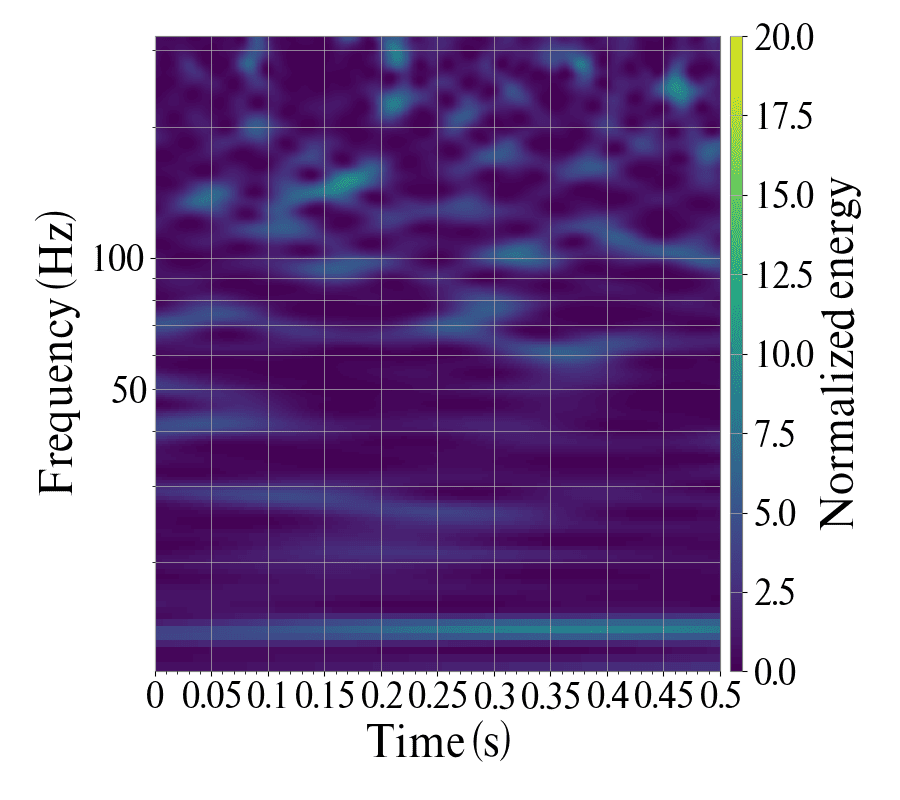}
    \par\smallskip
    {\scriptsize \scalebox{0.8}[1.0]{\textit{K1:VIS-ETMX\_MN\_PSDAMP\_Y\_IN1\_DQ}}\\ Apr 07, 2020	20:43:08	UTC	(GPS: 1270327406)}
    \end{minipage}
\hfill
\vrule width 0.5pt
\hfill    
\begin{minipage}[t]{0.45\textwidth}
    \centering
    \includegraphics[width=0.48\textwidth]{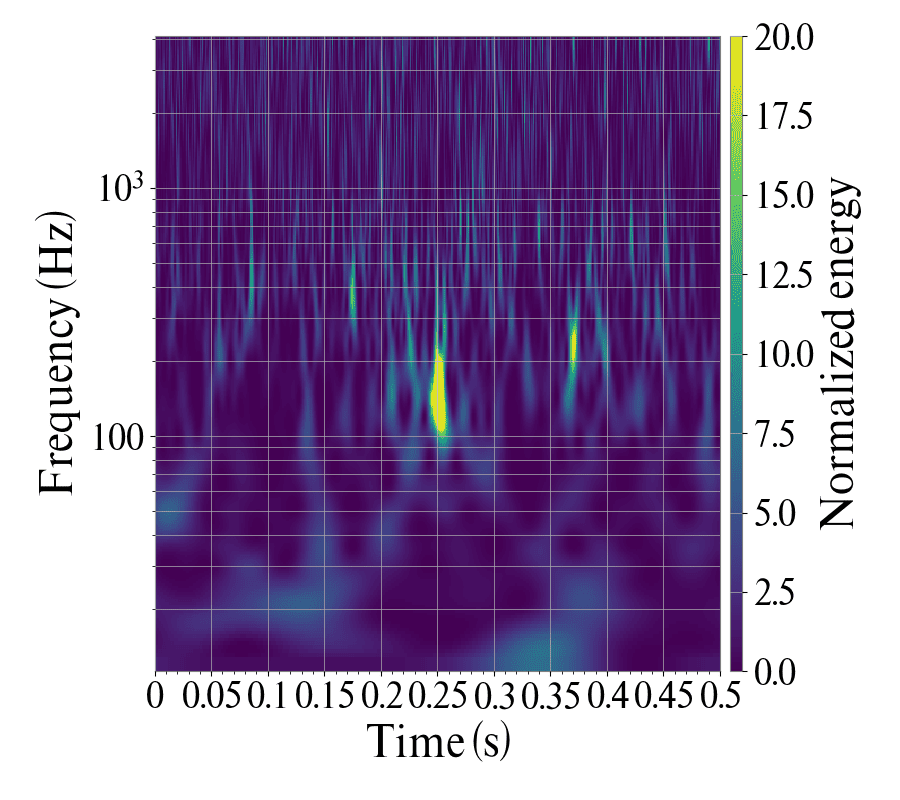}
    \includegraphics[width=0.48\textwidth]{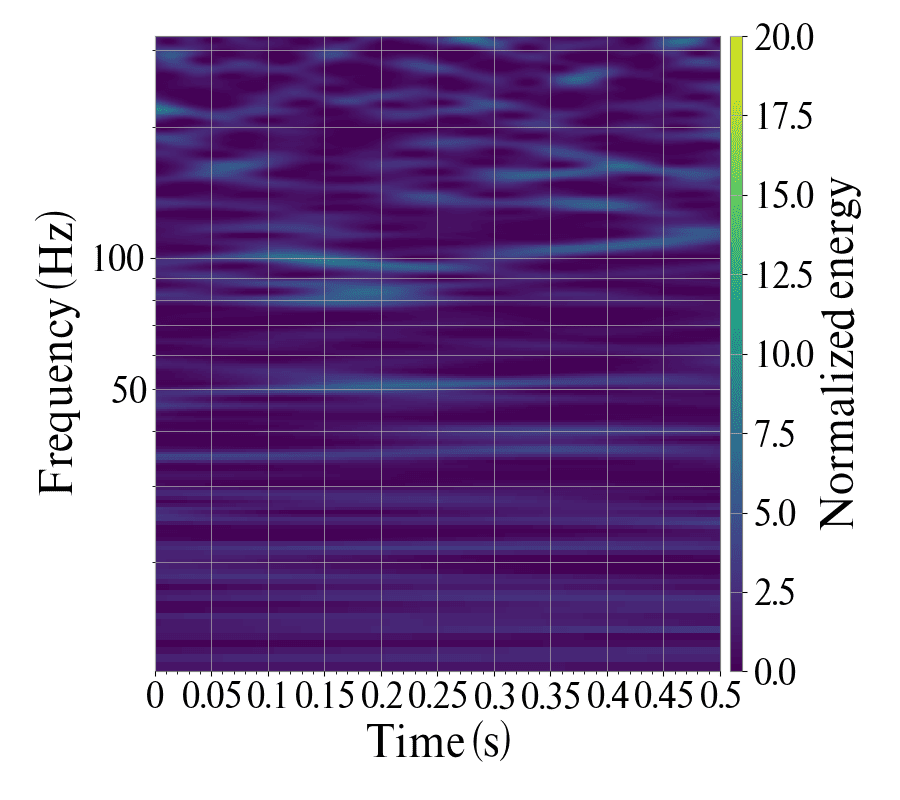}
    \par\smallskip
    {\scriptsize \scalebox{0.8}[1.0]{\textit{K1:VIS-ETMY\_MN\_PSDAMP\_Y\_IN1\_DQ}}\\ Apr 19, 2020	05:38:32	UTC	(GPS: 1271309930)}
    \end{minipage}
}

\resizebox{!}{0.136\textheight}
{%
\begin{minipage}[t]{0.45\textwidth}
    \centering
    \includegraphics[width=0.48\textwidth]{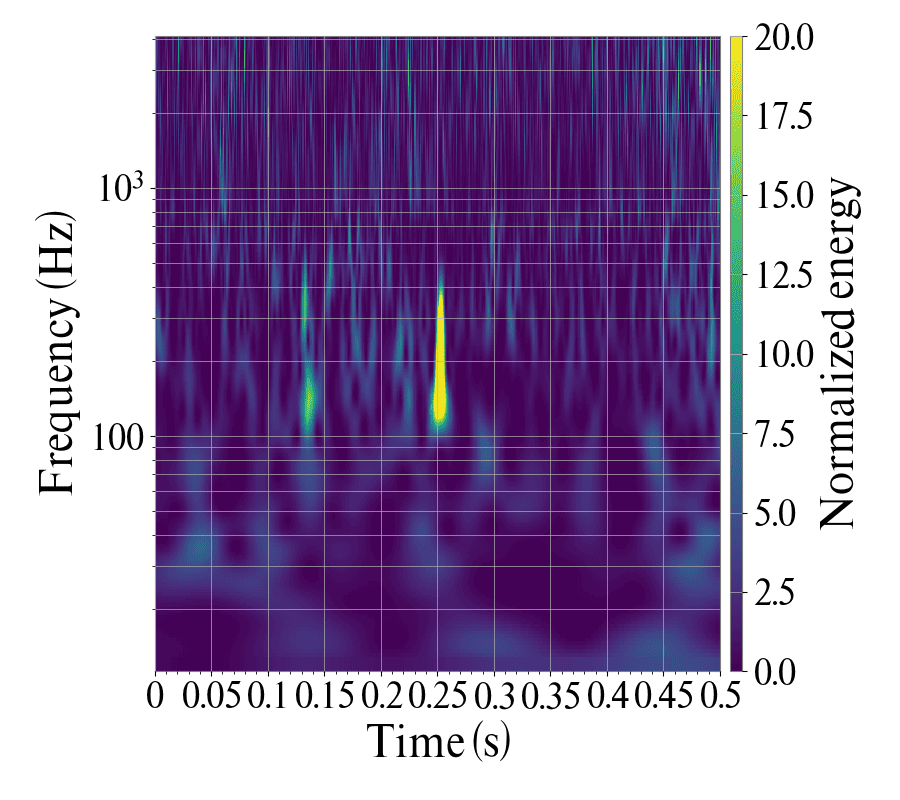}
    \includegraphics[width=0.48\textwidth]{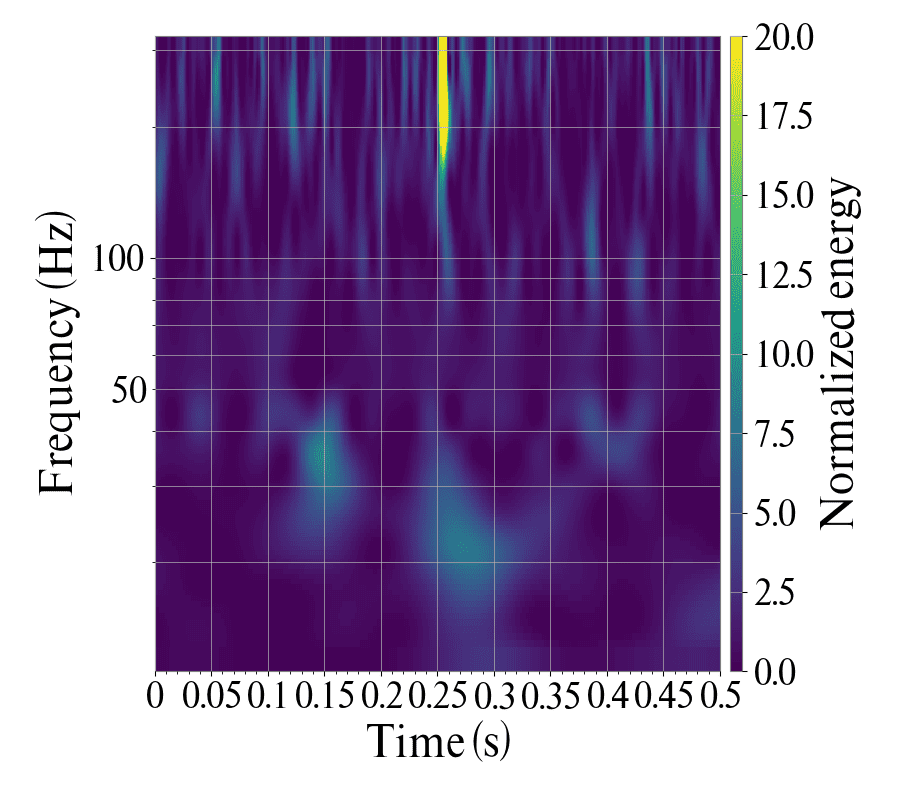}
    \par\smallskip
    {\scriptsize \scalebox{0.8}[1.0]{\textbf{K1:VIS-ITMY\_IM\_PSDAMP\_R\_IN1\_DQ}}\\ Apr 07, 2020	14:45:26	UTC	(GPS: 1270305944)}
    \end{minipage}
\hfill
\vrule width 0.5pt
\hfill
\begin{minipage}[t]{0.45\textwidth}
    \centering
    \includegraphics[width=0.48\textwidth]{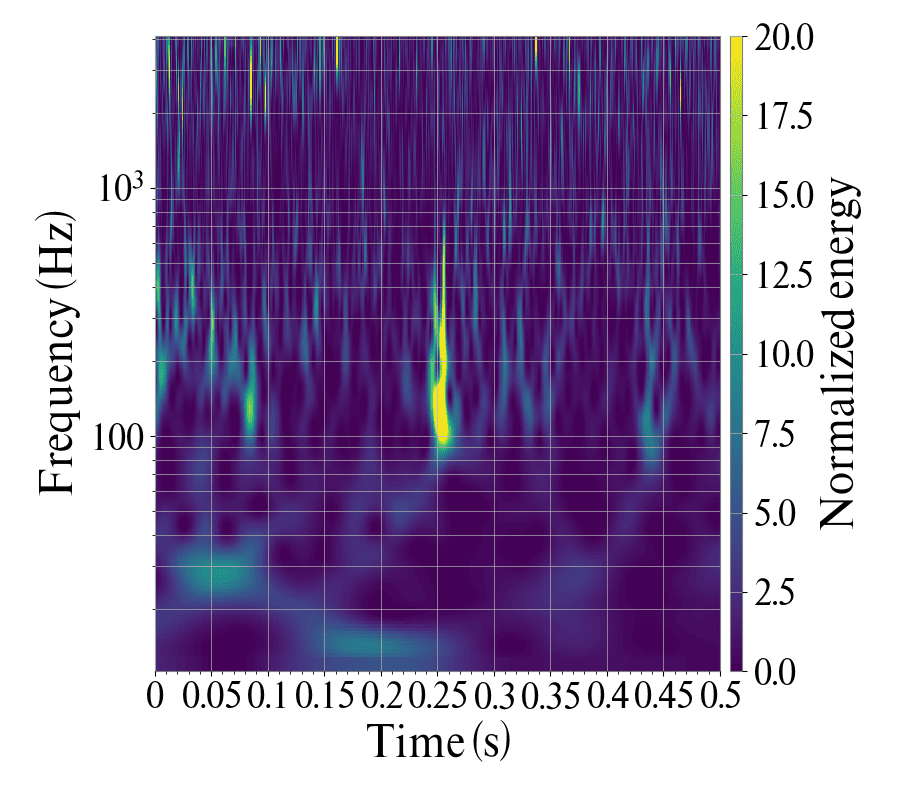}
    \includegraphics[width=0.48\textwidth]{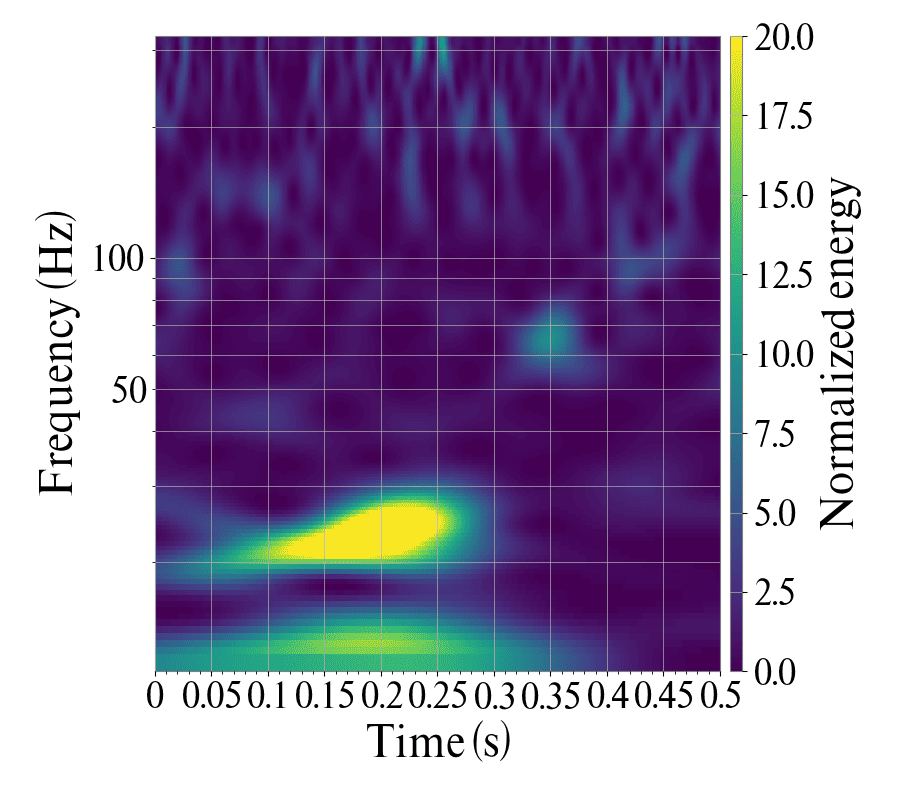}
    \par\smallskip
    {\scriptsize \scalebox{0.8}[1.0]{\textit{K1:VIS-ITMY\_MN\_OPLEV\_TILT\_YAW\_OUT\_DQ}}\\ Apr 09, 2020	01:58:55	UTC	(GPS: 1270432753)}
    \end{minipage}
}

\resizebox{!}{0.136\textheight}
{%
\begin{minipage}[t]{0.45\textwidth}
    \centering
    \includegraphics[width=0.48\textwidth]{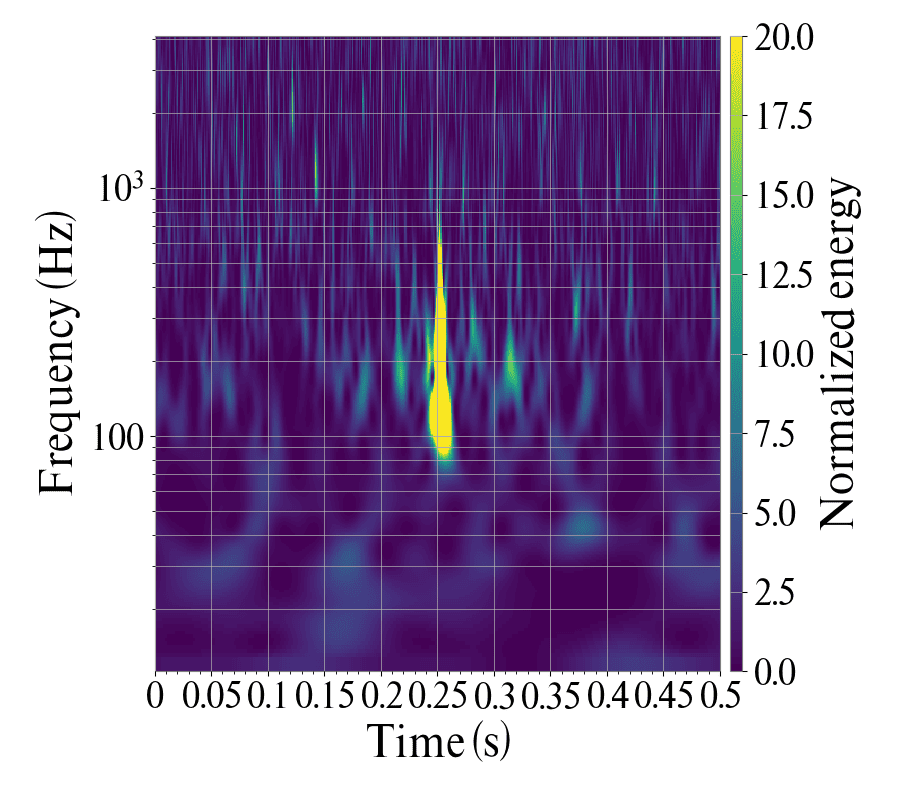}
    \includegraphics[width=0.48\textwidth]{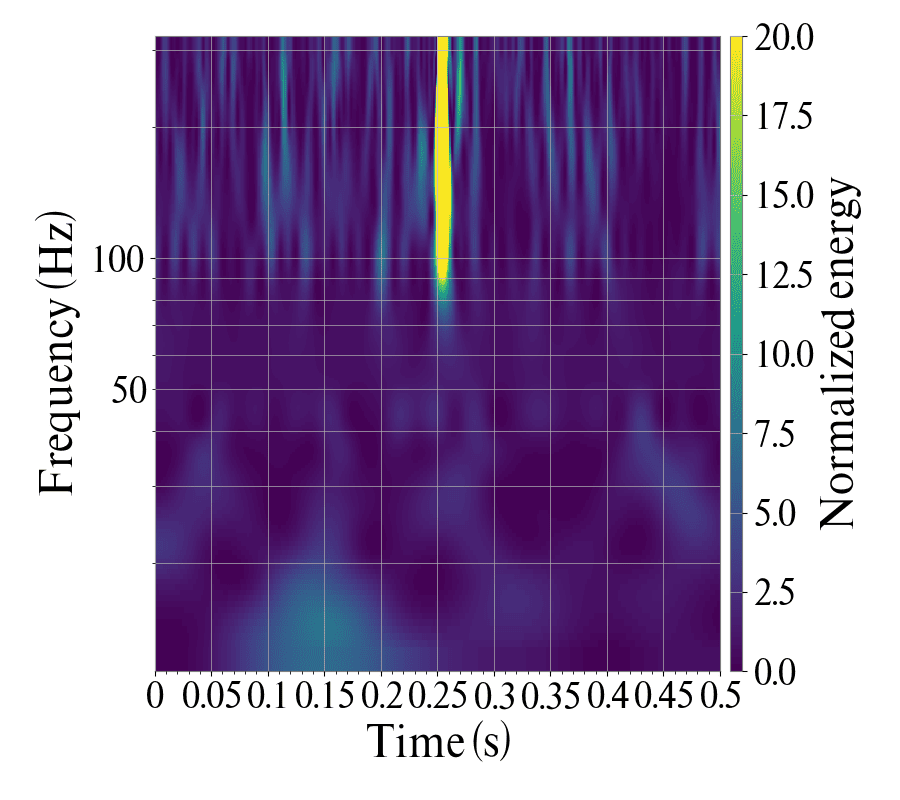}
    \par\smallskip
    {\scriptsize \scalebox{0.8}[1.0]{\textbf{K1:VIS-ITMY\_MN\_PSDAMP\_L\_IN1\_DQ}}\\ Apr 17, 2020, 12:59:06 UTC (GPS: 1271163564)}
    \end{minipage}
\hfill
\vrule width 0.5pt
\hfill
\begin{minipage}[t]{0.45\textwidth}
    \centering
    \includegraphics[width=0.48\textwidth]{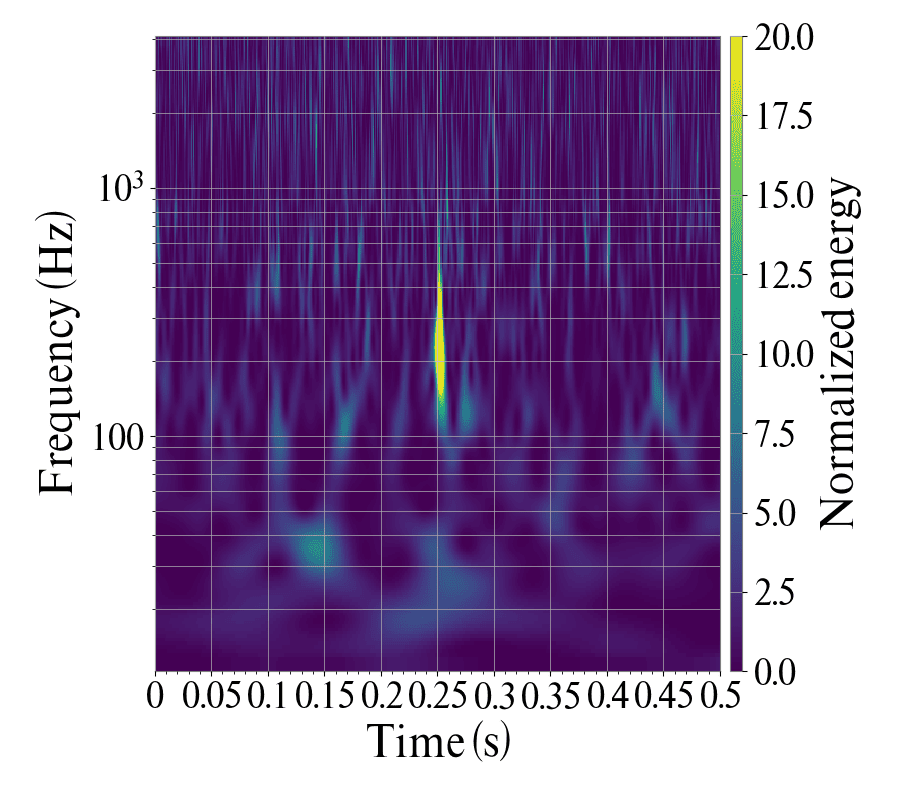}
    \includegraphics[width=0.48\textwidth]{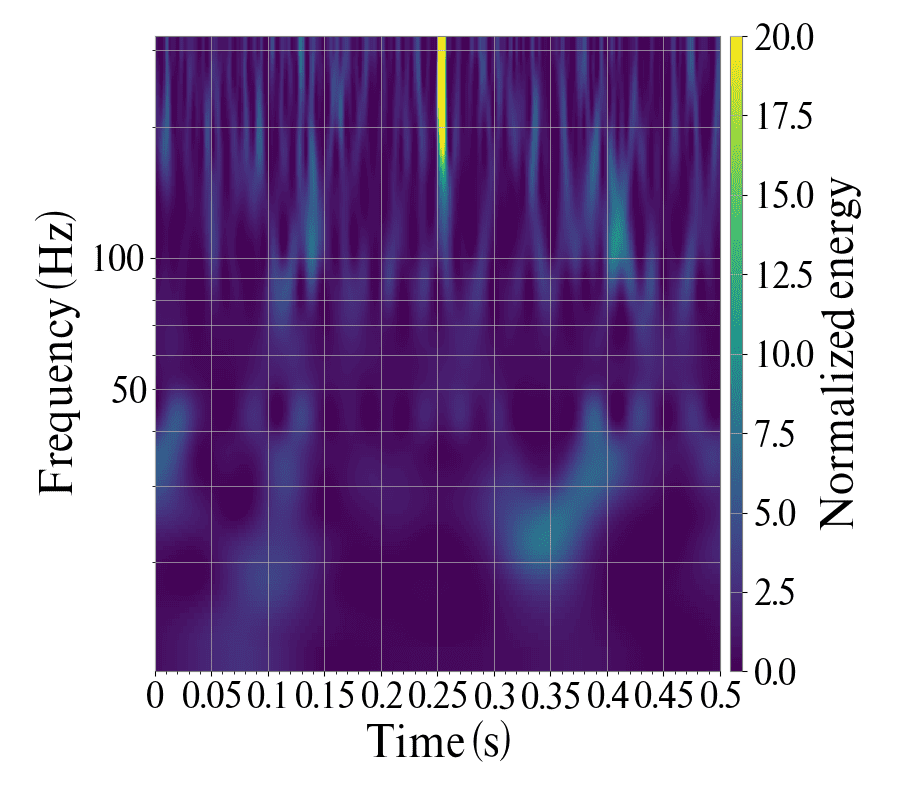}
    \par\smallskip
    {\scriptsize \scalebox{0.8}[1.0]{\textbf{K1:VIS-ITMY\_MN\_PSDAMP\_Y\_IN1\_DQ}}\\ Apr 10, 2020	19:53:03	UTC (GPS: 1270583601)}
    \end{minipage}
    }

\resizebox{!}{0.134\textheight}
{%
\begin{minipage}[t]{0.45\textwidth}
    \centering
    \includegraphics[width=0.48\textwidth]{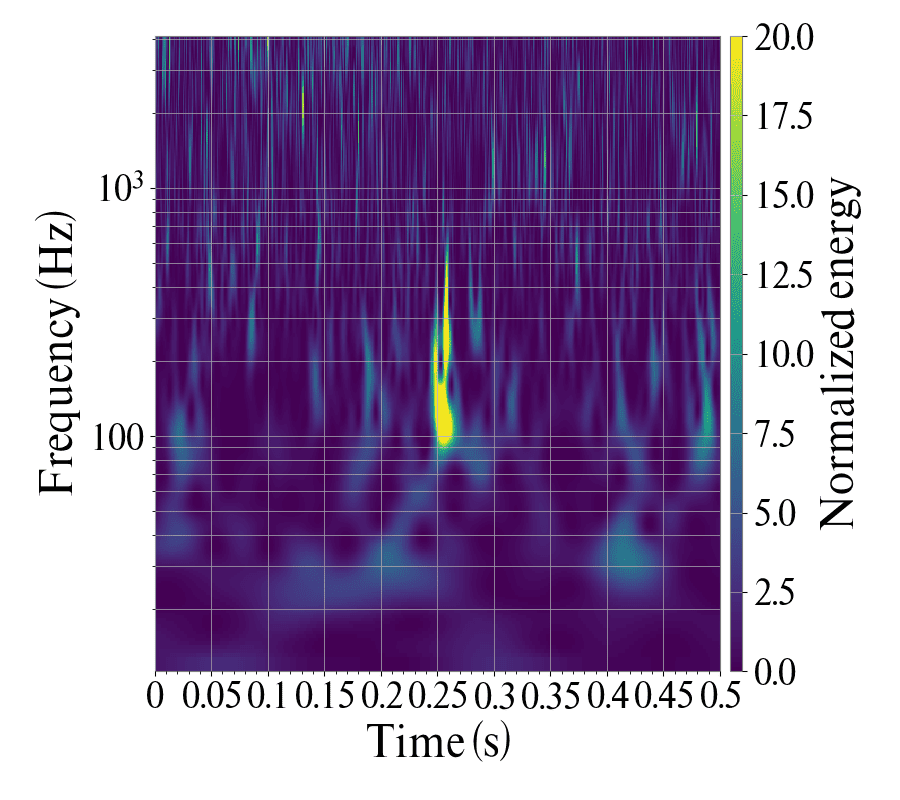}
    \includegraphics[width=0.48\textwidth]{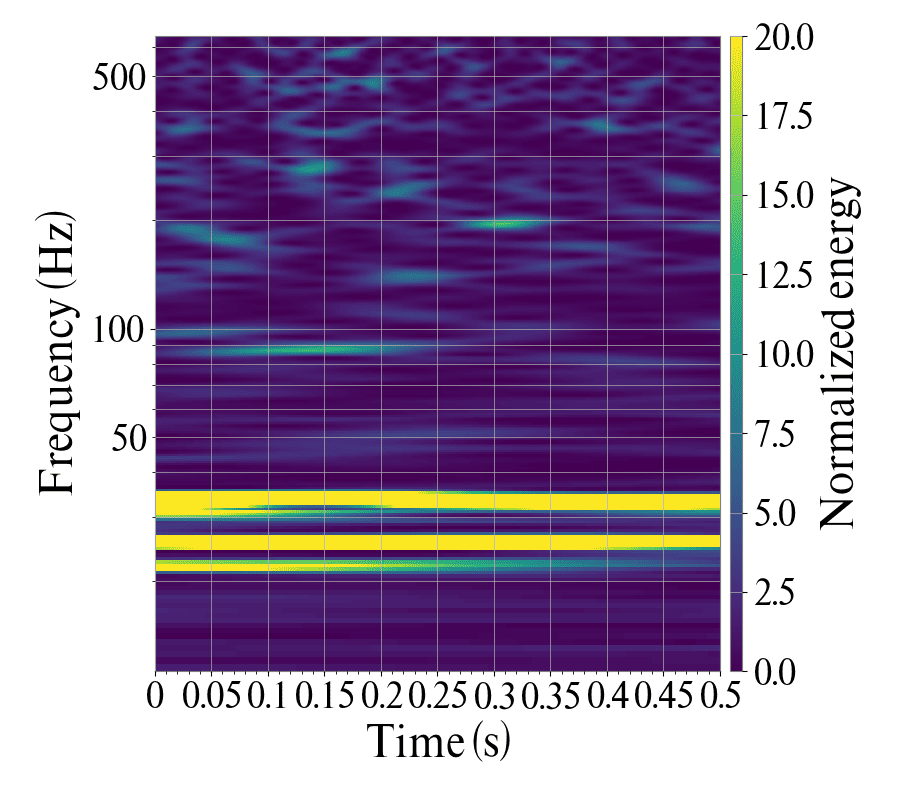}
    \par\smallskip
    {\scriptsize \scalebox{0.8}[1.0]{\textit{K1:VIS-OMMT1\_TM\_OPLEV\_PIT\_OUT\_DQ}}\\ Apr 08, 2020	22:00:18	UTC (GPS: 1270418436)}
    \end{minipage}
\hfill
\vrule width 0.5pt
\hfill
\begin{minipage}[t]{0.45\textwidth}
    \centering
    \includegraphics[width=0.48\textwidth]{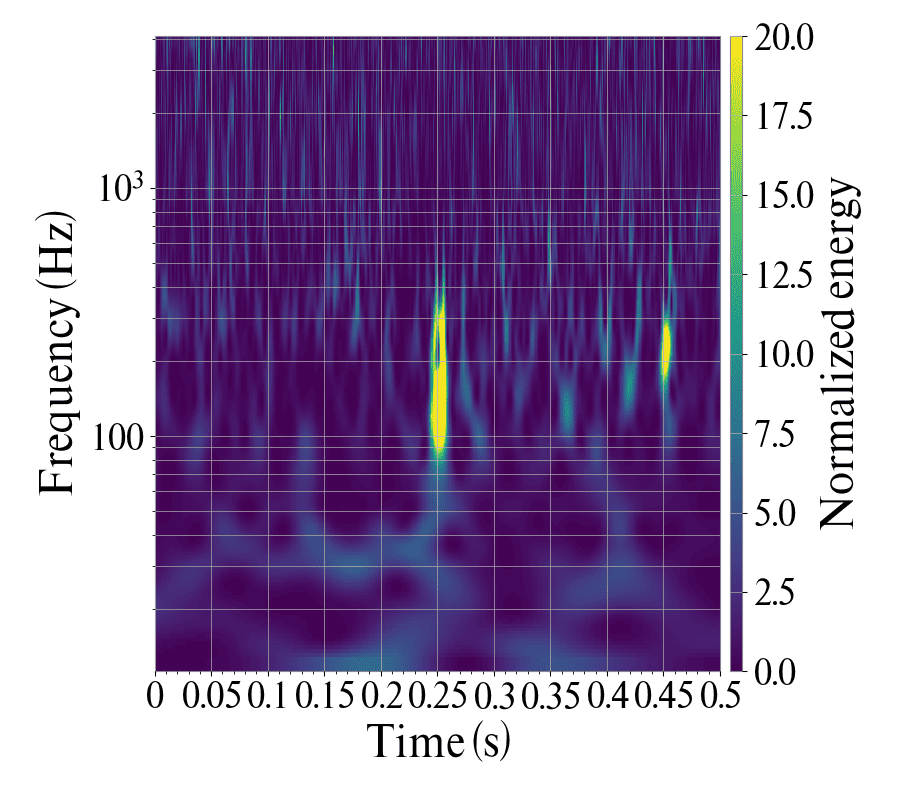}
    \includegraphics[width=0.48\textwidth]{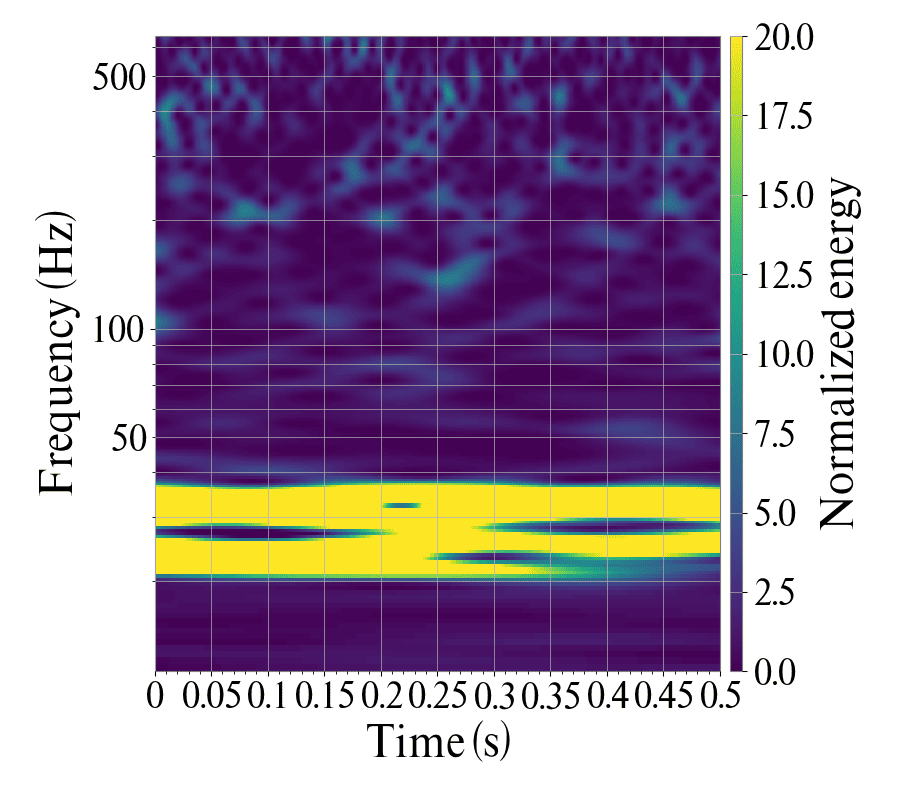}
    \par\smallskip
    {\scriptsize \scalebox{0.8}[1.0]{\textit{K1:VIS-OMMT1\_TM\_OPLEV\_YAW\_OUT\_DQ}}\\ Apr 18, 2020	18:29:21	UTC (GPS: 1271269779)}
    \end{minipage}
}
\end{figure}
\clearpage

\begin{figure*}[h]
\centering
\resizebox{!}{0.134\textheight}
{%
\begin{minipage}[t]{0.45\textwidth}
    \centering
    \includegraphics[width=0.48\textwidth]{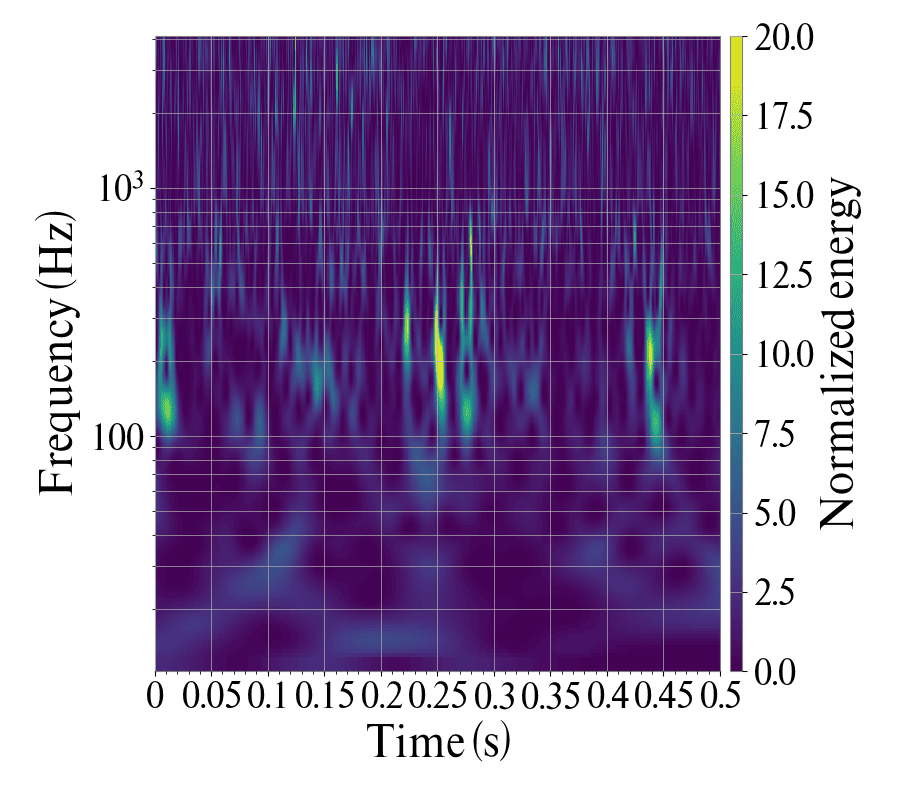}
    \includegraphics[width=0.48\textwidth]{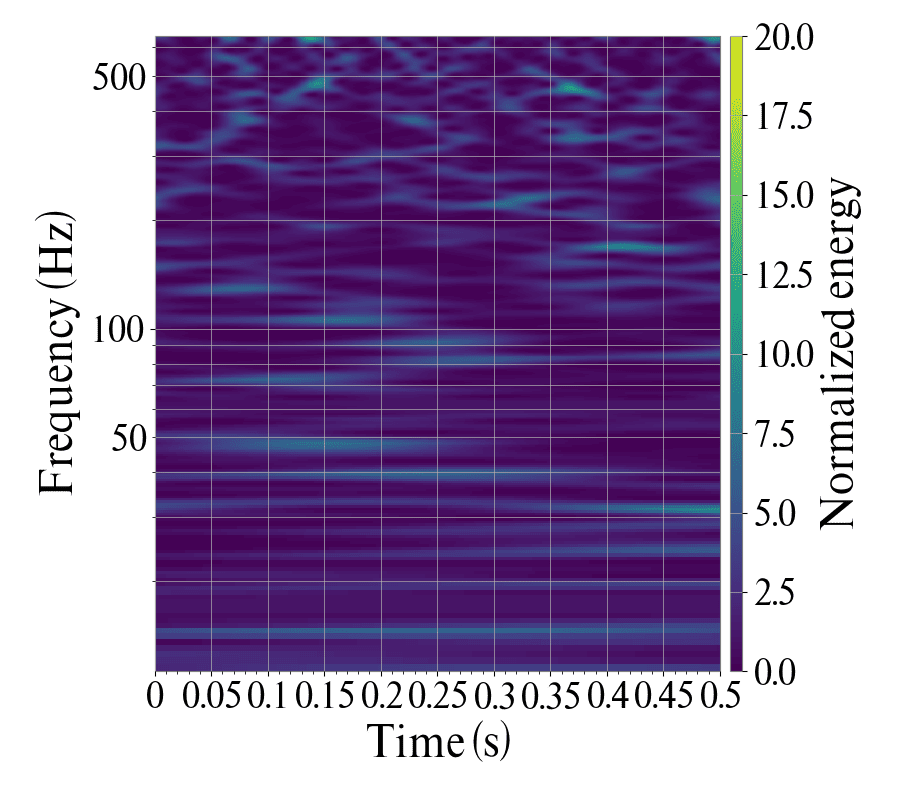}
    \par\smallskip
    {\scriptsize \scalebox{0.8}[1.0]{\textit{K1:VIS-OSTM\_TM\_OPLEV\_YAW\_OUT\_DQ}}\\ Apr 07, 2020	23:36:02	UTC (GPS: 1270337780)}
    \end{minipage}
\hfill
\vrule width 0.5pt
\hfill    
\begin{minipage}[t]{0.45\textwidth}
    \centering
    \includegraphics[width=0.48\textwidth]{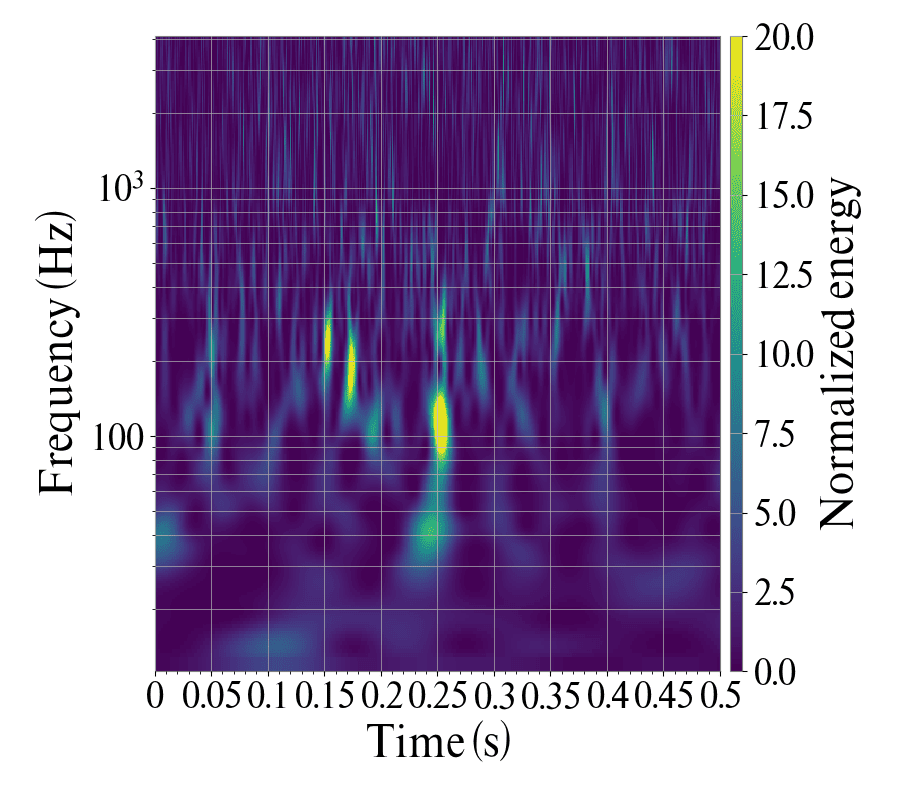}
    \includegraphics[width=0.48\textwidth]{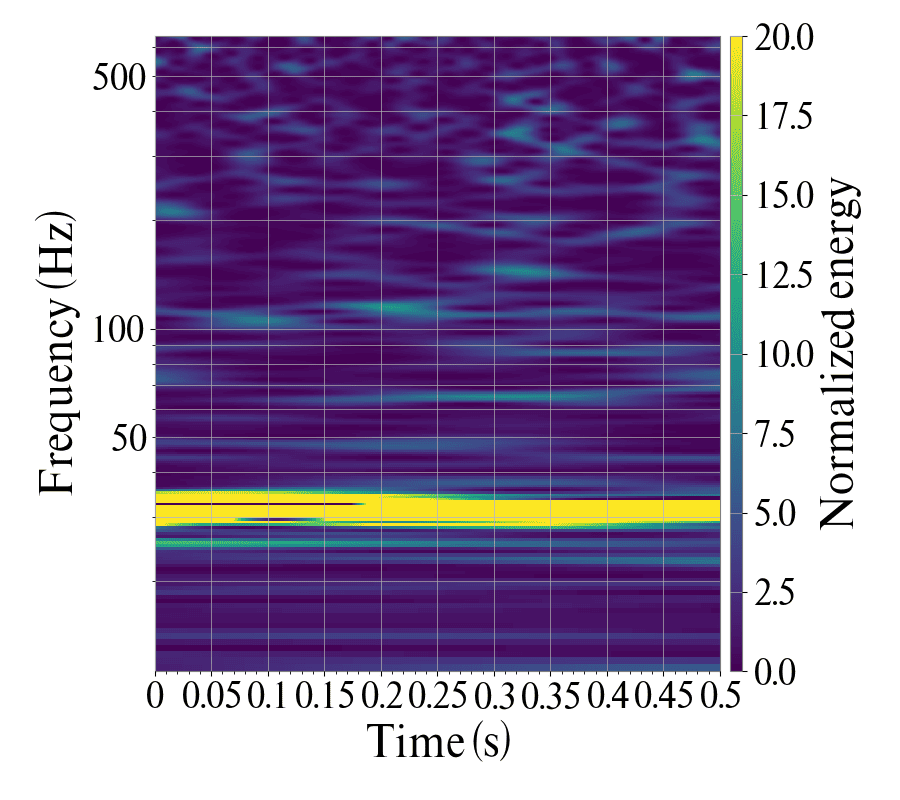}
    \par\smallskip
    {\scriptsize \scalebox{0.8}[1.0]{\textit{K1:VIS-SR3\_TM\_OPLEV\_TILT\_YAW\_OUT\_DQ}}\\ Apr 11, 2020	22:23:58	UTC (GPS: 1270679056)}
  \end{minipage}
}

\resizebox{!}{0.134\textheight}
{%
\begin{minipage}[t]{0.45\textwidth}
    \centering
    \includegraphics[width=0.48\textwidth]{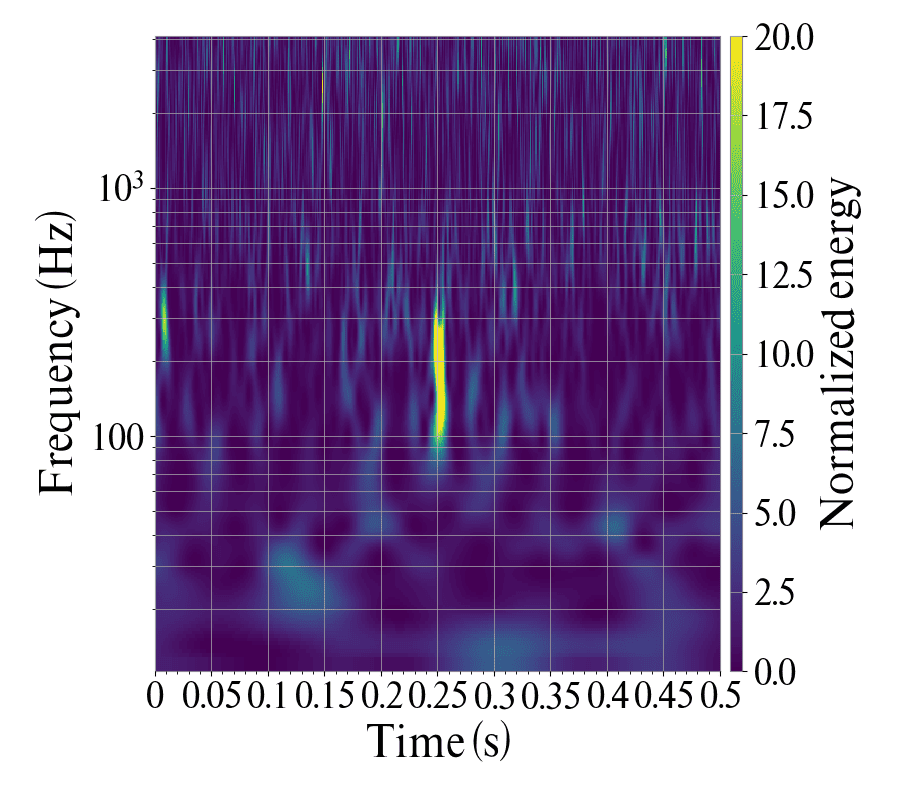}
    \includegraphics[width=0.48\textwidth]{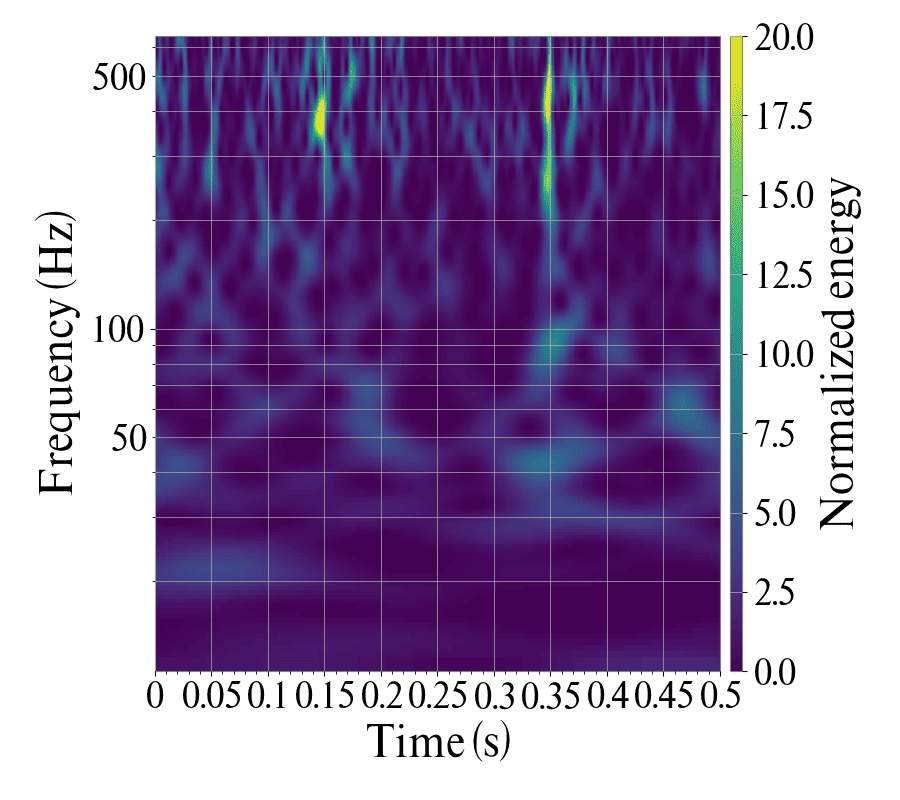}
    \par\smallskip
    {\scriptsize \scalebox{0.8}[1.0]{\textit{K1:VIS-TMSY\_DAMP\_R\_IN1\_DQ}}\\ Apr 20, 2020	11:46:07	UTC (GPS: 1271418385)}
  \end{minipage}
  }
  \caption{An O3GK KAGRA blip glitch found. The left and right panels show the spectrogram of the main channel and an auxiliary channel, respectively.
The auxiliary channels are marked with \textbf{bold font} (\textit{italic font}), because the glitch shape in the main channel spectrogram (left panel) is \textbf{similar} (\textit{nonsimilar}) to that in the auxiliary channel spectrogram (right panel). Each figure also includes, below the panels, the name of the auxiliary channel and the UTC time (GPS time) when the glitch was found. The same convention is used for all the following spectrogram figures.
In some examples, the auxiliary channel glitch may not be visually clear, as the selection focused on main channel features.
The y-axis is shown in logarithmic scale, with a different range in each panel, while the x-axis is in linear scale and has the same range in both panels. In the subsequent figures, the y-axis range for each auxiliary channel spectrogram in the right panel may differ, reflecting the fact that each auxiliary channel records a distinct frequency range.}
  \label{fig:Blipfigs}
\end{figure*}
\vspace{-0.5cm}
Most O3GK KAGRA blip glitches (572 out of 1,294) are related to the VIS subsystem, although 12 different auxiliary channels belonging to this subsystem are involved. These channels are distributed in multiple suspension locations and no unique physical cause has been identified so far. Our analysis focuses on statistical correlations, which may provide useful input for future noise mitigation efforts. In fact, such statistical information has been successfully utilized in previous work to aid noise subtraction or source identification~\cite{Davis_2022, Zhao_2021}.

We show examples of the O3GK KAGRA blip glitch in Figure \ref{fig:Blipfigs}.
It is necessary to mention that the type of glitch is determined by its shape in the main channel spectrogram, but in Figure \ref{fig:Blipfigs}, the spectrograms of the main channel (left panel) and the auxiliary channel of the round winner (right panel) are shown for comparison. In Figure \ref{fig:Blipfigs}, the glitch in the auxiliary channel marked with \textbf{bold font looks similar} to that in the main channel, while the glitch in the auxiliary channel marked with \textit{italic font looks not similar} from that on the main channel.

As mentioned earlier, similarity (or not similarity) in the glitch shape on the spectrogram between the main channel and the round winner auxiliary channel is related to the correlation between these two channels.  
For this reason, we indicate the round winner auxiliary channels that show glitch shapes similar to the main channel using the bold font in Table \ref{table:BlipGlitches}, while the italic font is used for the auxiliary channels that show different glitch shapes from the main channel. We use the same convention as in Tables \ref{table:DotGlitches} to \ref{table:scattered light Glitches} when we show the results for the other types of glitches in the following subsections. 
Later in the discussion section (Section \ref{sec:discussion}), 
we will discuss in more detail whether the similarity in the glitch shapes, judged by visual examination, is consistent with the true (or quantitatively measured) correlation between the main channel and the round winner auxiliary channel.  
It is worth mentioning that no auxiliary channel in the AOS and IMC subsystems of KAGRA shows blip glitches that are similar to those in the main channel. 
In contrast, the auxiliary channel in the LAS subsystem is indicated in bold font in Table \ref{table:BlipGlitches}, showing that blip glitches found in this subsystem look similar to those in the main channel. 

The spectrogram of K1:PEM-MAG\_BS\_BOOTH\_BS\_Y\_OUT\_DQ in Figure~\ref{fig:Blipfigs} shows another example of a blip glitch observed in the O3GK KAGRA data. Although the spectral features below 90 Hz in the auxiliary channel do not appear to be related to the blip glitch in the main channel, we consider this auxiliary channel to exhibit a relevant trend because the spectrogram shows a morphology similar to the main channel **in the frequency range above 100 Hz**. We note that this threshold does not correspond to a strict instrumental cut-off but rather reflects the typical frequency range where O3GK KAGRA blip glitches appear. Although these glitches may come from broadband noise sources, the whitening process - based on the detector's power spectral density - renders them with finite-frequency features in the spectrogram. Therefore, channels exhibiting similar morphology in the relevant frequency range are indicated in bold font in Table~\ref{table:BlipGlitches}.

\begin{table*}[!h]
\centering
\caption{\label{table:DotGlitches} O3GK KAGRA dot glitches}
\small
\resizebox{0.98\textwidth}{!}
{%
\begin{tabular}{|P{0.07\textwidth}|P{0.58\textwidth}|P{0.35\textwidth}|}
\hline
\begin{tabular}[c]{@{}c@{}}Sub-\\ System\end{tabular} & \begin{tabular}[c]{@{}c@{}}Round Winner\\ Auxiliary Channel\end{tabular} & \begin{tabular}[c]{@{}c@{}}Vetoed Date in April\\ (\# of Vetoed Events)\end{tabular} \\ \hline

\begin{tabular}[c]{@{}c@{}}AOS \\ (4) \end{tabular}
& \textit{K1:AOS-TMSX\_IR\_PDA1\_OUT\_DQ}
& \begin{tabular}[c]{@{}c@{}}10th (1), 16th (2), 18th (1)\end{tabular} \\ \hline

\multirow{2}{*}{\begin{tabular}[c]{@{}c@{}} IMC \\ (16) \end{tabular}}
& \begin{tabular}[c]{@{}c@{}}\textit{K1:IMC-IMMT1}\\ \textit{\_TRANS\_QPDA1\_DC\_PIT\_OUT\_DQ}\end{tabular}  
& \begin{tabular}[c]{@{}c@{}}8th (2), 9th (6),\\ 10th (2), 11th (1)\end{tabular}\\ \cline{2-3} 
& \begin{tabular}[c]{@{}c@{}}\textit{K1:IMC-IMMT1}\\ \textit{\_TRANS\_QPDA1\_DC\_YAW\_OUT\_DQ}\end{tabular} 
& 19th (5) \\ \hline

\begin{tabular}[c]{@{}c@{}}LSC \\ (14) \end{tabular}
& \begin{tabular}[c]{@{}c@{}}\textit{K1:LSC-ALS\_DARM\_OUT\_DQ}\end{tabular}
& 12th (11), 15th (2), 16th (1) 
\\ \hline

\multirow{2}{*}{\begin{tabular}[c]{@{}c@{}} PEM \\ (3) \end{tabular}}
& \textbf{K1:PEM-ACC\_OMC\_TABLE\_AS\_Z\_OUT\_DQ} 
& 17th (1)
\\ \cline{2-3}
& \textbf{K1:PEM-SEIS\_IXV\_GND\_EW\_IN1\_DQ} 
& 7th (2)
\\ \hline

\multirow{8}{*}{\begin{tabular}[c]{@{}c@{}}VIS \\ (13) \end{tabular}}
& \textit{K1:VIS-ETMX\_MN\_PSDAMP\_Y\_IN1\_DQ}
& 7th (1)
\\ \cline{2-3}
& \textit{K1:VIS-ETMY\_MN\_PSDAMP\_Y\_IN1\_DQ}
& 19th (1)
\\ \cline{2-3}
& \textbf{K1:VIS-ITMY\_IM\_PSDAMP\_R\_IN1\_DQ}
& 7th (1), 9th (3)
\\ \cline{2-3}
& \textbf{K1:VIS-ITMY\_MN\_PSDAMP\_L\_IN1\_DQ}
& 17th (1)
\\ \cline{2-3}
& \textit{K1:VIS-OMMT1\_TM\_OPLEV\_PIT\_OUT\_DQ}
&  12th (1)
\\ \cline{2-3}
& \begin{tabular}[c]{@{}c@{}}\textbf{K1:VIS-OMMT1}\\ \textbf{\_TM\_OPLEV\_YAW\_OUT\_DQ}\end{tabular}
& 11th (3)
\\ \cline{2-3}
& \textbf{K1:VIS-OSTM\_TM\_OPLEV\_PIT\_OUT\_DQ}
& 15th(1)
\\ \cline{2-3}
& \textit{K1:VIS-TMSY\_DAMP\_R\_IN1\_DQ}
& 20th (1)
\\ \hline

\end{tabular} }
\end{table*}

\subsubsection{Dot Glitch}
Dot glitch is named after their dot-like (or circular blob-like) shape on the spectrogram. 
During the O3GK period, 50 dot glitches were identified with Hveto in the five KAGRA subsystems (see Tables \ref{table:Glitchtypes} and \ref{table:DotGlitches}). 
Unlike the blip glitches, which were found mostly between $\sim 100$~Hz and $\sim 1000$~Hz on the main channel spectrogram, there is no typical frequency range at which the O3GK KAGRA dot glitches appeared on the main channel spectrogram. 

\begin{figure*}[h]
\centering
\resizebox{!}{0.135\textheight}
{%
\begin{minipage}[t]{0.45\textwidth}
    \centering
    \includegraphics[width=0.48\textwidth]{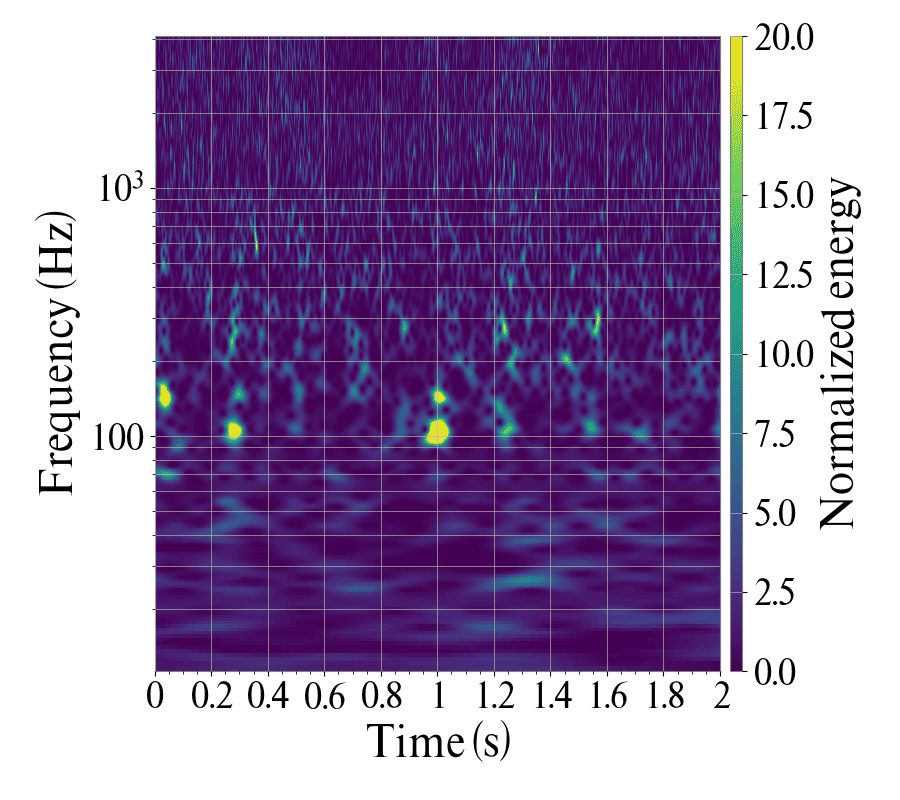}
    \includegraphics[width=0.48\textwidth]{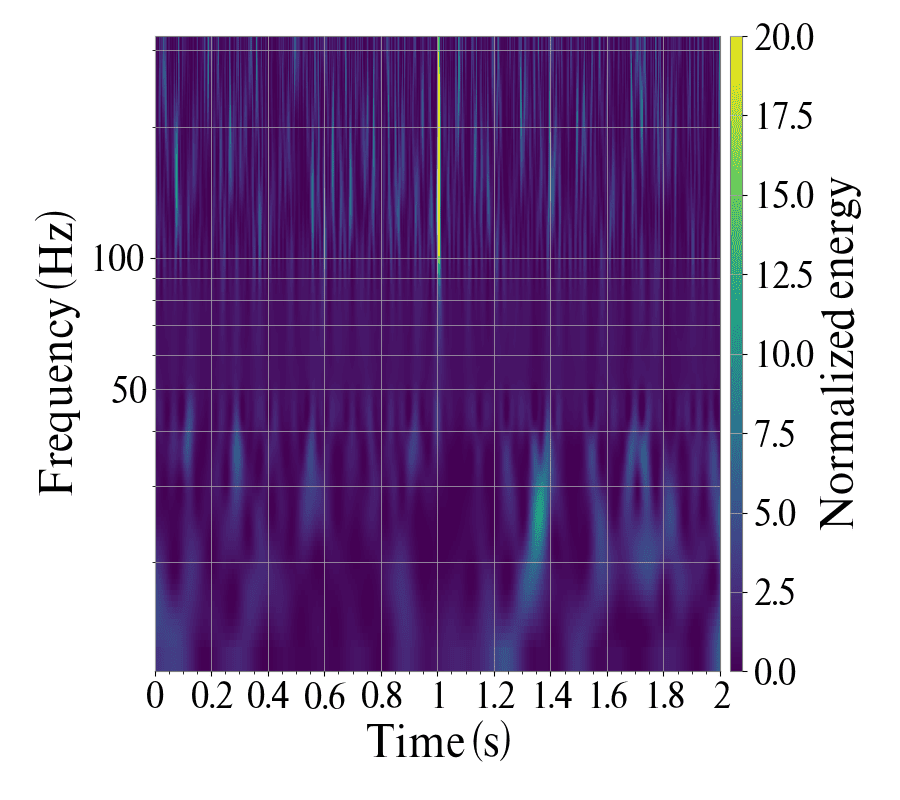}
    \par\smallskip
    {\scriptsize \scalebox{0.8}[1.0]{\textbf{K1:VIS-ITMY\_IM\_PSDAMP\_R\_IN1\_DQ}}\\ Apr 09, 2020	07:19:28	UTC	(GPS: 1270451986)}
    \end{minipage}
\hfill
\vrule width 0.5pt
\hfill 
\begin{minipage}[t]{0.45\textwidth}
    \centering
    \includegraphics[width=0.48\textwidth]{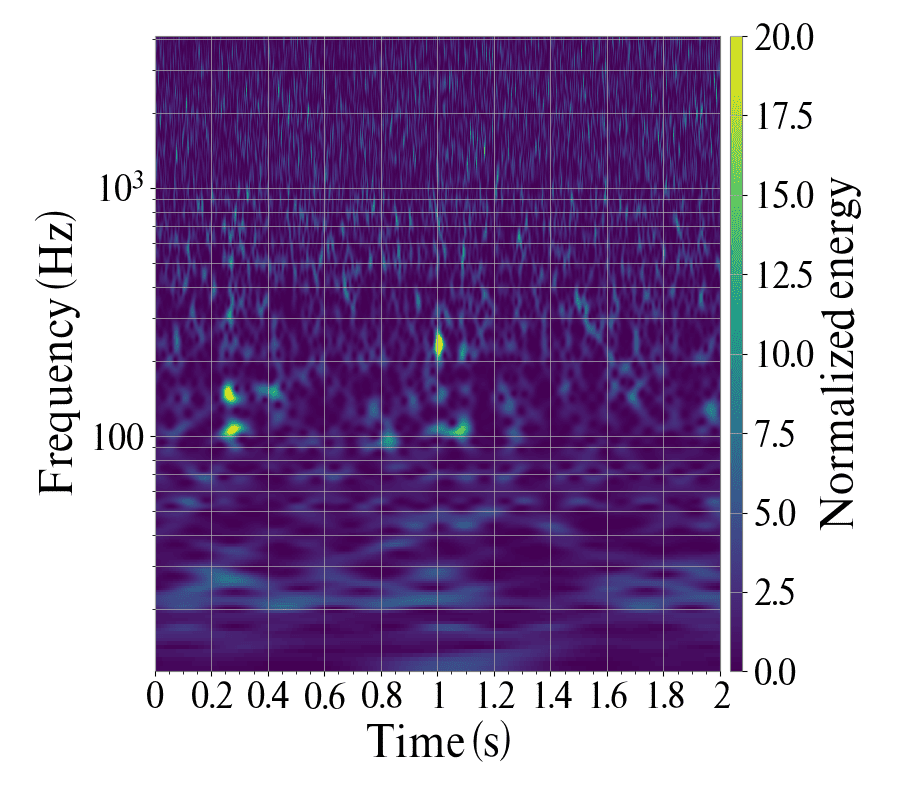}
    \includegraphics[width=0.48\textwidth]{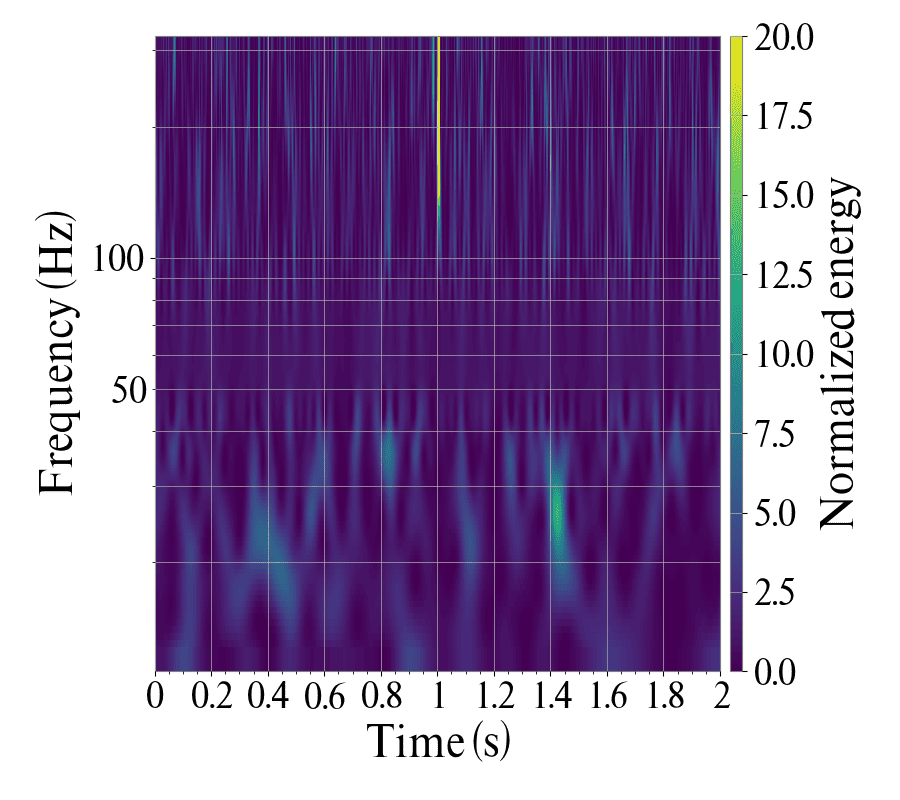}
    \par\smallskip
    {\scriptsize \scalebox{0.8}[1.0]{\textbf{K1:VIS-ITMY\_MN\_PSDAMP\_L\_IN1\_DQ}}\\ Apr 17, 2020	02:44:26	UTC (GPS: 1271126684)}
    \end{minipage}
    }

\resizebox{!}{0.135\textheight}
{%
\begin{minipage}[t]{0.45\textwidth}
    \centering
    \includegraphics[width=0.48\textwidth]{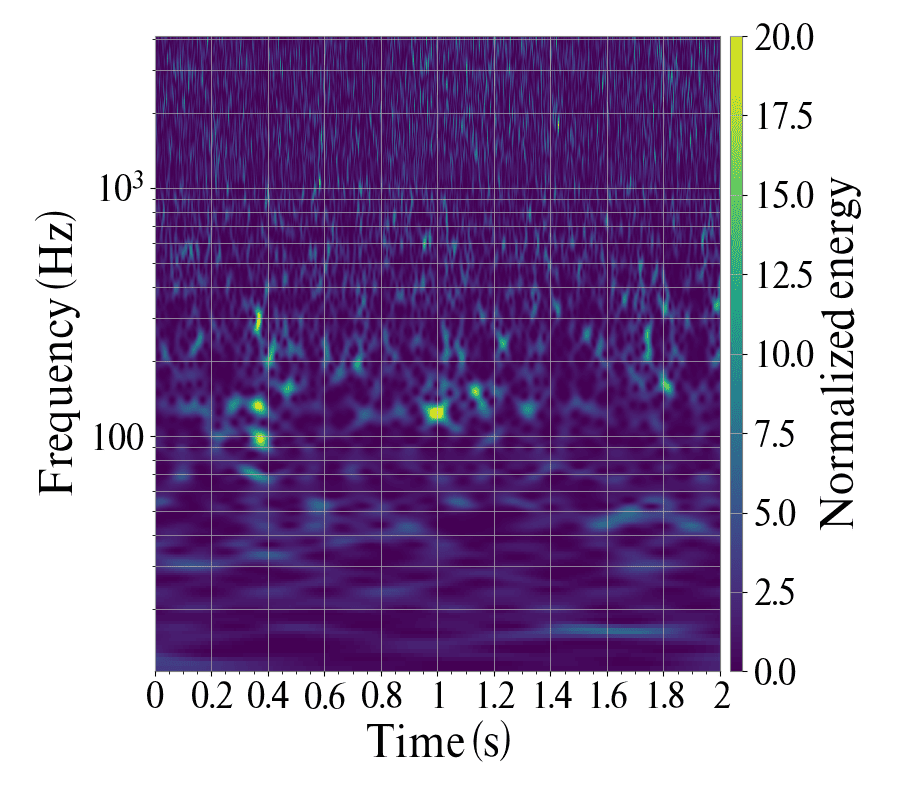}
    \includegraphics[width=0.48\textwidth]{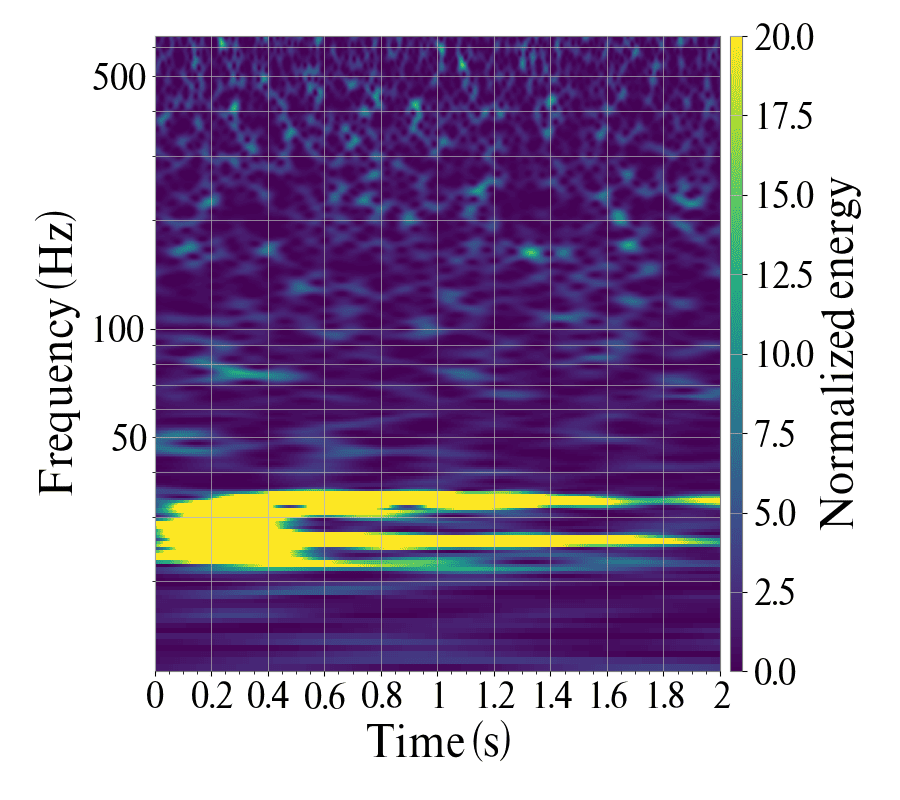}
    \par\smallskip
    {\scriptsize \scalebox{0.8}[1.0]{\textit{K1:VIS-OMMT1\_TM\_OPLEV\_PIT\_OUT\_DQ}}\\ Apr 12, 2020	05:00:08	UTC (GPS: 1270702826)}
    \end{minipage}
\hfill
\vrule width 0.5pt
\hfill 
\begin{minipage}[t]{0.45\textwidth}
    \centering
    \includegraphics[width=0.48\textwidth]{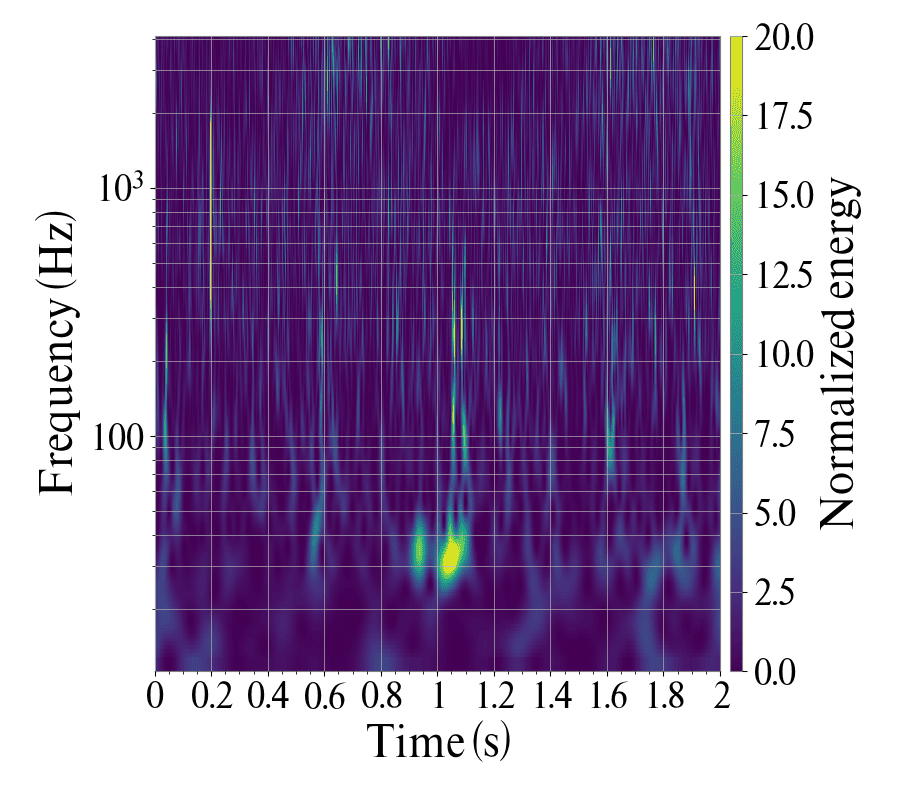}
    \includegraphics[width=0.48\textwidth]{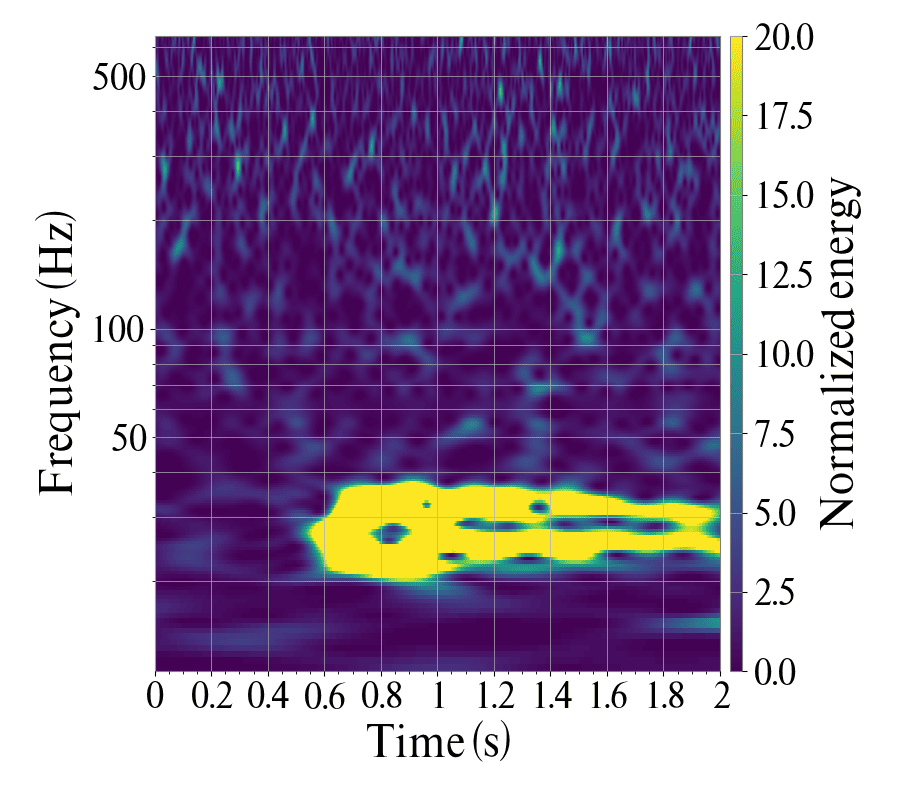}
    \par\smallskip
    {\scriptsize \scalebox{0.8}[1.0]{\textbf{K1:VIS-OMMT1\_TM\_OPLEV\_YAW\_OUT\_DQ}}\\ Apr 11, 2020	11:39:50	UTC (GPS: 1270640408)}
    \end{minipage}
}

\resizebox{!}{0.135\textheight}
{%
\begin{minipage}[t]{0.45\textwidth}
    \centering
    \includegraphics[width=0.48\textwidth]{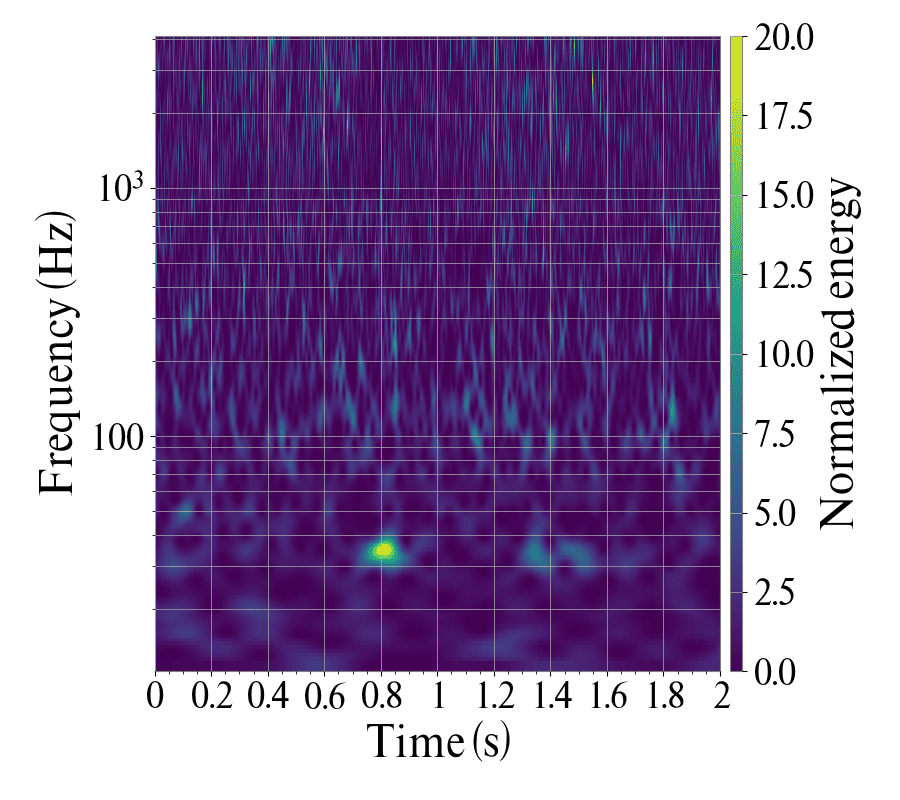}
    \includegraphics[width=0.48\textwidth]{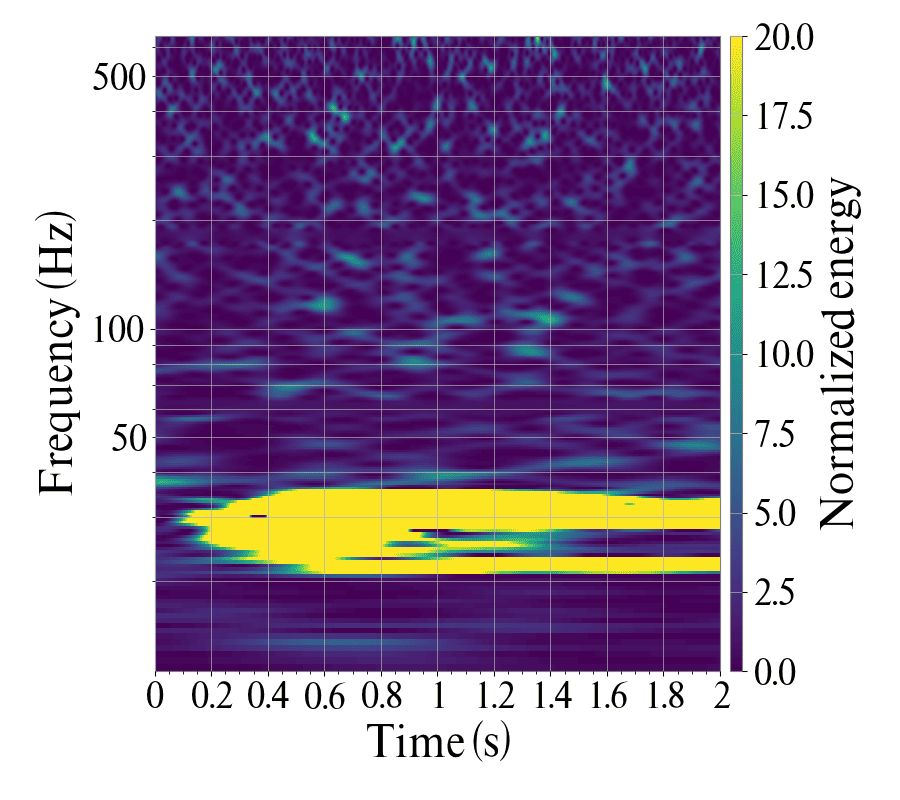}
    \par\smallskip
    {\scriptsize \scalebox{0.8}[1.0]{\textbf{K1:VIS-OSTM\_TM\_OPLEV\_YAW\_OUT\_DQ}}\\ Apr 15, 2020	11:10:41	UTC (GPS: 1270984259)}
    \end{minipage}
\hfill
\vrule width 0.5pt
\hfill     
\begin{minipage}[t]{0.45\textwidth}
    \centering
    \includegraphics[width=0.48\textwidth]{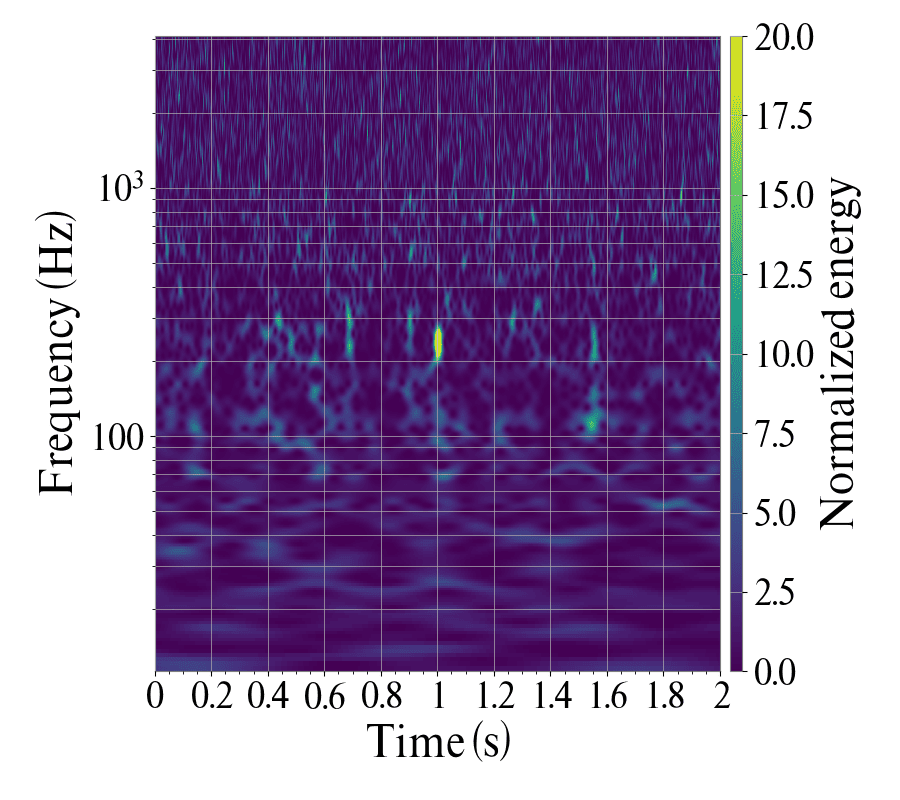}
    \includegraphics[width=0.48\textwidth]{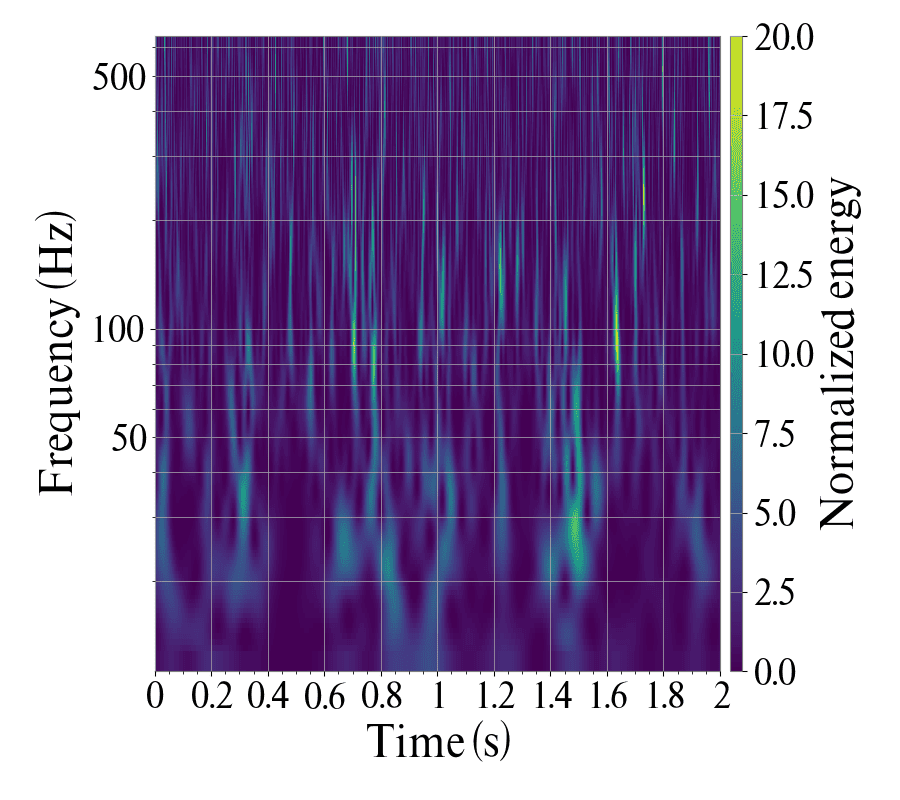}
    \par\smallskip
    {\scriptsize \scalebox{0.8}[1.0]{\textit{K1:VIS-TMSY\_DAMP\_R\_IN1\_DQ}}\\ Apr 20, 2020	07:17:25	UTC (GPS: 1271402263)}
  \end{minipage}
  }
  \caption{An O3GK KAGRA dot glitch found. The left and right panels show the spectrogram of the main channel and an auxiliary channel, respectively. Each figure also includes, below the panels, the name of the auxiliary channel and the UTC time (GPS time) when the glitch was found.}
  \label{fig:Dotfigs}
\end{figure*}

Figure \ref{fig:Dotfigs} shows examples of O3GK KAGRA dot glitches. As mentioned in the previous subsection, the auxiliary channel that shows a similar (different) shape to the main channel on the spectrogram is indicated in bold (italic) font.
As shown in the spectrogram of K1:VIS-OSTM\_TM\_OPLEV\_PIT\_OUT\_DQ in Figure \ref{fig:Dotfigs}, we determined that the auxiliary channel is correlated with the main channel (that is, marked in bold font) when the high-energy region of the auxiliary channel spectrogram covers the dot glitch in the main channel spectrogram.

\begin{figure}[h]
\centering
\resizebox{!}{0.131\textheight}
{%
\begin{minipage}[t]{0.45\textwidth}
    \centering
    \includegraphics[width=0.48\textwidth]{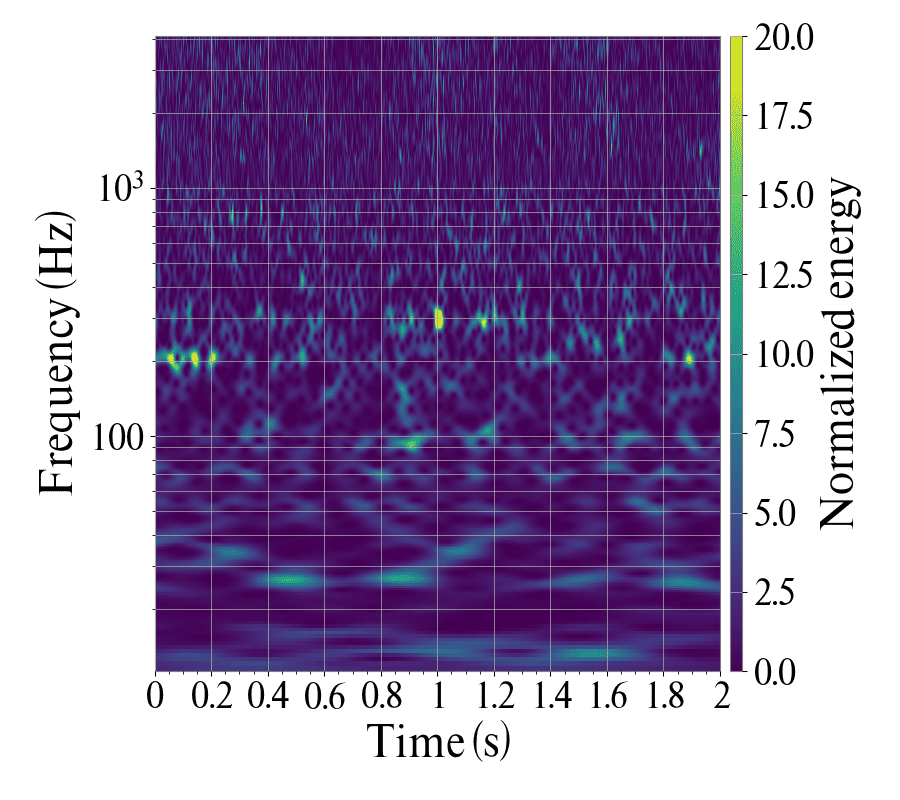}
    \includegraphics[width=0.48\textwidth]{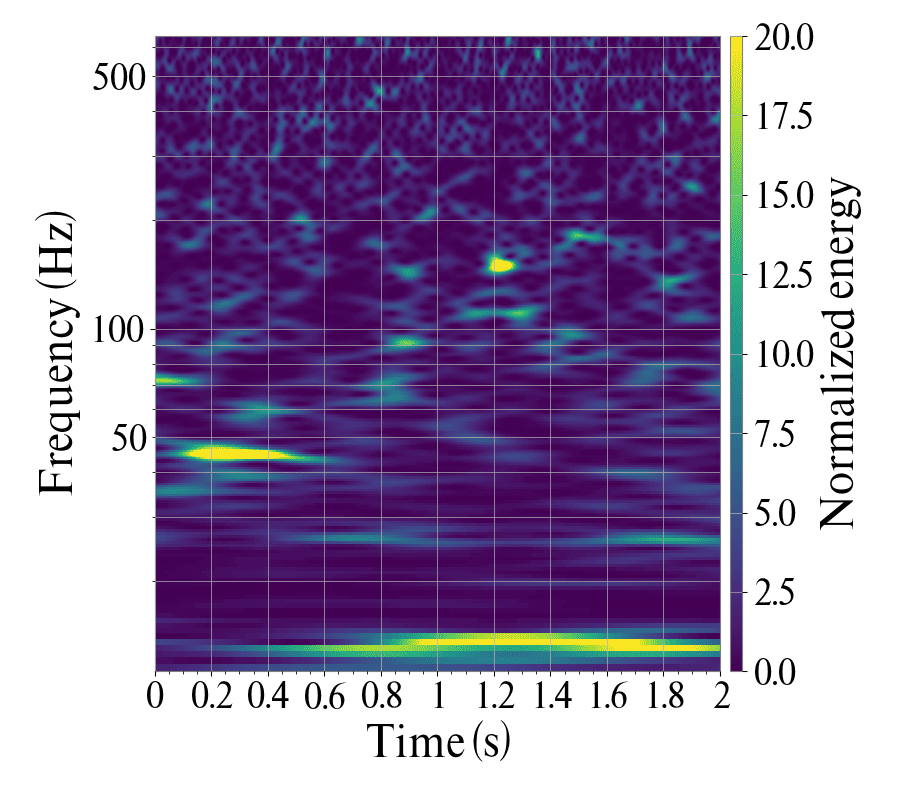}
    \par\smallskip
    {\scriptsize \scalebox{0.8}[1.0]{\textit{K1:AOS-TMSX\_IR\_PDA1\_OUT\_DQ}} \\ Apr 10, 2020	17:56:55	UTC (GPS: 1270576633)}
    \end{minipage}
\hfill
\vrule width 0.5pt
\hfill
  \begin{minipage}[t]{0.45\textwidth}
    \centering
    \includegraphics[width=0.48\textwidth]{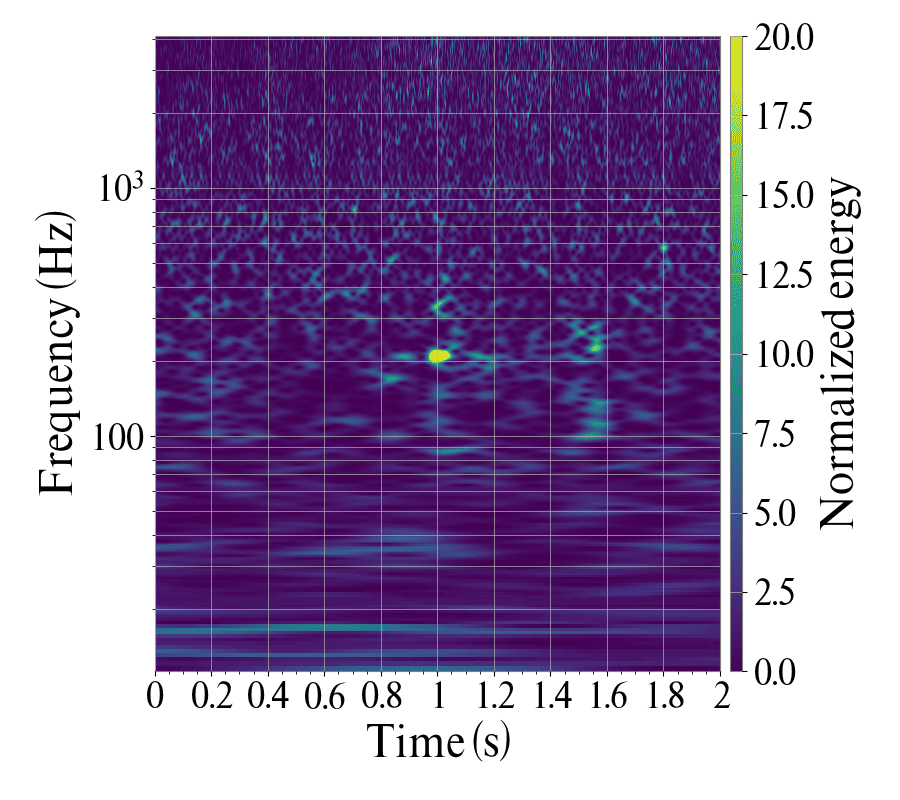}
    \includegraphics[width=0.48\textwidth]{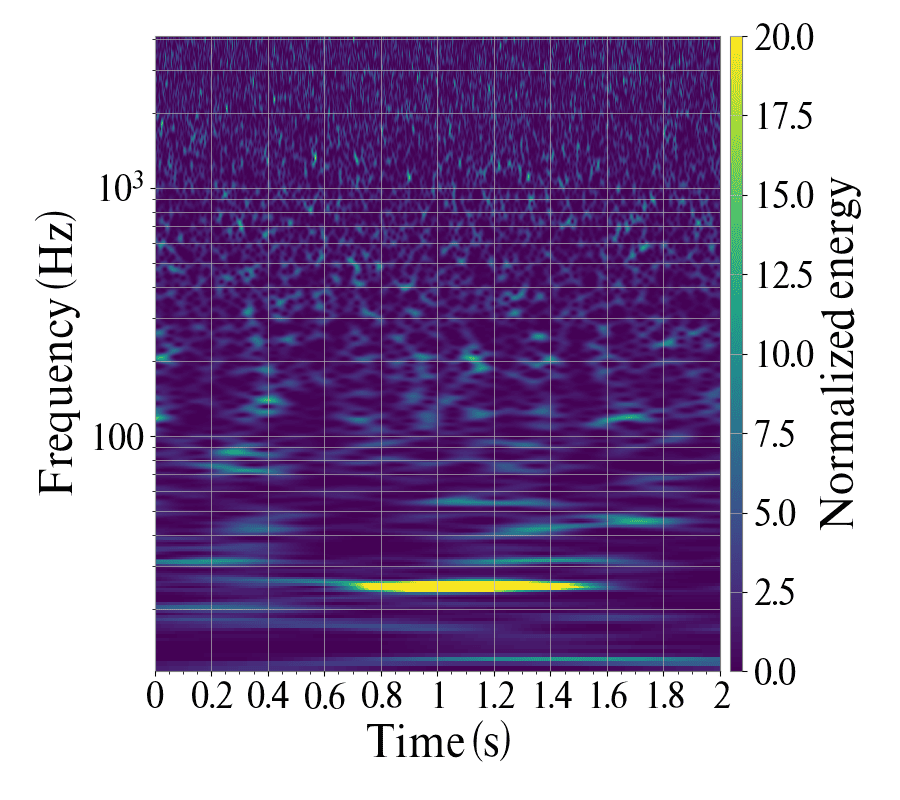}
    \par\smallskip
    {\scriptsize \scalebox{0.8}[1.0]{\textit{K1:IMC-IMMT1\_TRANS\_QPDA1\_DC\_PIT\_OUT\_DQ}} \\ Apr 08, 2020	17:55:38	UTC (GPS: 1270403756)}
    \end{minipage}
}

\resizebox{!}{0.131\textheight}
{%
\begin{minipage}[t]{0.45\textwidth}
    \centering
    \includegraphics[width=0.48\textwidth]{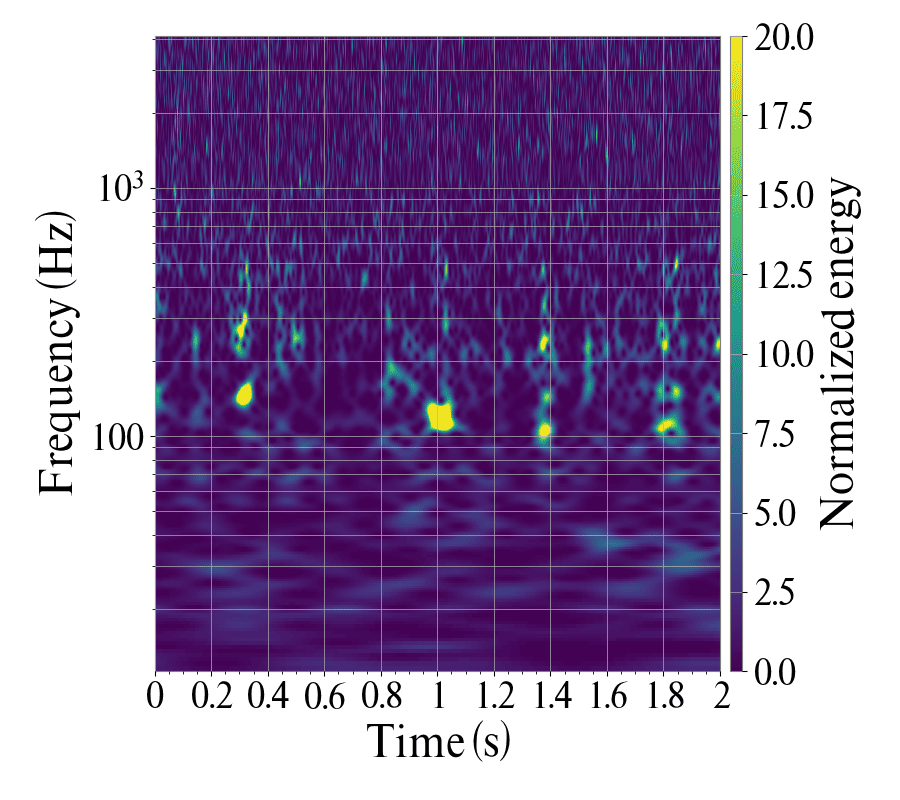}
    \includegraphics[width=0.48\textwidth]{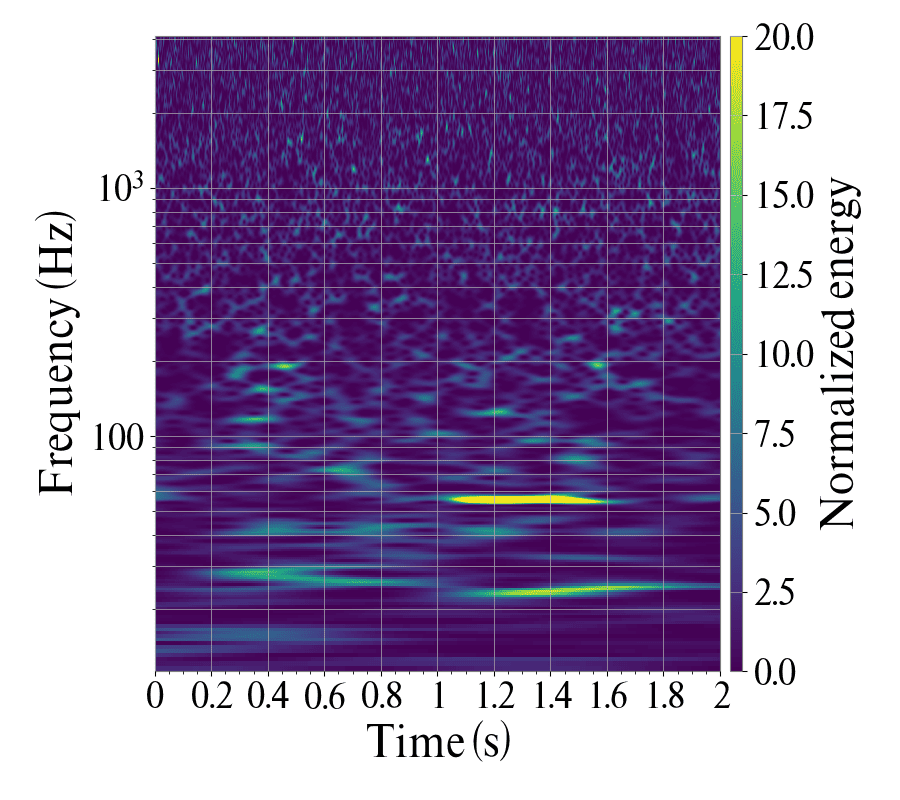}
    \par\smallskip
    {\scriptsize \scalebox{0.8}[1.0]{\textit{K1:IMC-IMMT1\_TRANS\_QPDA1\_DC\_YAW\_OUT\_DQ}} \\ Apr 19, 2020	02:27:16	UTC (GPS: 1271298454)}
    \end{minipage}
\hfill
\vrule width 0.5pt
\hfill
\begin{minipage}[t]{0.45\textwidth}
    \centering
    \includegraphics[width=0.48\textwidth]{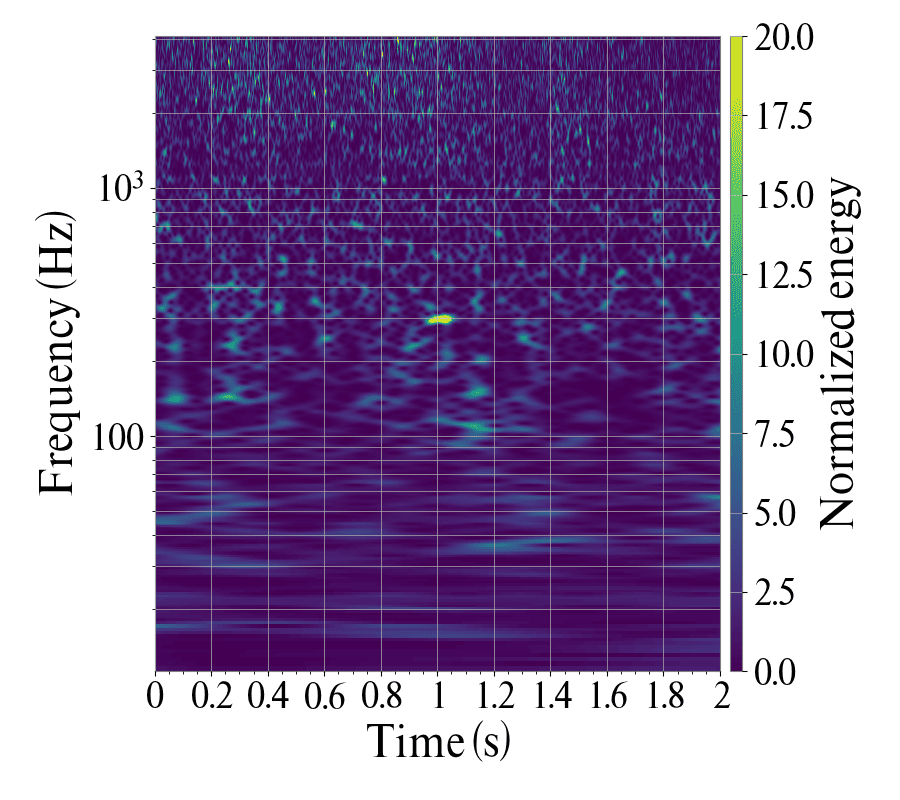}
    \includegraphics[width=0.48\textwidth]{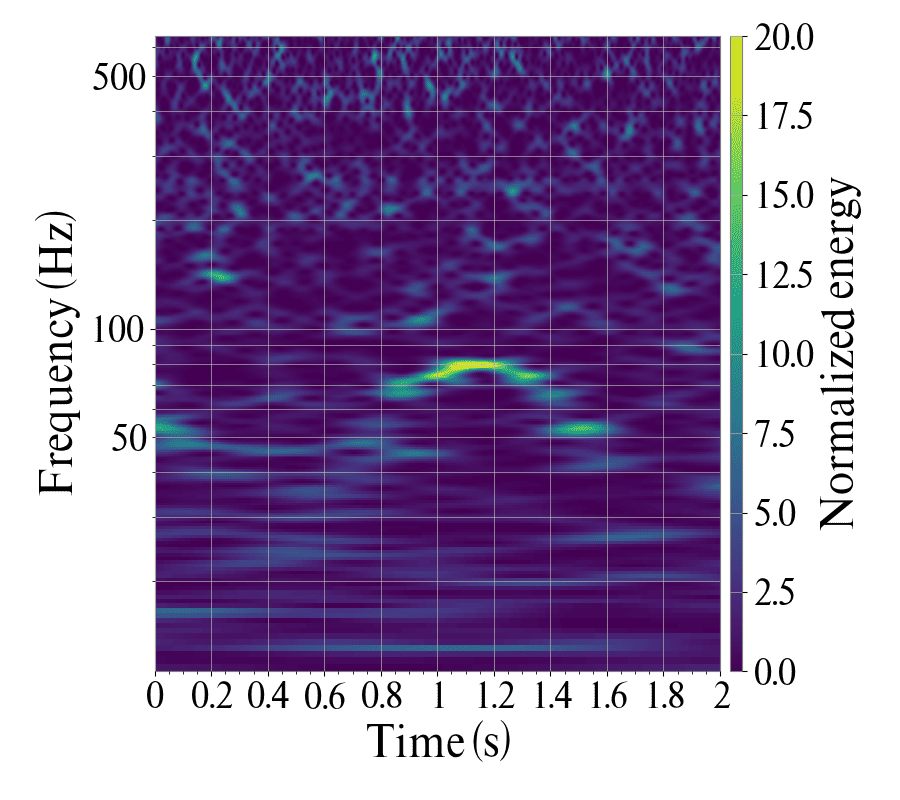}
    \par\smallskip
    {\scriptsize \scalebox{0.8}[1.0]{\textit{K1:LSC-ALS\_DARM\_OUT\_DQ}} \\ Apr 12, 2020	20:37:53	UTC (GPS: 1270759091)}
    \end{minipage}
}

\resizebox{!}{0.131\textheight}
{%
\begin{minipage}[t]{0.45\textwidth}
    \centering
    \includegraphics[width=0.48\textwidth]{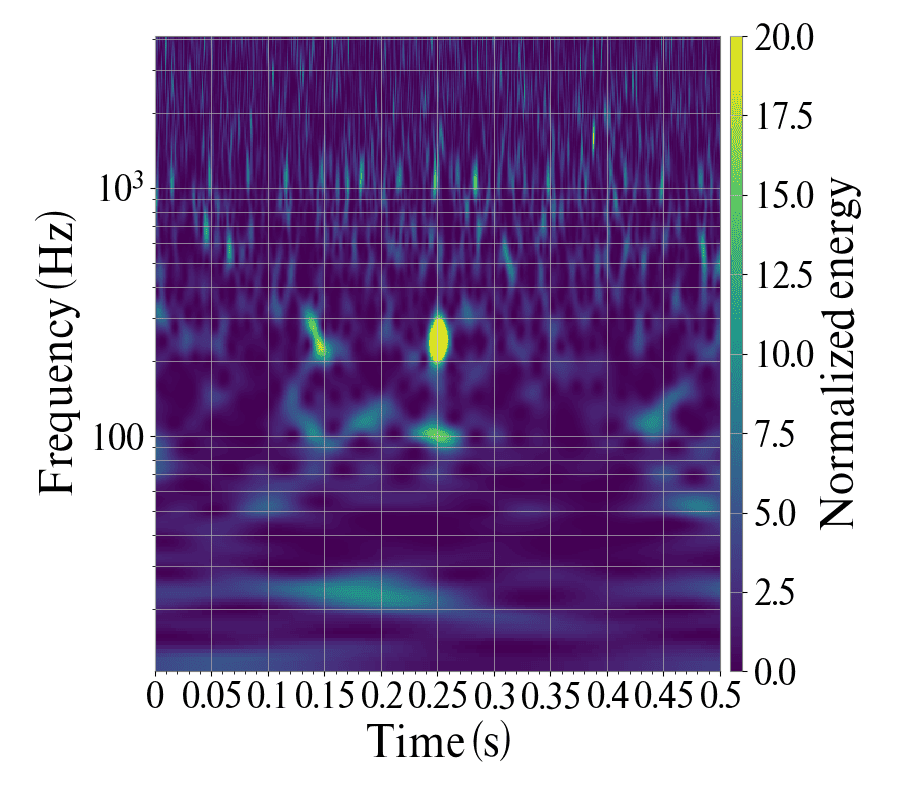}
    \includegraphics[width=0.48\textwidth]{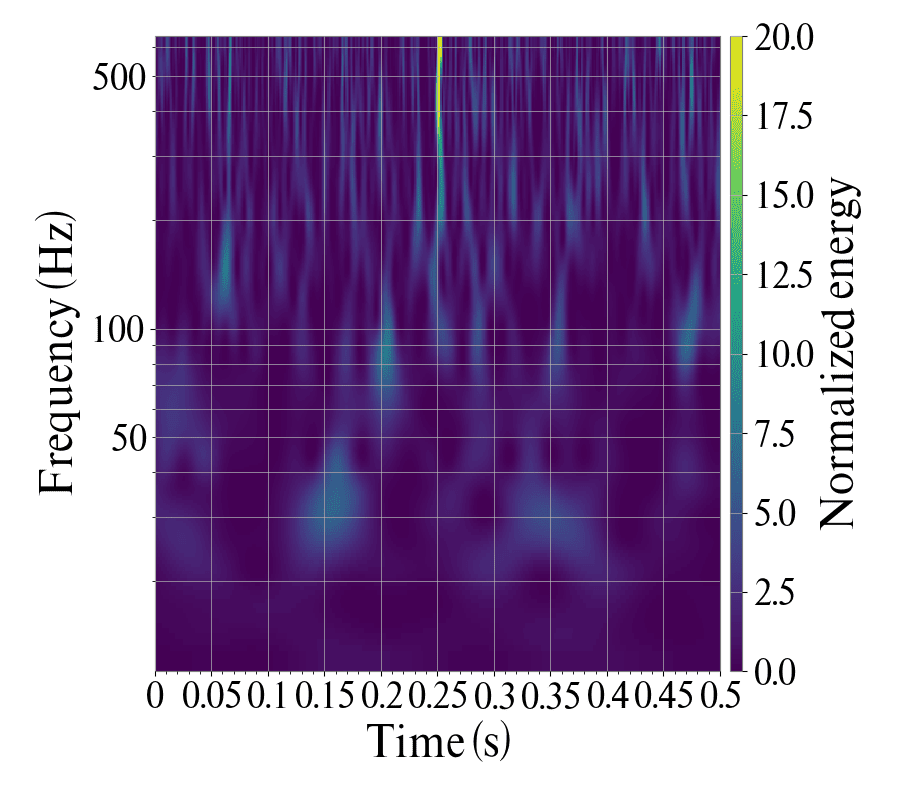}
    \par\smallskip
    {\scriptsize \scalebox{0.8}[1.0]{\textbf{K1:PEM-ACC\_OMC\_TABLE\_AS\_Z\_OUT\_DQ}}\\ Apr 17, 2020	13:14:27	UTC	(GPS: 1271164485)}
    \end{minipage}
\hfill
\vrule width 0.5pt
\hfill    
\begin{minipage}[t]{0.45\textwidth}
    \centering
    \includegraphics[width=0.48\textwidth]{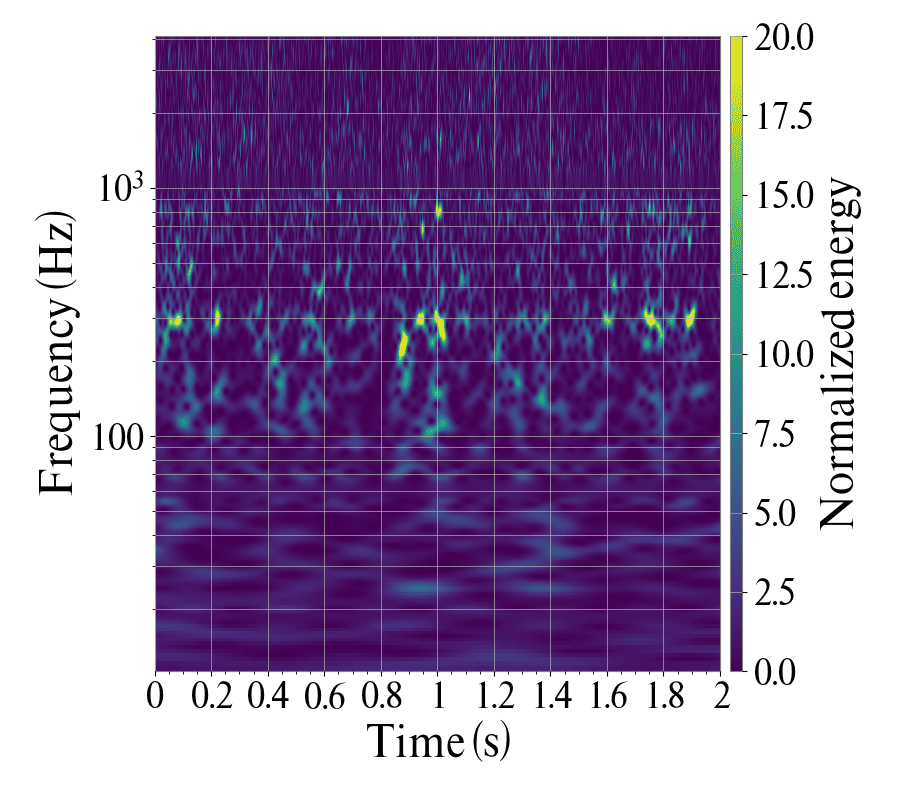}
    \includegraphics[width=0.48\textwidth]{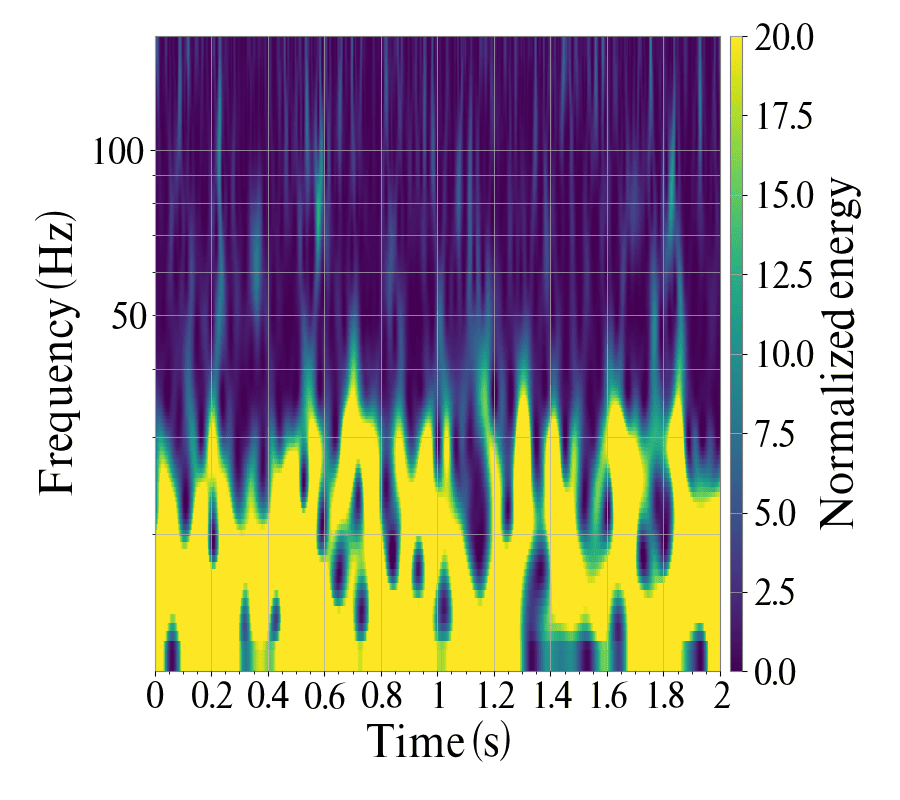}
    \par\smallskip
    {\scriptsize \scalebox{0.8}[1.0]{\textbf{K1:PEM-SEIS\_IXV\_GND\_EW\_IN1\_DQ}}\\ Apr 07, 2020	12:34:08	UTC	(GPS: 1270298066)}
    \end{minipage}
}

\resizebox{!}{0.131\textheight}
{%
\begin{minipage}[t]{0.45\textwidth}
    \centering
    \includegraphics[width=0.48\textwidth]{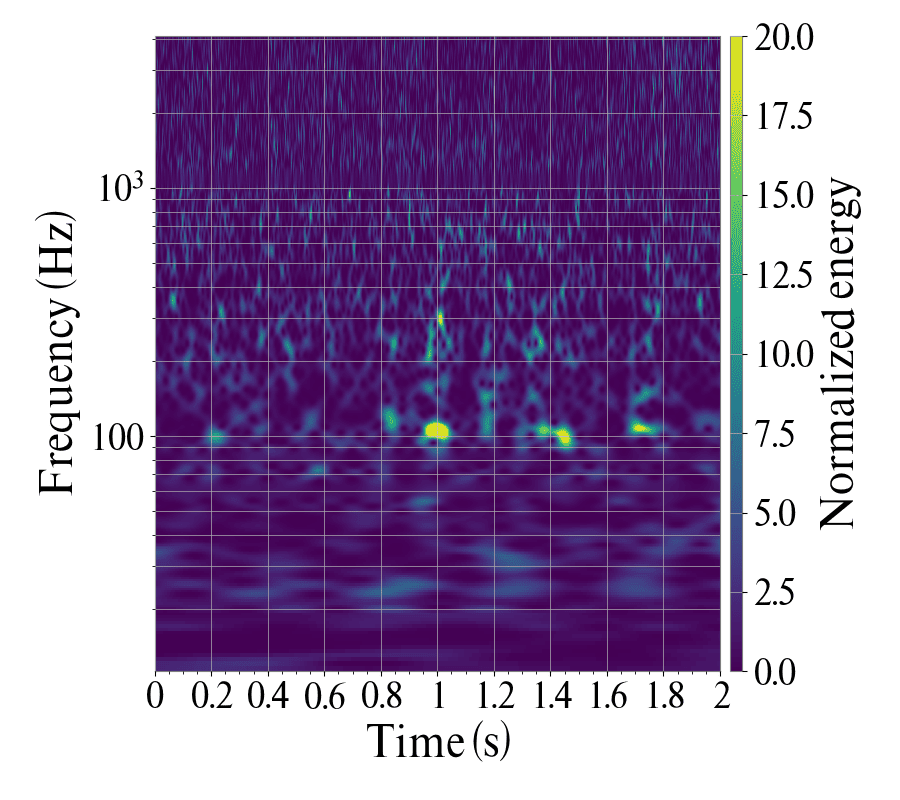}
    \includegraphics[width=0.48\textwidth]{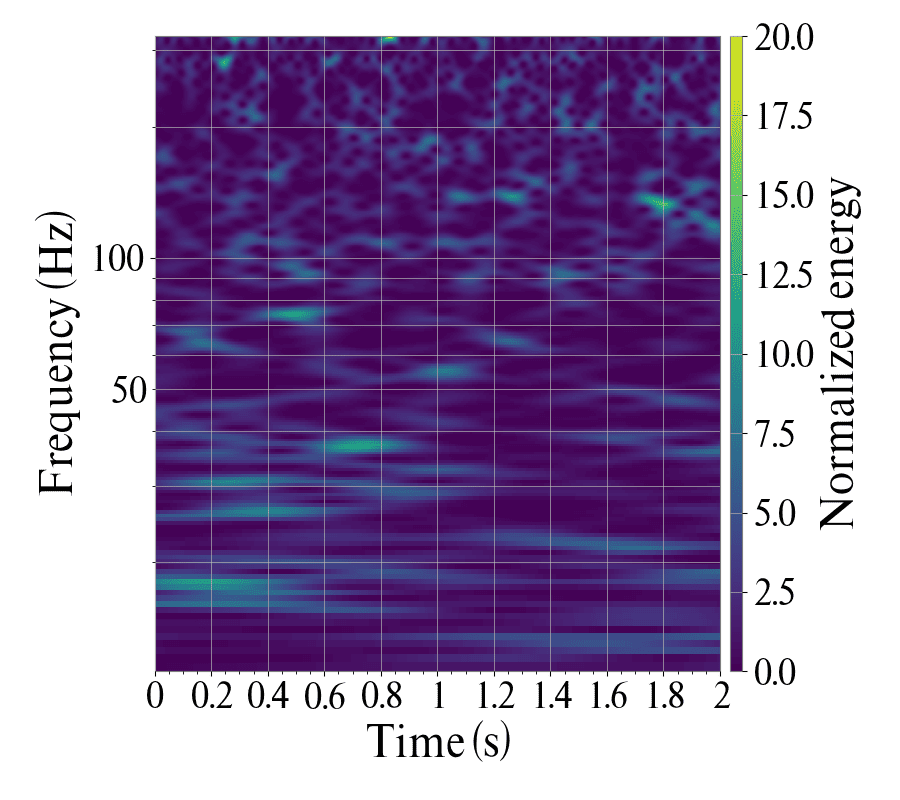}
    \par\smallskip
    {\scriptsize \scalebox{0.8}[1.0]{\textit{K1:VIS-ETMX\_MN\_PSDAMP\_Y\_IN1\_DQ}}\\ Apr 07, 2020	20:39:59	UTC	(GPS: 1270327217)}
    \end{minipage}
\hfill
\vrule width 0.5pt
\hfill 
\begin{minipage}[t]{0.45\textwidth}
    \centering
    \includegraphics[width=0.48\textwidth]{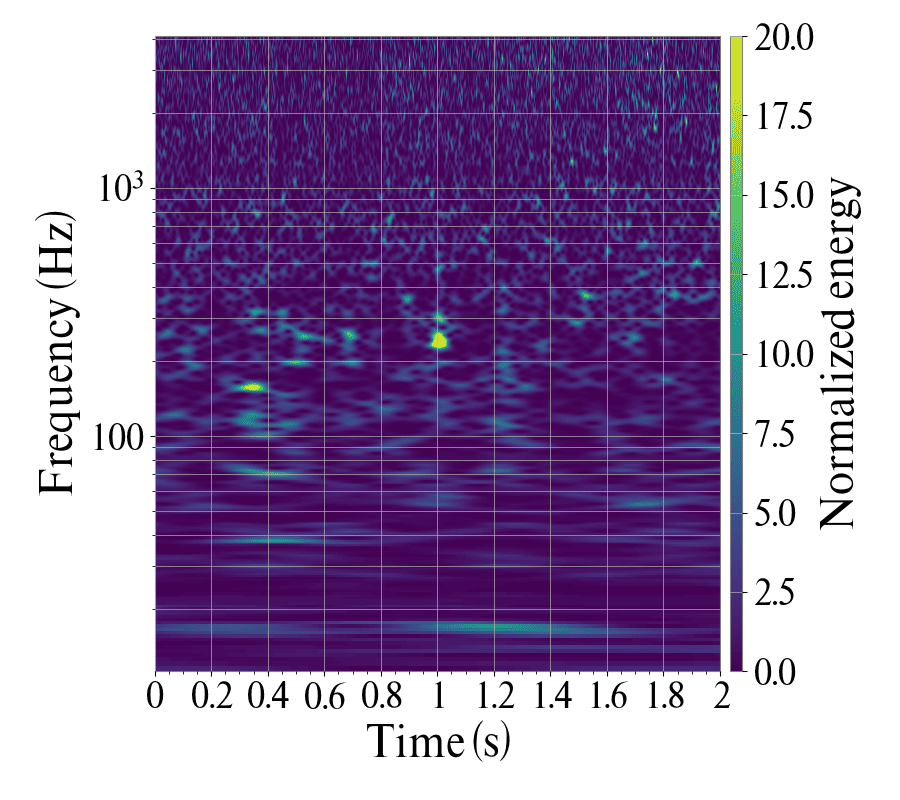}
    \includegraphics[width=0.48\textwidth]{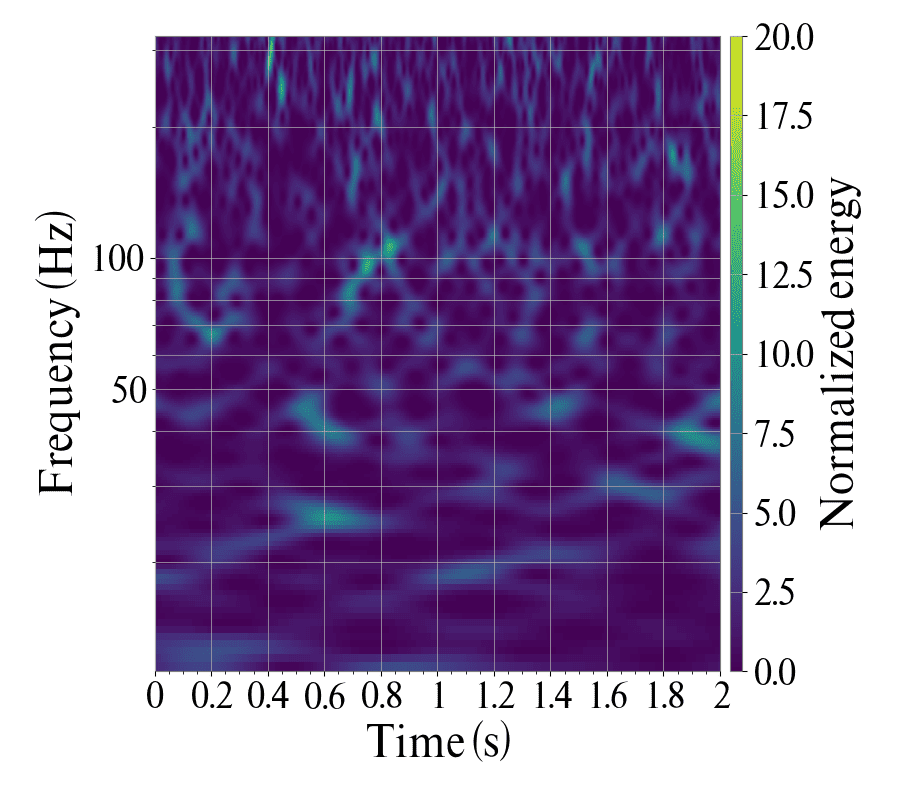}
    \par\smallskip
    {\scriptsize \scalebox{0.8}[1.0]{\textit{K1:VIS-ETMY\_MN\_PSDAMP\_Y\_IN1\_DQ}}\\ Apr 19, 2020	06:19:35	UTC	(GPS: 1271312393)}
    \end{minipage}
}
\end{figure} 
\clearpage

\subsubsection{Helix Glitch}

\begin{figure*}[h]
\centering
\resizebox{!}{0.134\textheight}
{%
\begin{minipage}[t]{0.45\textwidth}
    \centering
    \includegraphics[width=0.48\textwidth]{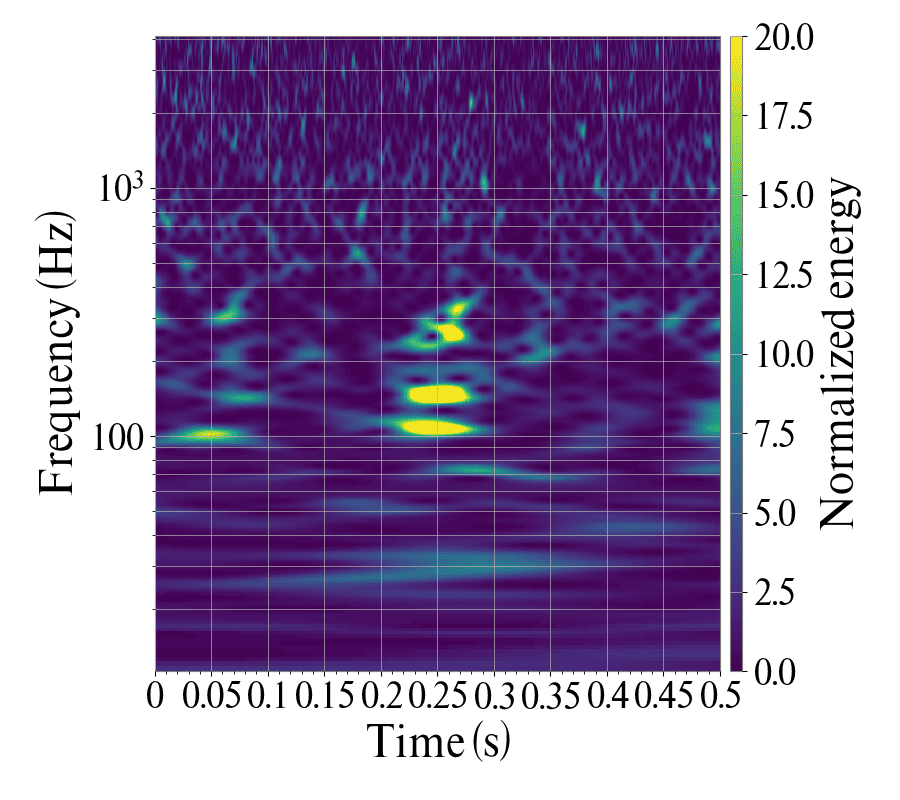}
    \includegraphics[width=0.48\textwidth]{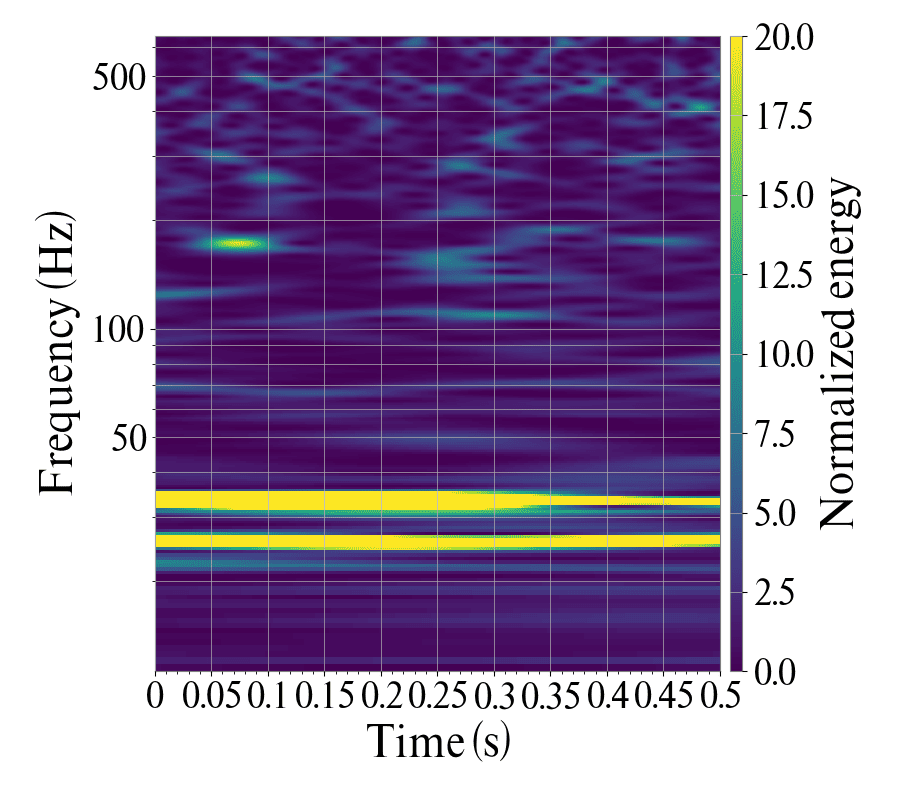}
    \par\smallskip
    {\scriptsize \scalebox{0.8}[1.0]{\textit{K1:VIS-OMMT1\_TM\_OPLEV\_PIT\_OUT\_DQ}}\\ Apr 18, 2020	22:06:30	UTC (GPS: 1271282808)}
    \end{minipage}
\hfill
\vrule width 0.5pt
\hfill
\begin{minipage}[t]{0.45\textwidth}
    \centering
    \includegraphics[width=0.48\textwidth]{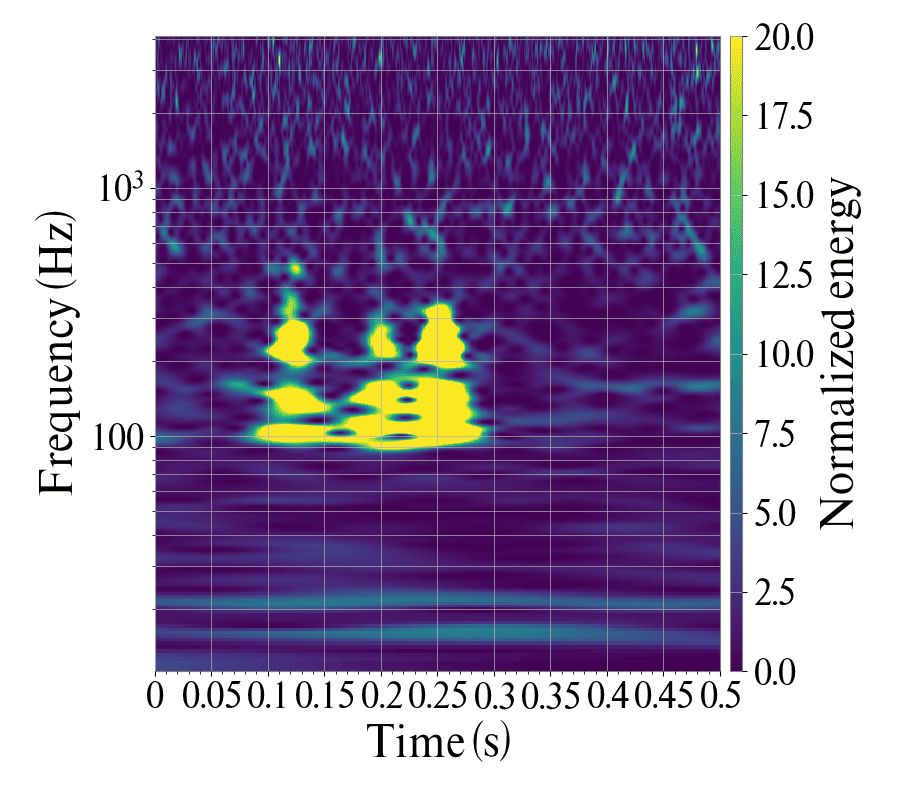}
    \includegraphics[width=0.48\textwidth]{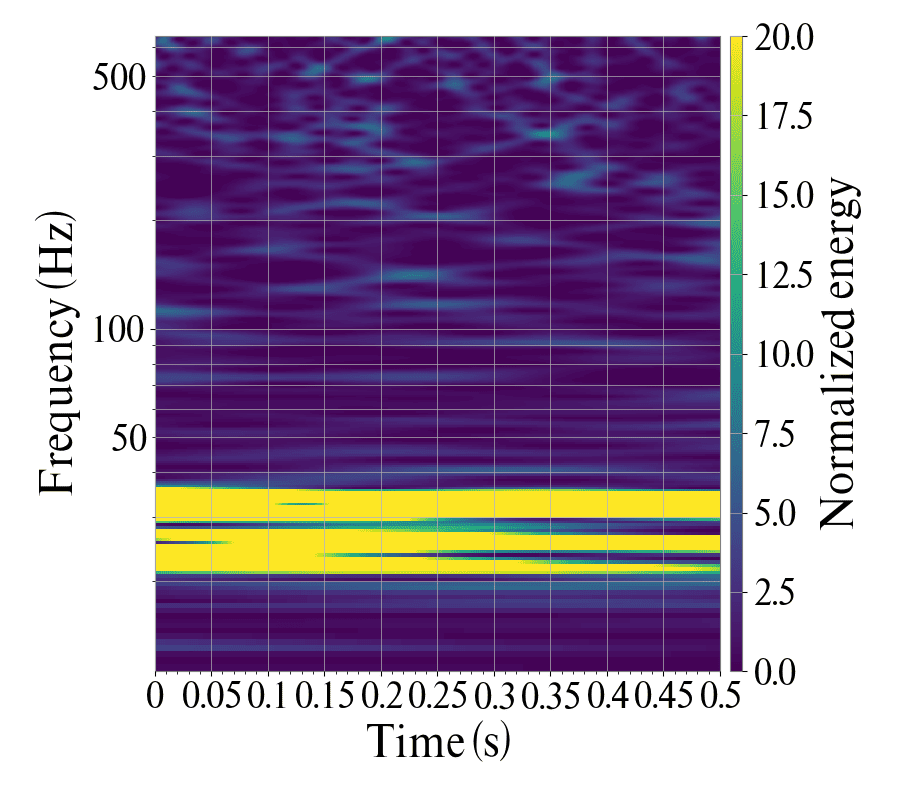}
    \par\smallskip
    {\scriptsize \scalebox{0.8}[1.0]{\textit{K1:VIS-OMMT1\_TM\_OPLEV\_YAW\_OUT\_DQ}}\\ Apr 17, 2020	11:24:59	UTC (GPS: 1271157917)}
    \end{minipage}
}

\resizebox{!}{0.134\textheight}
{%
\begin{minipage}[t]{0.45\textwidth}
    \centering
    \includegraphics[width=0.48\textwidth]{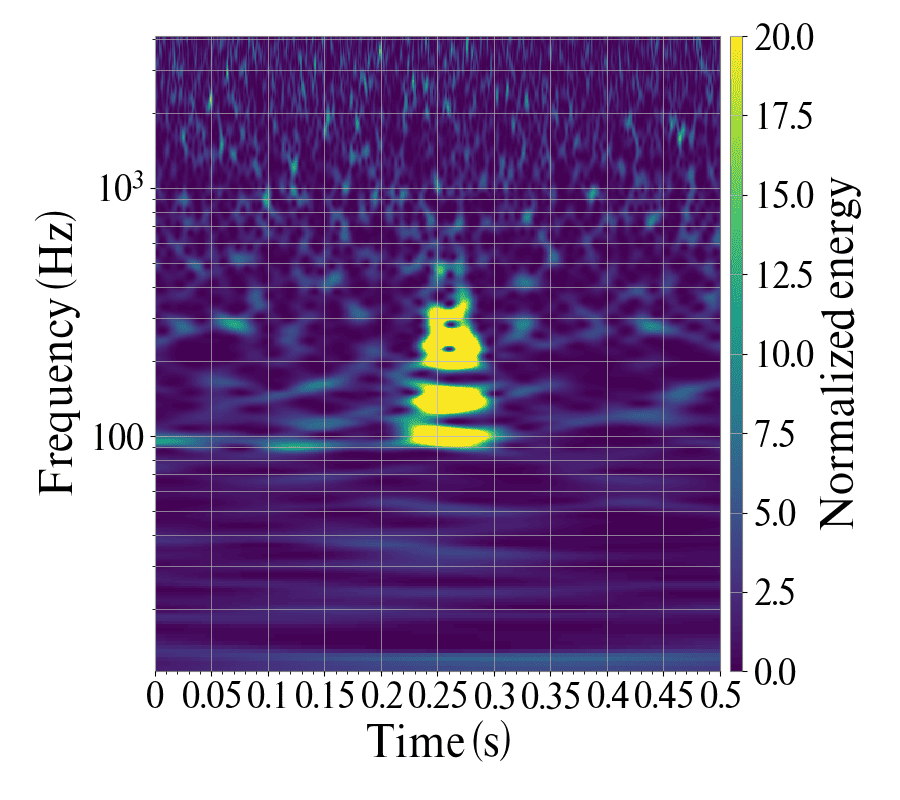}
    \includegraphics[width=0.48\textwidth]{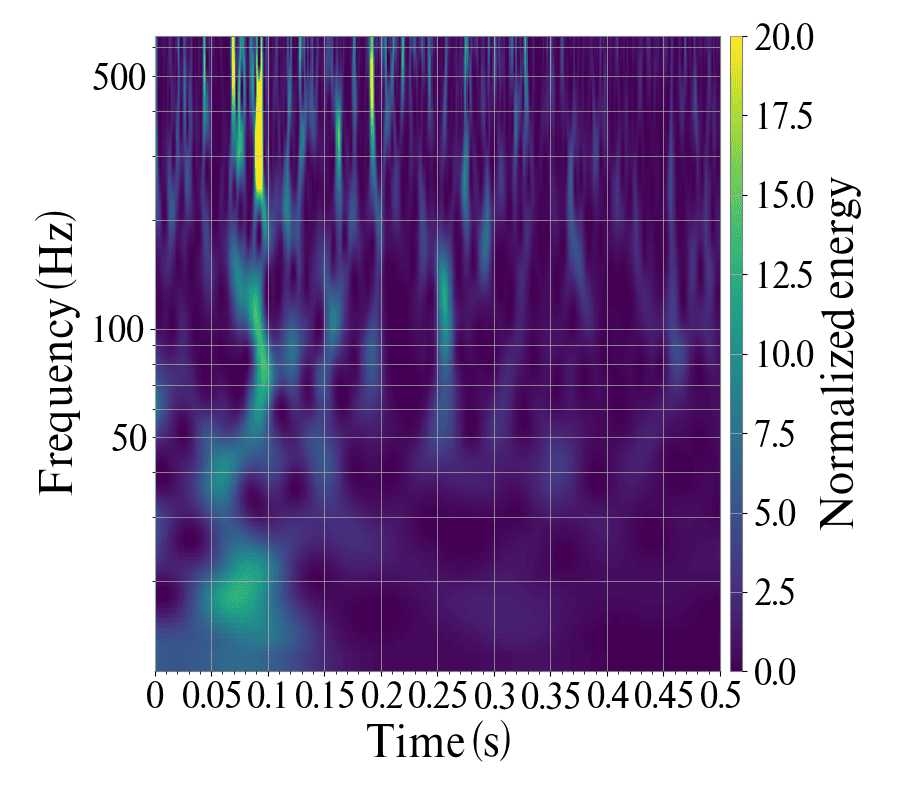}
    \par\smallskip
    {\scriptsize \scalebox{0.8}[1.0]{\textit{K1:VIS-TMSY\_DAMP\_R\_IN1\_DQ}}\\ Apr 20, 2020	09:21:06	UTC (GPS: 1271409684)}
  \end{minipage}
  }
  \caption{An O3GK KAGRA helix glitch found. The left and right panels show the spectrogram of the main channel and an auxiliary channel, respectively. Each figure also includes, below the panels, the name of the auxiliary channel and the UTC time (GPS time) when the glitch was found.}
  \label{fig:Helixfigs}
\end{figure*}

The helix glitch is named after its spiral upward shape. In other words, it typically appears as multiple dots that pile up in a triangular shape on the spectrogram. 
During the O3GK period, 320 Helix glitches were identified in all six subsystems (see Tables \ref{table:Glitchtypes} and \ref{table:HelixGlitches}).
Figure \ref{fig:Helixfigs} shows examples of the O3GK KAGRA helix glitch, which show strong and weak correlations between the main channel and the auxiliary channel, respectively. 

Similarly to the dot glitch (Figure \ref{fig:Dotfigs}), the correlation is determined to be strong (i.e., this auxiliary channel is indicated in bold font) when the helix glitch in the main channel spectrogram coincides (even partially) with the high-energy region in the auxiliary channel spectrogram, as the spectrogram of K1:VIS-ITMY\_IM\_PSDAMP\_R\_IN1\_DQ in Figure \ref{fig:Helixfigs}. In contrast, the spectrogram of K1:AOS-TMSX\_IR\_PDA1\_OUT\_DQ is marked in italic font, because the region of coincidence between the main channel and the auxiliary channel is insignificant. 

\begin{table*}[h]
\centering
\caption{\label{table:HelixGlitches} O3GK KAGRA helix glitches}
\small
\resizebox{0.98\textwidth}{!}
{%
\begin{tabular}{|P{0.07\textwidth}|P{0.58\textwidth}|P{0.35\textwidth}|}
\hline
\begin{tabular}[c]{@{}c@{}}Sub-\\ System\end{tabular} & \begin{tabular}[c]{@{}c@{}}Round Winner\\ Auxiliary Channel\end{tabular} & \begin{tabular}[c]{@{}c@{}}Vetoed Date in April\\ (\# of Vetoed Events)\end{tabular} \\ \hline

\begin{tabular}[c]{@{}c@{}}AOS \\ (92) \end{tabular}
& \textit{K1:AOS-TMSX\_IR\_PDA1\_OUT\_DQ}
& \begin{tabular}[c]{@{}c@{}}10th (1), 16th (17), \\ 17th (23), 18th (51)\end{tabular} \\ \hline

\multirow{2}{*}{\begin{tabular}[c]{@{}c@{}} IMC \\ (96) \end{tabular}}
& \begin{tabular}[c]{@{}c@{}}\textit{K1:IMC-IMMT1\_TRANS\_QPDA1\_DC\_}\\ \textit{PIT\_OUT\_DQ}\end{tabular}  
& \begin{tabular}[c]{@{}c@{}}8th (1), 9th (11),\\ 10th (6), 11th (2)\end{tabular}\\ \cline{2-3} 
& \begin{tabular}[c]{@{}c@{}}\textit{K1:IMC-IMMT1\_TRANS\_QPDA1\_DC\_}\\  \textit{YAW\_OUT\_DQ}\end{tabular} 
& 9th (1), 19th (75)                                                          \\ \hline

\begin{tabular}[c]{@{}c@{}}LAS \\ (7) \end{tabular}
& \textbf{K1:LAS-POW\_FIB\_OUT\_DQ}
& 7th (7) \\ \hline

\multirow{2}{*}{\begin{tabular}[c]{@{}c@{}}LSC \\ (39) \end{tabular}}
& \begin{tabular}[c]{@{}c@{}}\textit{K1:LSC-ALS\_CARM\_OUT\_DQ}\end{tabular}
& 16th (2) 
\\ \cline{2-3}
& \begin{tabular}[c]{@{}c@{}}\textbf{K1:LSC-ALS\_DARM\_OUT\_DQ}\end{tabular}
& 12th (22), 15th (10), 16th (5) 
\\ \hline

\multirow{5}{*}{\begin{tabular}[c]{@{}c@{}}PEM \\ (9) \end{tabular}}
& \textbf{K1:PEM-ACC\_OMC\_TABLE\_AS\_Z\_OUT\_DQ}
& 17th (2) 
\\ \cline{2-3}
& \textit{K1:PEM-MAG\_BS\_BOOTH\_BS\_X\_OUT\_DQ}
& 16th (1) 
\\ \cline{2-3}
& \textbf{K1:PEM-MAG\_BS\_BOOTH\_BS\_Z\_OUT\_DQ}
& 12th (2), 18th (1) 
\\ \cline{2-3}
& \textbf{K1:PEM-MIC\_SR\_BOOTH\_SR\_Z\_OUT\_DQ}
& 16th (1)
\\ \cline{2-3}
& \textit{K1:PEM-VOLT\_REFL\_TABLE\_GND\_OUT\_DQ}
& 11th (2) 
\\ \hline

\multirow{8}{*}{\begin{tabular}[c]{@{}c@{}}VIS \\ (77) \end{tabular}}
& \textit{K1:VIS-ETMY\_MN\_PSDAMP\_Y\_IN1\_DQ}
& 19th (1)
\\ \cline{2-3}
& \textbf{K1:VIS-ITMY\_IM\_PSDAMP\_R\_IN1\_DQ}
& \begin{tabular}[c]{@{}c@{}}8th (3), 9th (3),\\ 11th (2), 18th (14)\end{tabular}
\\ \cline{2-3}
& \textit{K1:VIS-ITMY\_MN\_OPLEV\_TILT\_YAW\_OUT\_DQ}
& 9th(1)
\\ \cline{2-3}
& \textbf{K1:VIS-ITMY\_MN\_PSDAMP\_L\_IN1\_DQ}
& 16th (7), 17th (2), 19th (6)
\\ \cline{2-3}
& \textbf{K1:VIS-ITMY\_MN\_PSDAMP\_Y\_IN1\_DQ}
& 10th (2)
\\ \cline{2-3}
& \textit{K1:VIS-OMMT1\_TM\_OPLEV\_PIT\_OUT\_DQ}
& 12th (1), 18th (3)
\\ \cline{2-3}
& \textit{K1:VIS-OMMT1\_TM\_OPLEV\_YAW\_OUT\_DQ}
& 17th (1)
\\ \cline{2-3}
& \textit{K1:VIS-TMSY\_DAMP\_R\_IN1\_DQ}
& 20th (31)
\\ \hline

\end{tabular} }
\end{table*}

\begin{figure}[h]
\centering
\resizebox{!}{0.134\textheight}
{%
\begin{minipage}[t]{0.45\textwidth}
    \centering
    \includegraphics[width=0.48\textwidth]{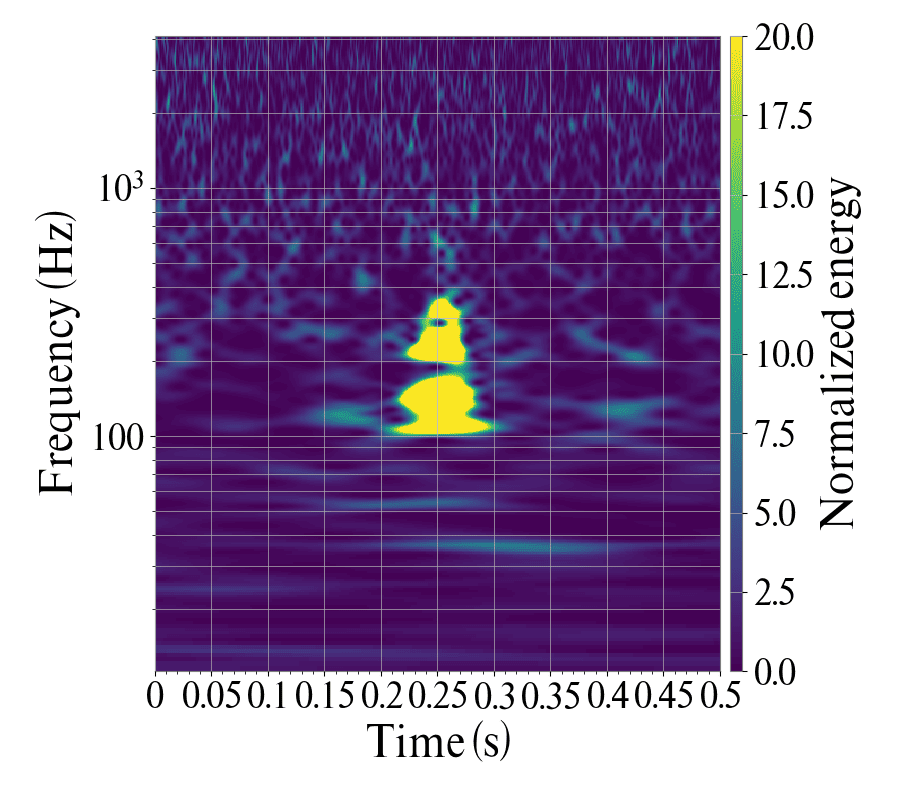}
    \includegraphics[width=0.48\textwidth]{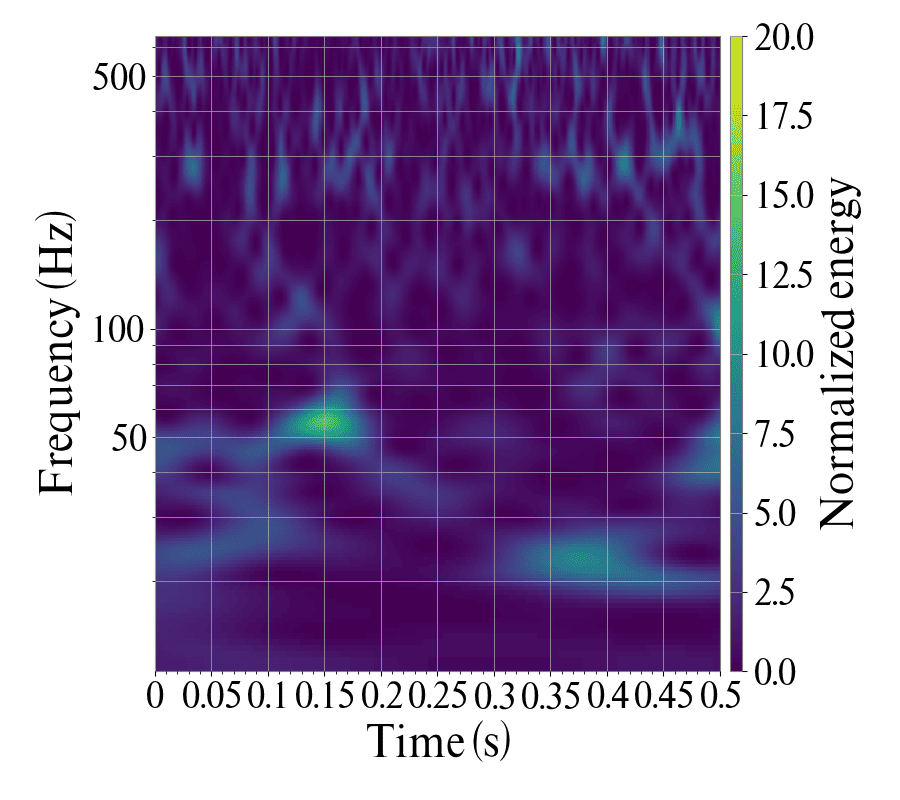}
    \par\smallskip
    {\scriptsize \scalebox{0.8}[1.0]{\textit{K1:AOS-TMSX\_IR\_PDA1\_OUT\_DQ}} \\ Apr 17, 2020	05:02:03	UTC (GPS: 1271134941)}
    \end{minipage}
\hfill
\vrule width 0.5pt
\hfill
  \begin{minipage}[t]{0.45\textwidth}
    \centering
    \includegraphics[width=0.48\textwidth]{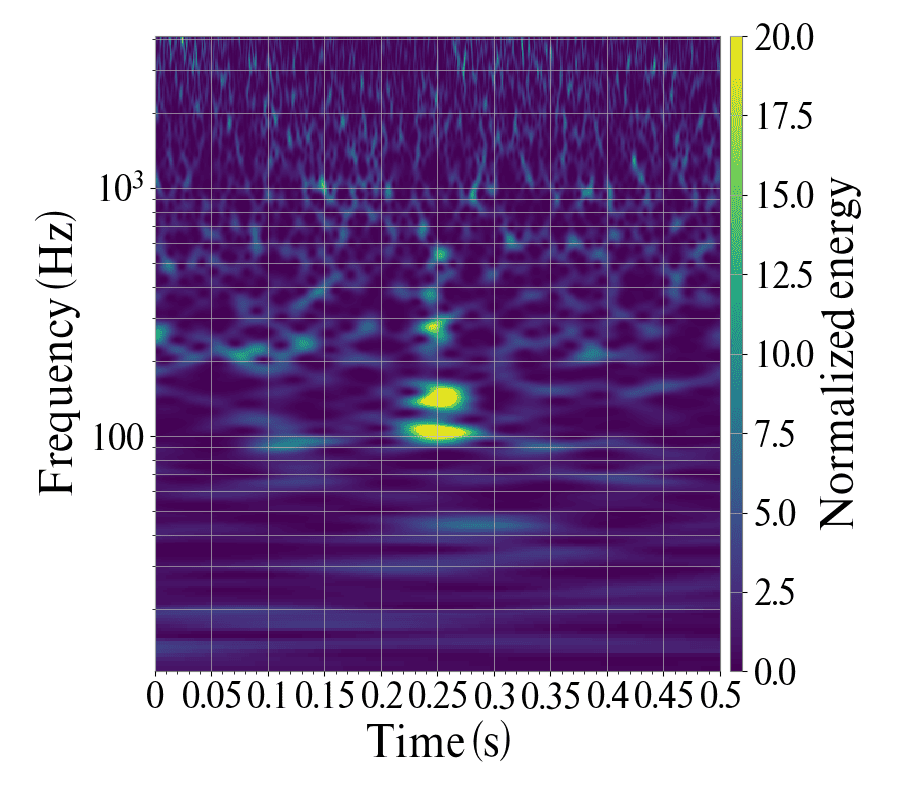}
    \includegraphics[width=0.48\textwidth]{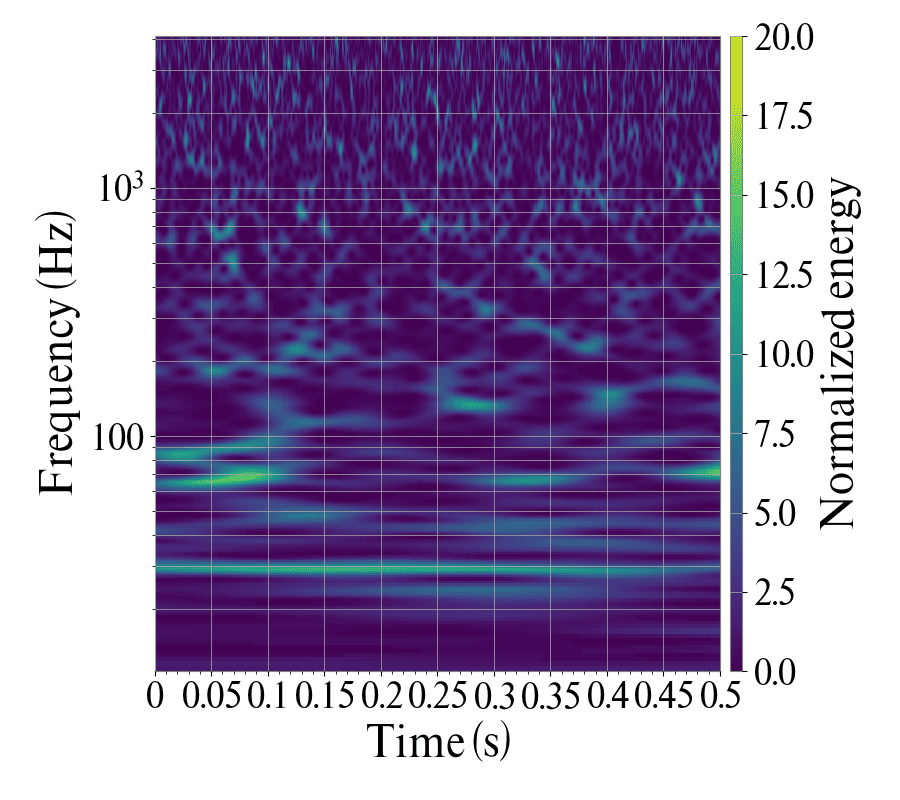}
    \par\smallskip
    {\scriptsize \scalebox{0.8}[1.0]{\textit{K1:IMC-IMMT1\_TRANS\_QPDA1\_DC\_PIT\_OUT\_DQ}} \\ Apr 10, 2020	07:52:56	UTC (GPS: 1270540394)}
    \end{minipage}
}

\resizebox{!}{0.134\textheight}
{%
\begin{minipage}[t]{0.45\textwidth}
    \centering
    \includegraphics[width=0.48\textwidth]{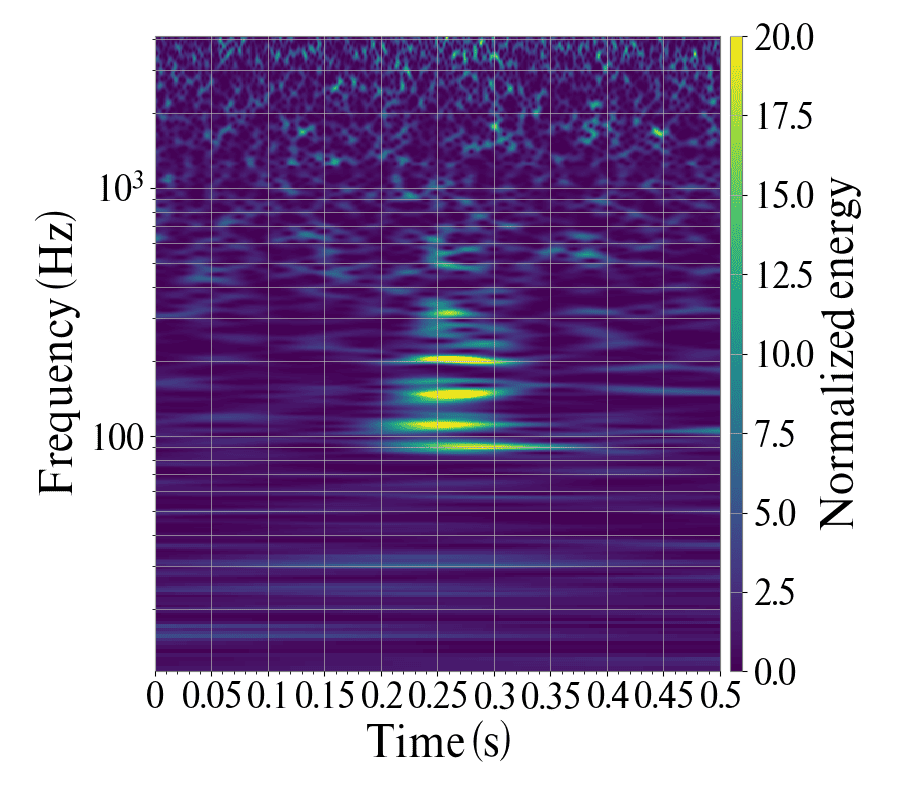}
    \includegraphics[width=0.48\textwidth]{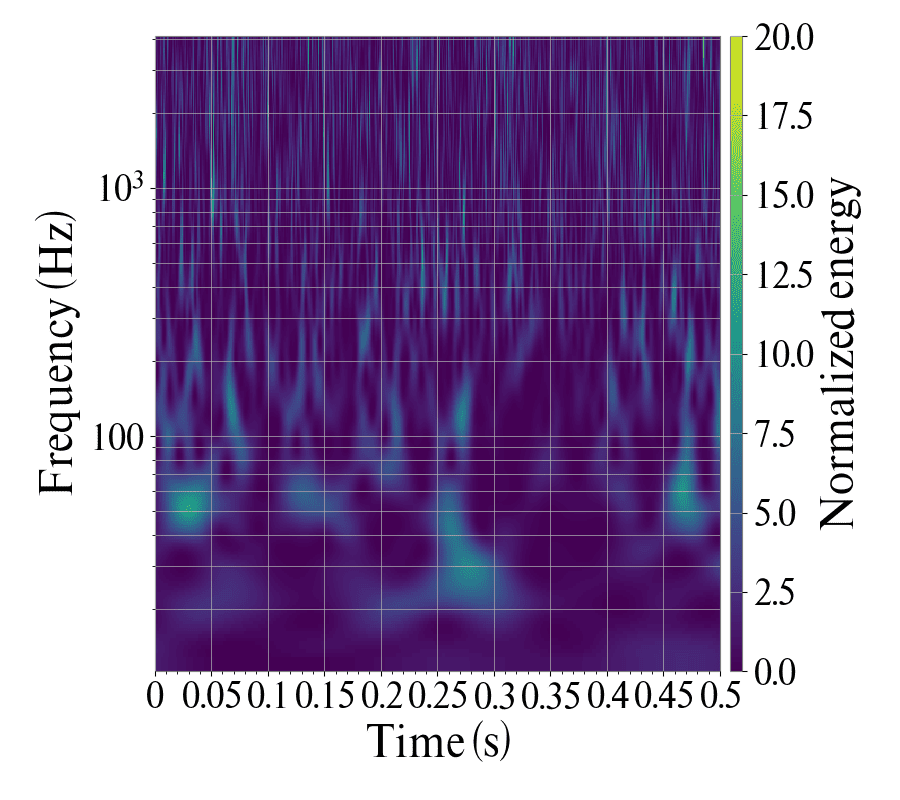}
    \par\smallskip
    {\scriptsize \scalebox{0.8}[1.0]{\textit{K1:IMC-IMMT1\_TRANS\_QPDA1\_DC\_YAW\_OUT\_DQ}} \\ Apr 19, 2020	01:37:00	UTC (GPS: 1271295438)}
    \end{minipage}
\hfill
\vrule width 0.5pt
\hfill
\begin{minipage}[t]{0.45\textwidth}   
    \centering
    \includegraphics[width=0.48\textwidth]{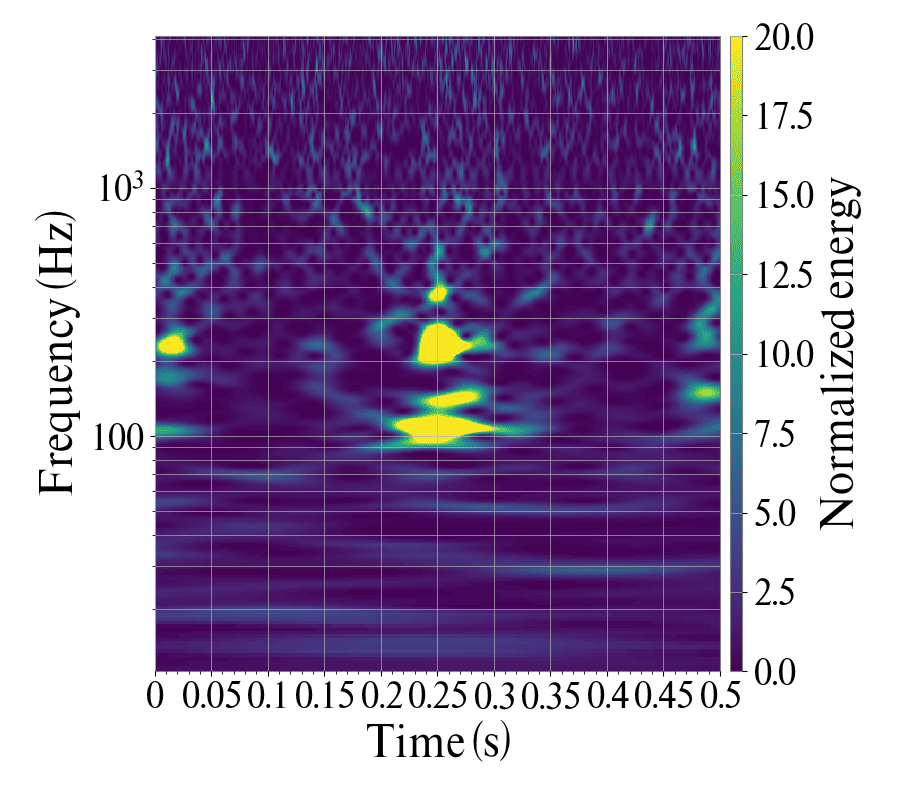}
    \includegraphics[width=0.48\textwidth]{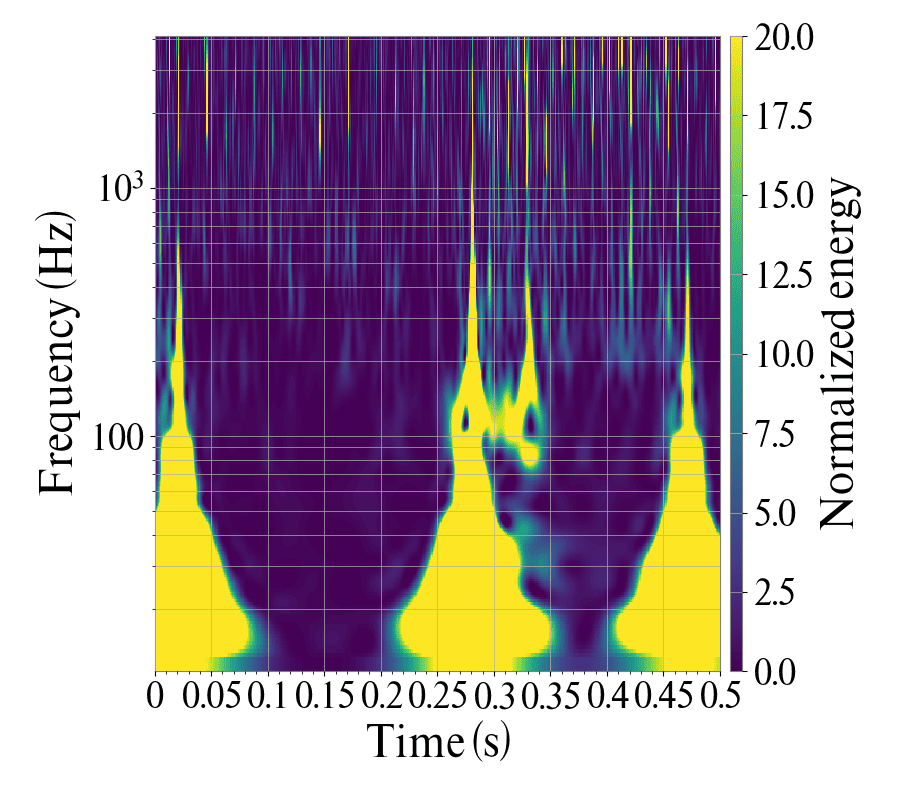}
    \par\smallskip
    {\scriptsize \scalebox{0.8}[1.0]{\textbf{K1:LAS-POW\_FIB\_OUT\_DQ}} \\ Apr 07, 2020	23:34:35	UTC (GPS: 1270337693)}
    \end{minipage}
}

\resizebox{!}{0.134\textheight}
{%
\begin{minipage}[t]{0.45\textwidth}
    \centering
    \includegraphics[width=0.48\textwidth]{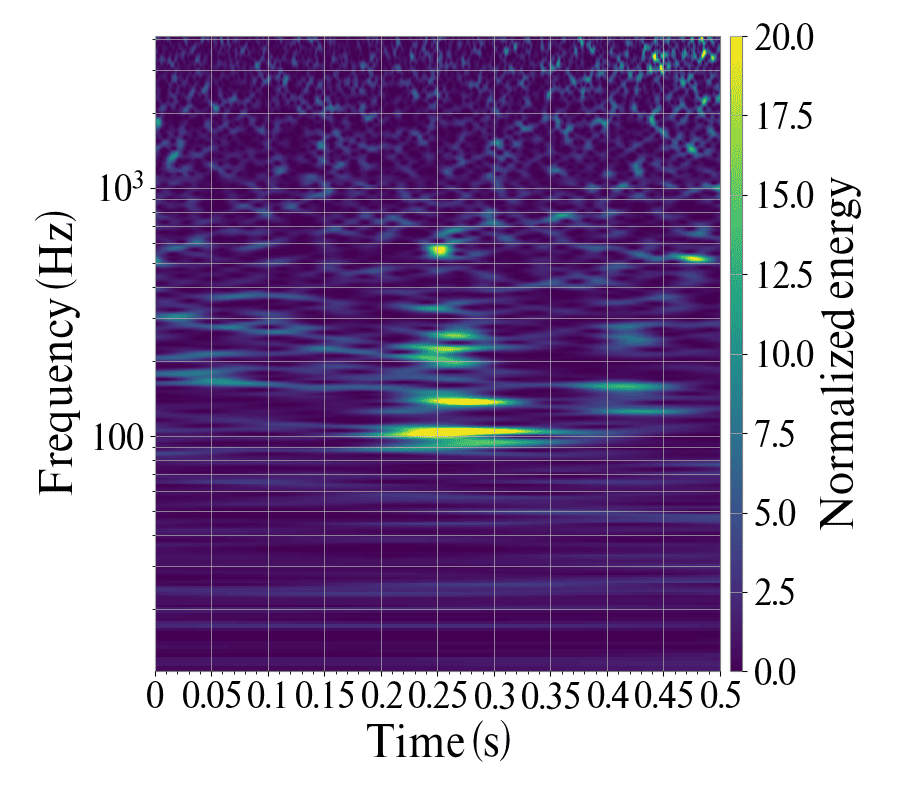}
    \includegraphics[width=0.48\textwidth]{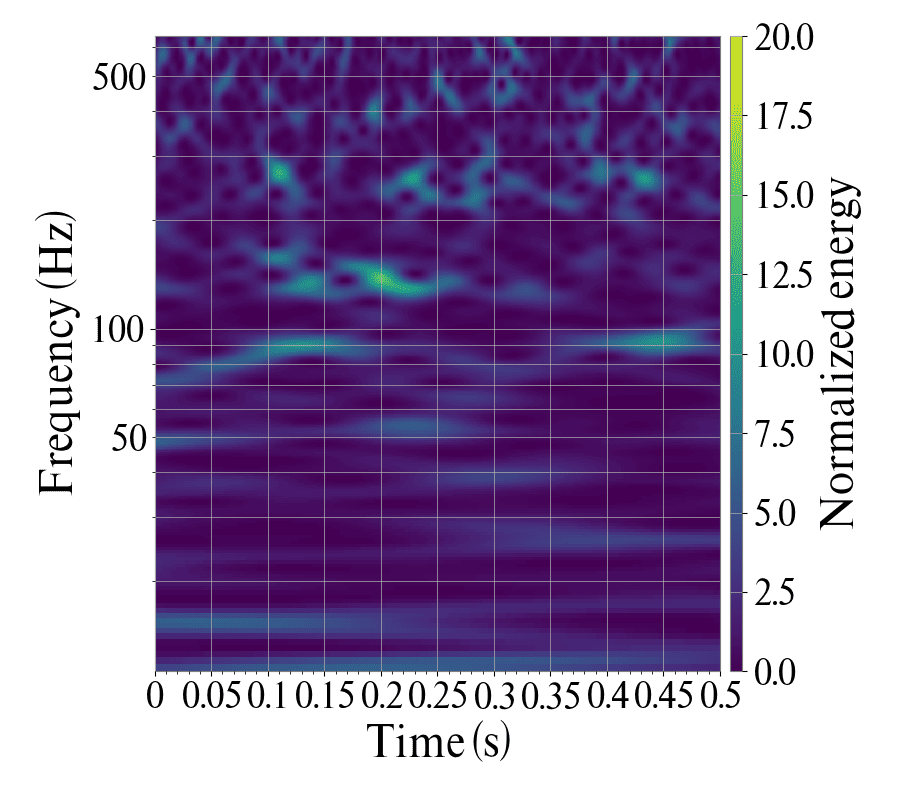}
    \par\smallskip
    {\scriptsize \scalebox{0.8}[1.0]{\textit{K1:LSC-ALS\_CARM\_OUT\_DQ}} \\ Apr 16, 2020	05:43:43	UTC (GPS: 1271051041)}
    \end{minipage}
\hfill
\vrule width 0.5pt
\hfill
\begin{minipage}[t]{0.45\textwidth}
    \centering
    \includegraphics[width=0.48\textwidth]{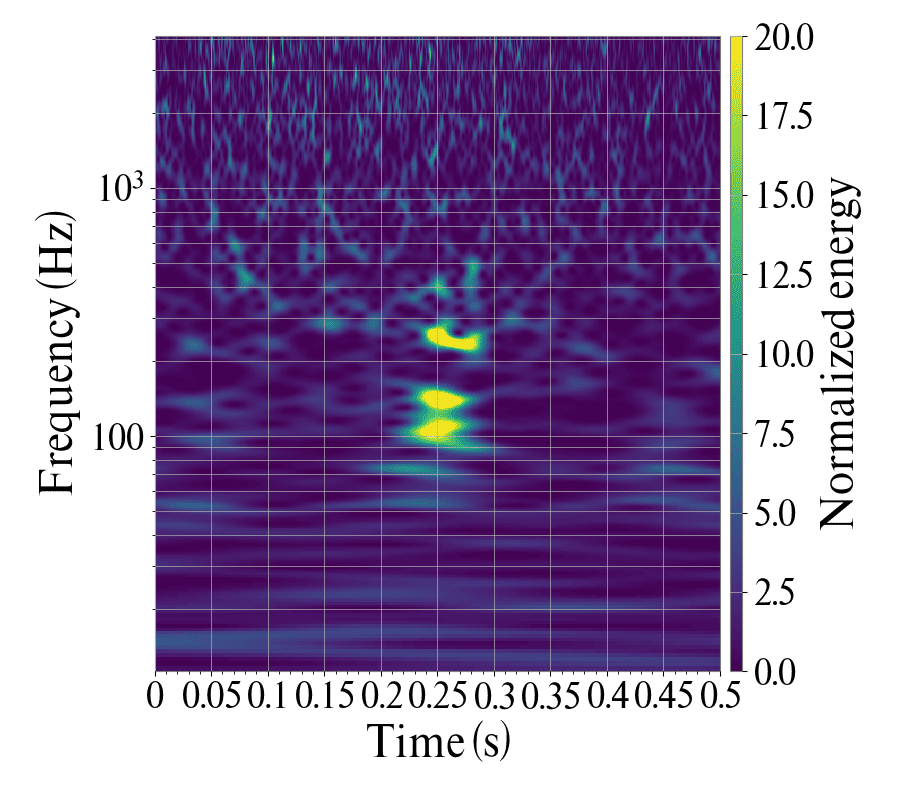}
    \includegraphics[width=0.48\textwidth]{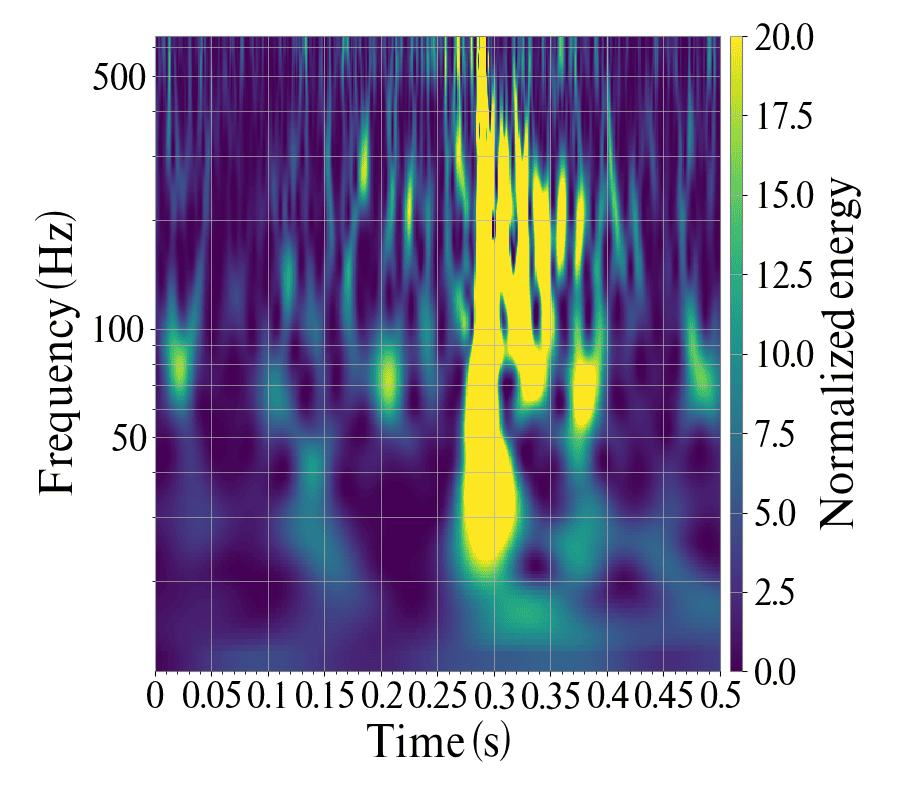}
    \par\smallskip
    {\scriptsize \scalebox{0.8}[1.0]{\textbf{K1:LSC-ALS\_DARM\_OUT\_DQ}} \\ Apr 15, 2020	10:02:13	UTC (GPS: 1270980151)}
    \end{minipage}
}
\end{figure}

\clearpage

\begin{figure}[h]
\resizebox{!}{0.136\textheight}
{%
\begin{minipage}[t]{0.45\textwidth}
    \centering
    \includegraphics[width=0.48\textwidth]{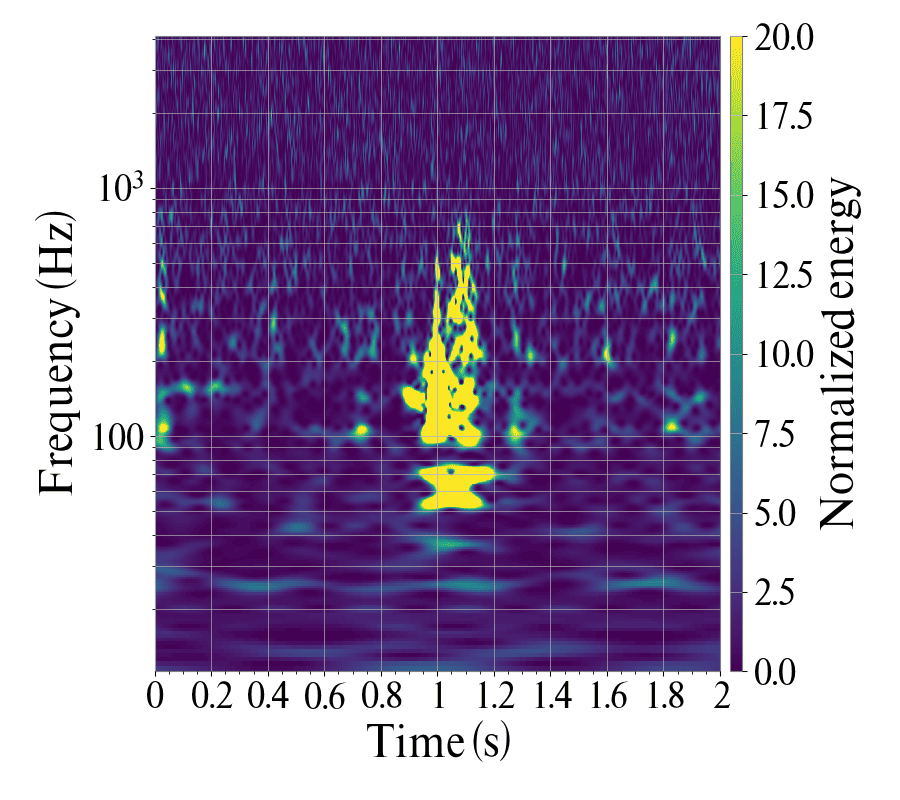}
    \includegraphics[width=0.48\textwidth]{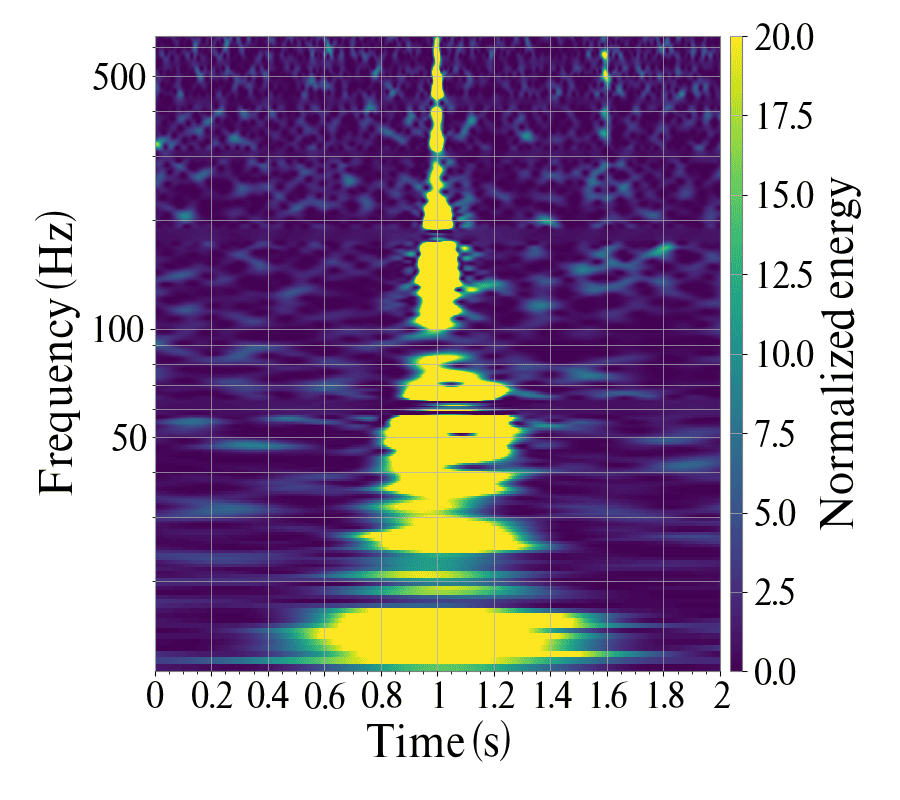}
    \par\smallskip
    {\scriptsize \scalebox{0.8}[1.0]{\textbf{K1:PEM-ACC\_OMC\_TABLE\_AS\_Z\_OUT\_DQ}} \\ Apr 17, 2020	01:11:37	UTC	(GPS: 1271121115)}
    \end{minipage}
\hfill
\vrule width 0.5pt
\hfill
\begin{minipage}[t]{0.45\textwidth}
    \centering
    \includegraphics[width=0.48\textwidth]{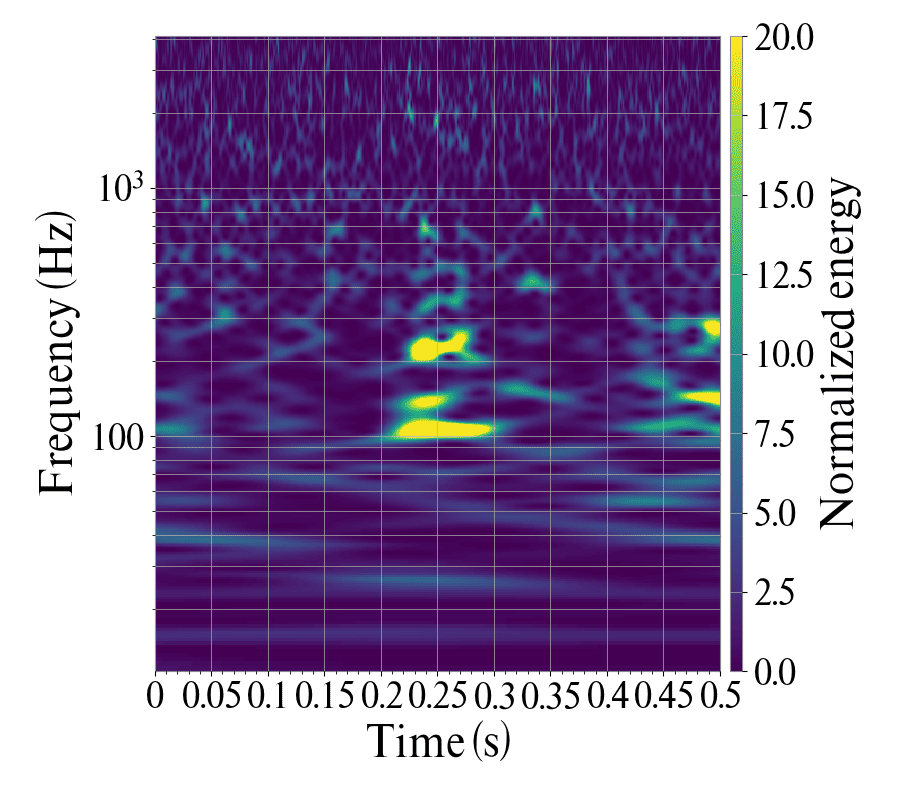}
    \includegraphics[width=0.48\textwidth]{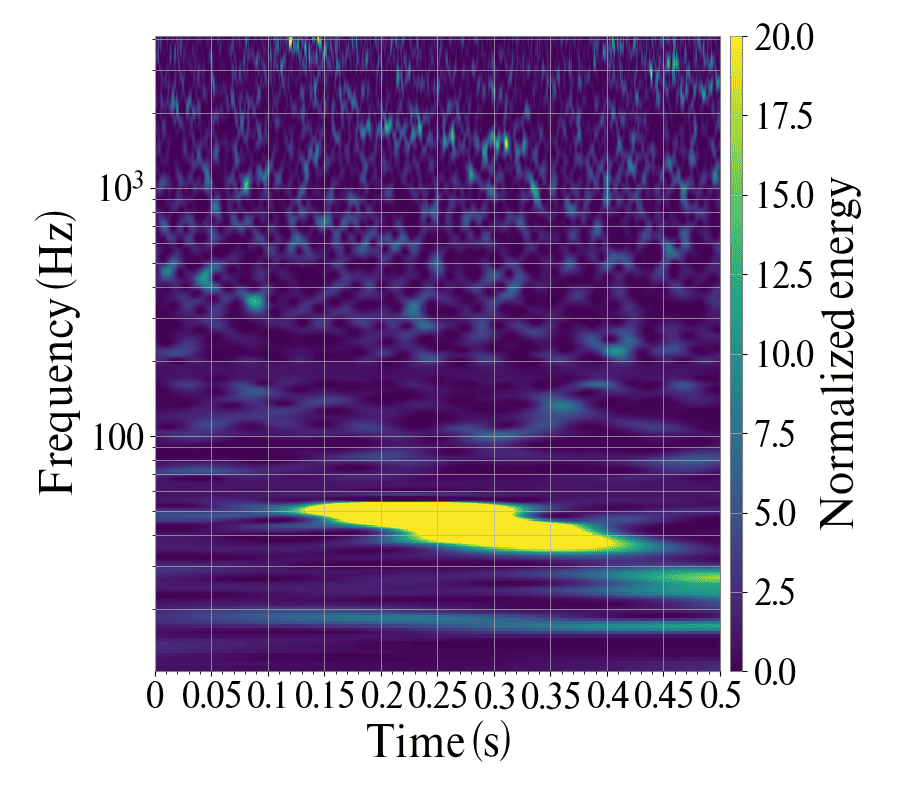}
    \par\smallskip
    {\scriptsize \scalebox{0.8}[1.0]{\textit{K1:PEM-MAG\_BS\_BOOTH\_BS\_X\_OUT\_DQ}} \\ Apr 16, 2020	20:29:41	UTC	(GPS: 1271104199)}
    \end{minipage}
    }

\resizebox{!}{0.136\textheight}
{%
\begin{minipage}[t]{0.45\textwidth}
    \centering
    \includegraphics[width=0.48\textwidth]{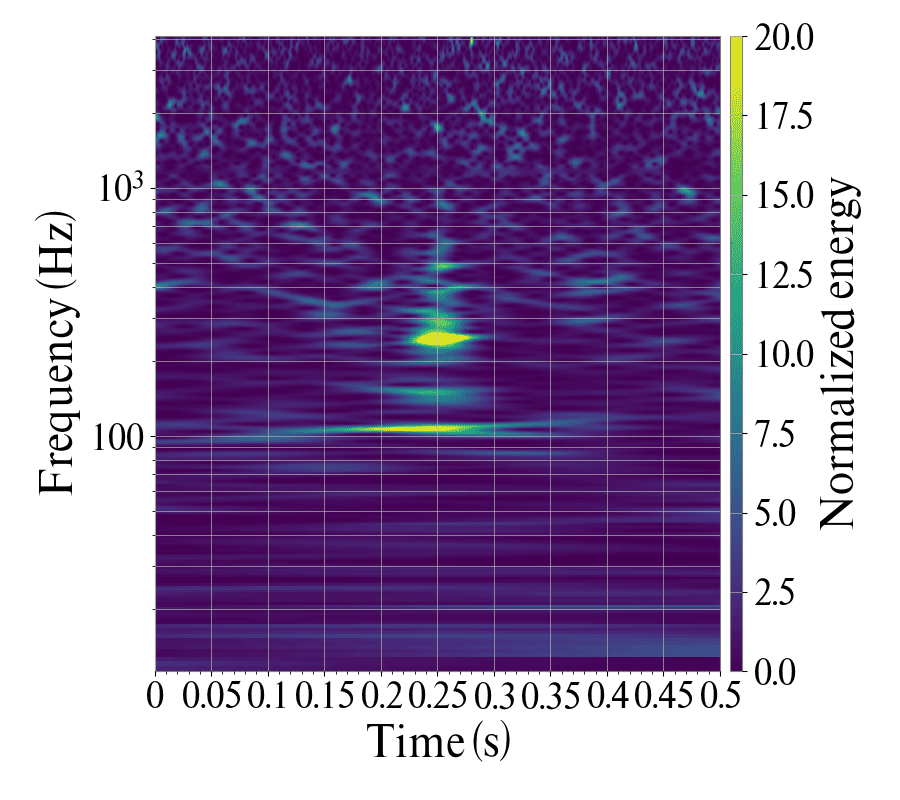}
    \includegraphics[width=0.48\textwidth]{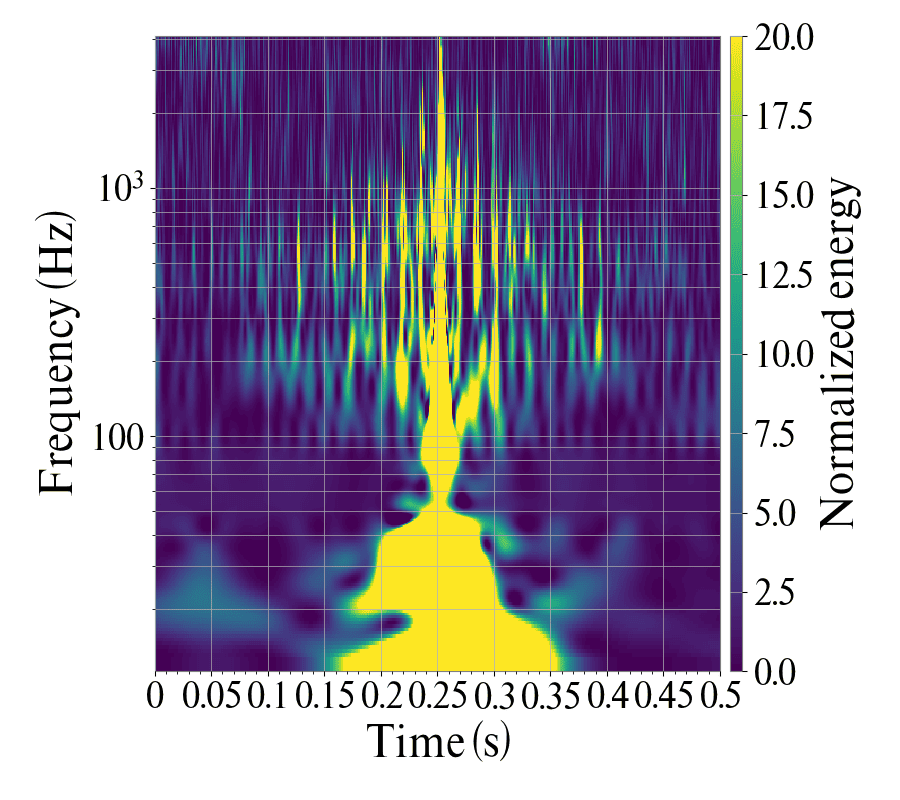}
    \par\smallskip
    {\scriptsize \scalebox{0.8}[1.0]{\textbf{K1:PEM-MAG\_BS\_BOOTH\_BS\_Z\_OUT\_DQ}} \\ Apr 12, 2020	19:08:31	UTC	(GPS: 1270753729)}
    \end{minipage}
\hfill
\vrule width 0.5pt
\hfill
\begin{minipage}[t]{0.45\textwidth}
    \centering
    \includegraphics[width=0.48\textwidth]{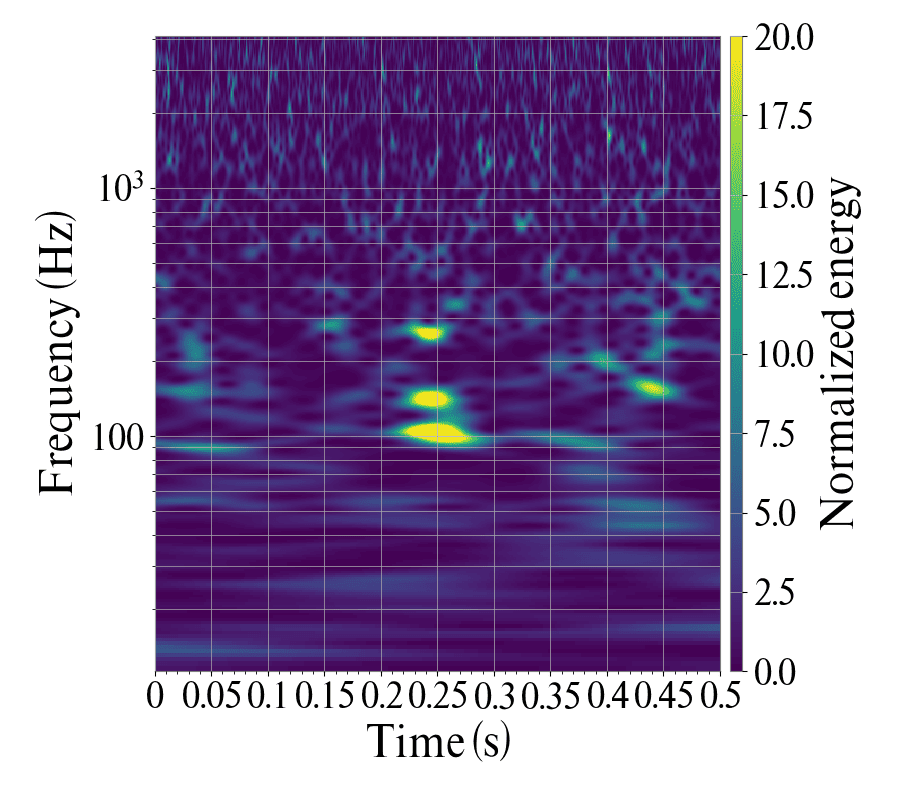}
    \includegraphics[width=0.48\textwidth]{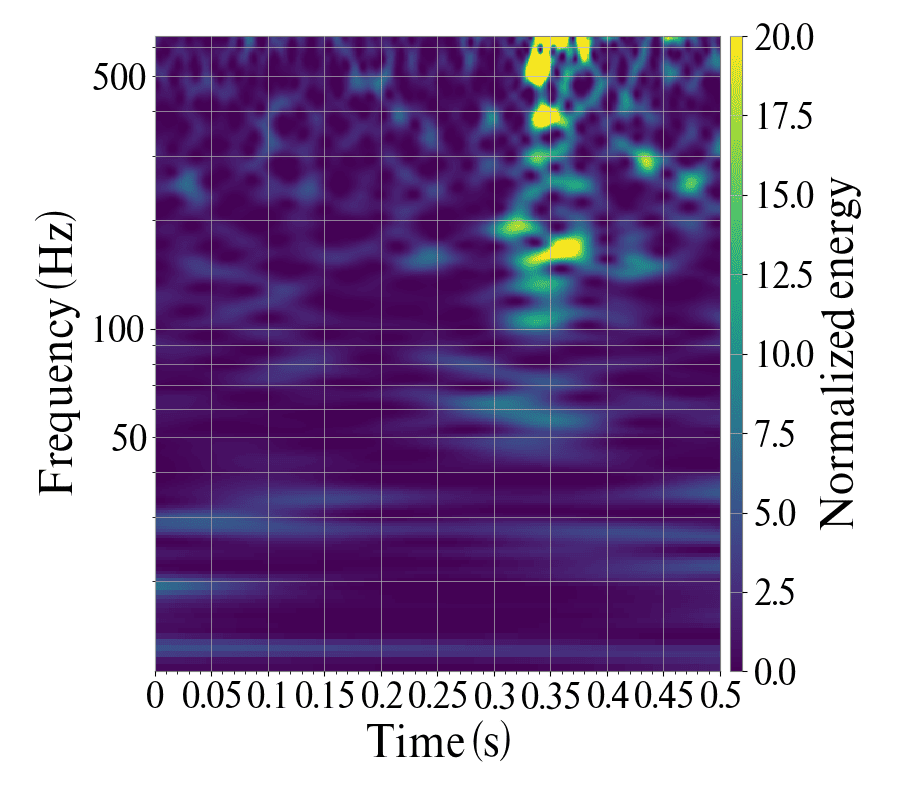}
    \par\smallskip
    {\scriptsize \scalebox{0.8}[1.0]{\textbf{K1:PEM-MIC\_SR\_BOOTH\_SR\_Z\_OUT\_DQ}}\\ Apr 16, 2020	20:00:58	UTC	(GPS: 1271102476)}
    \end{minipage}
    }

\resizebox{!}{0.136\textheight}
{%
\begin{minipage}[t]{0.45\textwidth}
    \centering
    \includegraphics[width=0.48\textwidth]{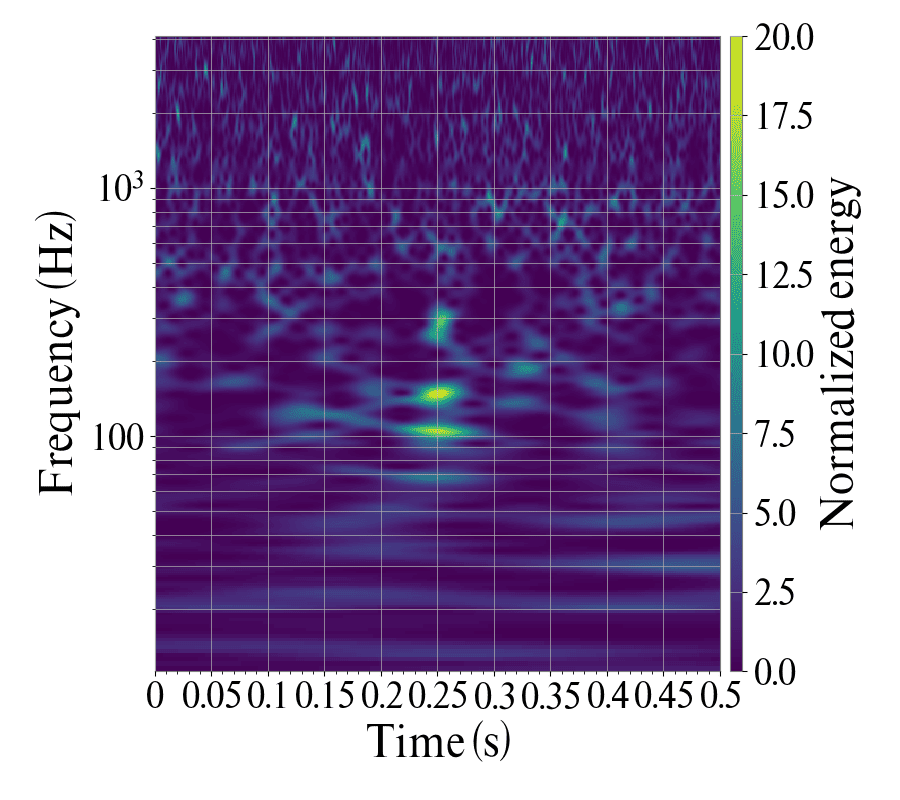}
    \includegraphics[width=0.48\textwidth]{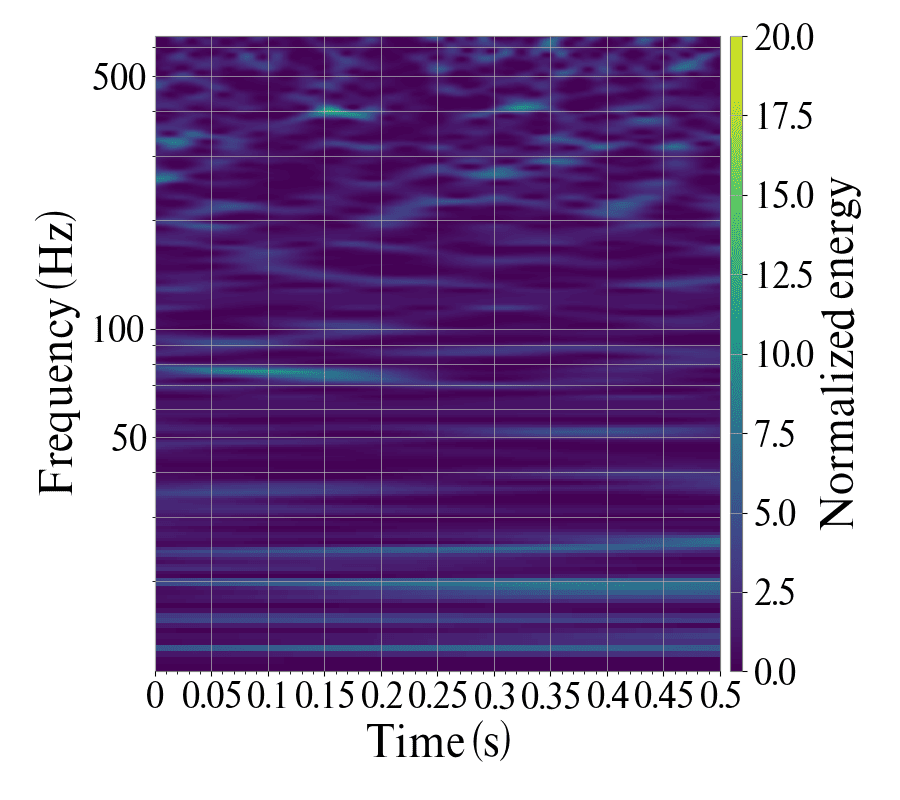}
    \par\smallskip
    {\scriptsize \scalebox{0.8}[1.0]{\textit{K1:PEM-VOLT\_REFL\_TABLE\_GND\_OUT\_DQ}}\\ Apr 11, 2020	00:40:51	UTC	(GPS: 1270600869)}
    \end{minipage}
\hfill
\vrule width 0.5pt
\hfill
\begin{minipage}[t]{0.45\textwidth}
    \centering
    \includegraphics[width=0.48\textwidth]{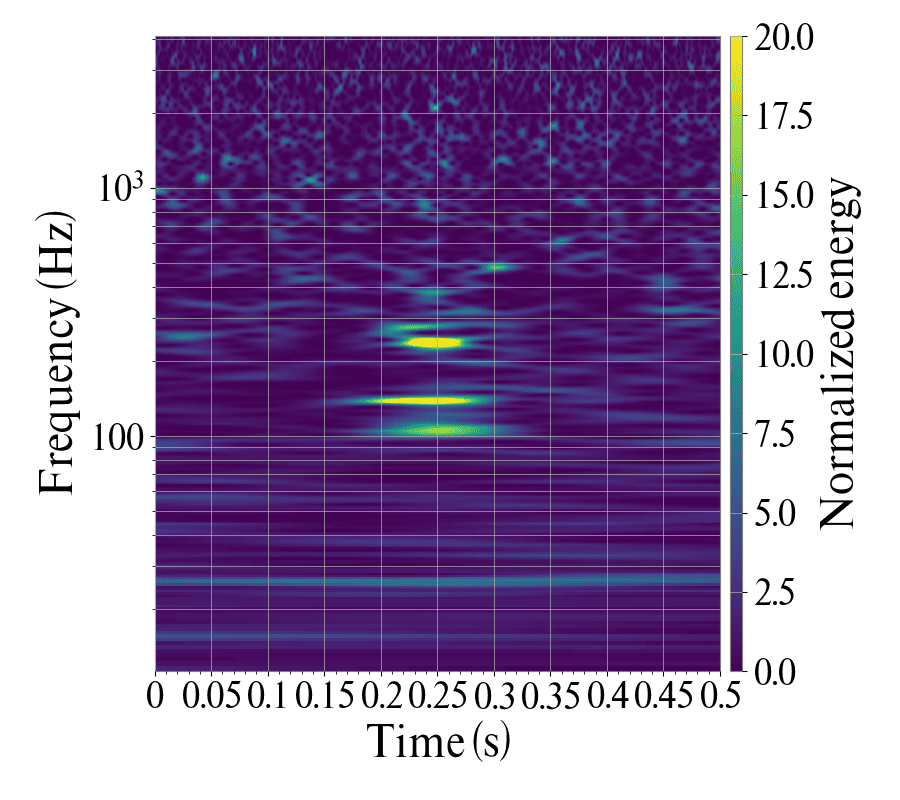}
    \includegraphics[width=0.48\textwidth]{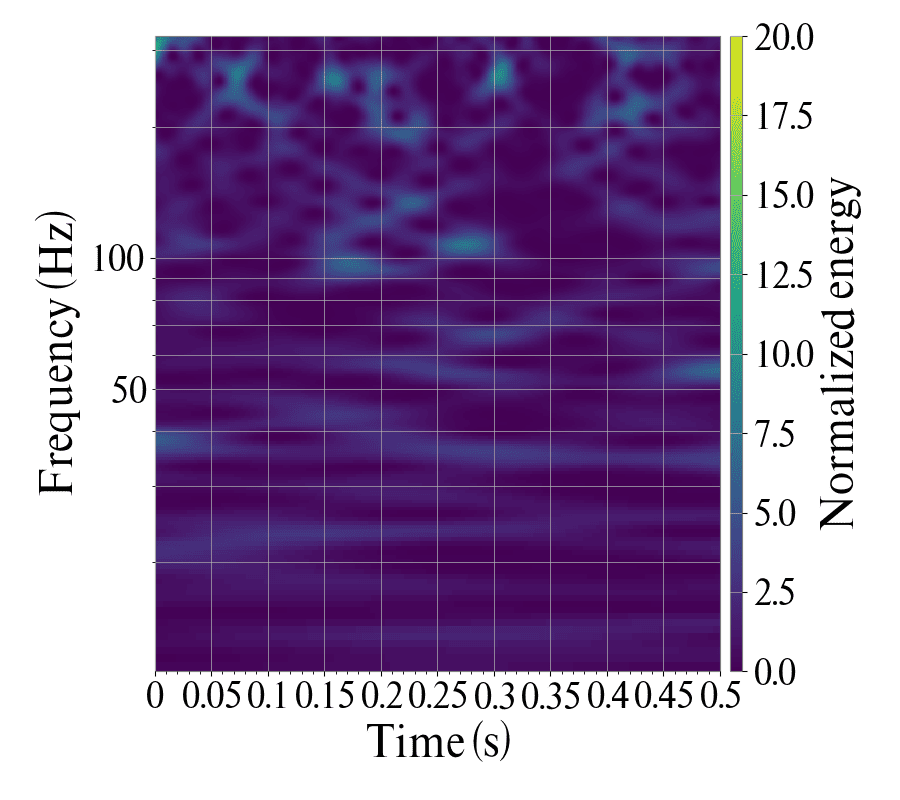}
    \par\smallskip
    {\scriptsize \scalebox{0.8}[1.0]{\textit{K1:VIS-ETMY\_MN\_PSDAMP\_Y\_IN1\_DQ}}\\ Apr 19, 2020	06:19:32	UTC	(GPS: 1271312390)}
    \end{minipage}
    }

\resizebox{!}{0.136\textheight}
{%
\begin{minipage}[t]{0.45\textwidth}
    \centering
    \includegraphics[width=0.48\textwidth]{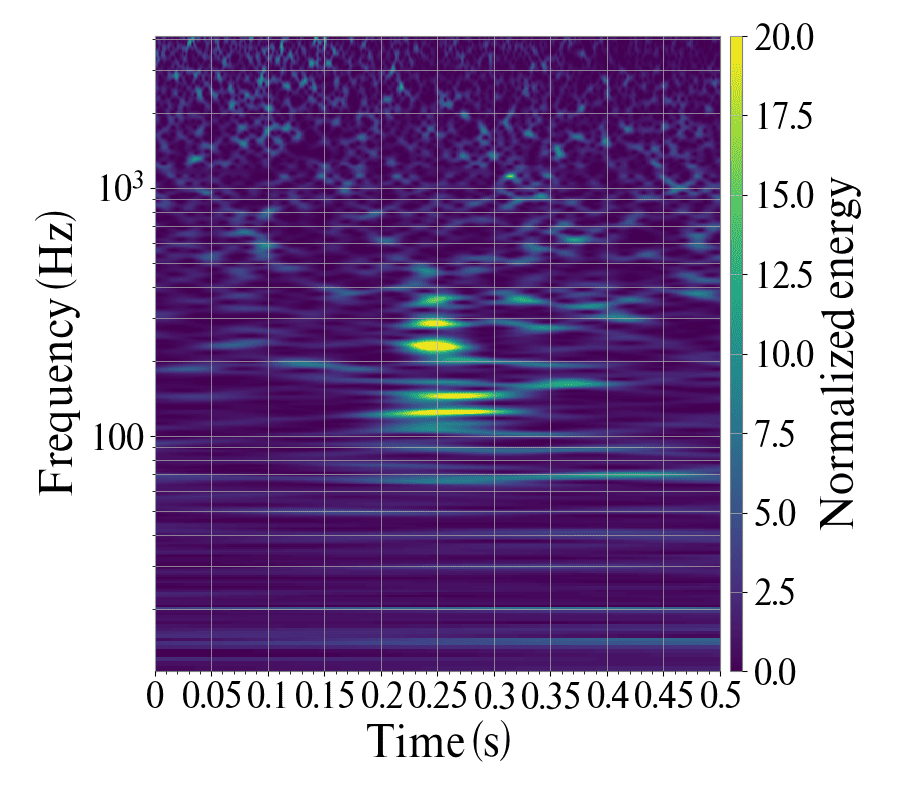}
    \includegraphics[width=0.48\textwidth]{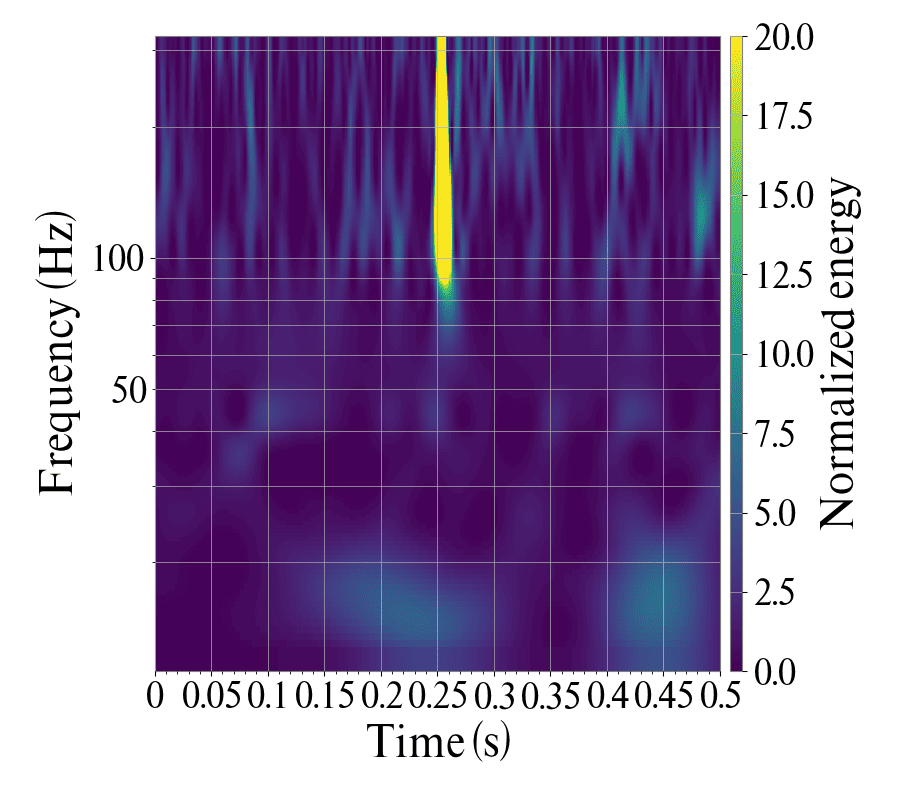}
    \par\smallskip
    {\scriptsize \scalebox{0.8}[1.0]{\textbf{K1:VIS-ITMY\_IM\_PSDAMP\_R\_IN1\_DQ}}\\ Apr 08, 2020	20:50:54	UTC	(GPS: 1270414272)}
    \end{minipage}
\hfill
\vrule width 0.5pt
\hfill
\begin{minipage}[t]{0.45\textwidth}
    \centering
    \includegraphics[width=0.48\textwidth]{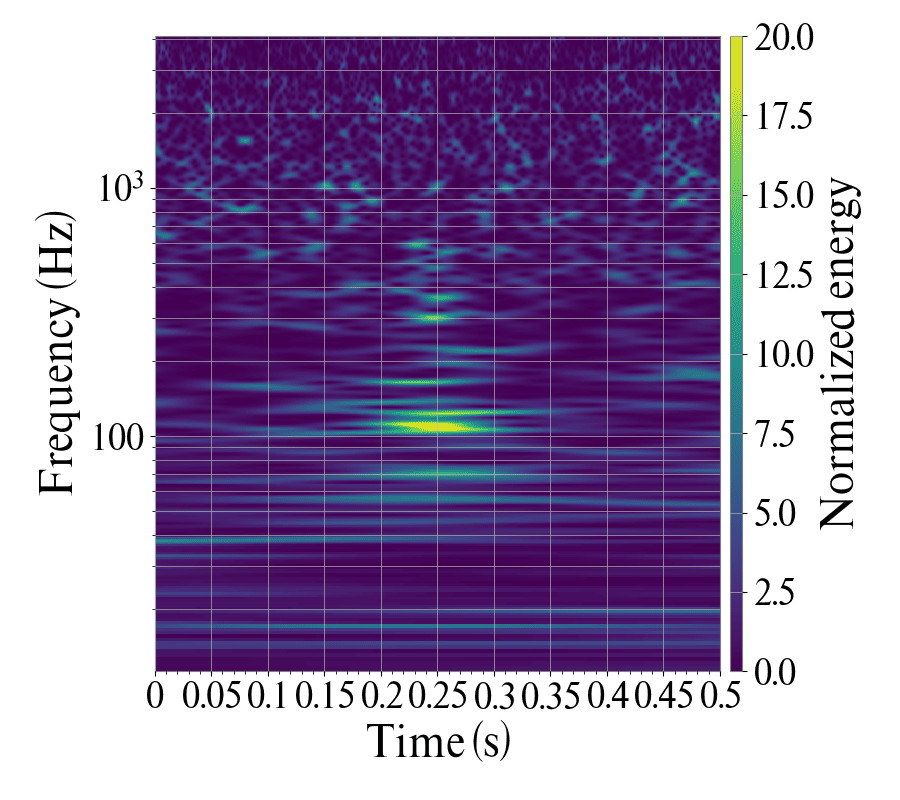}
    \includegraphics[width=0.48\textwidth]{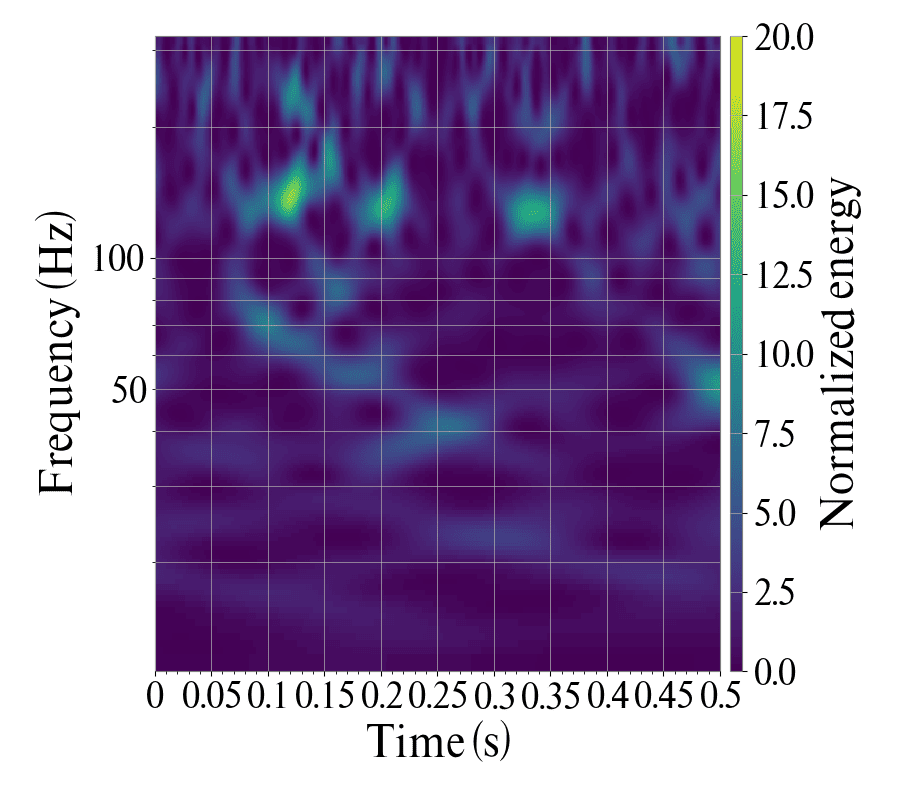}
    \par\smallskip
    {\scriptsize \scalebox{0.8}[1.0]{\textit{K1:VIS-ITMY\_MN\_OPLEV\_TILT\_YAW\_OUT\_DQ}}\\ Apr 09, 2020	12:52:28	UTC	(GPS: 1270471966)}
    \end{minipage}
}

\resizebox{!}{0.136\textheight}
{%
\begin{minipage}[t]{0.45\textwidth}
    \centering
    \includegraphics[width=0.48\textwidth]{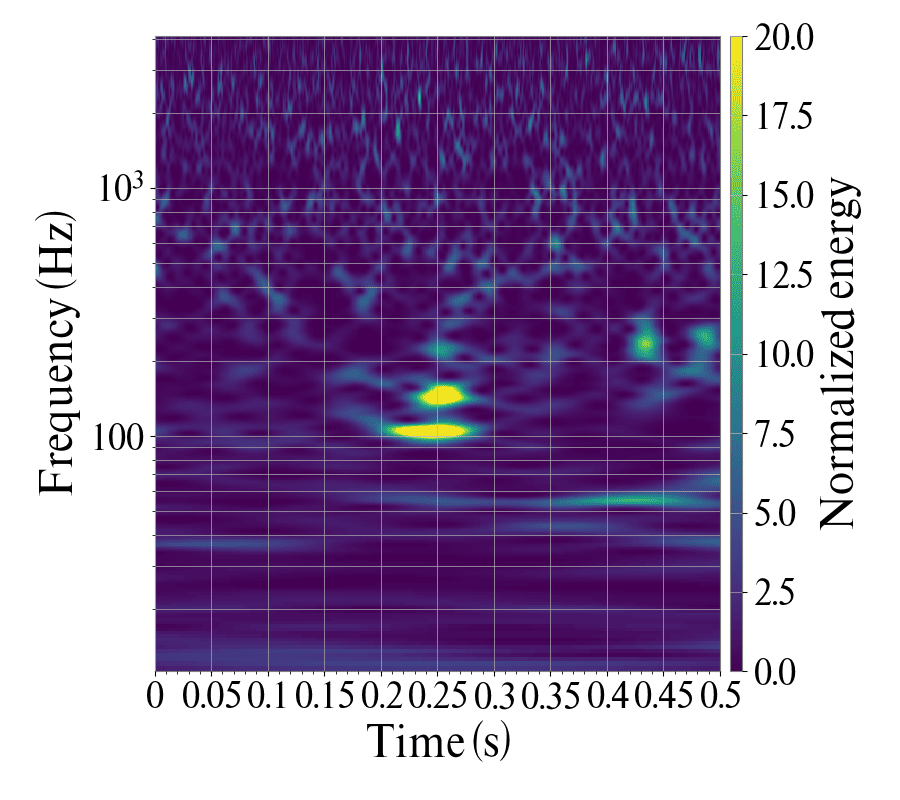}
    \includegraphics[width=0.48\textwidth]{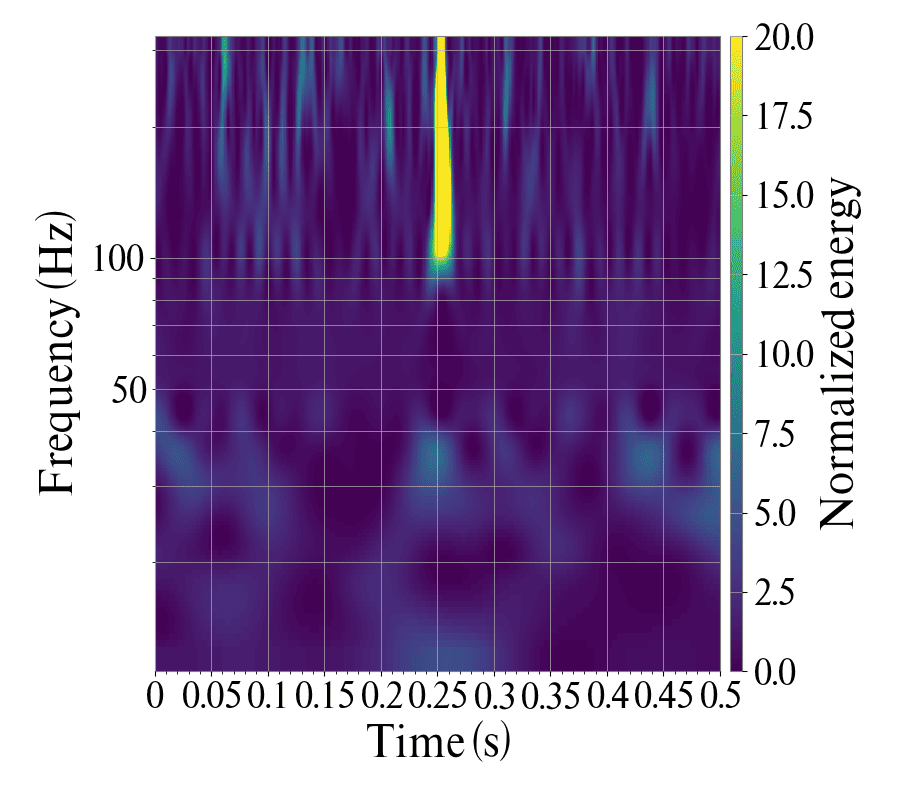}
    \par\smallskip
    {\scriptsize \scalebox{0.8}[1.0]{\textbf{K1:VIS-ITMY\_MN\_PSDAMP\_L\_IN1\_DQ}}\\ Apr 16, 2020	20:56:49	UTC (GPS: 1271105827)}
    \end{minipage}
\hfill
\vrule width 0.5pt
\hfill
\begin{minipage}[t]{0.45\textwidth}
    \centering
    \includegraphics[width=0.48\textwidth]{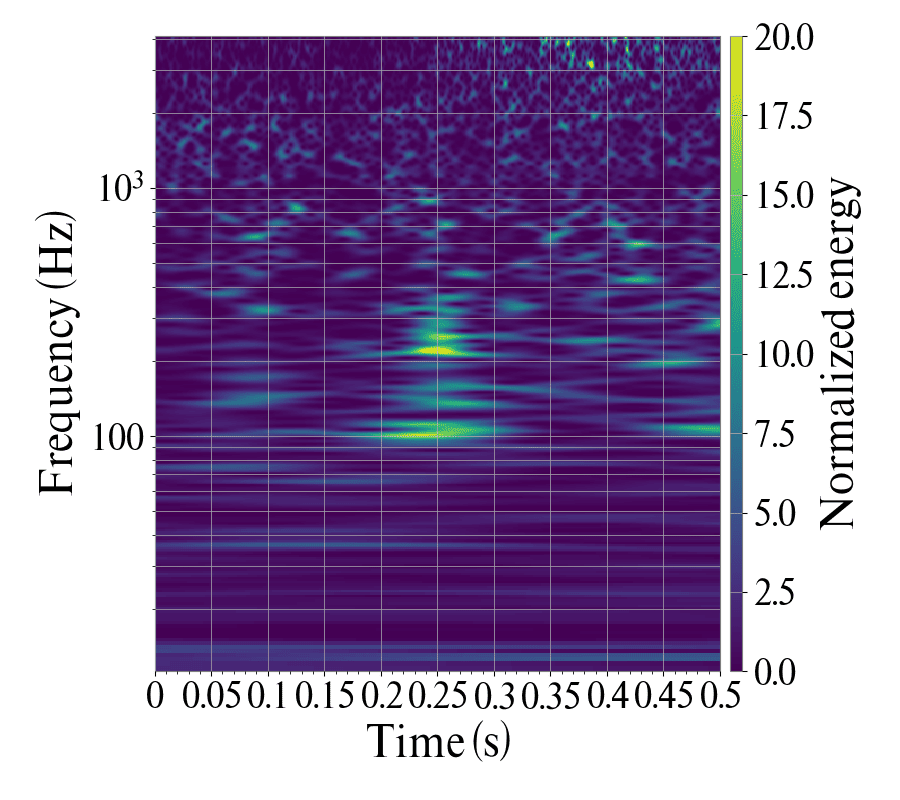}
    \includegraphics[width=0.48\textwidth]{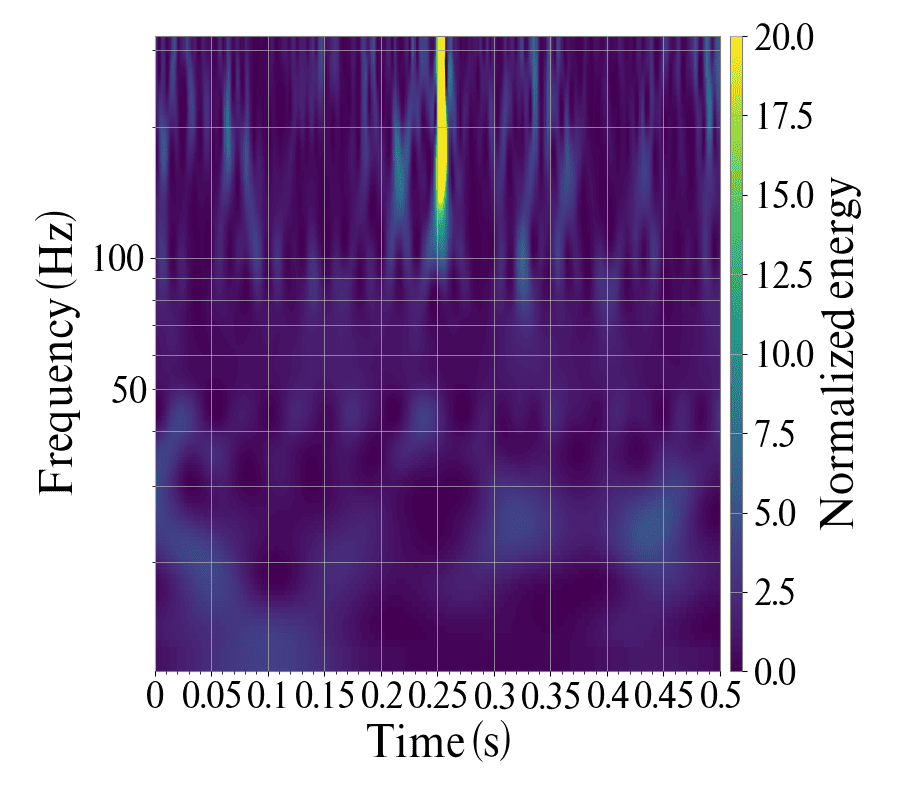}
    \par\smallskip
    {\scriptsize \scalebox{0.8}[1.0]{\textbf{K1:VIS-ITMY\_MN\_PSDAMP\_Y\_IN1\_DQ}}\\ Apr 10, 2020	03:56:35	UTC (GPS: 1270526213)}
    \end{minipage}
    }
\end{figure}
\clearpage

\vspace{-1cm}

\begin{table*}[!h]
\centering
\caption{\label{table:LineGlitches} O3GK KAGRA line glitches}
\small
\resizebox{0.98\textwidth}{!}
{%
\begin{tabular}{|P{0.07\textwidth}|P{0.58\textwidth}|P{0.35\textwidth}|}
\hline
\begin{tabular}[c]{@{}c@{}}Sub-\\ System\end{tabular} & \begin{tabular}[c]{@{}c@{}}Round Winner\\ Auxiliary Channel\end{tabular} & \begin{tabular}[c]{@{}c@{}}Vetoed Date in April\\ (\# of Vetoed Events)\end{tabular} \\ \hline

\begin{tabular}[c]{@{}c@{}}AOS \\ (82) \end{tabular}
& \textbf{K1:AOS-TMSX\_IR\_PDA1\_OUT\_DQ}
& \begin{tabular}[c]{@{}c@{}}10th (7), 11th (2), 16th (21), \\ 17th (16), 18th (36)\end{tabular} \\ \hline

\multirow{2}{*}{\begin{tabular}[c]{@{}c@{}} IMC \\ (122) \end{tabular}}
& \begin{tabular}[c]{@{}c@{}}\textbf{K1:IMC-IMMT1}\\ \textbf{\_TRANS\_QPDA1\_DC\_PIT\_OUT\_DQ}\end{tabular}  
& \begin{tabular}[c]{@{}c@{}}8th (7), 9th (26),\\ 10th (25), 11th (4)\end{tabular}\\ \cline{2-3} 
& \begin{tabular}[c]{@{}c@{}}\textbf{K1:IMC-IMMT1}\\  \textbf{\_TRANS\_QPDA1\_DC\_YAW\_OUT\_DQ}\end{tabular} 
& 9th (5), 11th (21), 19th (34)                                                          \\ \hline

\multirow{2}{*}{\begin{tabular}[c]{@{}c@{}}LSC \\ (65) \end{tabular}}
& \begin{tabular}[c]{@{}c@{}}\textbf{K1:LSC-ALS\_CARM\_OUT\_DQ}\end{tabular}
& 16th (8) 
\\ \cline{2-3}
& \begin{tabular}[c]{@{}c@{}}\textbf{K1:LSC-ALS\_DARM\_OUT\_DQ}\end{tabular}
& 12th (41), 15th (9), 16th (7) 
\\ \hline

\multirow{4}{*}{\begin{tabular}[c]{@{}c@{}}PEM \\ (10) \end{tabular}}
& \textbf{K1:PEM-ACC\_OMC\_TABLE\_AS\_Z\_OUT\_DQ}
& 17th (1) 
\\ \cline{2-3}
& \begin{tabular}[c]{@{}c@{}}\textbf{K1:PEM-MIC}\\ \textbf{\_MCF\_TABLE\_REFL\_Z\_OUT\_DQ}\end{tabular}
& 15th (5) 
\\ \cline{2-3}
& \textit{K1:PEM-MIC\_SR\_BOOTH\_SR\_Z\_OUT\_DQ}
& 18th (3)
\\ \cline{2-3}
& \textit{K1:PEM-VOLT\_REFL\_TABLE\_GND\_OUT\_DQ}
& 11th (1) 
\\ \hline

\multirow{8}{*}{\begin{tabular}[c]{@{}c@{}}VIS \\ (199) \end{tabular}}
& \textit{K1:VIS-ETMY\_MN\_PSDAMP\_Y\_IN1\_DQ}
& 19th (1)
\\ \cline{2-3}
& \textit{K1:VIS-ITMY\_MN\_OPLEV\_TILT\_YAW\_OUT\_DQ}
& 9th(1)
\\ \cline{2-3}
& \textit{K1:VIS-ITMY\_MN\_PSDAMP\_L\_IN1\_DQ}
& 19th (1)
\\ \cline{2-3}
& \textbf{K1:VIS-OMMT1\_TM\_OPLEV\_PIT\_OUT\_DQ}
& \begin{tabular}[c]{@{}c@{}}8th (22), 9th (17), 12th (9),\\ 16th (10), 18th (9)\end{tabular} 
\\ \cline{2-3}
& \begin{tabular}[c]{@{}c@{}}\textbf{K1:VIS-OMMT1}\\ \textbf{\_TM\_OPLEV\_YAW\_OUT\_DQ}\end{tabular}
& \begin{tabular}[c]{@{}c@{}}11th (36), 17th (5),\\ 18th (7), 19th (38)\end{tabular}
\\ \cline{2-3}
& \textbf{K1:VIS-OSTM\_TM\_OPLEV\_YAW\_OUT\_DQ}
& 7th (11), 15th (16), 20th (10)
\\ \cline{2-3}
& \begin{tabular}[c]{@{}c@{}}\textbf{K1:VIS-SR3}\\ \textbf{\_TM\_OPLEV\_TILT\_YAW\_OUT\_DQ}\end{tabular}
& 11th (3)
\\ \cline{2-3}
& \textit{K1:VIS-TMSY\_DAMP\_R\_IN1\_DQ}
& 20th (3)
\\ \hline

\end{tabular} }
\end{table*}

\clearpage

\begin{figure}[p]
\centering
\resizebox{!}{0.136\textheight}
{%
\begin{minipage}[t]{0.45\textwidth}
    \centering
    \includegraphics[width=0.48\textwidth]{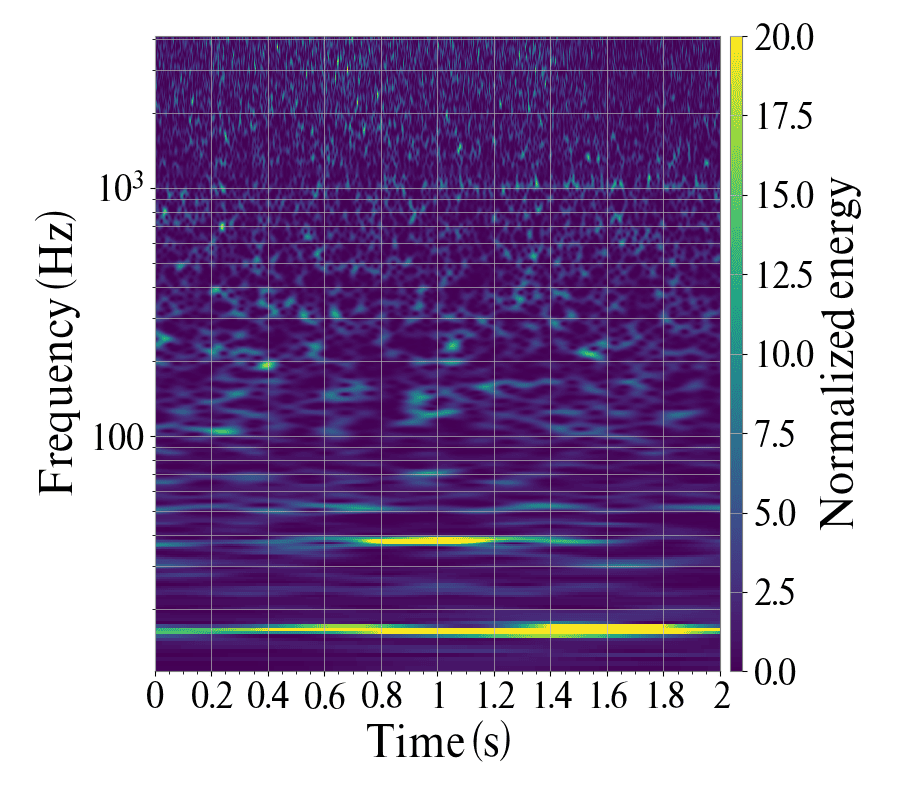}
    \includegraphics[width=0.48\textwidth]{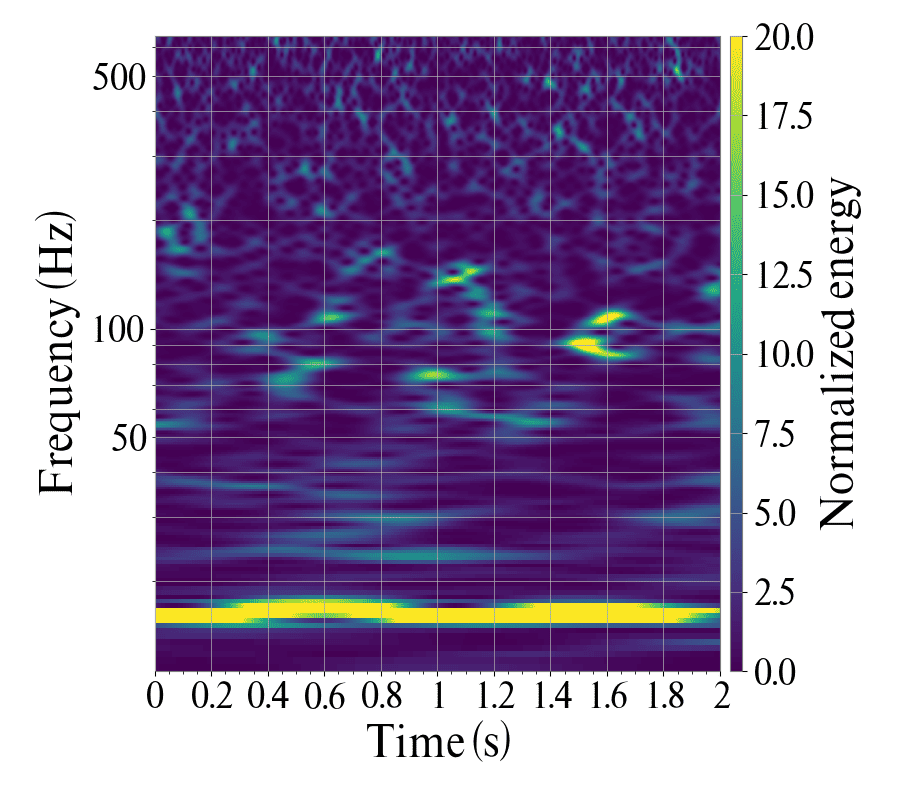}
    \par\smallskip
    {\scriptsize \scalebox{0.8}[1.0]{\textbf{K1:AOS-TMSX\_IR\_PDA1\_OUT\_DQ}} \\ Apr 10, 2020	22:21:58	UTC (GPS: 1270592536)}
    \end{minipage}
\hfill
\vrule width 0.5pt
\hfill
  \begin{minipage}[t]{0.45\textwidth}
    \centering
    \includegraphics[width=0.48\textwidth]{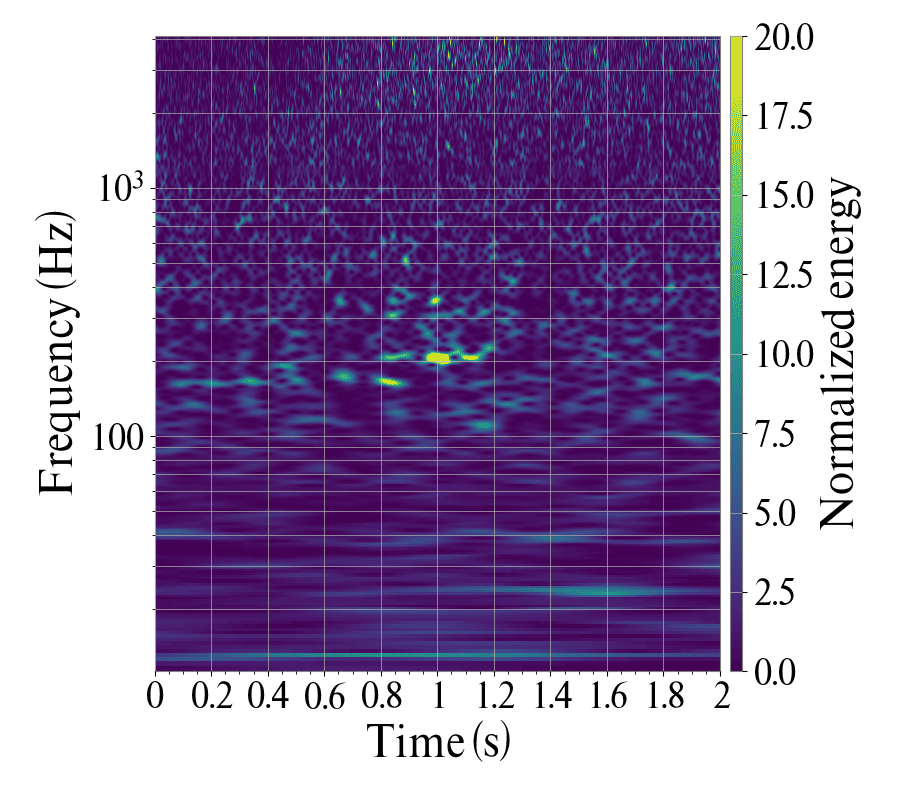}
    \includegraphics[width=0.48\textwidth]{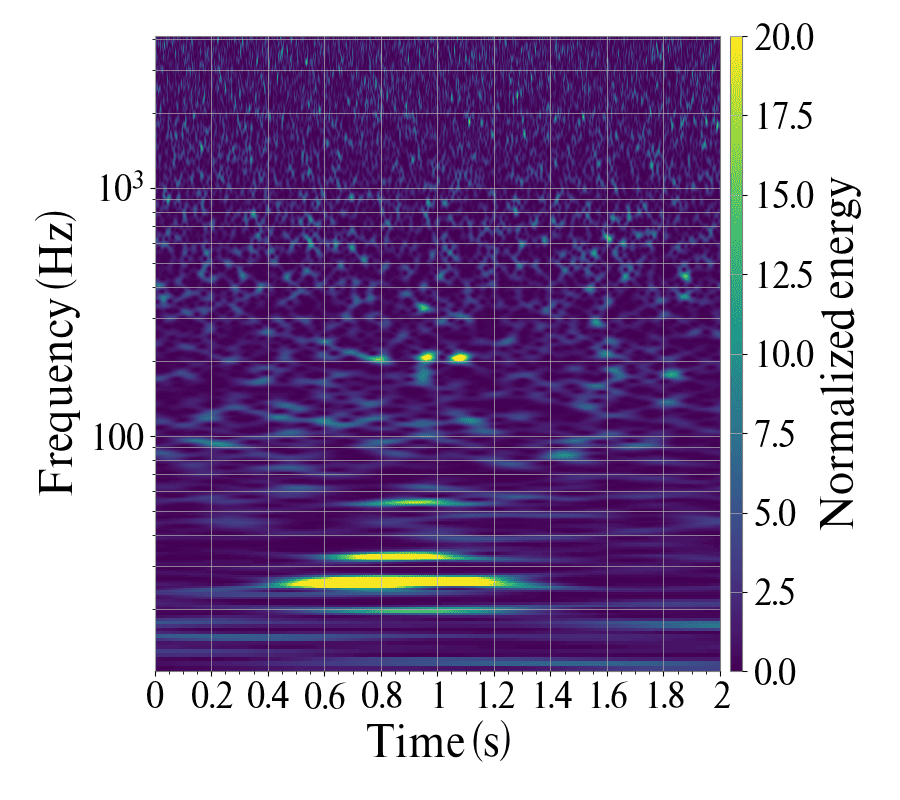}
    \par\smallskip
    {\scriptsize \scalebox{0.8}[1.0]{\textbf{K1:IMC-IMMT1\_TRANS\_QPDA1\_DC\_PIT\_OUT\_DQ}} \\ Apr 09, 2020	18:44:23	UTC (GPS: 1270493081)}
    \end{minipage}
    }

\resizebox{!}{0.136\textheight}
{%
\begin{minipage}[t]{0.45\textwidth}
    \centering
    \includegraphics[width=0.48\textwidth]{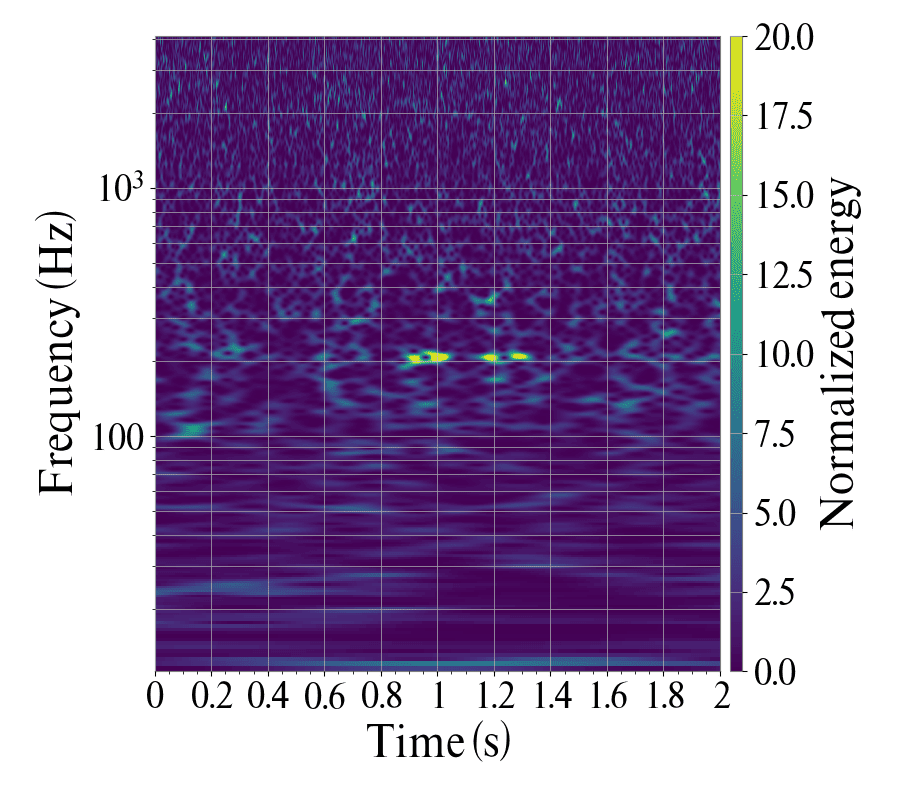}
    \includegraphics[width=0.48\textwidth]{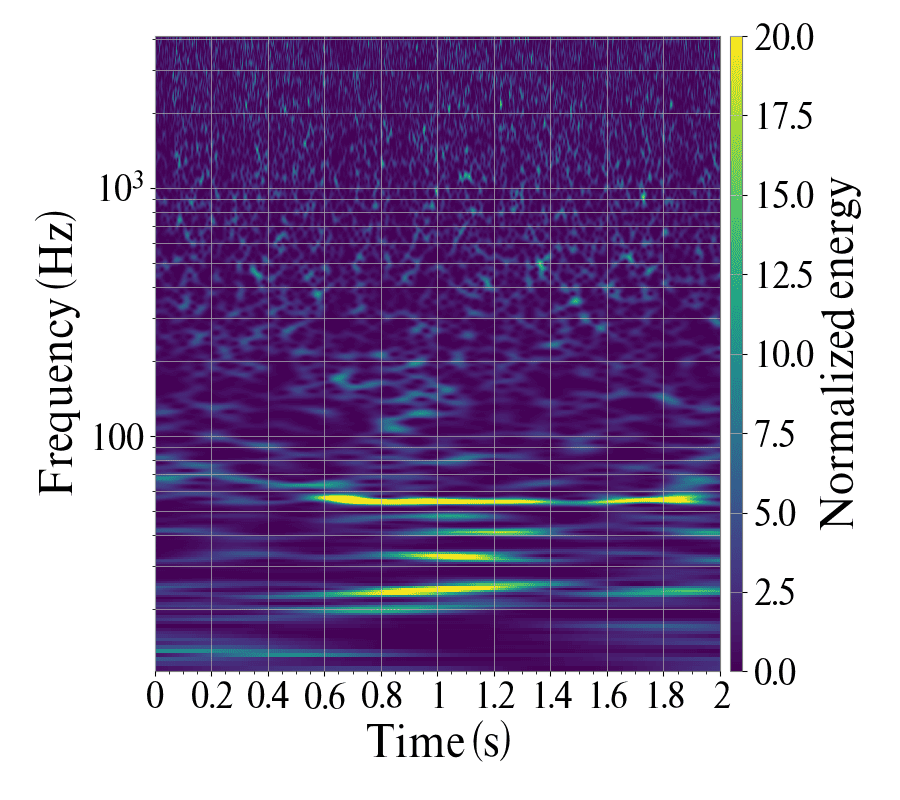}
    \par\smallskip
    {\scriptsize \scalebox{0.8}[1.0]{\textbf{K1:IMC-IMMT1\_TRANS\_QPDA1\_DC\_YAW\_OUT\_DQ}} \\ Apr 11, 2020	01:52:18	UTC (GPS: 1270605156)}
    \end{minipage}
\hfill
\vrule width 0.5pt
\hfill
\begin{minipage}[t]{0.45\textwidth}
    \centering
    \includegraphics[width=0.48\textwidth]{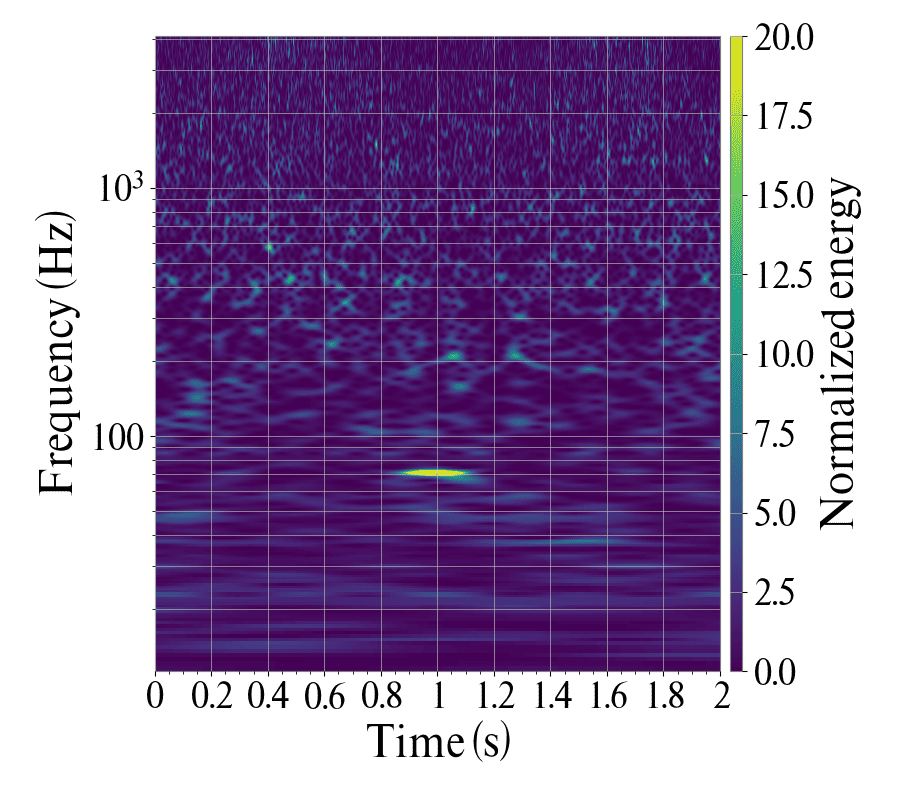}
    \includegraphics[width=0.48\textwidth]{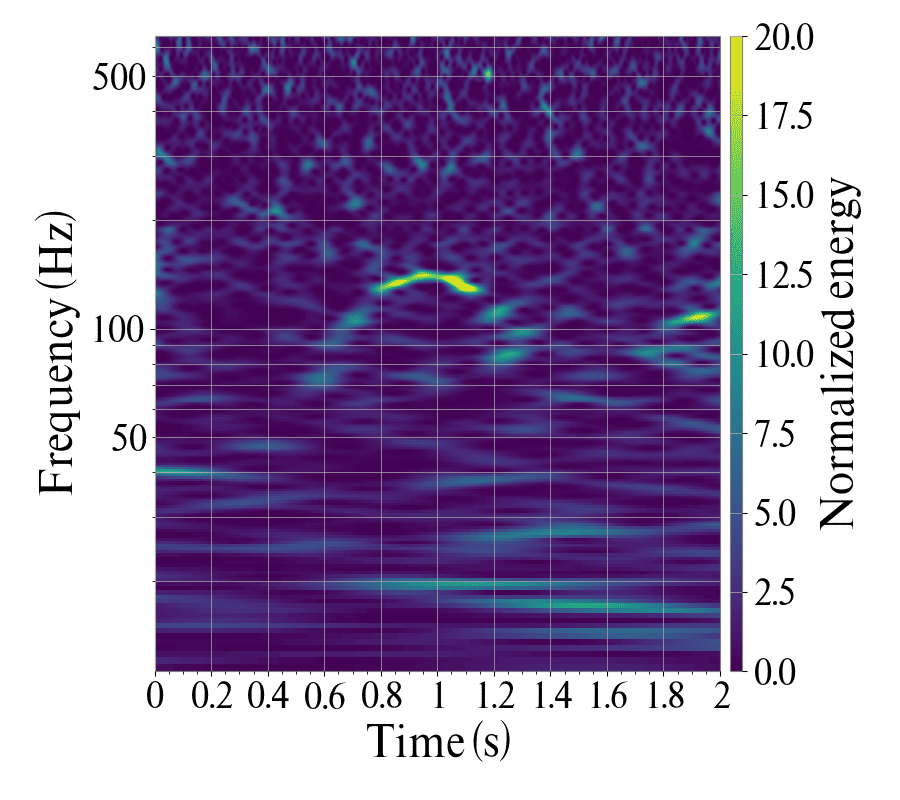}
    \par\smallskip
    {\scriptsize \scalebox{0.8}[1.0]{\textbf{K1:LSC-ALS\_CARM\_OUT\_DQ}} \\ Apr 16, 2020	05:51:06	UTC (GPS: 1271051484)}
    \end{minipage}
    }

\resizebox{!}{0.136\textheight}
{%
\begin{minipage}[t]{0.45\textwidth}
    \centering
    \includegraphics[width=0.48\textwidth]{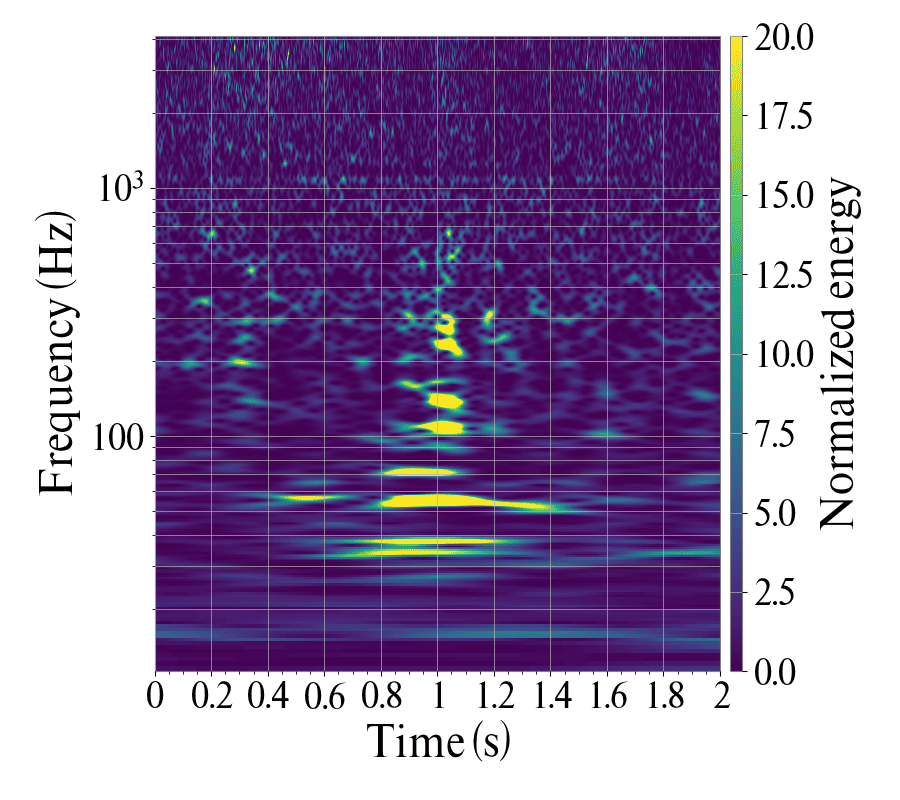}
    \includegraphics[width=0.48\textwidth]{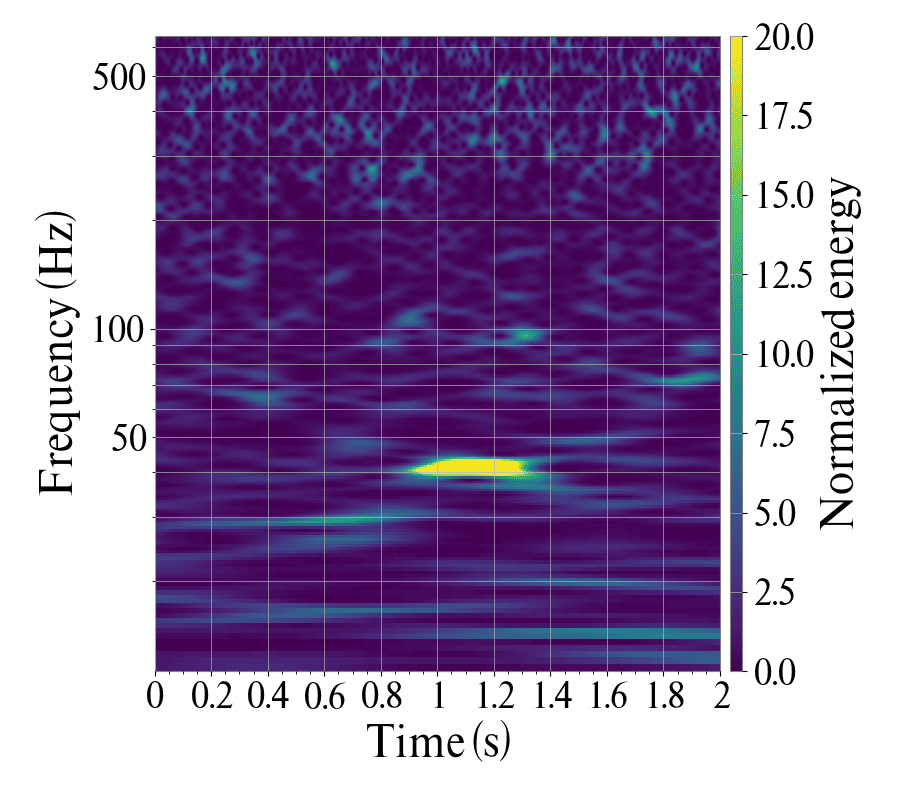}
    \par\smallskip
    {\scriptsize \scalebox{0.8}[1.0]{\textbf{K1:LSC-ALS\_DARM\_OUT\_DQ}} \\ Apr 12, 2020	19:19:17	UTC (GPS: 1270754375)}
    \end{minipage}
\hfill
\vrule width 0.5pt
\hfill    
\begin{minipage}[t]{0.45\textwidth}
    \centering
    \includegraphics[width=0.48\textwidth]{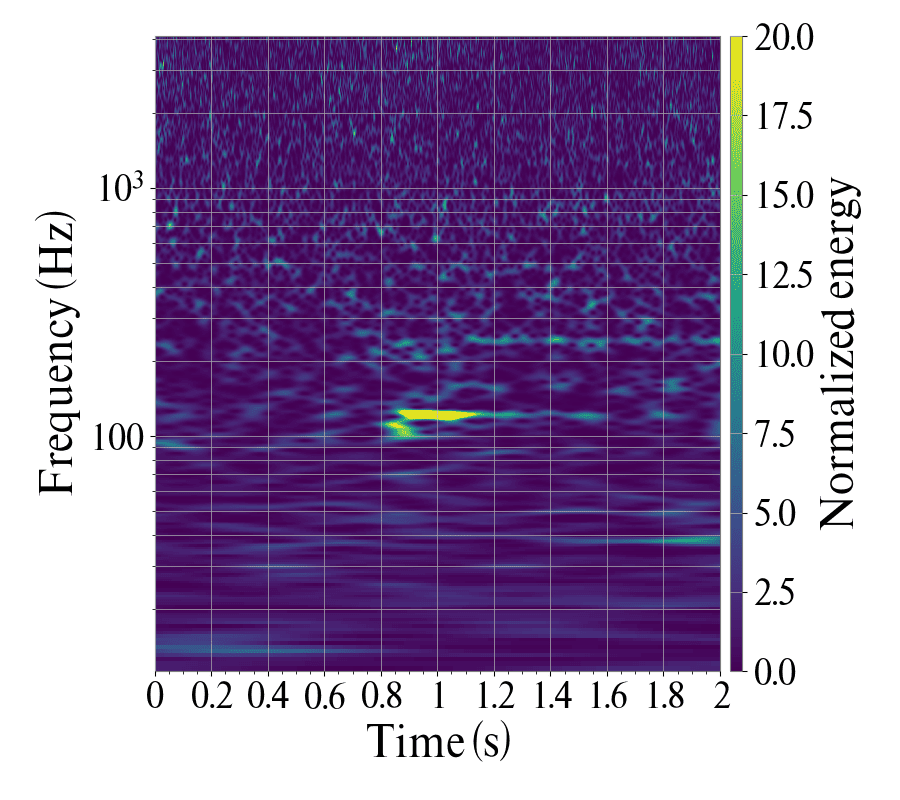}
    \includegraphics[width=0.48\textwidth]{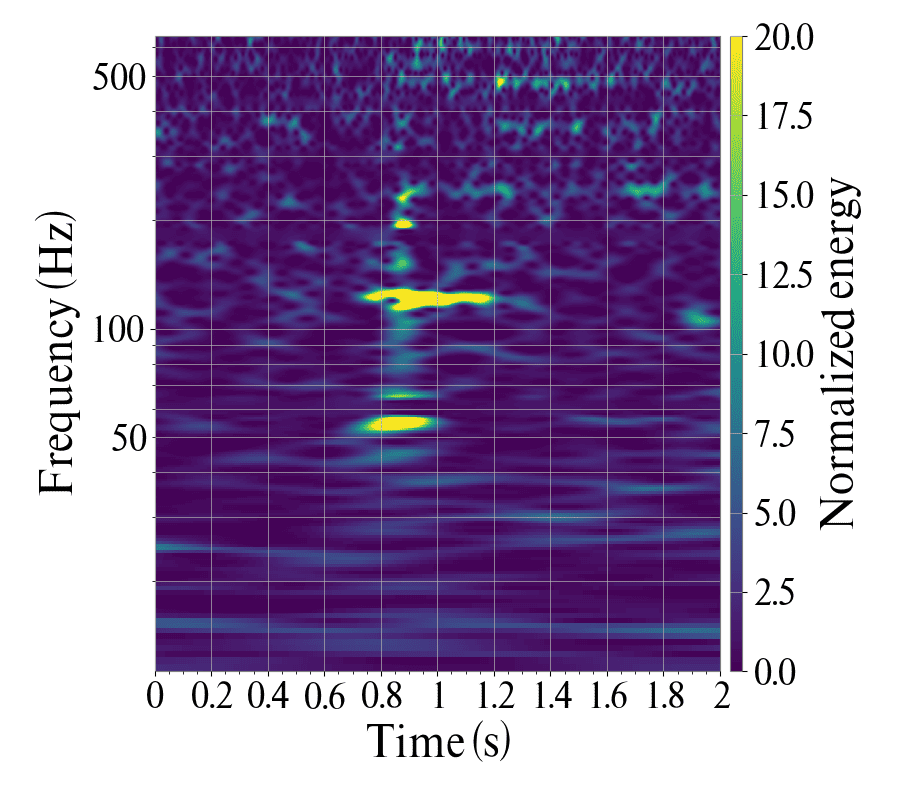}
    \par\smallskip
    {\scriptsize \scalebox{0.8}[1.0]{\textbf{K1:PEM-ACC\_OMC\_TABLE\_AS\_Z\_OUT\_DQ}} \\ Apr 17, 2020	07:33:46	UTC	(GPS: 1271144044)}
    \end{minipage}
    }

\resizebox{!}{0.136\textheight}
{%
\begin{minipage}[t]{0.45\textwidth}
    \centering
    \includegraphics[width=0.48\textwidth]{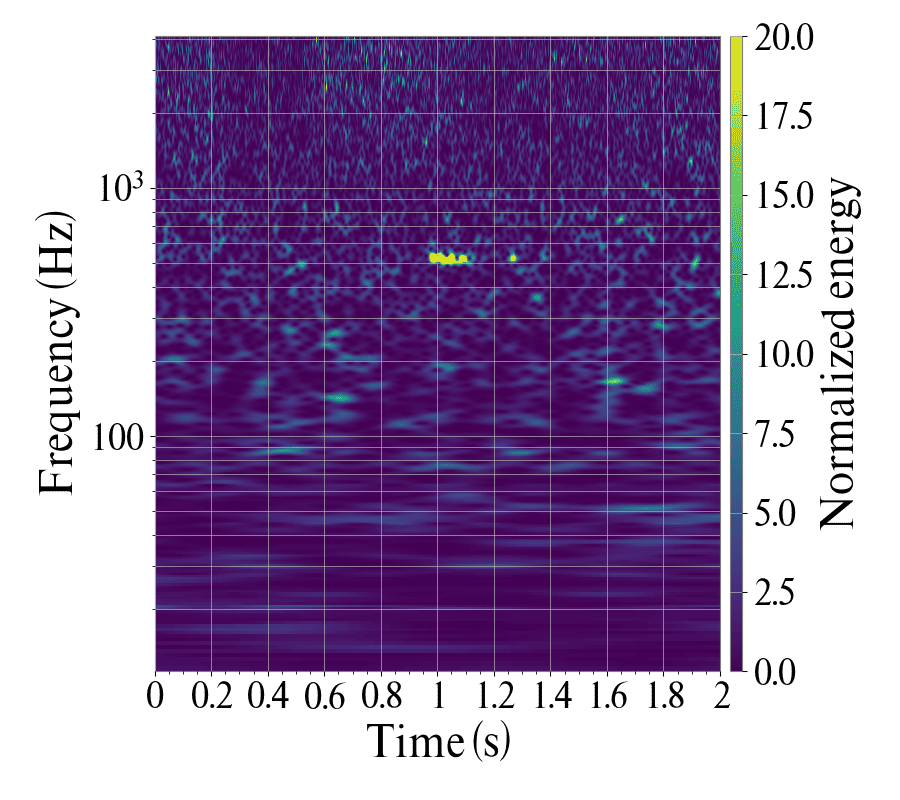}
    \includegraphics[width=0.48\textwidth]{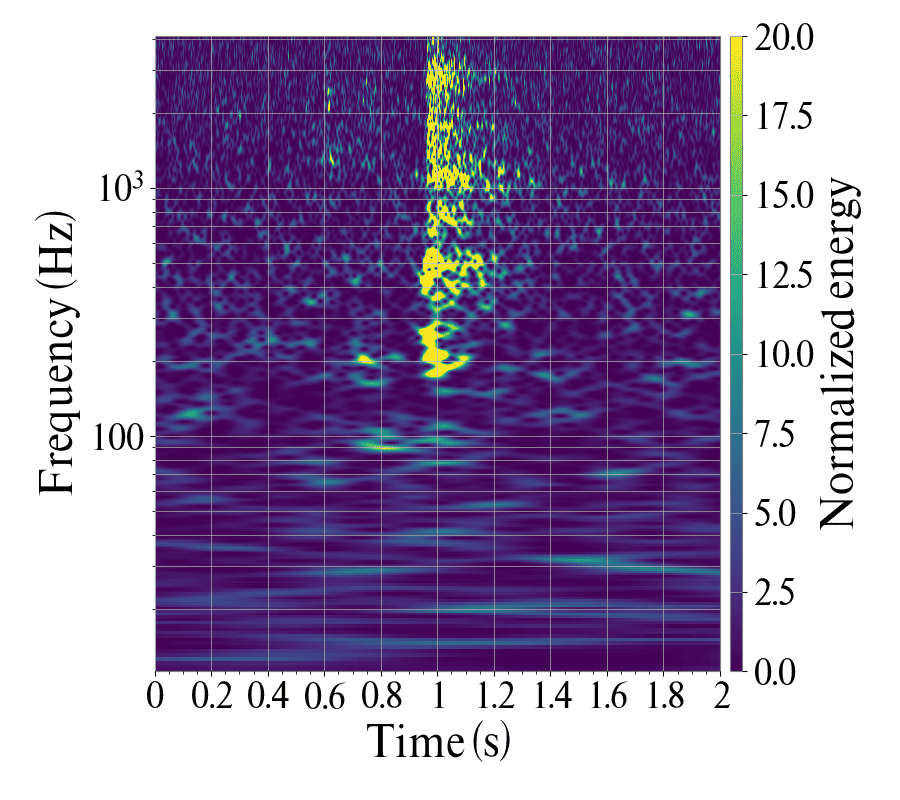}
    \par\smallskip
    {\scriptsize \scalebox{0.8}[1.0]{\textit{K1:PEM-MIC\_MCF\_TABLE\_REFL\_Z\_OUT\_DQ}}\\ Apr 15, 2020	08:22:30	UTC	(GPS: 1270974168)}
    \end{minipage}
\hfill
\vrule width 0.5pt
\hfill    
\begin{minipage}[t]{0.45\textwidth}
    \centering
    \includegraphics[width=0.48\textwidth]{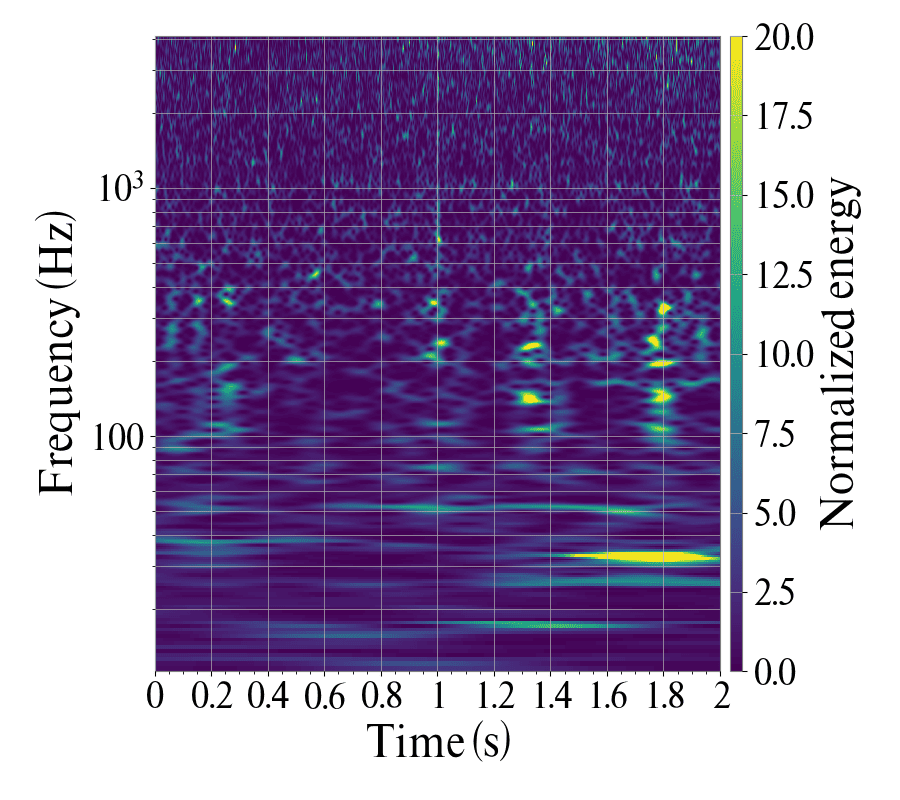}
    \includegraphics[width=0.48\textwidth]{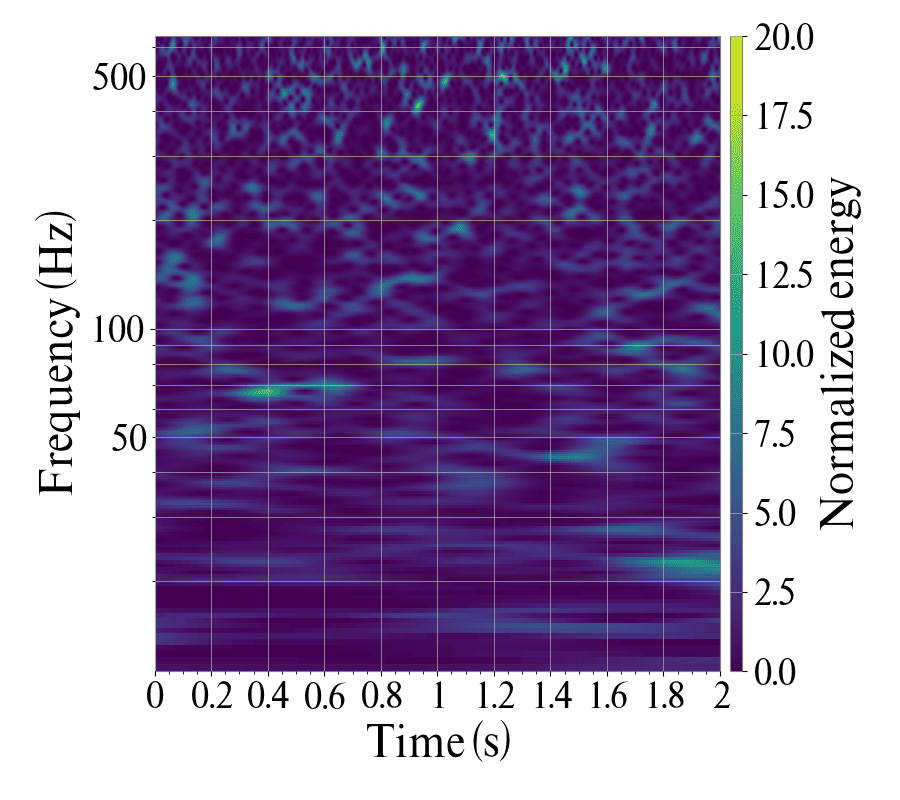}
    \par\smallskip
    {\scriptsize \scalebox{0.8}[1.0]{\textit{K1:PEM-MIC\_SR\_BOOTH\_SR\_Z\_OUT\_DQ}}\\ Apr 18, 2020	20:39:33	UTC	(GPS: 1271277591)}
    \end{minipage}
    }

\resizebox{!}{0.136\textheight}
{%
\begin{minipage}[t]{0.45\textwidth}
    \centering
    \includegraphics[width=0.48\textwidth]{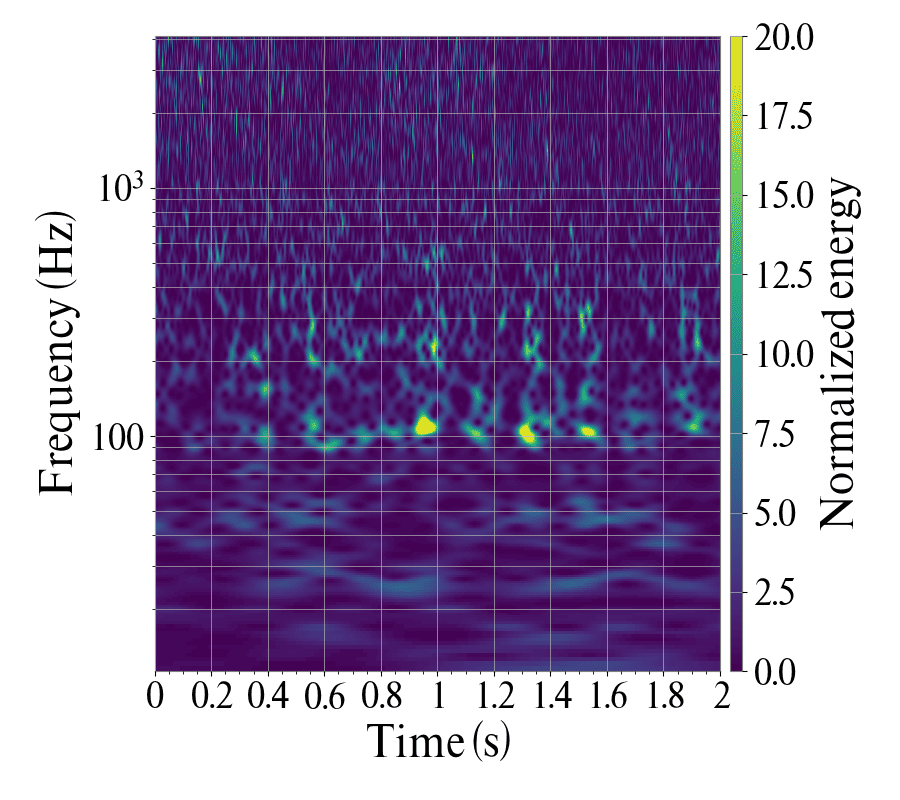}
    \includegraphics[width=0.48\textwidth]{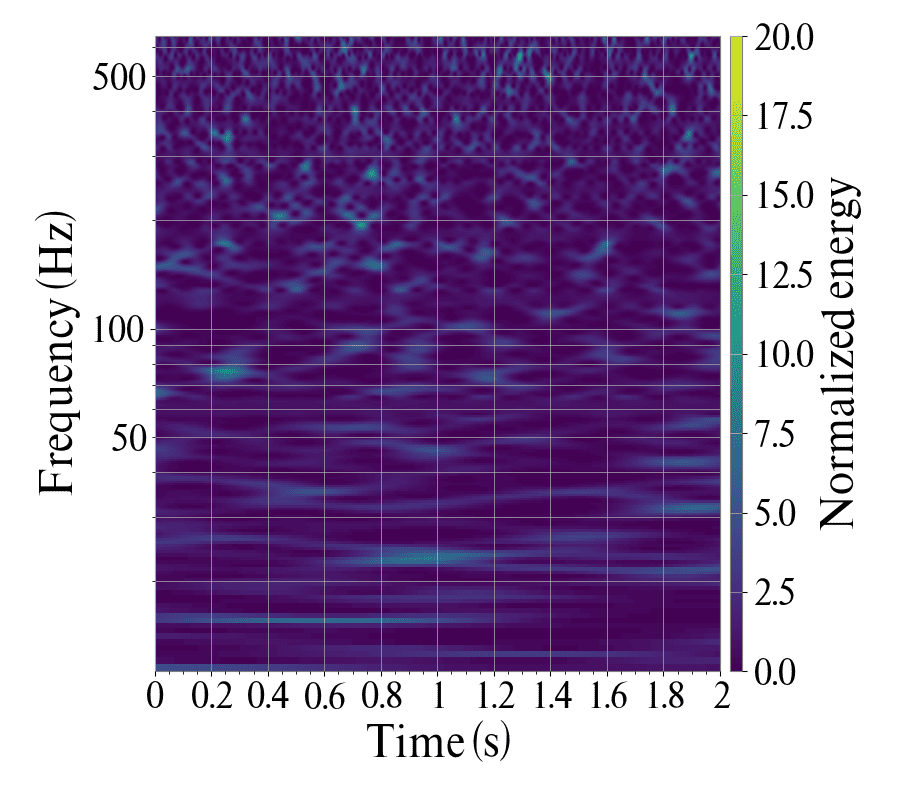}
    \par\smallskip
    {\scriptsize \scalebox{0.8}[1.0]{\textit{K1:PEM-VOLT\_REFL\_TABLE\_GND\_OUT\_DQ}}\\ Apr 11, 2020	01:41:58	UTC	(GPS: 1270604536)}
    \end{minipage}
\hfill
\vrule width 0.5pt
\hfill   
\begin{minipage}[t]{0.45\textwidth}
    \centering
    \includegraphics[width=0.48\textwidth]{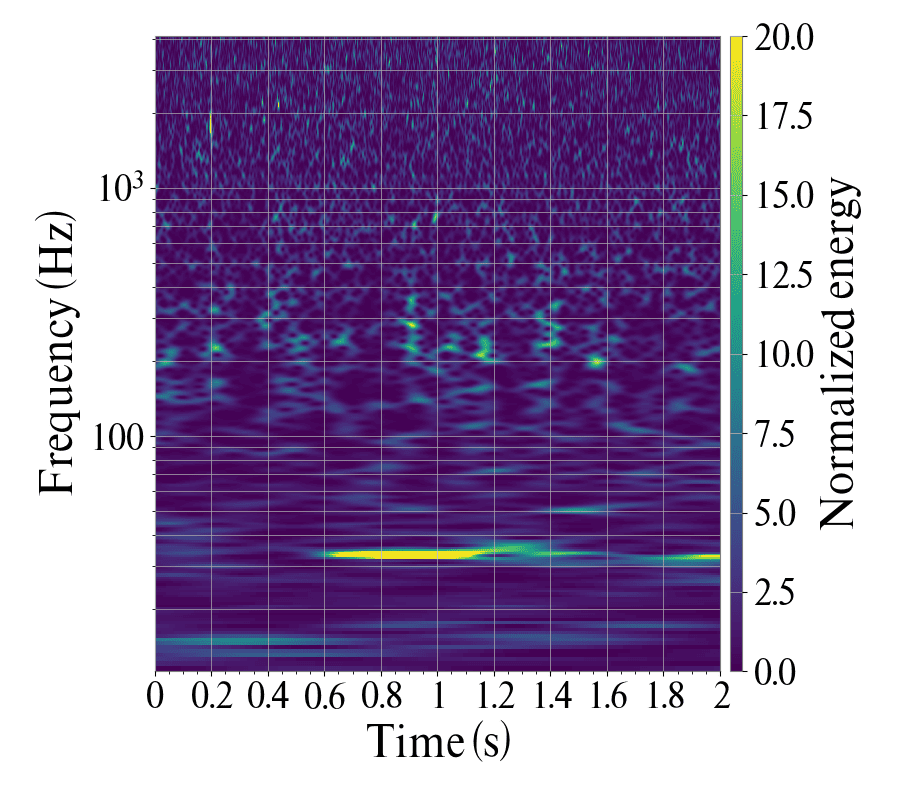}
    \includegraphics[width=0.48\textwidth]{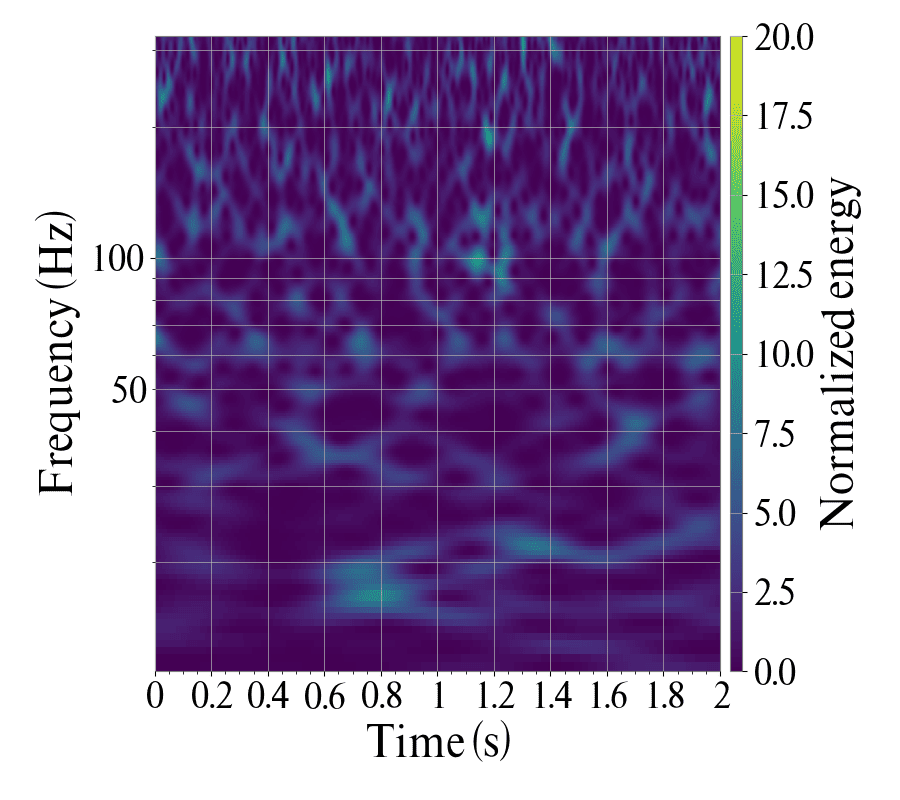}
    \par\smallskip
    {\scriptsize \scalebox{0.8}[1.0]{\textit{K1:VIS-ETMY\_MN\_PSDAMP\_Y\_IN1\_DQ}}\\ Apr 19, 2020	16:02:50	UTC	(GPS: 1271347388)}
    \end{minipage}
    }
\end{figure}

\clearpage

\begin{figure*}[h]
\centering
\resizebox{!}{0.136\textheight}
{%
\begin{minipage}[t]{0.45\textwidth}
    \centering
    \includegraphics[width=0.48\textwidth]{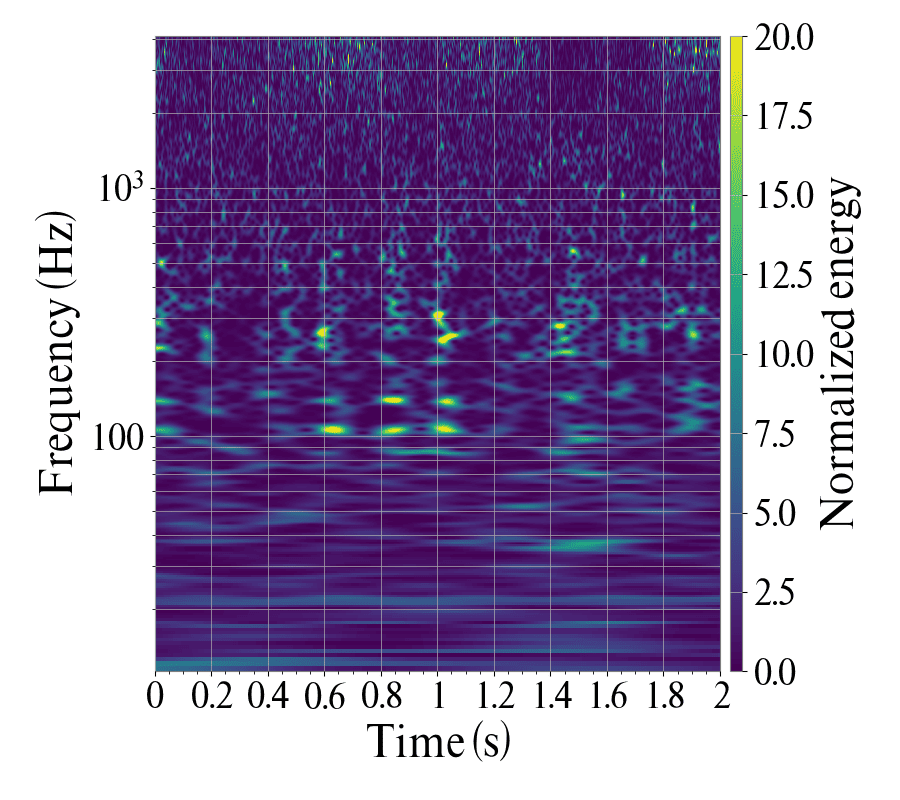}
    \includegraphics[width=0.48\textwidth]{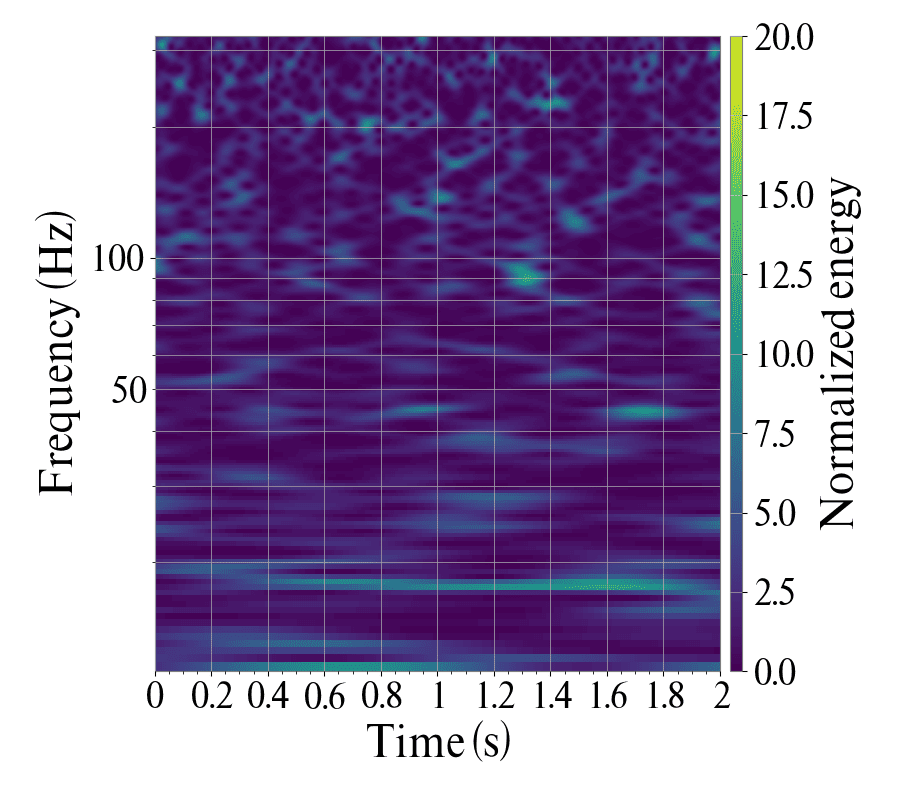}
    \par\smallskip
    {\scriptsize \scalebox{0.8}[1.0]{\textit{K1:VIS-ITMY\_MN\_OPLEV\_TILT\_YAW\_OUT\_DQ}}\\ Apr 09, 2020	10:35:09	UTC	(GPS: 1270463727)}
    \end{minipage}
\hfill
\vrule width 0.5pt
\hfill    
\begin{minipage}[t]{0.45\textwidth}
    \centering
    \includegraphics[width=0.48\textwidth]{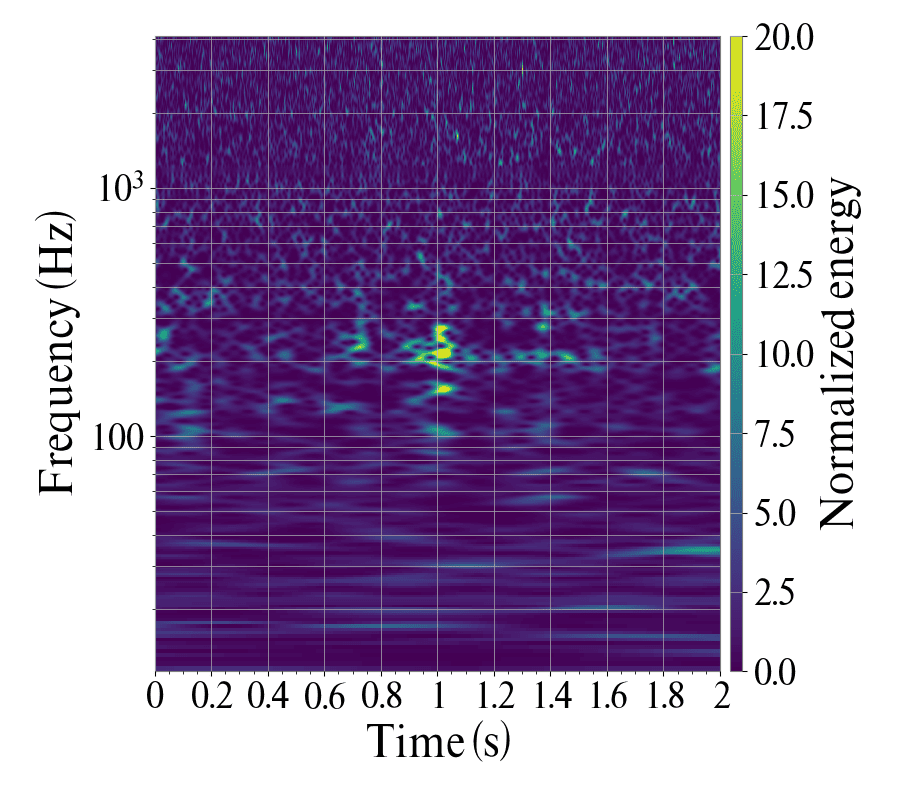}
    \includegraphics[width=0.48\textwidth]{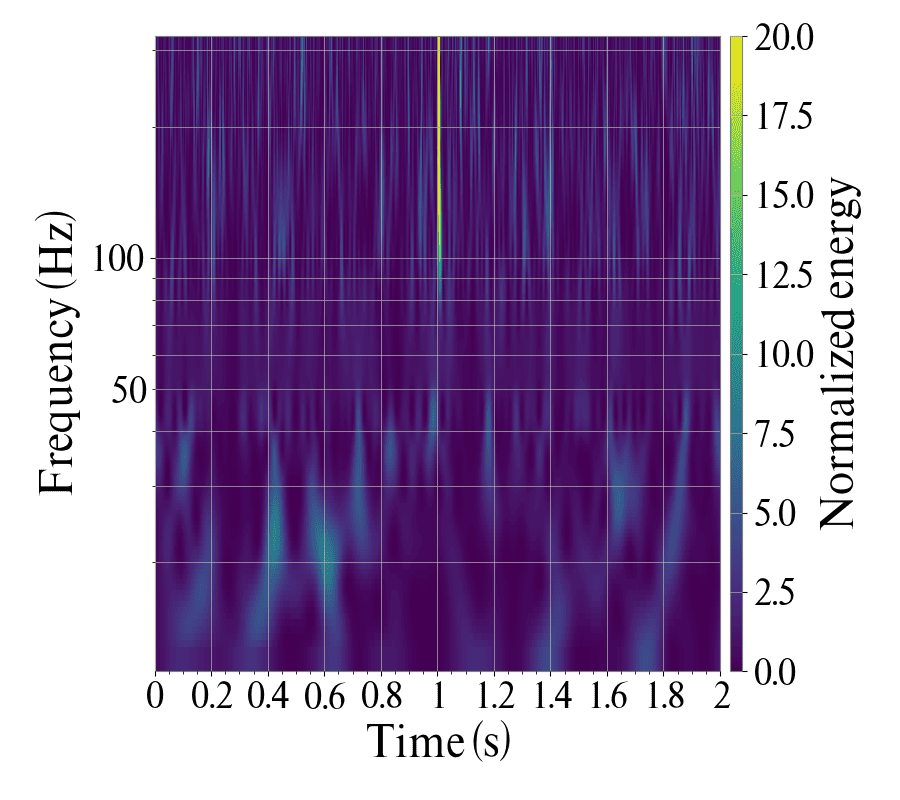}
    \par\smallskip
    {\scriptsize \scalebox{0.8}[1.0]{\textit{K1:VIS-ITMY\_MN\_PSDAMP\_L\_IN1\_DQ}}\\ Apr 19, 2020	01:03:04	UTC (GPS: 1271293402)}
    \end{minipage}
    }

\resizebox{!}{0.136\textheight}
{%
\begin{minipage}[t]{0.45\textwidth}
    \centering
    \includegraphics[width=0.48\textwidth]{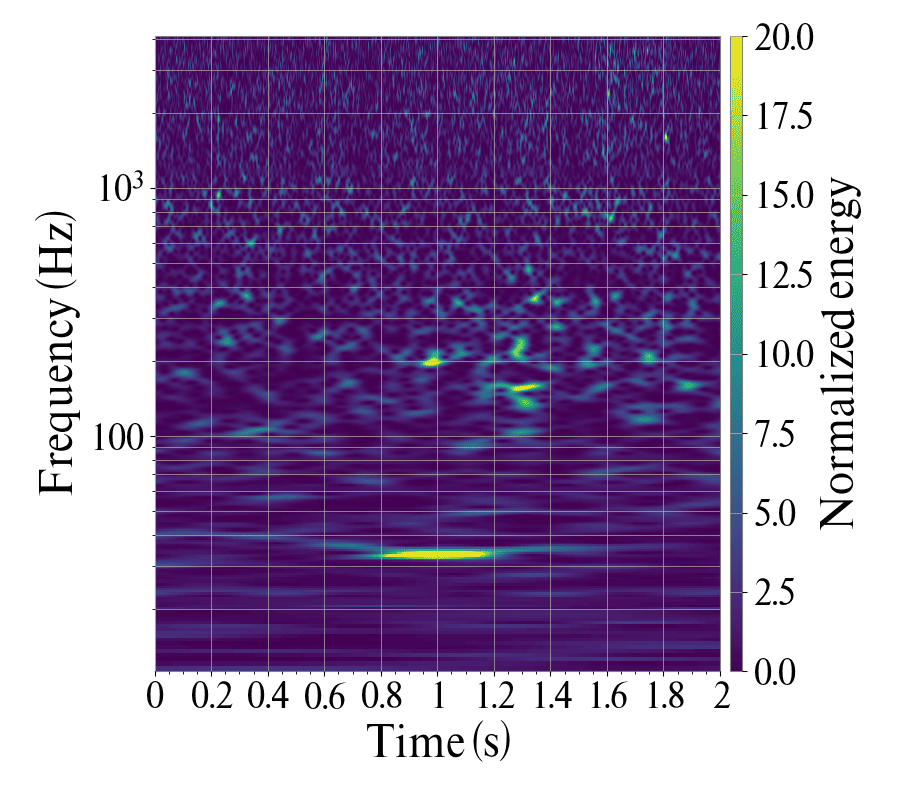}
    \includegraphics[width=0.48\textwidth]{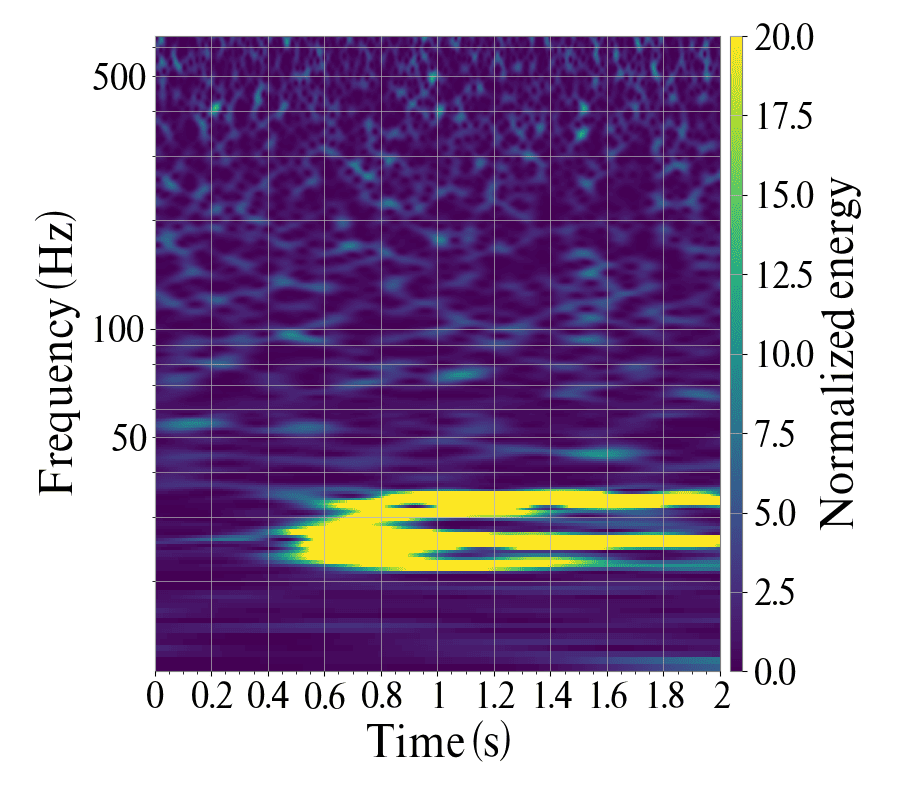}
    \par\smallskip
    {\scriptsize \scalebox{0.8}[1.0]{\textbf{K1:VIS-OMMT1\_TM\_OPLEV\_PIT\_OUT\_DQ}}\\ Apr 12, 2020	18:40:17	UTC (GPS: 1270752035)}
    \end{minipage}
\hfill
\vrule width 0.5pt
\hfill
\begin{minipage}[t]{0.45\textwidth}
    \centering
    \includegraphics[width=0.48\textwidth]{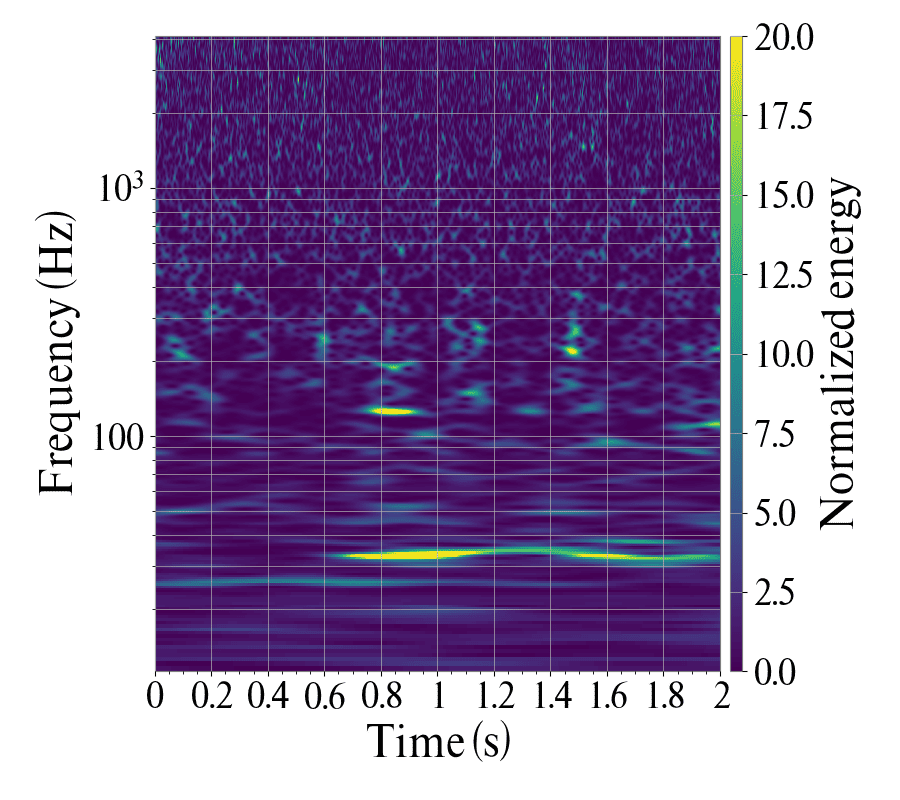}
    \includegraphics[width=0.48\textwidth]{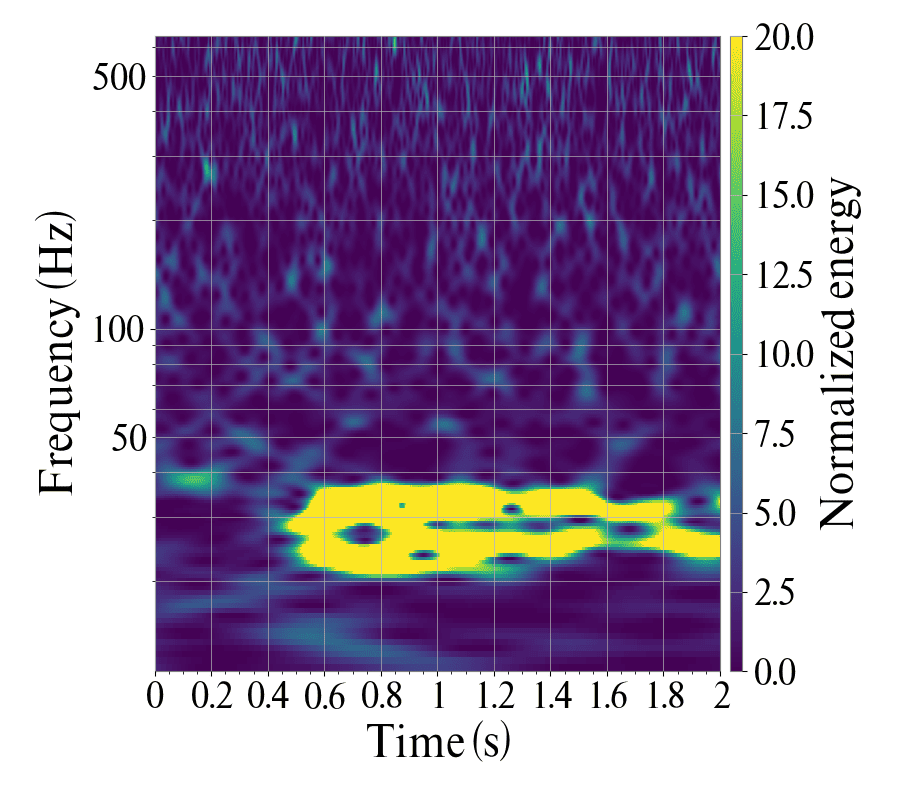}
    \par\smallskip
    {\scriptsize \scalebox{0.8}[1.0]{\textbf{K1:VIS-OMMT1\_TM\_OPLEV\_YAW\_OUT\_DQ}}\\ Apr 17, 2020	05:47:50	UTC (GPS: 1271137688)}
    \end{minipage}
    }

\resizebox{!}{0.136\textheight}
{%
\begin{minipage}[t]{0.45\textwidth}
    \centering
    \includegraphics[width=0.48\textwidth]{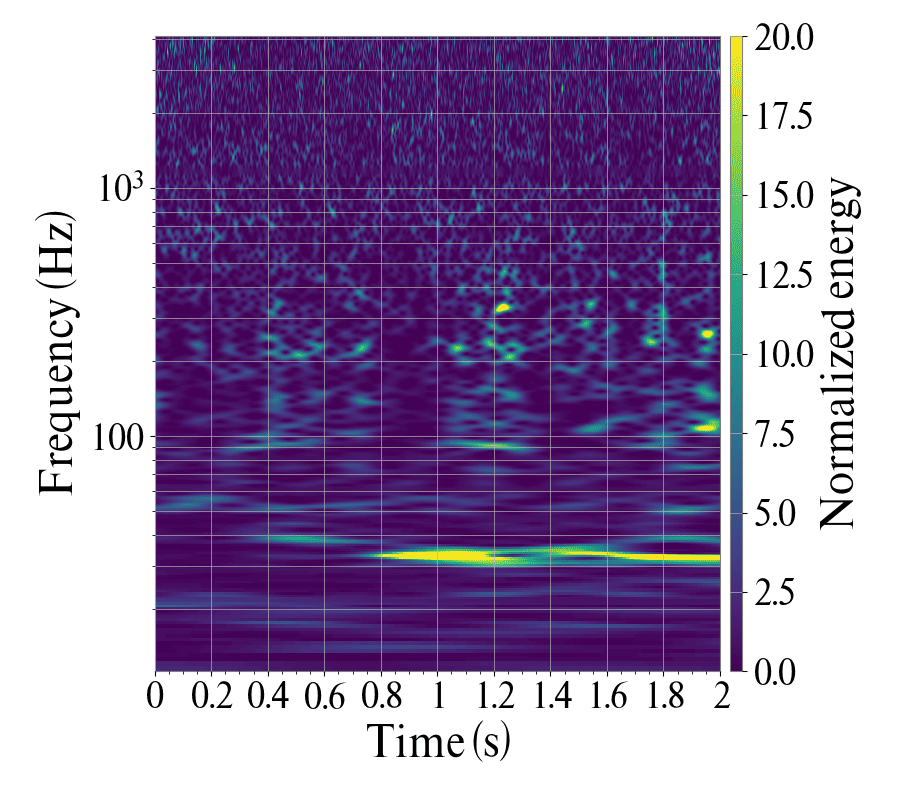}
    \includegraphics[width=0.48\textwidth]{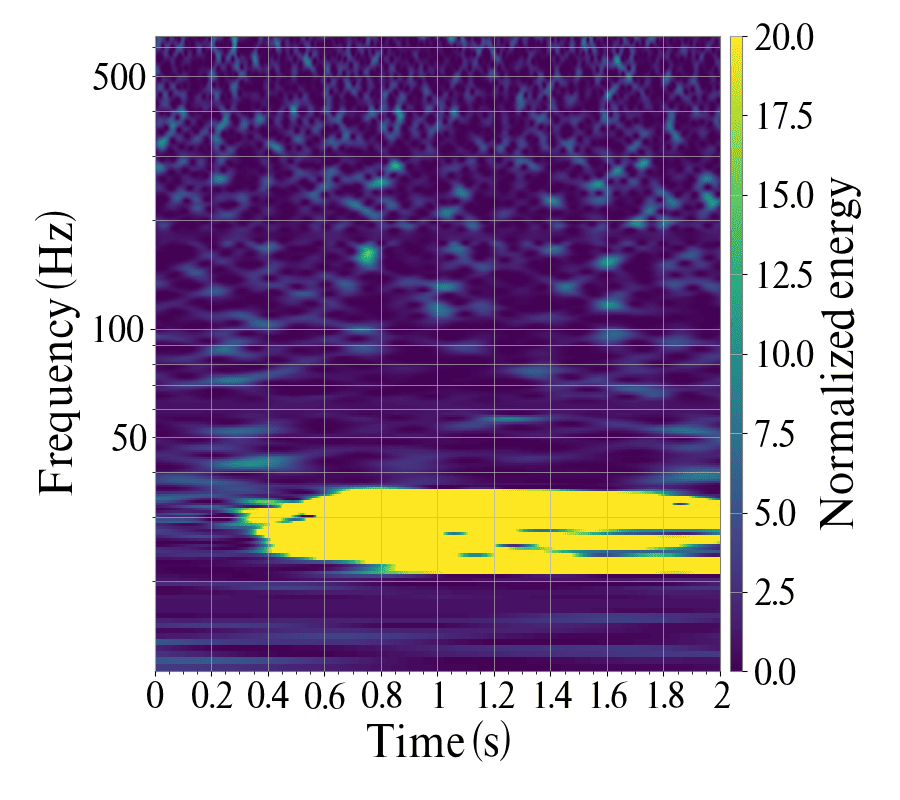}
    \par\smallskip
    {\scriptsize \scalebox{0.8}[1.0]{\textbf{K1:VIS-OSTM\_TM\_OPLEV\_YAW\_OUT\_DQ}}\\ Apr 20, 2020	14:13:47	UTC (GPS: 1271427245)}
    \end{minipage}
\hfill
\vrule width 0.5pt
\hfill    
\begin{minipage}[t]{0.45\textwidth}
    \centering
    \includegraphics[width=0.48\textwidth]{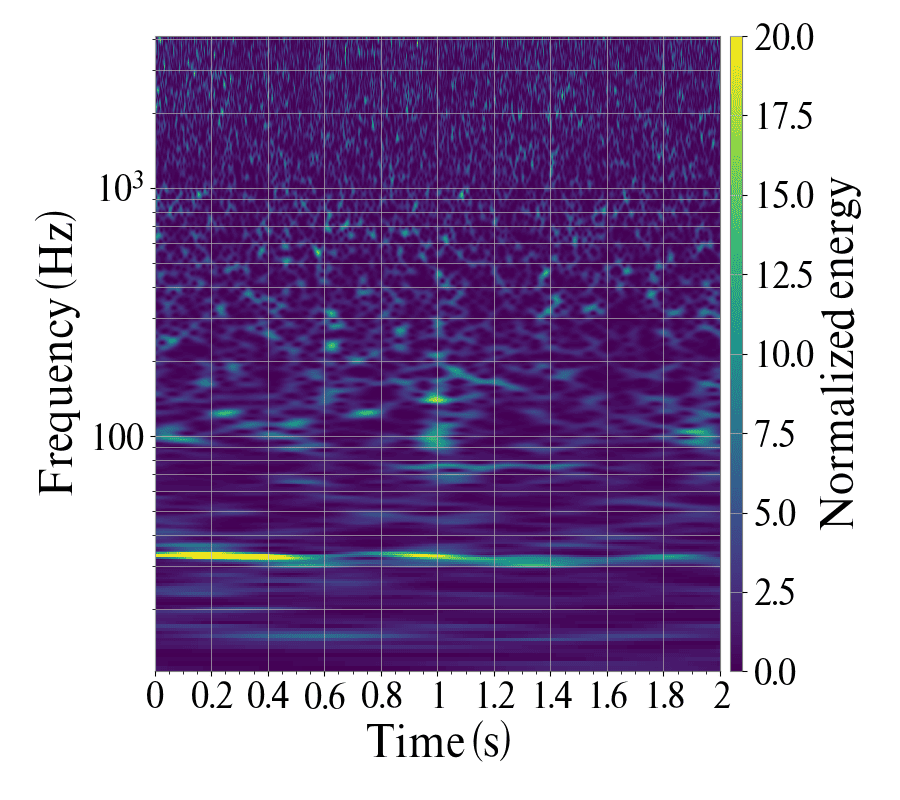}
    \includegraphics[width=0.48\textwidth]{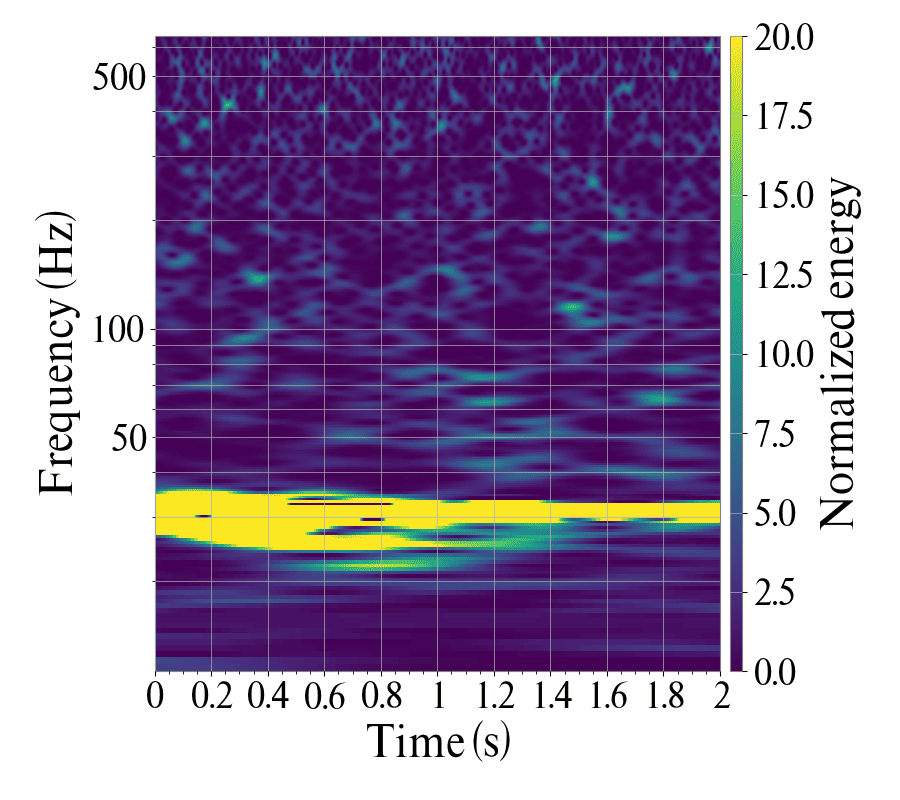}
    \par\smallskip
    {\scriptsize \scalebox{0.8}[1.0]{\textbf{K1:VIS-SR3\_TM\_OPLEV\_TILT\_YAW\_OUT\_DQ}}\\ Apr 11, 2020	09:22:38	UTC (GPS: 1270632176)}
  \end{minipage}
    }

\resizebox{!}{0.136\textheight}
{%
\begin{minipage}[t]{0.45\textwidth}
    \centering
    \includegraphics[width=0.48\textwidth]{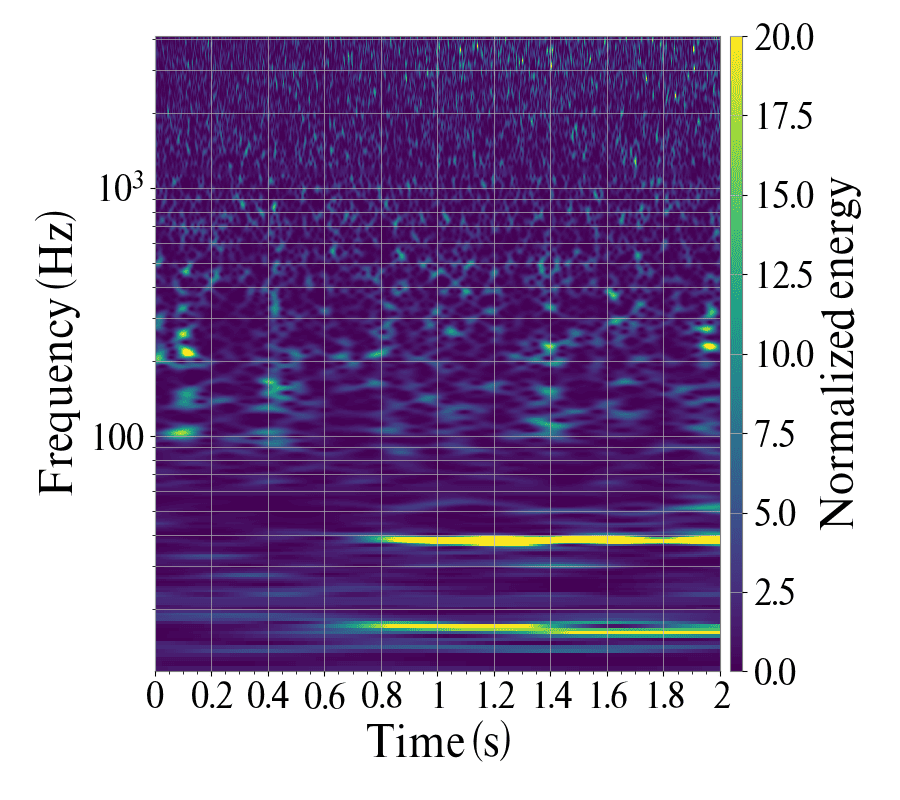}
    \includegraphics[width=0.48\textwidth]{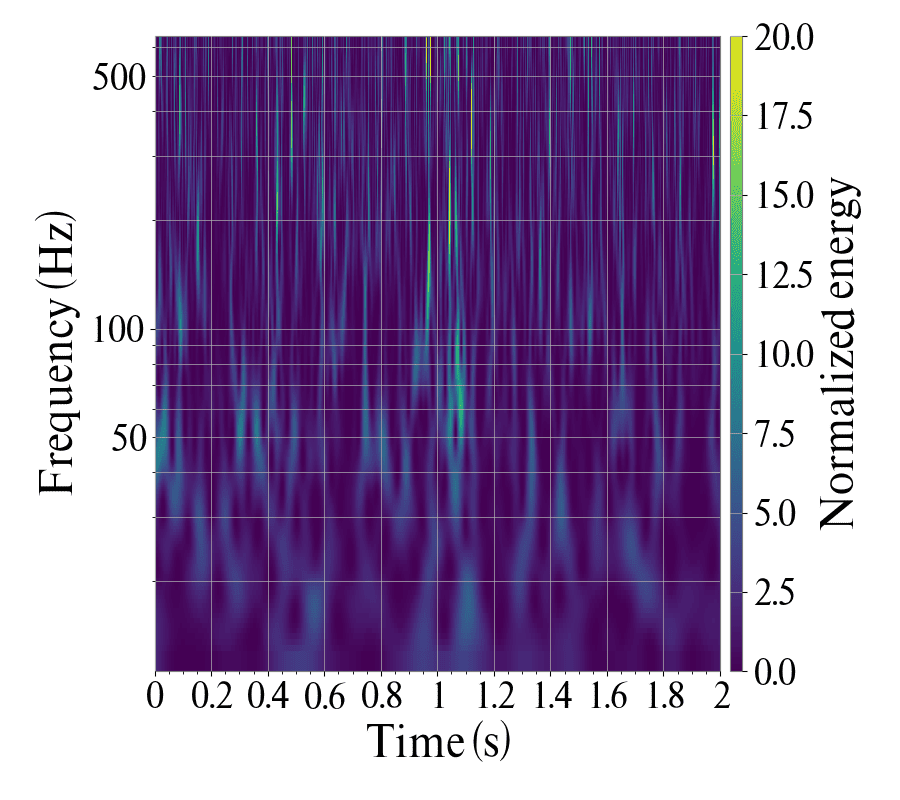}
    \par\smallskip
    {\scriptsize \scalebox{0.8}[1.0]{\textit{K1:VIS-TMSY\_DAMP\_R\_IN1\_DQ}}\\ Apr 20, 2020	12:07:43	UTC (GPS: 1271419681)}
  \end{minipage}
}
  \caption{An O3GK KAGRA line glitch found. The left and right panels show the spectrogram of the main channel and an auxiliary channel, respectively. Each figure also includes, below the panels, the name of the auxiliary channel and the UTC time (GPS time) when the glitch was found.}
  \label{fig:Linefigs}
\end{figure*}

\subsubsection{Line Glitch}
Line glitch appears as horizontal lines in the main channel spectrogram. A unique feature of the line glitch is that multiple horizontal lines can often be seen in a single spectrogram, which may be caused by resonance. 
During the O3GK period, a total of 478 line glitches were found in five KAGRA subsystems (see Tables \ref{table:Glitchtypes} and \ref{table:LineGlitches}). Note that neither dot nor line glitches were found in the LAS subsystem, while blip, helix, and scratchy glitches were found in all six subsystems.

Figure \ref{fig:Linefigs} shows examples of the KAGRA O3GK line glitch. 
The left panels of figure \ref{fig:Linefigs} (main channel spectrograms) show the resonance features of the line glitch, which occur at 16, 24, and 32~Hz. We consider the horizontal consecutive dots appearing in the left panels as the spectrogram of K1:IMC-IMMT1\_TRANS\_QPDA1\_DC\_YAW\_OUT\_DQ (near about 200 Hz) and K1:VIS-TMSY\_DAMP\_R\_IN1\_DQ (near about 100 and 200 Hz) line glitches. 
The auxiliary channels, such as K1:VIS-OSTM\_TM\_OPLEV\_YAW\_OUT\_DQ and K1:IMC-IMMT1\_TRANS\_QPDA1\_DC\_YAW\_OUT\_DQ, are marked with the bold font in Table \ref{table:LineGlitches} because the right panel covers the region of line glitches in the left panel or shows a pattern similar to the left panel. The auxiliary channel, such as K1:VIS-TMSY\_DAMP\_R\_IN1\_DQ is indicated in italics, because the auxiliary channel does not show correlation with the main channel.

\subsubsection{Scratchy Glitch}

\begin{figure*}[h]
\centering
\resizebox{!}{0.136\textheight}
{%
\begin{minipage}[t]{0.45\textwidth}
    \centering
    \includegraphics[width=0.48\textwidth]{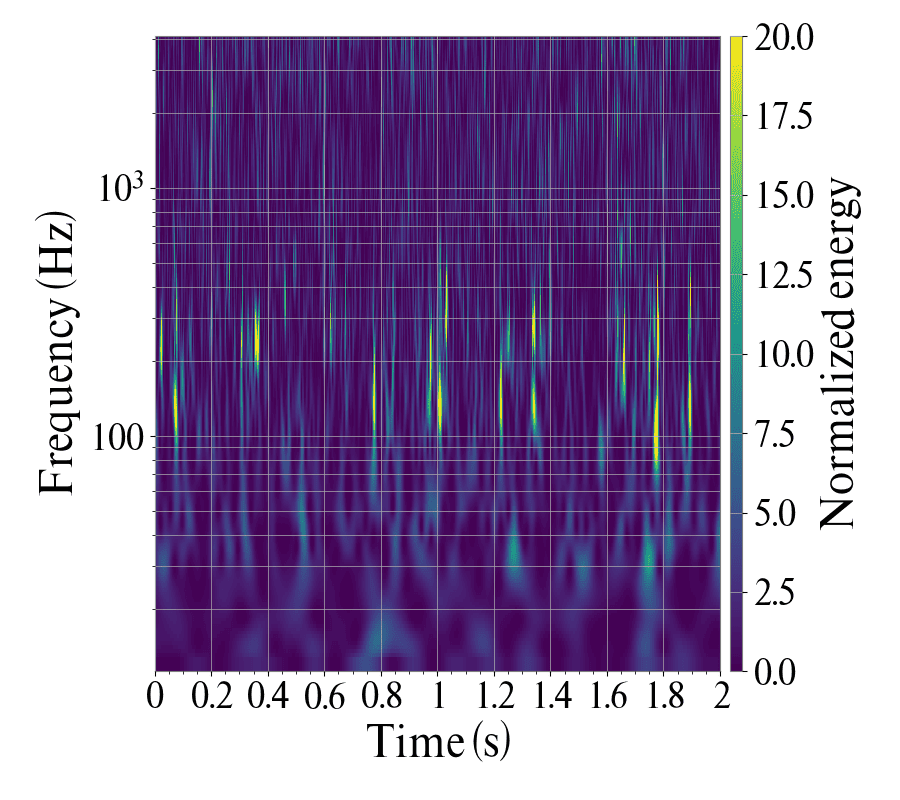}
    \includegraphics[width=0.48\textwidth]{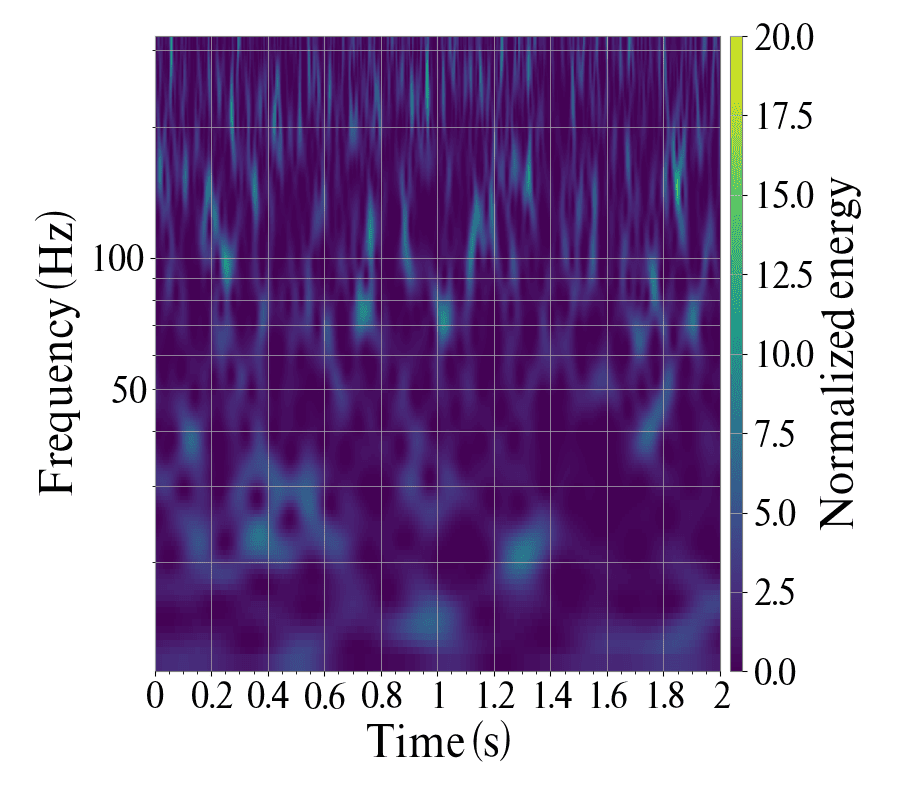}
    \par\smallskip
    {\scriptsize \scalebox{0.8}[1.0]{\textit{K1:VIS-ETMY\_MN\_PSDAMP\_Y\_IN1\_DQ}}\\ Apr 19, 2020	17:15:09	UTC	(GPS: 1271351727)}
    \end{minipage}
\hfill
\vrule width 0.5pt
\hfill
\begin{minipage}[t]{0.45\textwidth}
    \centering
    \includegraphics[width=0.48\textwidth]{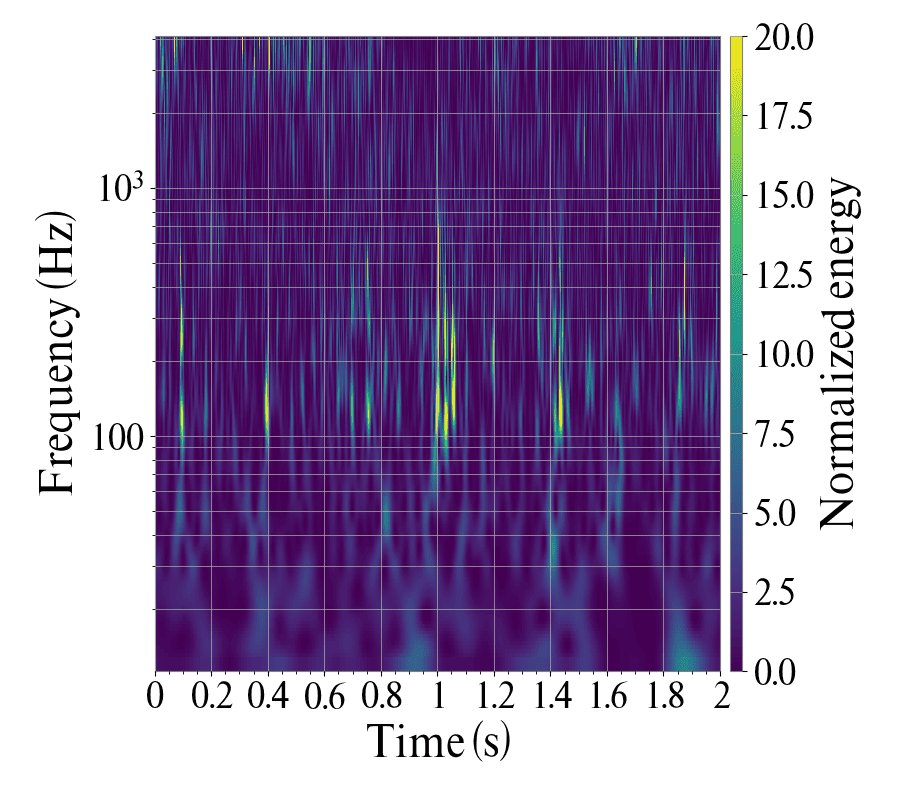}
    \includegraphics[width=0.48\textwidth]{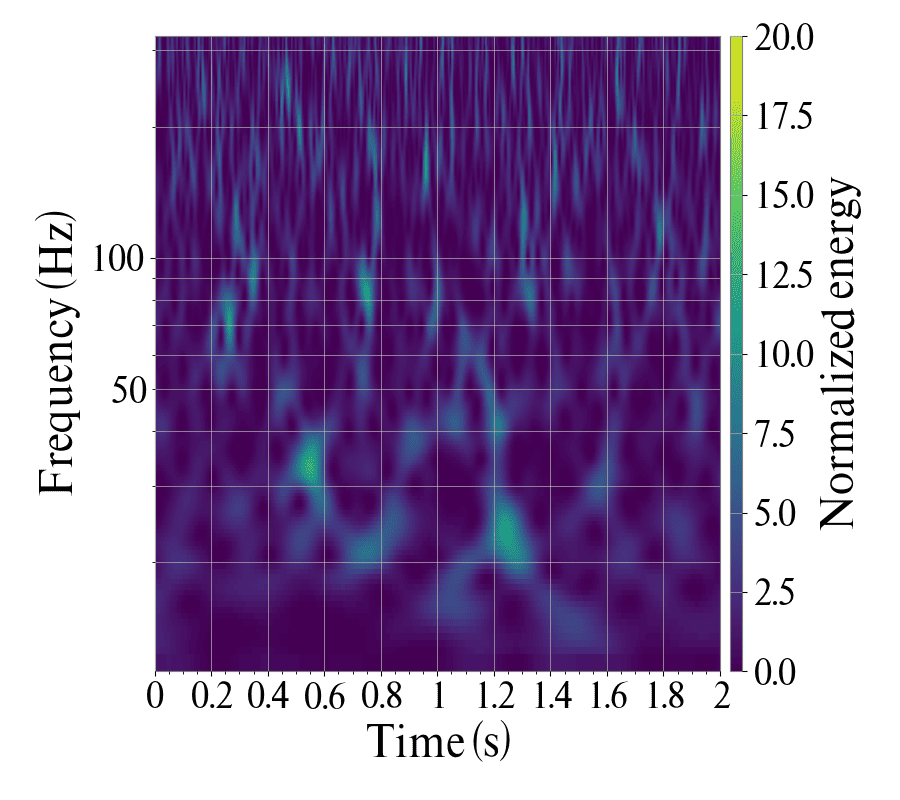}
    \par\smallskip
    {\scriptsize \scalebox{0.8}[1.0]{\textit{K1:VIS-ITMY\_MN\_OPLEV\_TILT\_YAW\_OUT\_DQ}}\\ Apr 09, 2020	10:50:18	UTC	(GPS: 1270464636)}
    \end{minipage}
    }

\resizebox{!}{0.136\textheight}
{%
\begin{minipage}[t]{0.45\textwidth}
    \centering
    \includegraphics[width=0.48\textwidth]{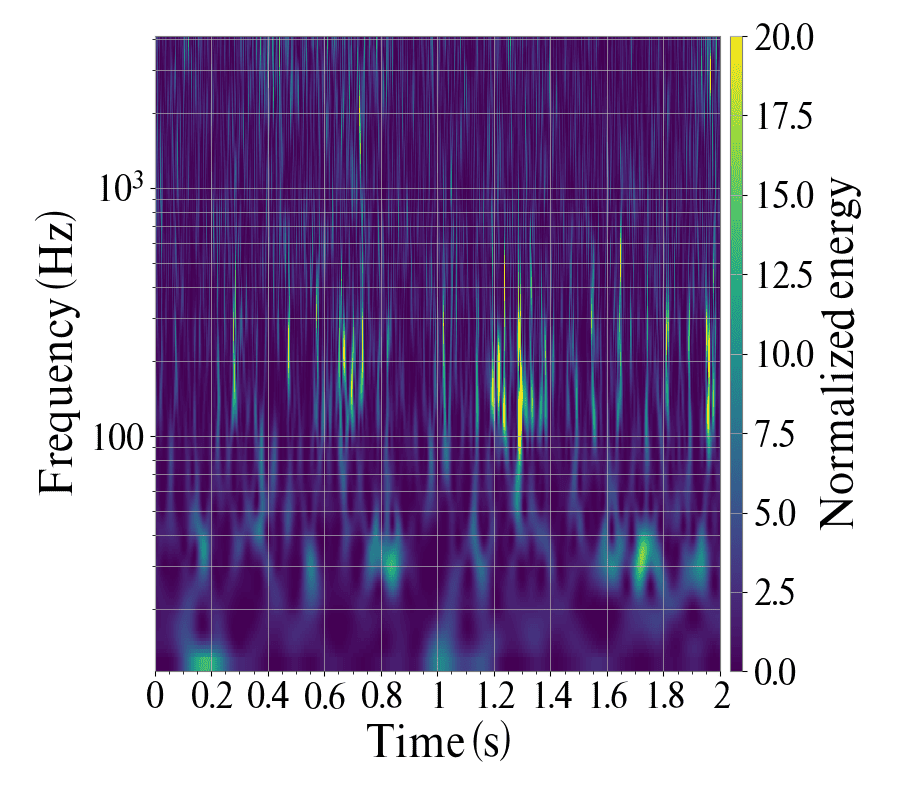}
    \includegraphics[width=0.48\textwidth]{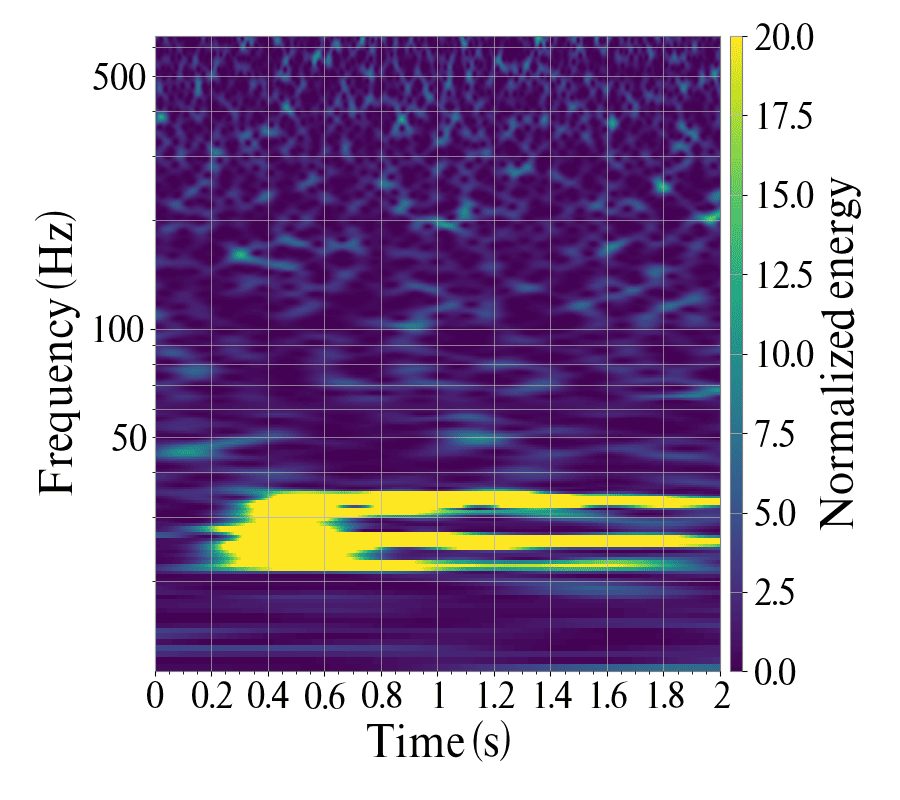}
    \par\smallskip
    {\scriptsize \scalebox{0.8}[1.0]{\textit{K1:VIS-OMMT1\_TM\_OPLEV\_PIT\_OUT\_DQ}}\\ Apr 08, 2020	22:25:58	UTC (GPS: 1270419976)}
    \end{minipage}
\hfill
\vrule width 0.5pt
\hfill
\begin{minipage}[t]{0.45\textwidth}
    \centering
    \includegraphics[width=0.48\textwidth]{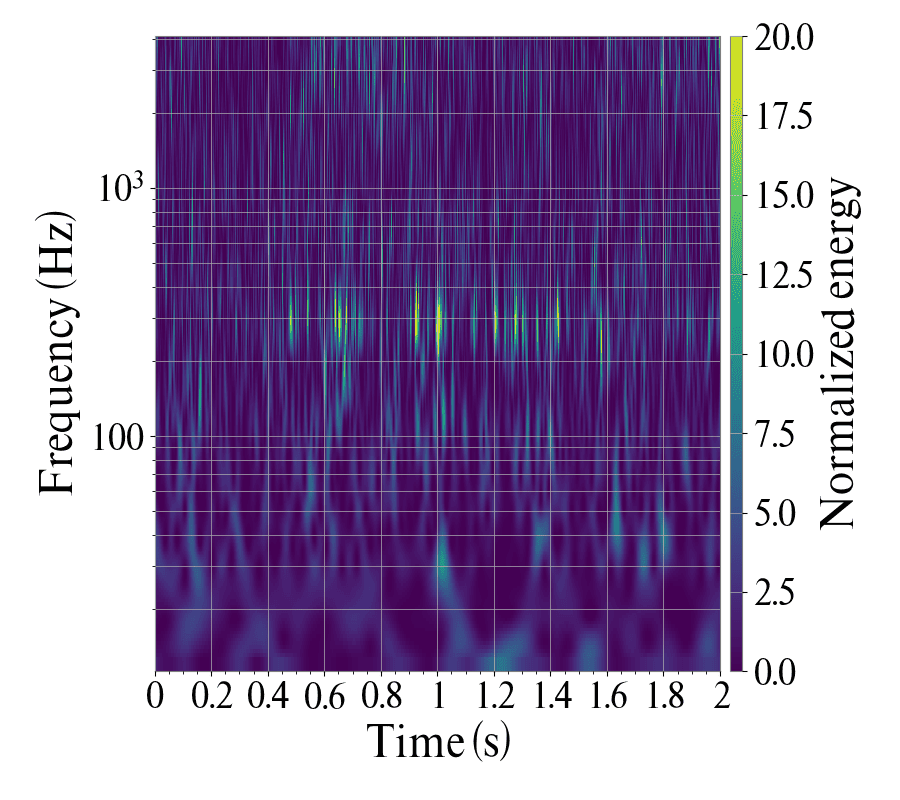}
    \includegraphics[width=0.48\textwidth]{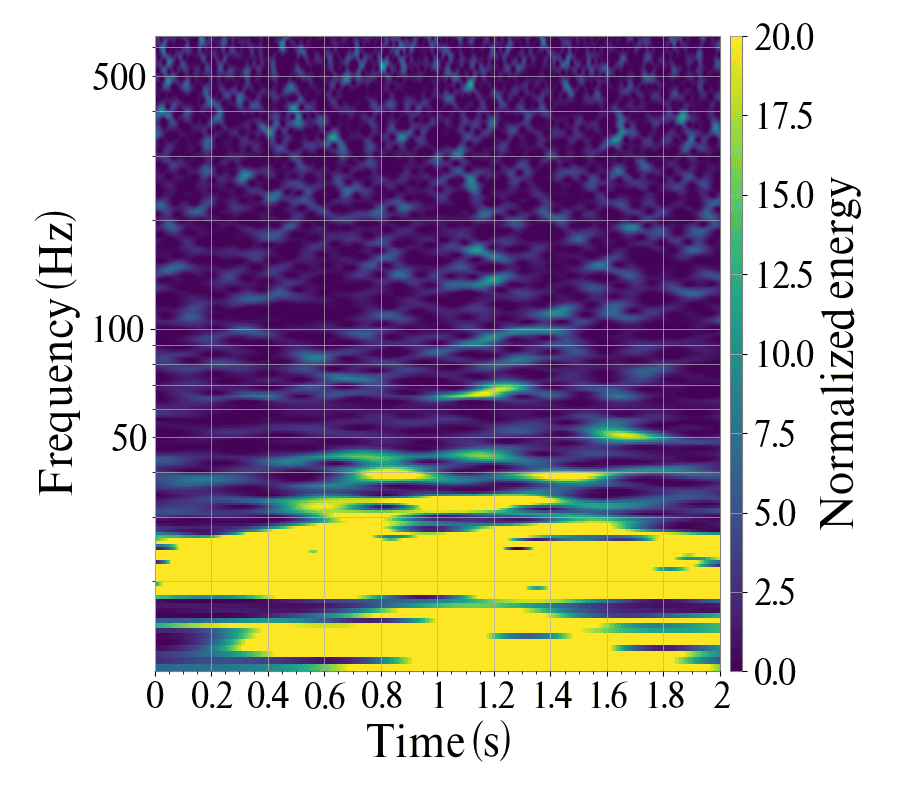}
    \par\smallskip
    {\scriptsize \scalebox{0.8}[1.0]{\textit{K1:VIS-OMMT1\_TM\_OPLEV\_YAW\_OUT\_DQ}}\\ Apr 11, 2020	20:26:31	UTC (GPS: 1270672009)}
    \end{minipage}
    }

\resizebox{!}{0.136\textheight}
{%
\begin{minipage}[t]{0.45\textwidth}
    \centering
    \includegraphics[width=0.48\textwidth]{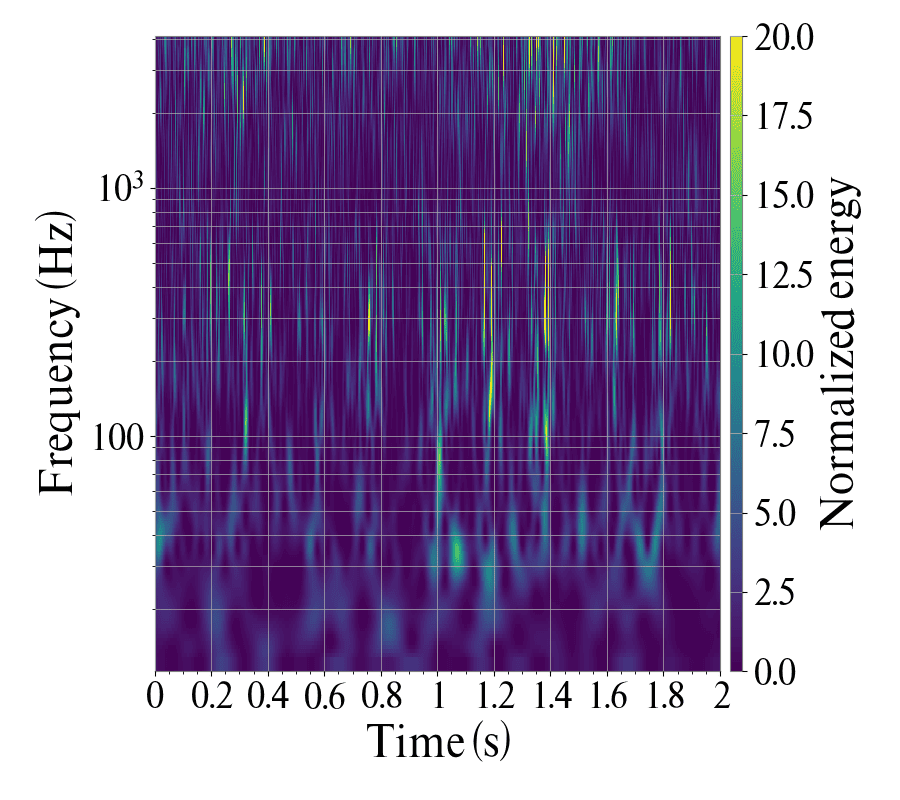}
    \includegraphics[width=0.48\textwidth]{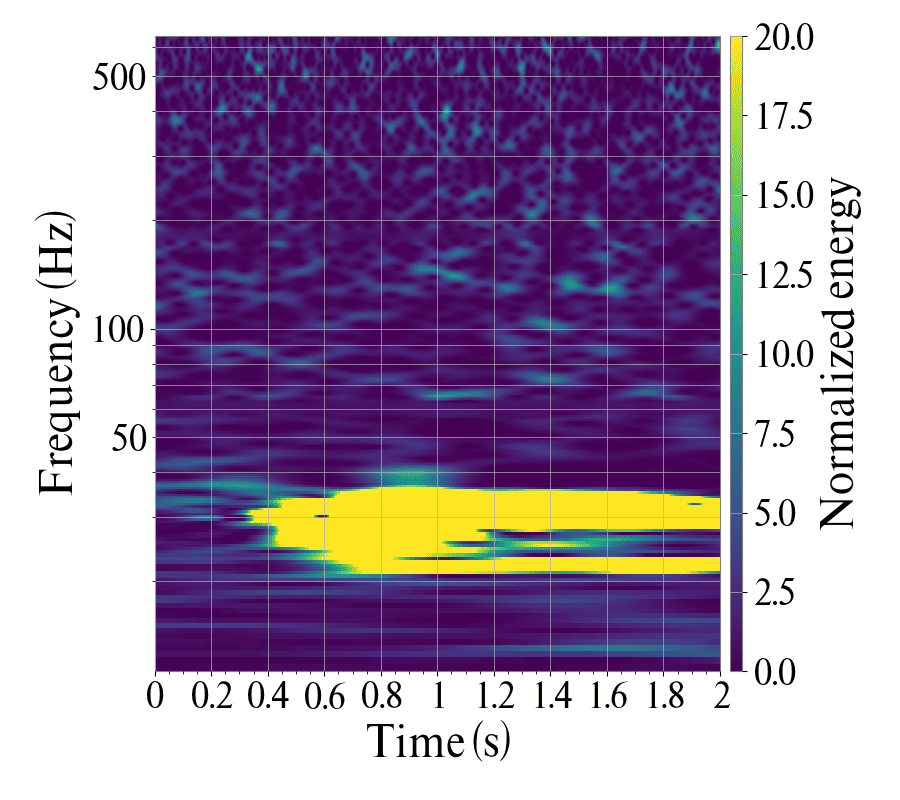}
    \par\smallskip
    {\scriptsize \scalebox{0.8}[1.0]{\textit{K1:VIS-OSTM\_TM\_OPLEV\_YAW\_OUT\_DQ}}\\ Apr 15, 2020	14:29:19	UTC (GPS: 1270996177)}
    \end{minipage}
  \hfill
  \vrule width 0.5pt
  \hfill
\begin{minipage}[t]{0.45\textwidth}
    \centering
    \includegraphics[width=0.48\textwidth]{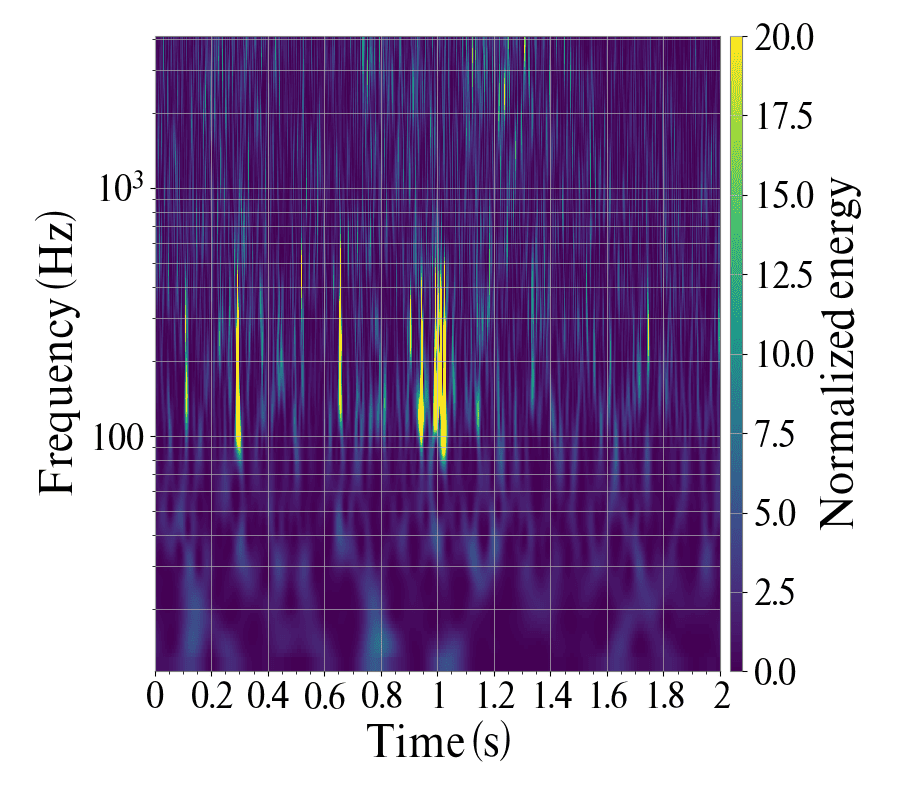}
    \includegraphics[width=0.48\textwidth]{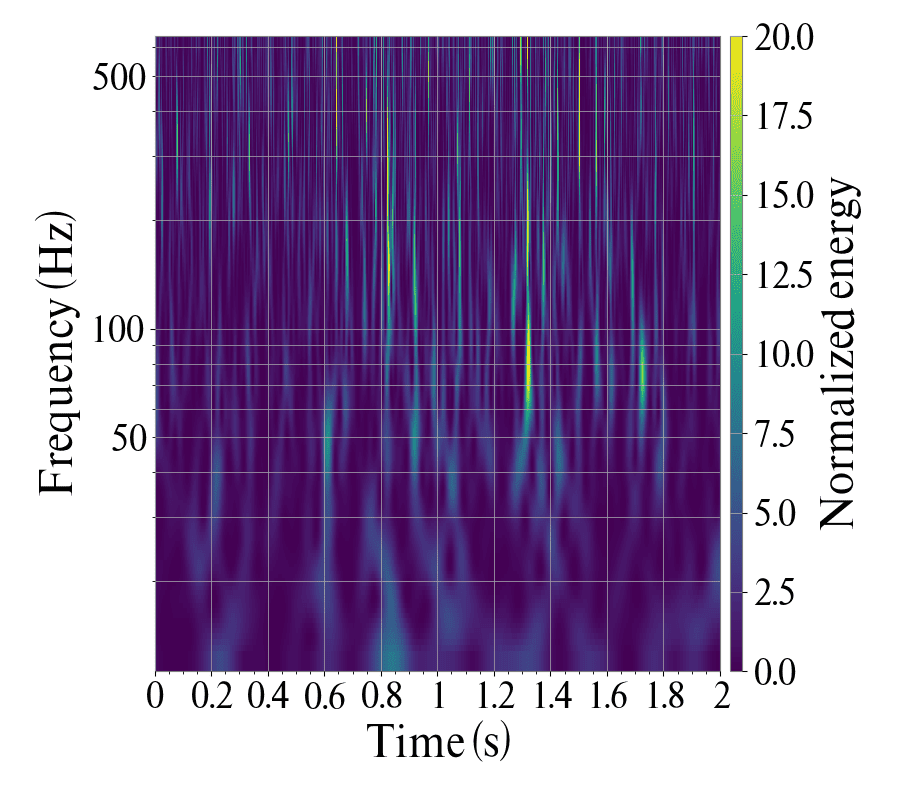}
    \par\smallskip
    {\scriptsize \scalebox{0.8}[1.0]{\textit{K1:VIS-TMSY\_DAMP\_R\_IN1\_DQ}}\\ Apr 20, 2020	09:20:29	UTC (GPS: 1271409647)}
  \end{minipage}
  }
  \caption{An O3GK KAGRA scratchy glitch found. The left and right panels show the spectrogram of the main channel and an auxiliary channel, respectively. Each figure also includes, below the panels, the name of the auxiliary channel and the UTC time (GPS time) when the glitch was found.}
  \label{fig:Scratchyfigs}
\end{figure*}

Scratchy glitch is composed of many vertical lines, which are confined within a certain region on the main channel spectrogram. Individual lines may look like a blip glitch, but the scratchy glitch is classified with multiple blip-like glitches over a relatively long period of time within a certain frequency range. Figure \ref{fig:Scratchyfigs} shows examples of the KAGRA O3GK scratchy glitch. Following convention, the spectrogram such as K1:LAS-POW\_FIB\_OUT\_DQ (K1:PEM-SEIS\_IXV\_GND\_EW\_IN1\_DQ) in Figure \ref{fig:Scratchyfigs}  is marked with a bold (italic) font in Table \ref{table:ScratchyGlitches} because the pattern in the main channel spectrogram is similar to (different from) that in the auxiliary channel. 
During the O3GK period, a total of 157 scratchy glitches were detected (see Tables \ref{table:Glitchtypes} and \ref{table:ScratchyGlitches}). As mentioned already, the O3GK KAGRA scratchy glitches were found in all six subsystems, as the blip and helix glitches.

\begin{table*}[h]
\centering
\caption{\label{table:ScratchyGlitches} O3GK KAGRA scratchy glitches}
\small
\resizebox{0.98\textwidth}{!}
{%
\begin{tabular}{|P{0.07\textwidth}|P{0.58\textwidth}|P{0.35\textwidth}|}
\hline
\begin{tabular}[c]{@{}c@{}}Sub-\\ System\end{tabular} & \begin{tabular}[c]{@{}c@{}}Round Winner\\ Auxiliary Channel\end{tabular} & \begin{tabular}[c]{@{}c@{}}Vetoed Date in April\\ (\# of Vetoed Events)\end{tabular} \\ \hline

\begin{tabular}[c]{@{}c@{}}AOS \\ (31) \end{tabular}
& \textit{K1:AOS-TMSX\_IR\_PDA1\_OUT\_DQ}
& \begin{tabular}[c]{@{}c@{}}10th (2), 16th (15), \\ 17th (5), 18th (9)\end{tabular} \\ \hline

\multirow{2}{*}{\begin{tabular}[c]{@{}c@{}} IMC \\ (44) \end{tabular}}
& \begin{tabular}[c]{@{}c@{}}\textit{K1:IMC-IMMT1}\\ \textit{\_TRANS\_QPDA1\_DC\_PIT\_OUT\_DQ}\end{tabular}  
& \begin{tabular}[c]{@{}c@{}}8th (7), 9th (13),\\ 10th (6), 11th (3)\end{tabular}\\ \cline{2-3} 
& \begin{tabular}[c]{@{}c@{}}\textit{K1:IMC-IMMT1}\\ \textit{\_TRANS\_QPDA1\_DC\_YAW\_OUT\_DQ}\end{tabular} 
& 9th (4), 11th (7), 19th (4)                                                          \\ \hline

\begin{tabular}[c]{@{}c@{}}LAS \\ (12) \end{tabular}
& \textbf{K1:LAS-POW\_FIB\_OUT\_DQ}
& 7th (12) \\ \hline

\begin{tabular}[c]{@{}c@{}}LSC \\ (31) \end{tabular}
& \begin{tabular}[c]{@{}c@{}}\textbf{K1:LSC-ALS\_DARM\_OUT\_DQ}\end{tabular}
& 12th (8), 15th (21), 16th (2) 
\\ \hline

\multirow{3}{*}{\begin{tabular}[c]{@{}c@{}}PEM \\ (5) \end{tabular}}
& \textit{K1:PEM-MAG\_BS\_BOOTH\_BS\_X\_OUT\_DQ}
& 16th (1) 
\\ \cline{2-3}
& \textit{K1:PEM-MIC\_SR\_BOOTH\_SR\_Z\_OUT\_DQ}
& 16th (1)
\\ \cline{2-3}
& \textit{K1:PEM-SEIS\_IXV\_GND\_EW\_IN1\_DQ}
& 7th (3) 
\\ \hline

\multirow{8}{*}{\begin{tabular}[c]{@{}c@{}}VIS \\ (34) \end{tabular}}
& \textit{K1:VIS-ETMX\_MN\_PSDAMP\_R\_IN1\_DQ}
& 7th (5)
\\ \cline{2-3}
& \textit{K1:VIS-ETMX\_MN\_PSDAMP\_Y\_IN1\_DQ}
& 7th (3)
\\ \cline{2-3}
& \textit{K1:VIS-ETMY\_MN\_PSDAMP\_Y\_IN1\_DQ}
& 19th (3)
\\ \cline{2-3}
& \textit{K1:VIS-ITMY\_MN\_OPLEV\_TILT\_YAW\_OUT\_DQ}
& 9th(1)
\\ \cline{2-3}
& \textit{K1:VIS-OMMT1\_TM\_OPLEV\_PIT\_OUT\_DQ}
& \begin{tabular}[c]{@{}c@{}}8th (3), 9th (1), \\ 12th (1), 18th (3)\end{tabular} 
\\ \cline{2-3}
& \textit{K1:VIS-OMMT1\_TM\_OPLEV\_YAW\_OUT\_DQ}
& 11th (5), 18th (1)
\\ \cline{2-3}
& \textit{K1:VIS-OSTM\_TM\_OPLEV\_YAW\_OUT\_DQ}
& 7th (2), 15th(3)
\\ \cline{2-3}
& \textit{K1:VIS-TMSY\_DAMP\_R\_IN1\_DQ}
& 20th (3)
\\ \hline

\end{tabular} }
\end{table*}

\begin{figure}[htpb]
\resizebox{!}{0.136\textheight}
{%
\centering
\begin{minipage}[t]{0.45\textwidth}
    \centering
    \includegraphics[width=0.48\textwidth]{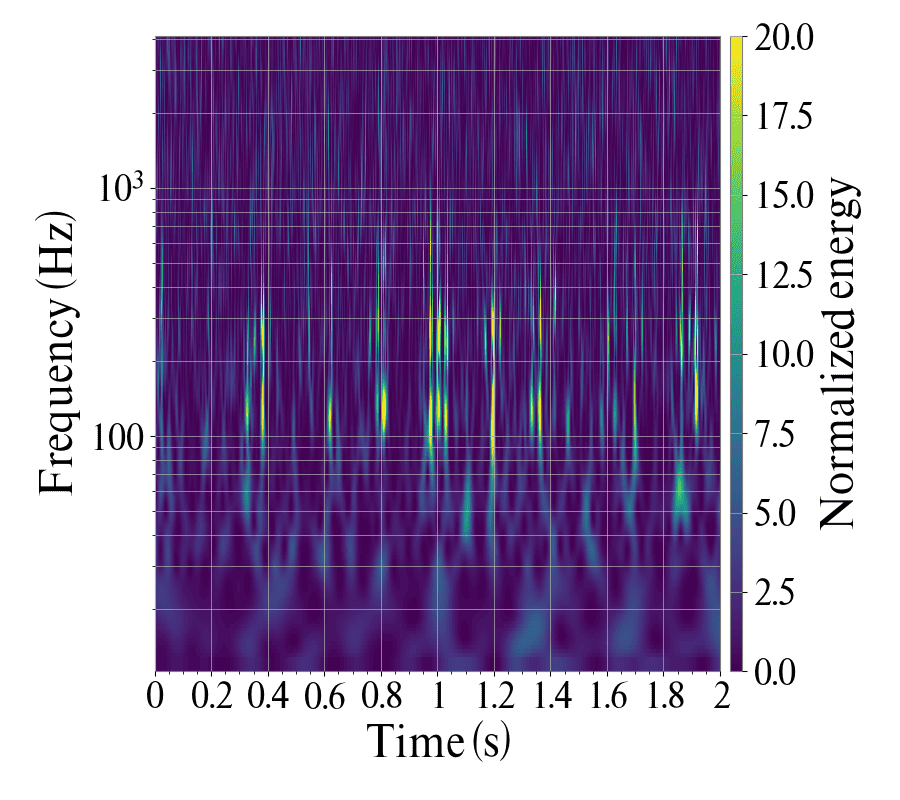}
    \includegraphics[width=0.48\textwidth]{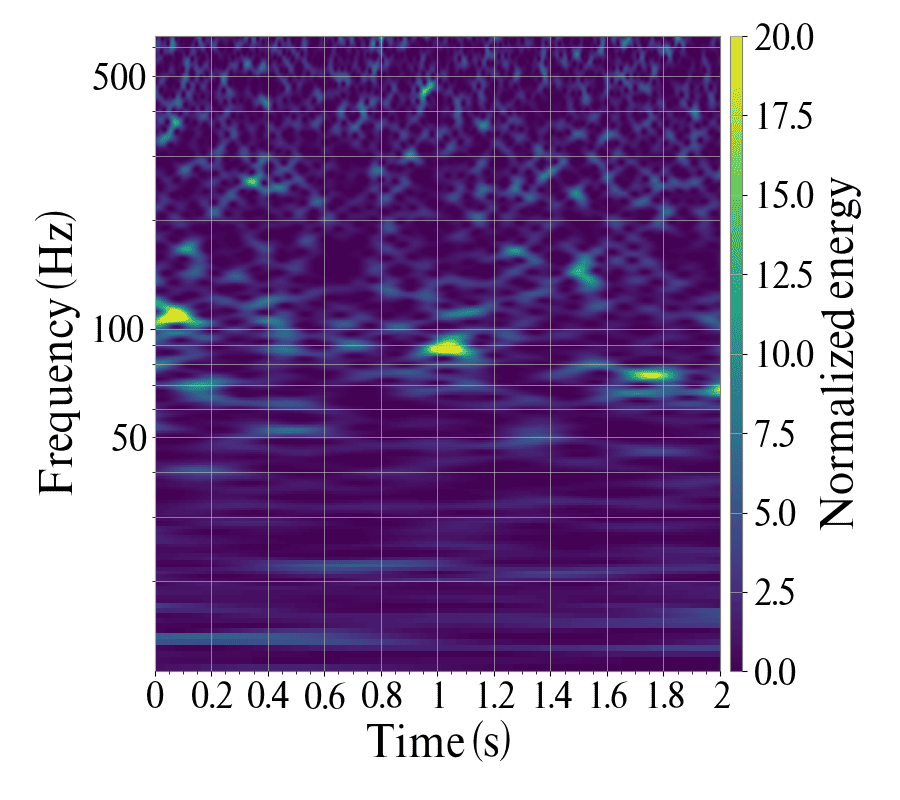}
    \par\smallskip
    {\scriptsize \scalebox{0.8}[1.0]{\textit{K1:AOS-TMSX\_IR\_PDA1\_OUT\_DQ}} \\ Apr 16, 2020	20:33:57	UTC (GPS: 1271104455)}
    \end{minipage}
\hfill
\vrule width 0.5pt
\hfill
  \begin{minipage}[t]{0.45\textwidth}
    \centering
    \includegraphics[width=0.48\textwidth]{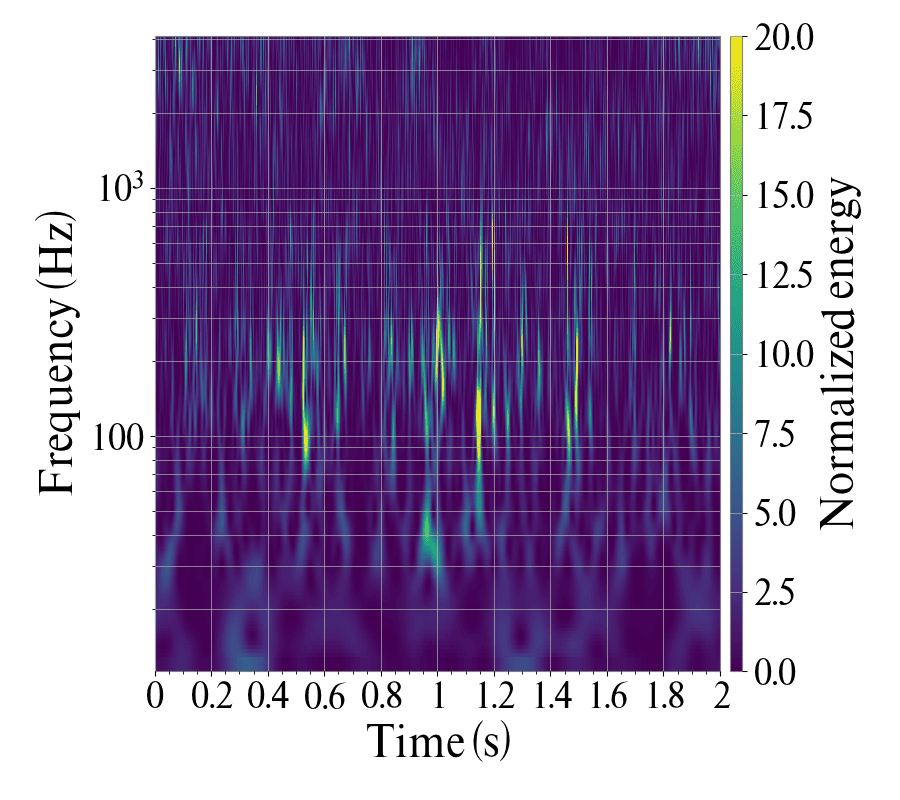}
    \includegraphics[width=0.48\textwidth]{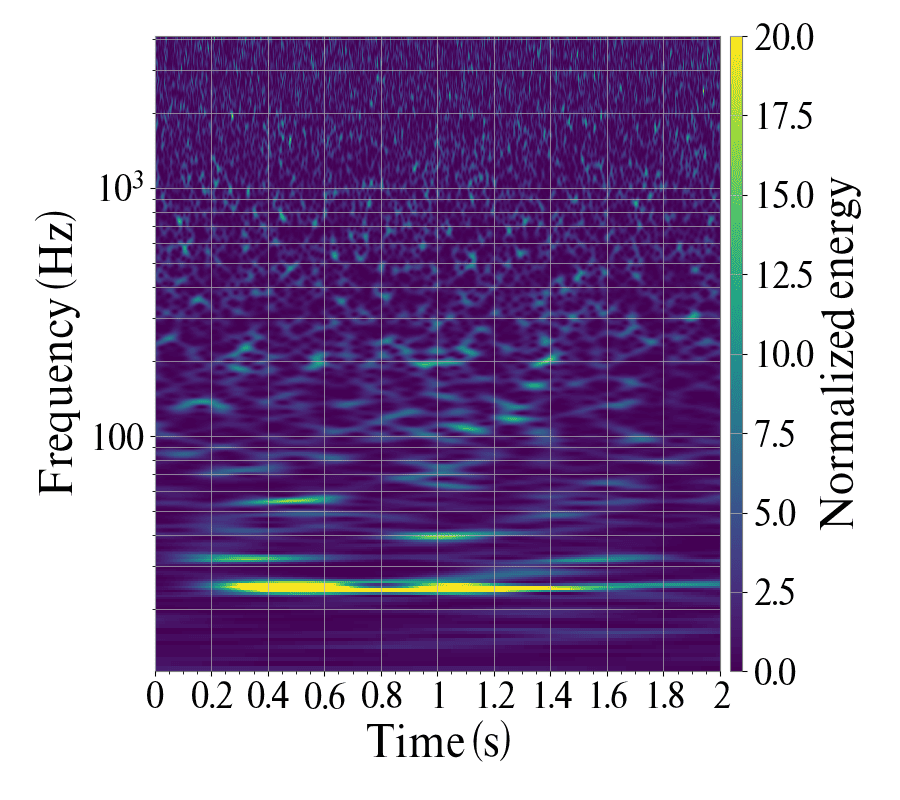}
    \par\smallskip
    {\scriptsize \scalebox{0.8}[1.0]{\textit{K1:IMC-IMMT1\_TRANS\_QPDA1\_DC\_PIT\_OUT\_DQ}} \\ Apr 09, 2020	00:13:38	UTC (GPS: 1270426436)}
    \end{minipage}
    }

\resizebox{!}{0.136\textheight}
{%
\begin{minipage}[t]{0.45\textwidth}
    \centering
    \includegraphics[width=0.48\textwidth]{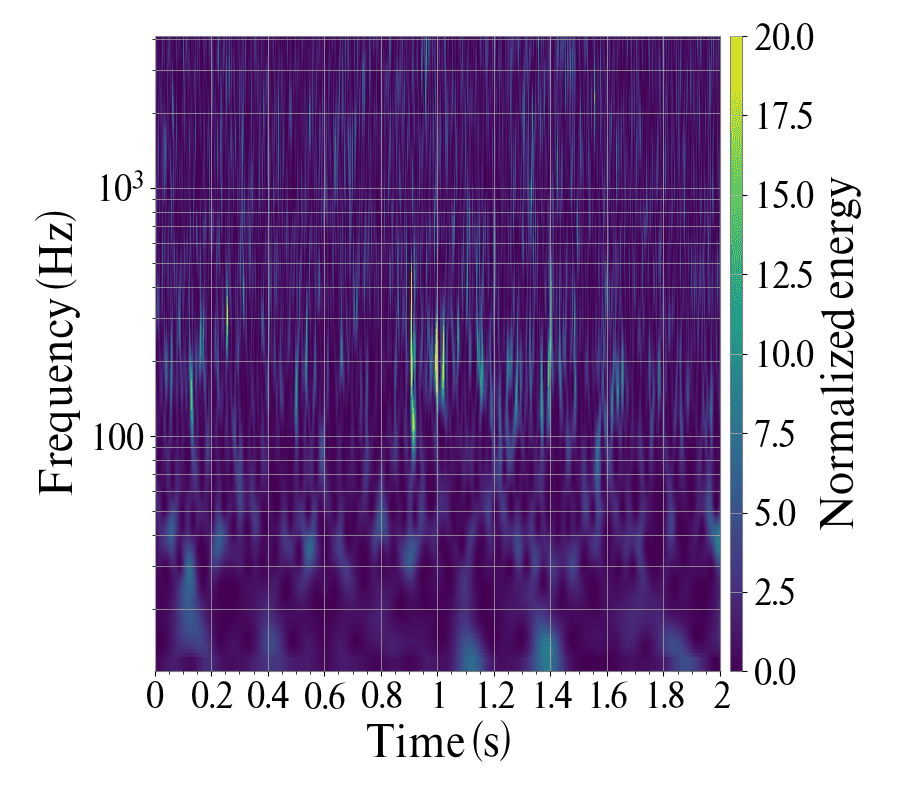}
    \includegraphics[width=0.48\textwidth]{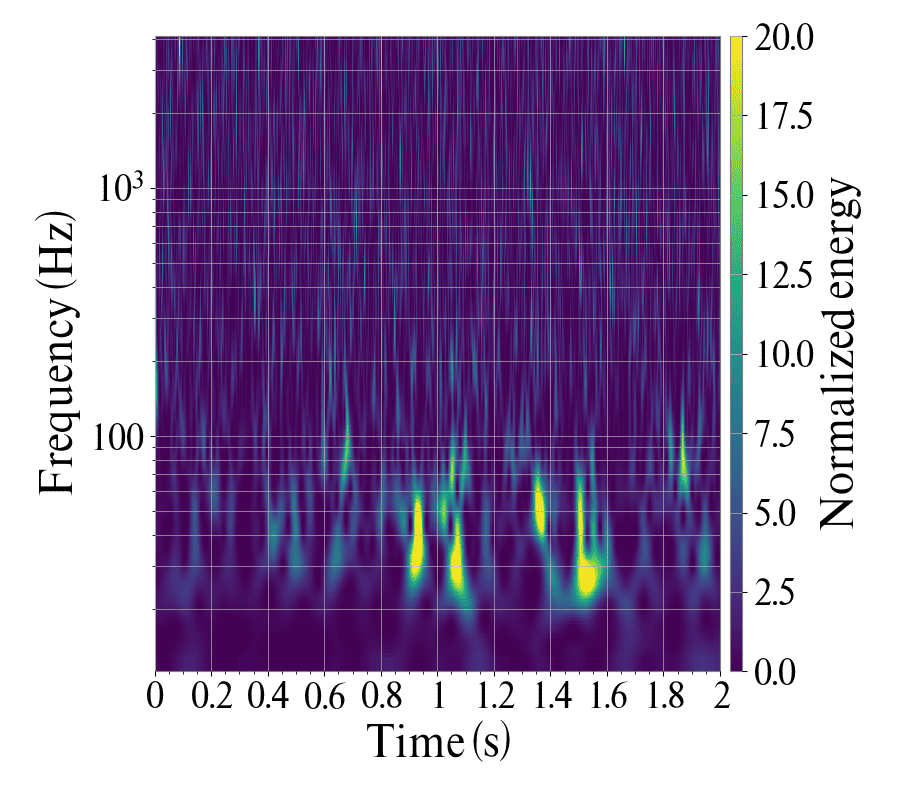}
    \par\smallskip
    {\scriptsize \scalebox{0.8}[1.0]{\textit{K1:IMC-IMMT1\_TRANS\_QPDA1\_DC\_YAW\_OUT\_DQ}} \\ Apr 11, 2020	05:24:26	UTC (GPS: 1270617884)}
    \end{minipage}
\hfill
\vrule width 0.5pt
\hfill
\begin{minipage}[t]{0.45\textwidth}   
    \centering
    \includegraphics[width=0.48\textwidth]{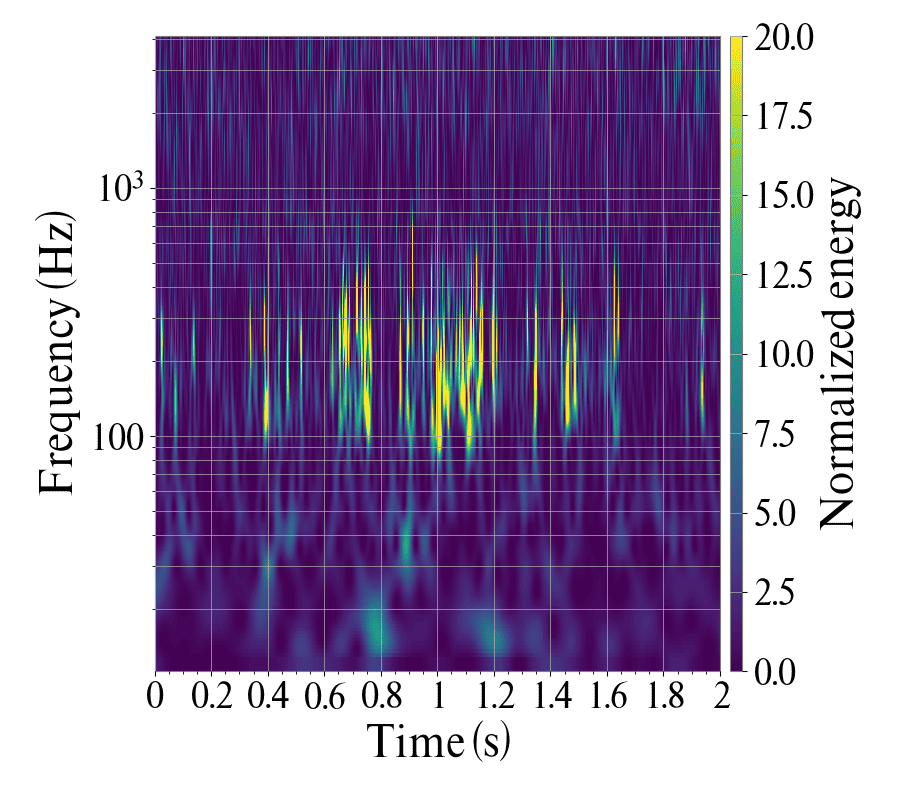}
    \includegraphics[width=0.48\textwidth]{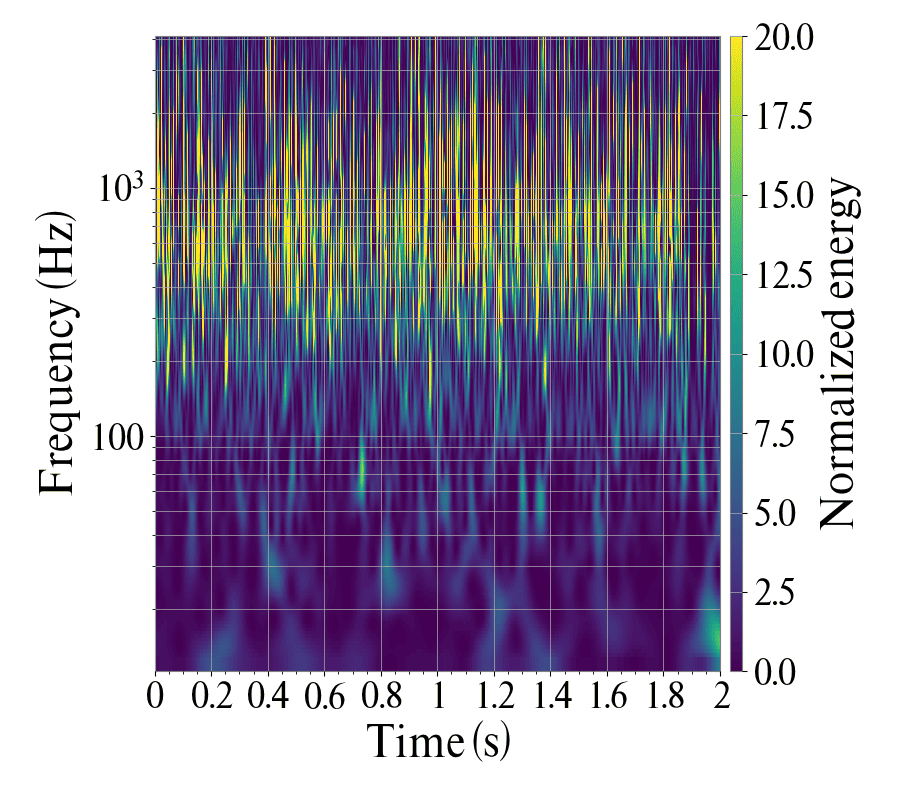}
    \par\smallskip
    {\scriptsize \scalebox{0.8}[1.0]{\textbf{K1:LAS-POW\_FIB\_OUT\_DQ}} \\ Apr 07, 2020	22:29:30	UTC (GPS: 1270333788)}
    \end{minipage}
    }

\resizebox{!}{0.136\textheight}
{%
\begin{minipage}[t]{0.45\textwidth}
    \centering
    \includegraphics[width=0.48\textwidth]{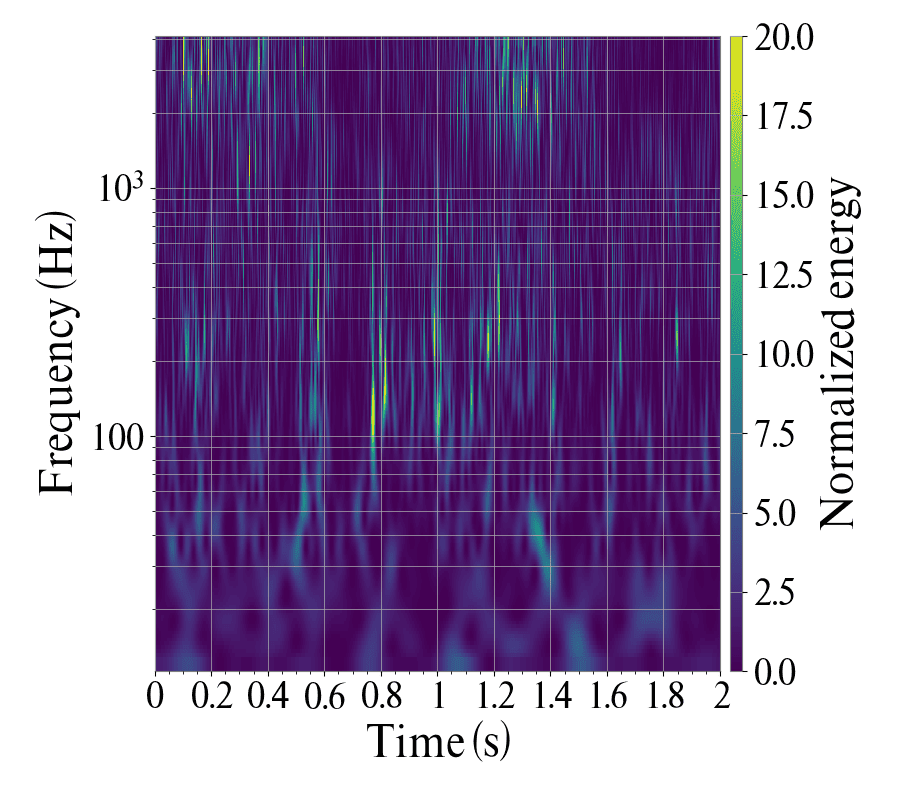}
    \includegraphics[width=0.48\textwidth]{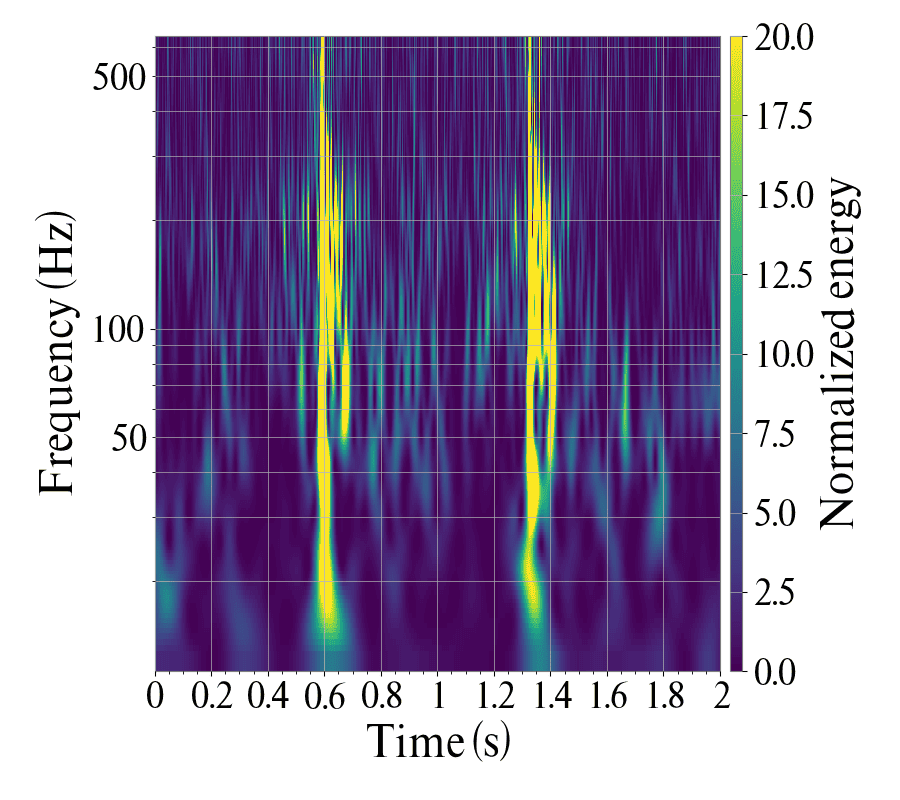}
    \par\smallskip
    {\scriptsize \scalebox{0.8}[1.0]{\textbf{K1:LSC-ALS\_DARM\_OUT\_DQ}} \\ Apr 12, 2020	21:23:06	UTC (GPS: 1270761804)}
    \end{minipage}  
\hfill
\vrule width 0.5pt
\hfill
\begin{minipage}[t]{0.45\textwidth}
    \centering
    \includegraphics[width=0.48\textwidth]{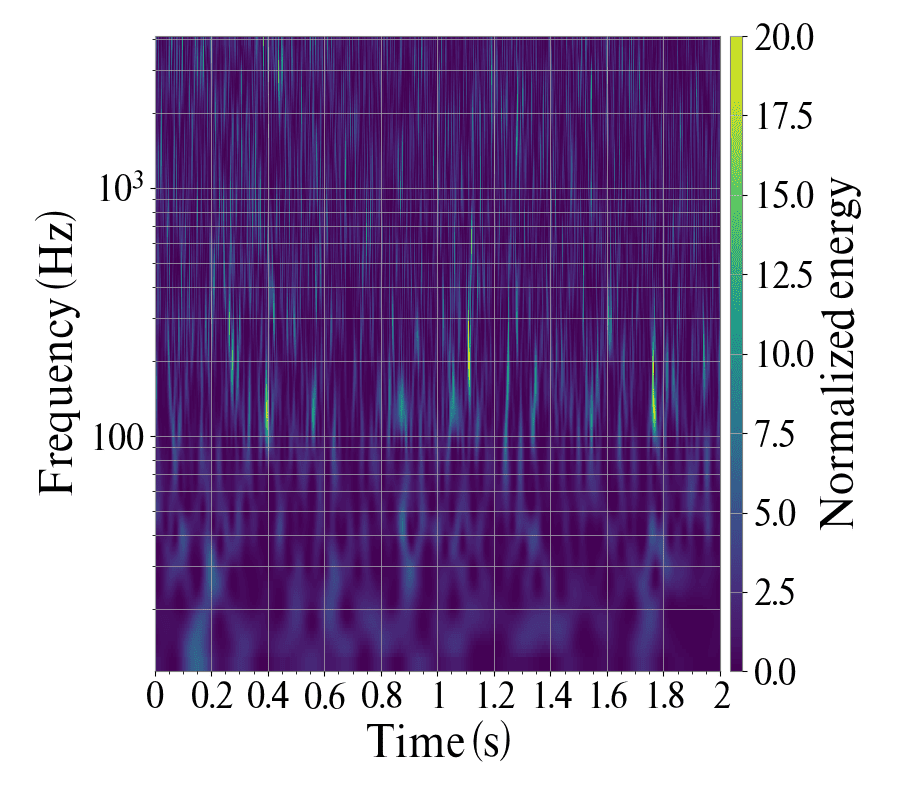}
    \includegraphics[width=0.48\textwidth]{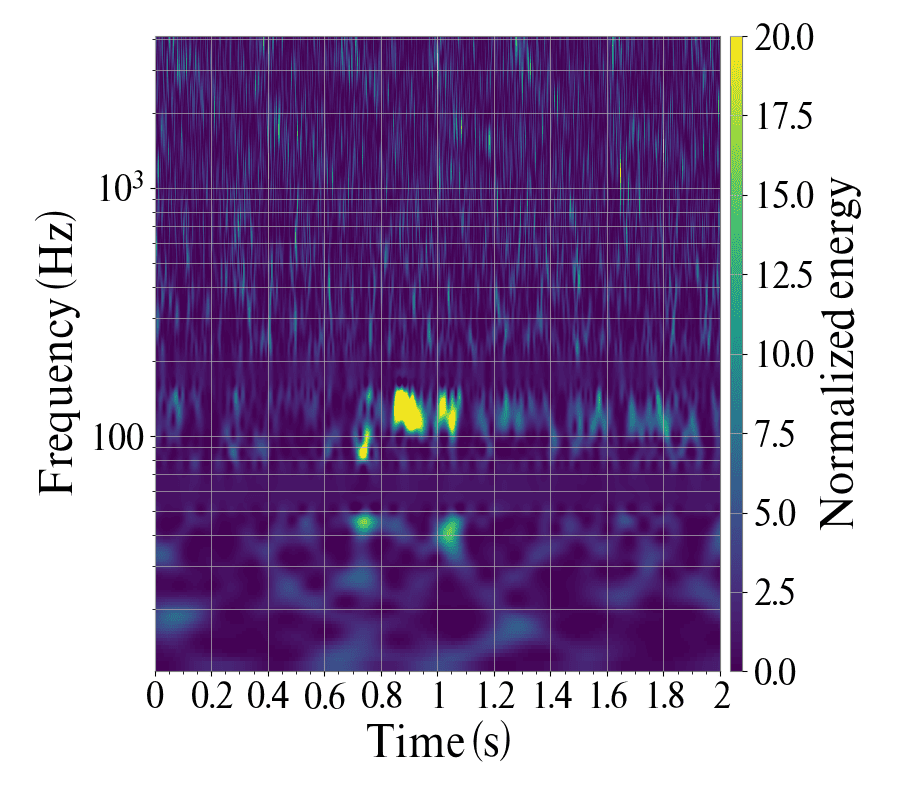}
    \par\smallskip
    {\scriptsize \scalebox{0.8}[1.0]{\textit{K1:PEM-MAG\_BS\_BOOTH\_BS\_X\_OUT\_DQ}} \\ Apr 16, 2020	15:57:36	UTC	(GPS: 1271087874)}
    \end{minipage}
    }

\resizebox{!}{0.136\textheight}
{%
\begin{minipage}[t]{0.45\textwidth}
    \centering
    \includegraphics[width=0.48\textwidth]{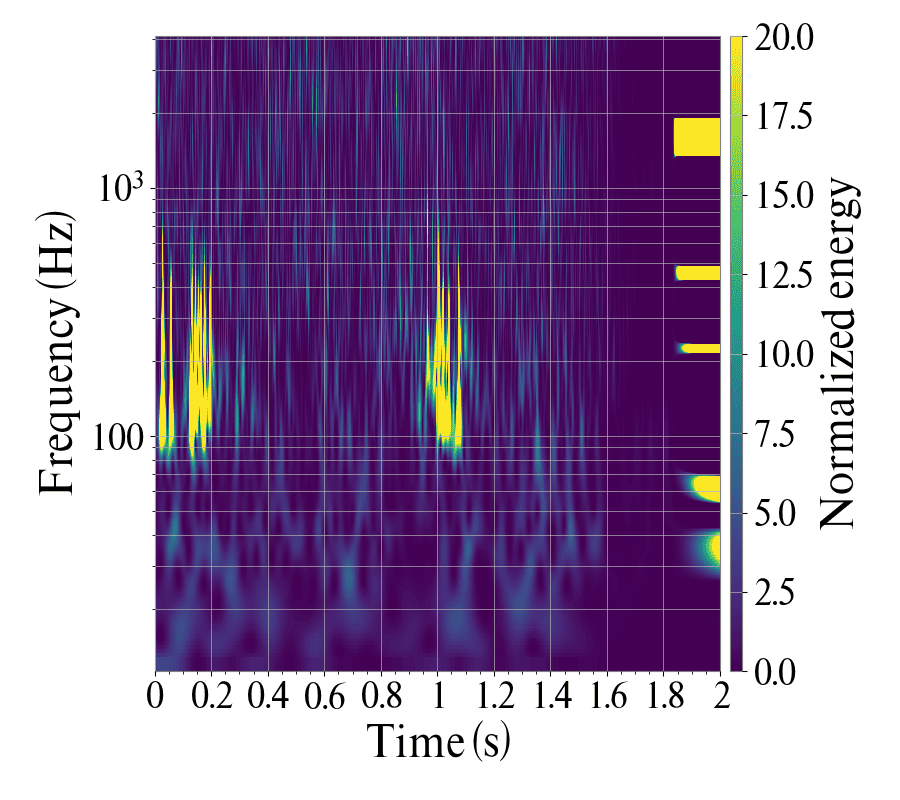}
    \includegraphics[width=0.48\textwidth]{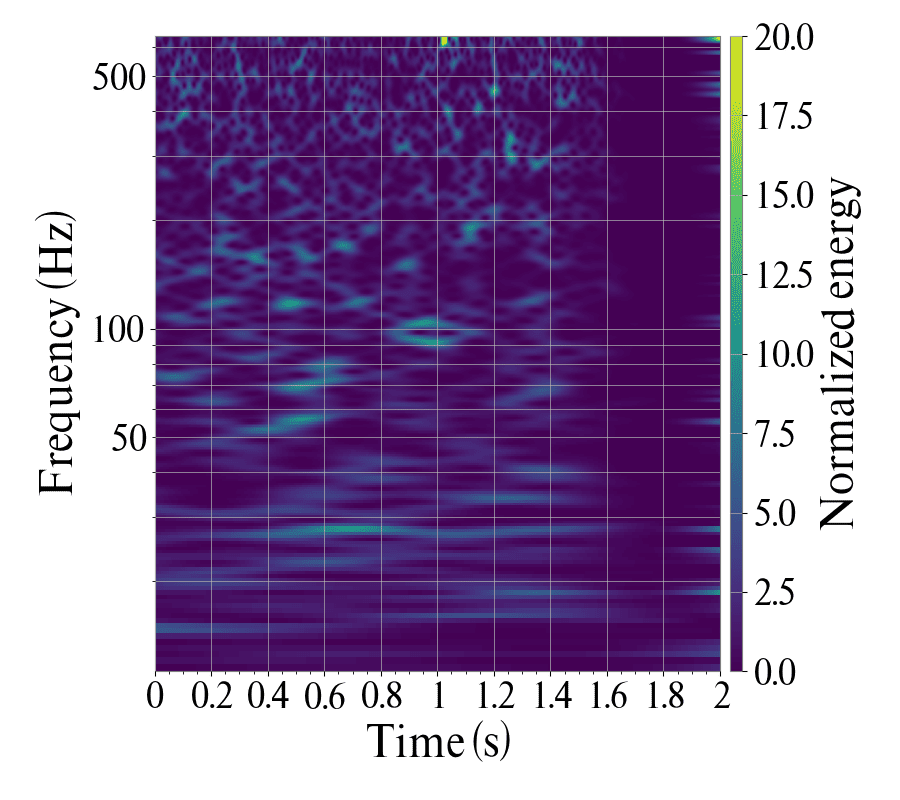}
    \par\smallskip
    {\scriptsize \scalebox{0.8}[1.0]{\textit{K1:PEM-MIC\_SR\_BOOTH\_SR\_Z\_OUT\_DQ}}\\ Apr 16, 2020	15:14:52	UTC	(GPS: 1271085310)}
    \end{minipage}
\hfill
\vrule width 0.5pt
\hfill    
\begin{minipage}[t]{0.45\textwidth}
    \centering
    \includegraphics[width=0.48\textwidth]{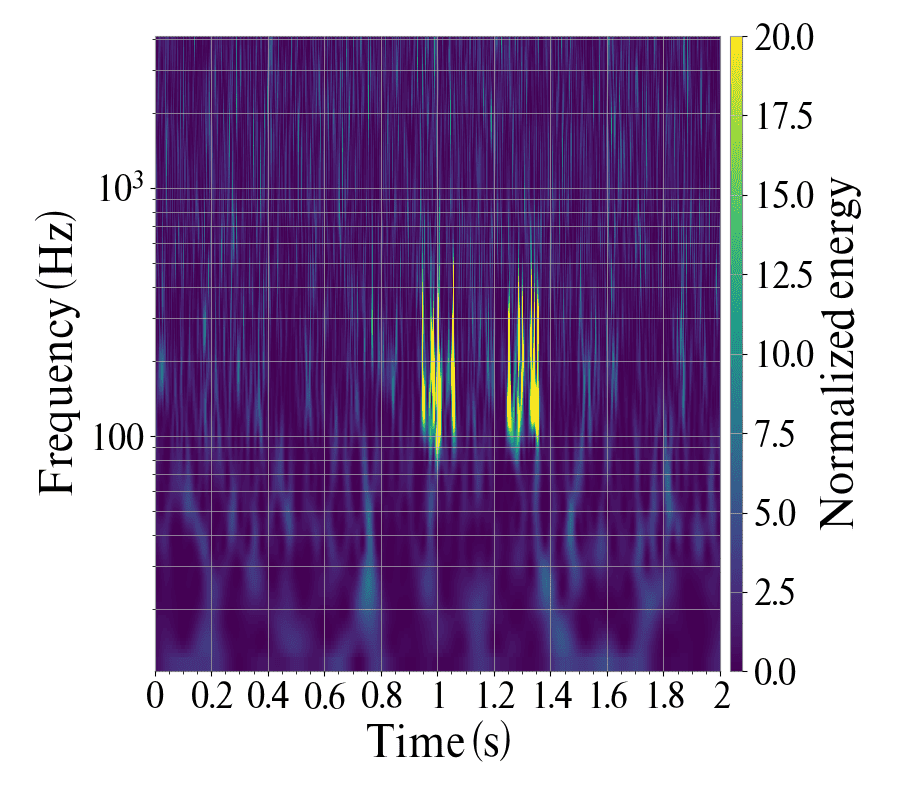}
    \includegraphics[width=0.48\textwidth]{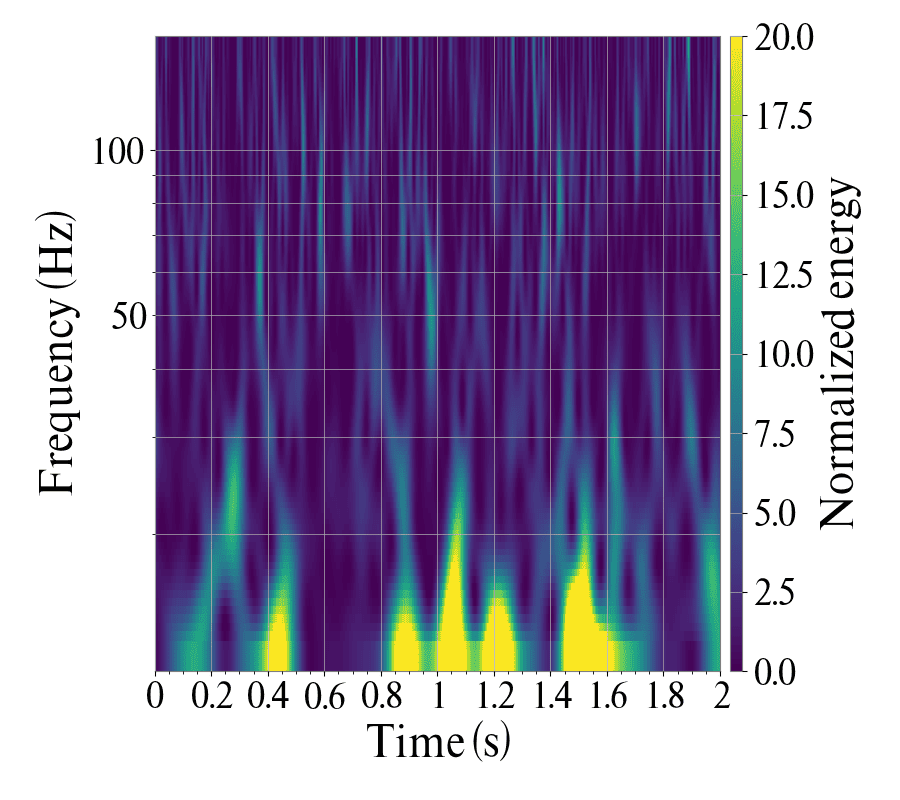}
    \par\smallskip
    {\scriptsize \scalebox{0.8}[1.0]{\textit{K1:PEM-SEIS\_IXV\_GND\_EW\_IN1\_DQ}}\\ Apr 07, 2020	22:17:27	UTC	(GPS: 1270333065)}
    \end{minipage}
    }

\resizebox{!}{0.136\textheight}
{%
\begin{minipage}[t]{0.45\textwidth}
    \centering
    \includegraphics[width=0.48\textwidth]{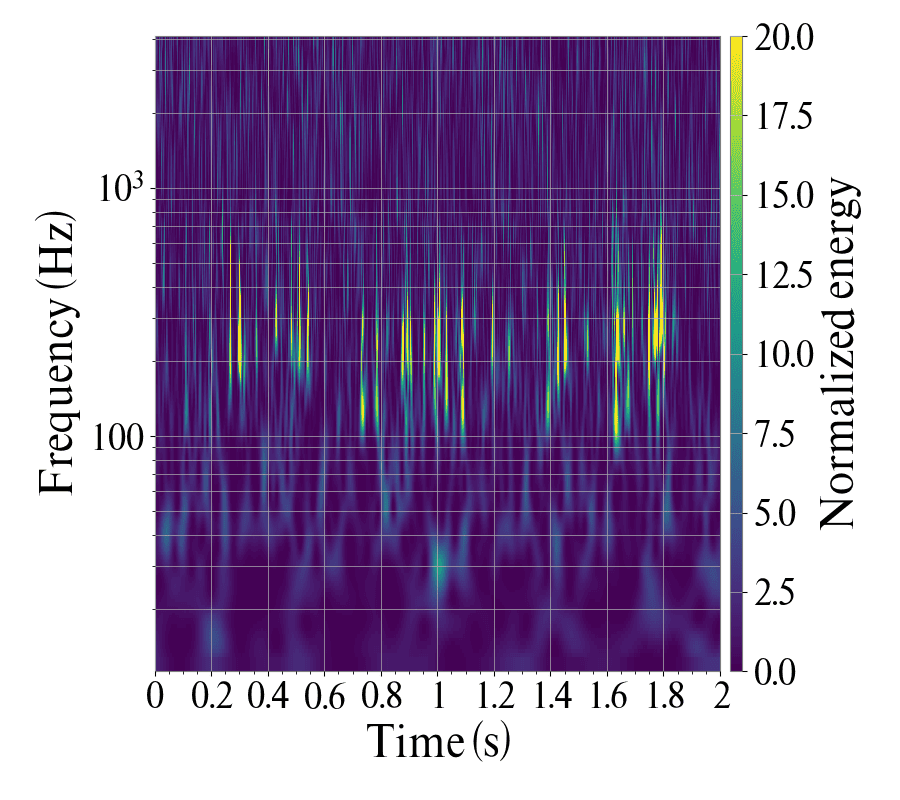}
    \includegraphics[width=0.48\textwidth]{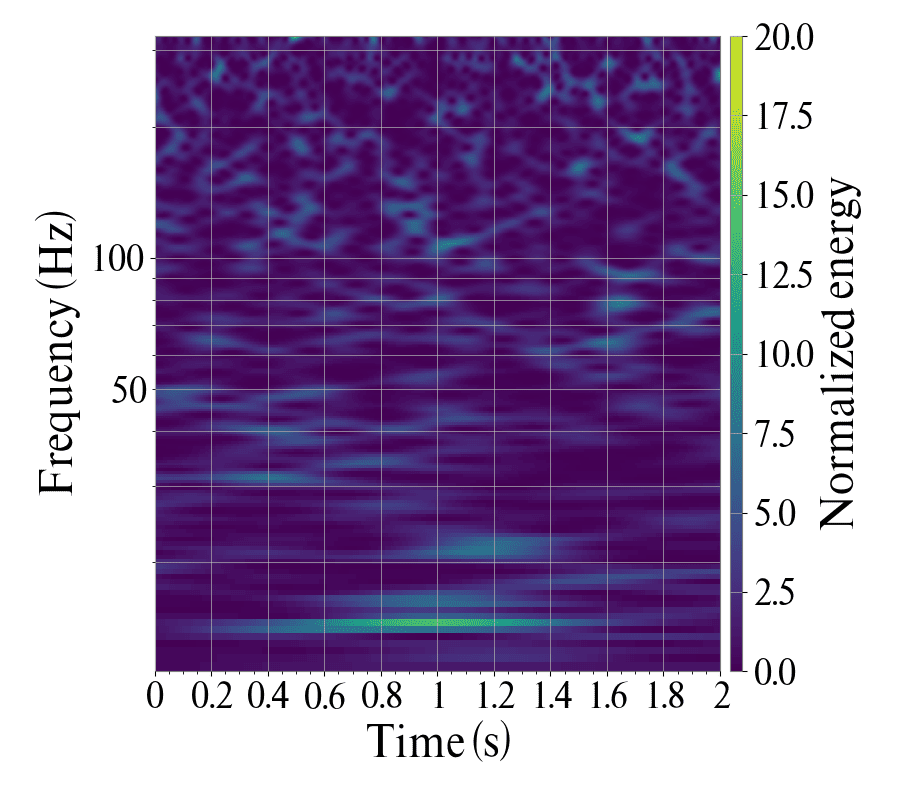}
    \par\smallskip
    {\scriptsize \scalebox{0.8}[1.0]{\textit{K1:VIS-ETMX\_MN\_PSDAMP\_R\_IN1\_DQ}}\\ Apr 07, 2020	22:29:20	UTC	(GPS: 1270333778)}
    \end{minipage}
\hfill
\vrule width 0.5pt
\hfill
\begin{minipage}[t]{0.45\textwidth}
    \centering
    \includegraphics[width=0.48\textwidth]{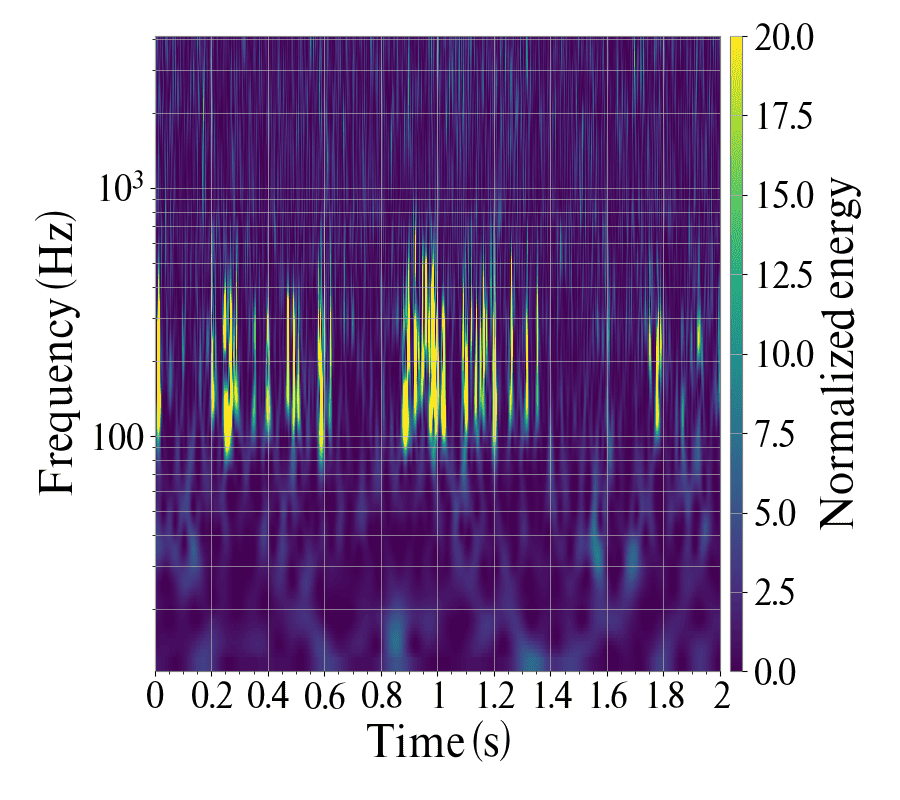}
    \includegraphics[width=0.48\textwidth]{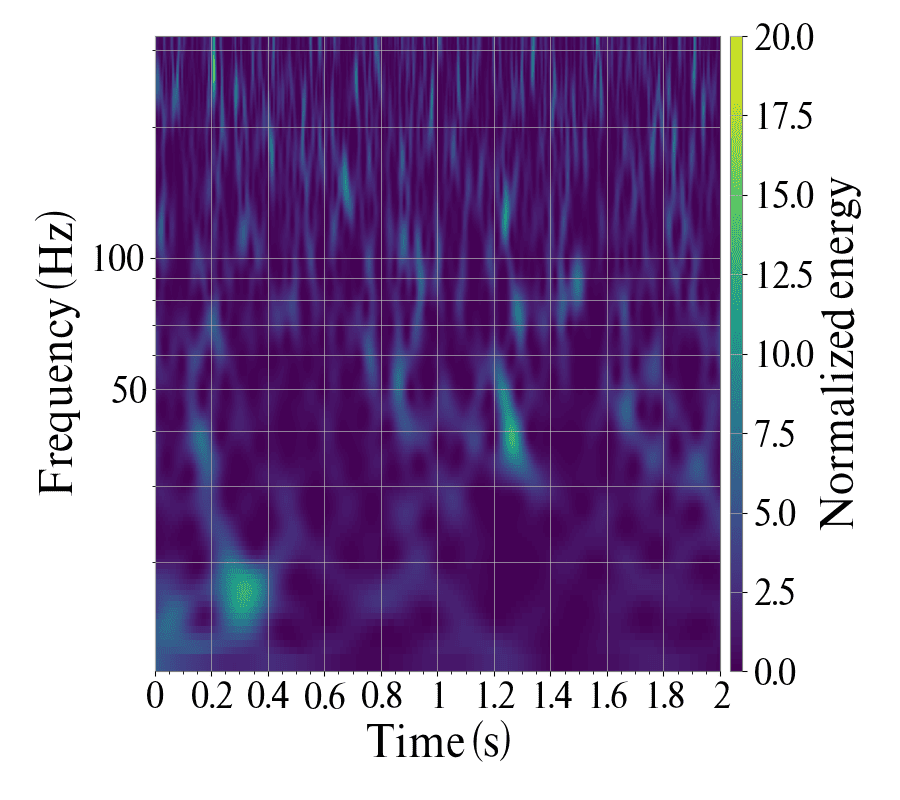}
    \par\smallskip
    {\scriptsize \scalebox{0.8}[1.0]{\textit{K1:VIS-ETMX\_MN\_PSDAMP\_Y\_IN1\_DQ}}\\ Apr 07, 2020	22:28:52	UTC	(GPS: 1270333750)}
    \end{minipage}
    }
\end{figure}

\clearpage

\subsubsection{Scattered Light Glitch}
scattered light glitch appears as repeated arches on the main channel spectrogram.
During the O3GK period, a total of 55 scattered light glitches were discovered (see Tables \ref{table:Glitchtypes} and \ref{table:scattered light Glitches}). Unlike the other types of glitch, the KAGRA O3GK scattered light glitches were found in only two subsystems (AOS and LSC) of KAGRA.  
\begin{table*}[!h]
\centering
\caption{\label{table:scattered light Glitches} O3GK KAGRA scattered light glitches}
\small
\resizebox{0.98\textwidth}{!}
{%
\begin{tabular}{|P{0.07\textwidth}|P{0.58\textwidth}|P{0.35\textwidth}|}
\hline
\begin{tabular}[c]{@{}c@{}}Sub-\\ System\end{tabular} & \begin{tabular}[c]{@{}c@{}}Round Winner\\ Auxiliary Channel\end{tabular} & \begin{tabular}[c]{@{}c@{}}Vetoed Date in April\\ (\# of Vetoed Events)\end{tabular} \\ \hline

\begin{tabular}[c]{@{}c@{}}AOS \\ (2) \end{tabular}
& \textit{K1:AOS-TMSX\_IR\_PDA1\_OUT\_DQ}
& 16th (2) \\ \hline

\multirow{2}{*}{\begin{tabular}[c]{@{}c@{}}LSC \\ (53) \end{tabular}}
& \begin{tabular}[c]{@{}c@{}}\textbf{K1:LSC-ALS\_CARM\_OUT\_DQ}\end{tabular}
& 16th (18) 
\\ \cline{2-3}
& \begin{tabular}[c]{@{}c@{}}\textbf{K1:LSC-ALS\_DARM\_OUT\_DQ}\end{tabular}
& 12th (1), 16th (34) 
\\ \hline

\end{tabular} }
\end{table*}

\begin{figure*}[h]
\centering
\resizebox{!}{0.136\textheight}
{%
\begin{minipage}[t]{0.45\textwidth}
    \centering
    \includegraphics[width=0.48\textwidth]{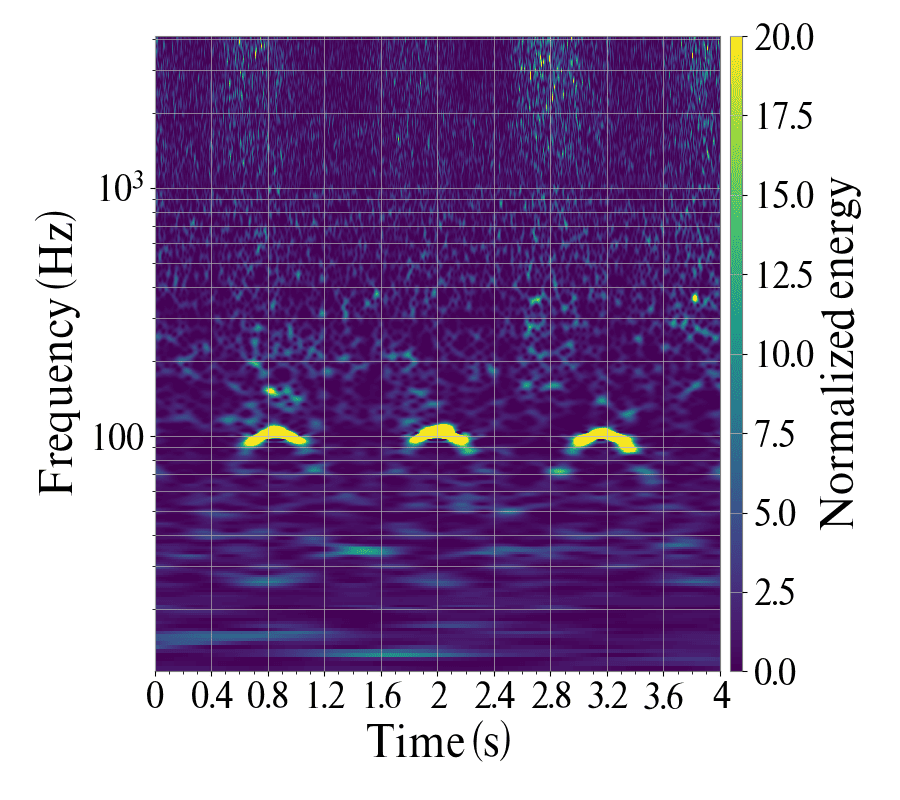}
    \includegraphics[width=0.48\textwidth]{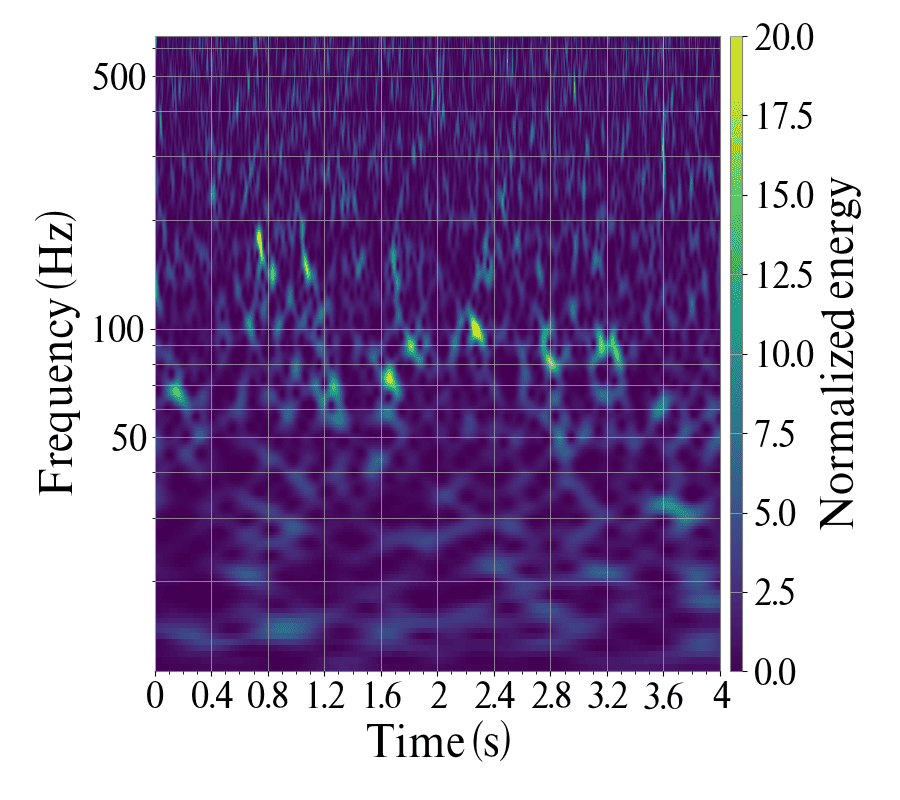}
    \par\smallskip
    {\scriptsize \scalebox{0.8}[1.0]{\textit{K1:AOS-TMSX\_IR\_PDA1\_OUT\_DQ}} \\ Apr 16, 2020, 05:38:33 UTC (GPS:1271050731)}
    \end{minipage}
\hfill
\vrule width 0.5pt
\hfill
\begin{minipage}[t]{0.45\textwidth}
    \centering
    \includegraphics[width=0.48\textwidth]{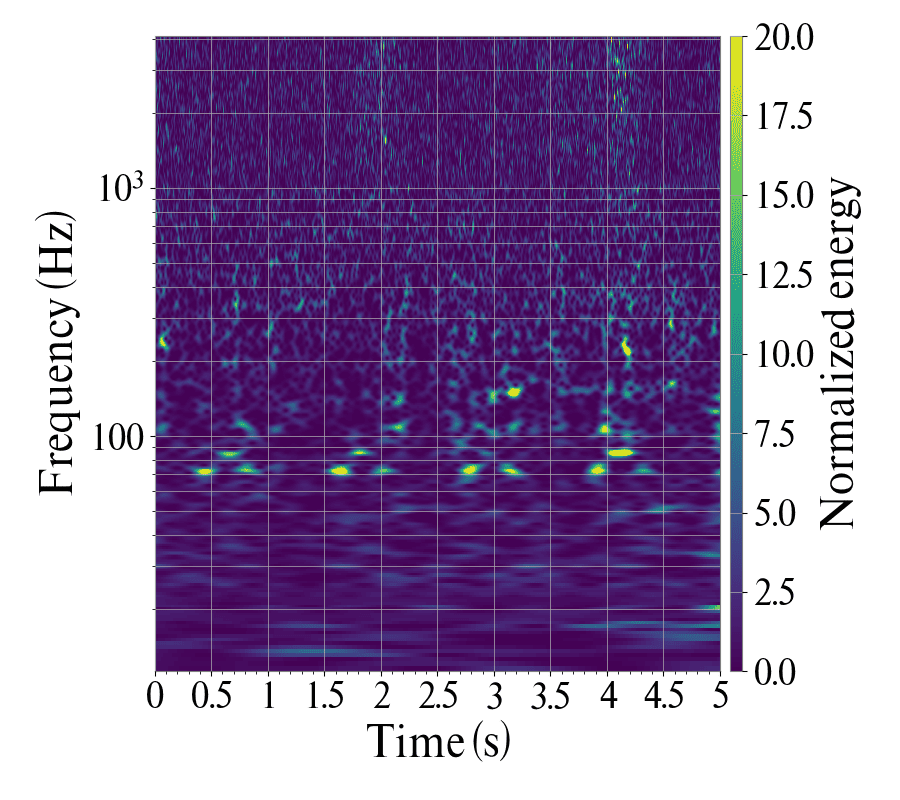}
    \includegraphics[width=0.48\textwidth]{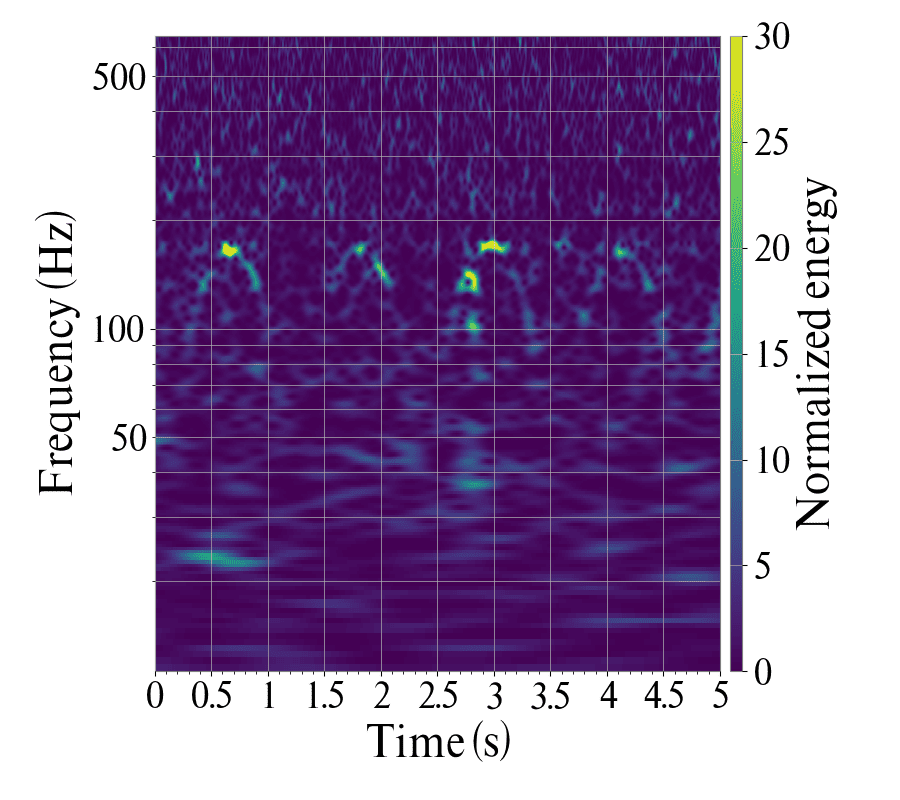}
    \par\smallskip
    {\scriptsize \scalebox{0.8}[1.0]{\textbf{K1:LSC-ALS\_CARM\_OUT\_DQ}} \\ Apr 16, 2020	11:41:29	UTC (GPS: 1271072507)}
    \end{minipage}
    }
    
\resizebox{!}{0.136\textheight}
{%
\begin{minipage}[t]{0.45\textwidth}
    \centering
    \includegraphics[width=0.48\textwidth]{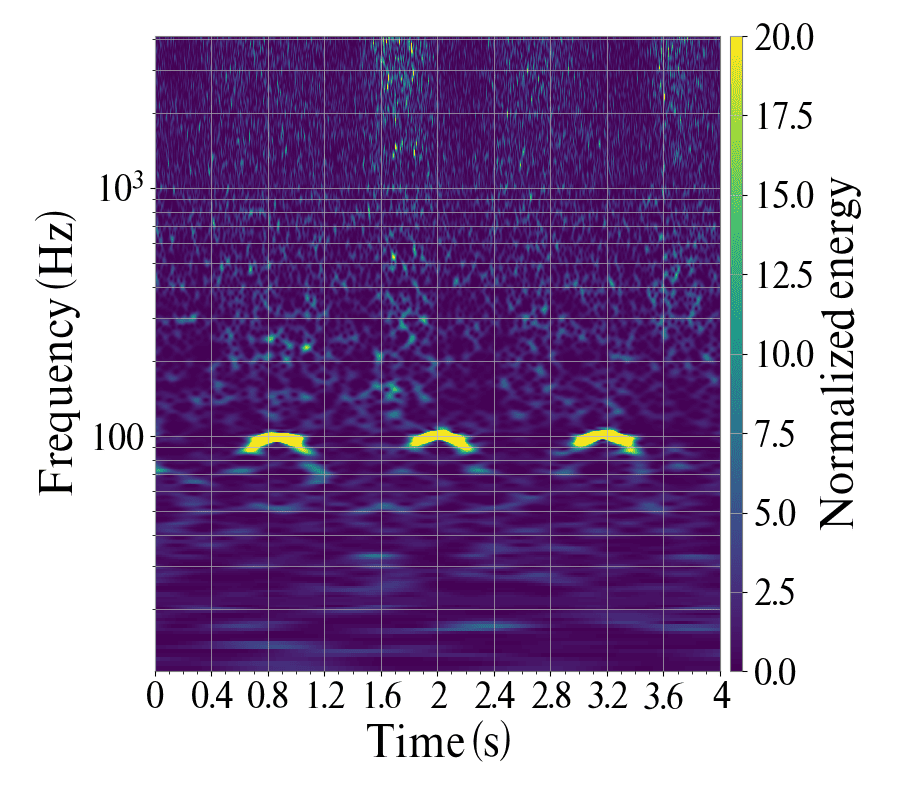}
    \includegraphics[width=0.48\textwidth]{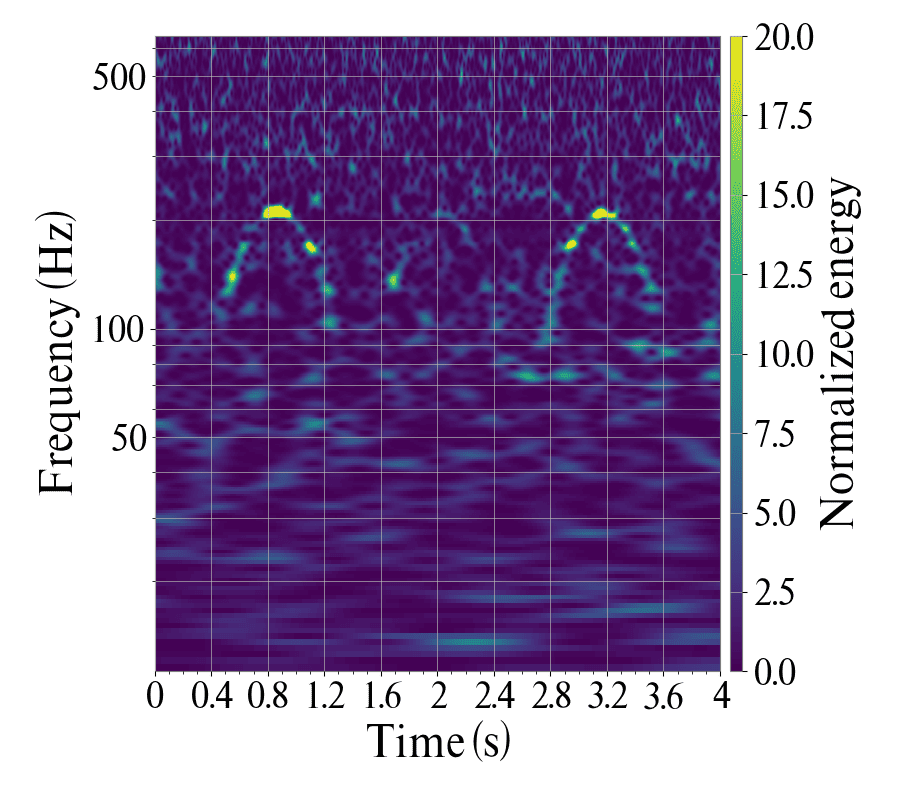}
    \par\smallskip
    {\scriptsize \scalebox{0.8}[1.0]{\textbf{K1:LSC-ALS\_DARM\_OUT\_DQ}} \\ Apr 16, 2020	05:38:28	UTC (GPS:1271050726)}
    \end{minipage}
    }
    \caption{An O3GK KAGRA scattered light glitch found. The left and right panels show the spectrogram of the main channel and an auxiliary channel, respectively. Each figure also includes, below the panels, the name of the auxiliary channel and the UTC time (GPS time) when the glitch was found.}
  \label{fig:scattered light figs}
\end{figure*}

Figure \ref{fig:scattered light figs} shows examples of the KAGRA O3GK scattered light glitch. The spectrogram of K1:LSC-ALS\_CARM\_OUT\_DQ is indicated in bold in Table \ref{table:scattered light Glitches} because the auxiliary channel spectrogram also shows the scattered light glitch pattern as in the main channel spectrogram, although the frequency ranges are not identical. We need to mention that this auxiliary channel is determined to be correlated to the main channel, based upon a similar pattern that has a tight time coincidence. For example, the peak of each arch is located at a similar time in both the auxiliary and the main-channel spectrograms. In contrast, the spectrogram of K1:AOS-TMSX\_IR\_PDA1\_OUT\_DQ is indicated in italic font in Table \ref{table:scattered light Glitches} because there is no arch-like pattern on the auxiliary channel spectrogram (right panel). 

\section{Discussion}\label{sec:discussion}

In this work, we searched for the statistical time correlation between main channel and auxiliary channels by applying Hveto to the O3GK KAGRA data and tried to figure out their further correlation that might be caused by physical reason by investigating similarity or difference, in the glitch shape on the spectrogram. Our investigation was presented by bold (auxiliary channel that shows similar glitch pattern to the main channel) or italic (auxiliary channel that shows different glitch pattern from the main channel) font in Tables \ref{table:BlipGlitches} to \ref{table:scattered light Glitches}. 

GWpy provides a module to calculate coherence between two channels (\url{https://gwpy.github.io/docs/1.0.0/examples/frequencyseries/coherence.html}), and we used this module to calculate coherence for O3GK KAGRA data. Coherence is a quantitative measure that indicates the linear correlation between two sets of time-series data, and provides insight into the frequency-domain correlation between channels. The correlated auxiliary channel identified through coherence analysis was previously used for noise subtraction in LIGO data~\cite{coherence_article_Derek}. In addition, successful subtraction of certain glitch classes has been demonstrated using auxiliary channels, particularly for persistent or repeated noise sources~\cite{Davis_2022}. However, the glitches analyzed in this study are short-duration transient events, for which direct subtraction is generally more difficult. We therefore focus on identifying statistically significant correlations to support vetoing or flagging of noisy segments. Therefore, our analysis emphasizes the identification of statistically significant correlations between glitches and auxiliary channels, aiming to support vetoing or flagging of noisy segments rather than performing explicit subtraction.

In our study, coherence analysis can be further utilized to assess the measure of frequency correlation between the glitches observed in the main channel and the round-winner auxiliary channels identified by Hveto. Since auxiliary channels in the detector are originally designed to monitor noise within specific frequency bands, the presence of a glitch in the same frequency range as that of an auxiliary channel suggests effective noise detection by the channel. Conversely, if a glitch appears in a different frequency range, it may suggest that the coincidence is accidental rather than causal.

In this section, we discuss the validity of our qualitative findings on the correlation between the main channel and the auxiliary channels, which were initially determined by visual inspection. We selected two representative cases and compared our qualitative results with the quantitative coherence measurements.

\begin{figure}[h]
\centering
\includegraphics[width=0.9\linewidth]{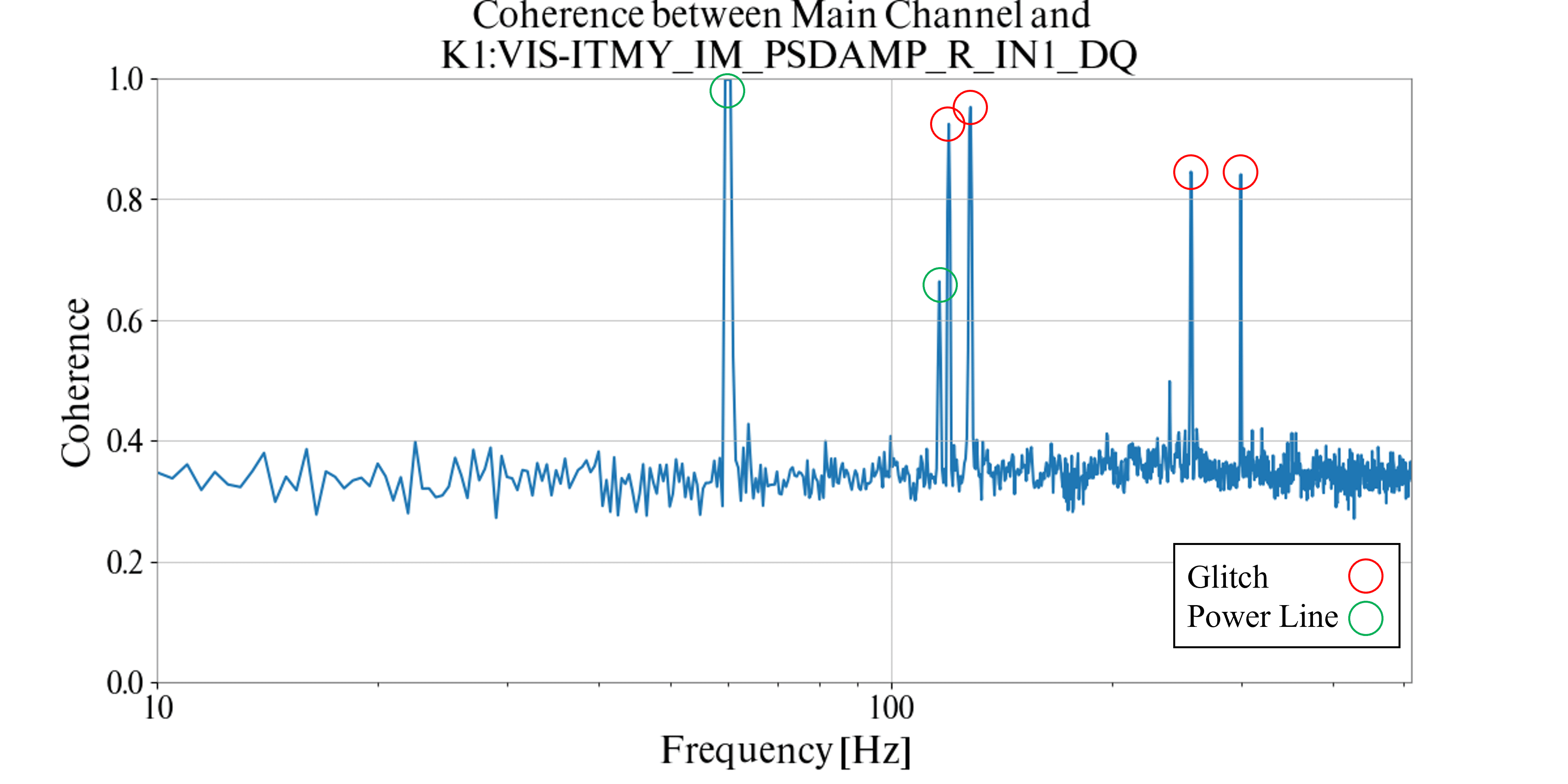}
\caption{Coherence plot between the main channel and the auxiliary channel K1:VIS-ITMY\_IM\_PSDAMP\_R\_IN1\_DQ.}
\label{fig:coherence_ITMY}
\end{figure}

Figure \ref{fig:coherence_ITMY} presents the average coherence plot between the main channel and the auxiliary channel K1:VIS-ITMY\_IM\_PSDAMP\_R\_IN1\_DQ. During the O3GK observation period, this channel was responsible for vetoing 193 events (Table \ref{table3:Vetoedchannels}), of which 187 events were classified into each type of glitch by visual examination. The glitches were classified as \textbf{ blip} (161 events), \textbf{dot} (4 events), and \textbf{helix} (22 events) (Tables \ref{table:BlipGlitches} to \ref{table:HelixGlitches}: bold font indicating that this auxiliary channel shows a pattern of glitch similar to the main channel). 
The average coherence plot in Figure \ref{fig:coherence_ITMY} was drawn using the following steps. For each event vetoed by the auxiliary channel K1:VIS-ITMY\_IM\_PSDAMP\_R\_IN1\_DQ, we chose the 4 second time window ($\pm2$ seconds around each event) in the time series data of both the main channel and the auxiliary channel. We then calculated the coherence values over the 4 second time window for all vetoed events and averaged them. However, for some of the vetoed events, the 4 second time window is contaminated with non-glitch noises. We excluded these events in the calculation of the average coherence plot after verifying the contamination with visual examination of the spectrogram. For example, in the case of Figure \ref{fig:coherence_ITMY}, a total of 169 out of the 193 vetoed events are chosen for the coherence calculation.

In Figure \ref{fig:coherence_ITMY}, high coherence values are observed near 60 Hz and at four distinct points between 100 Hz and 300 Hz (red), indicating that this auxiliary channel detects noise at specific frequencies where coherence is elevated. The well-known coherence peaks at 60 Hz and 120 Hz are associated with the power line frequency and its harmonics (green). The coherence peak near 300 Hz corresponds to the frequency region around 295 Hz identified by Hveto as the band in which this auxiliary channel significantly affects the main channel. In this frequency band, blip and helix glitches are predominantly observed (e.g., see Figure \ref{fig:Helixfigs}). Although the occurrence of dot glitches is relatively low, they are observed around 243 Hz, which also shows a prominent peak in the coherence plot. These observations suggest that future studies using coherence analysis to identify the characteristic frequency bands monitored by auxiliary channels could enhance the efficiency of glitch classification in the main channel, particularly in regions exhibiting high coherence.

\begin{figure}[h]
\centering
\includegraphics[width=0.9\linewidth]{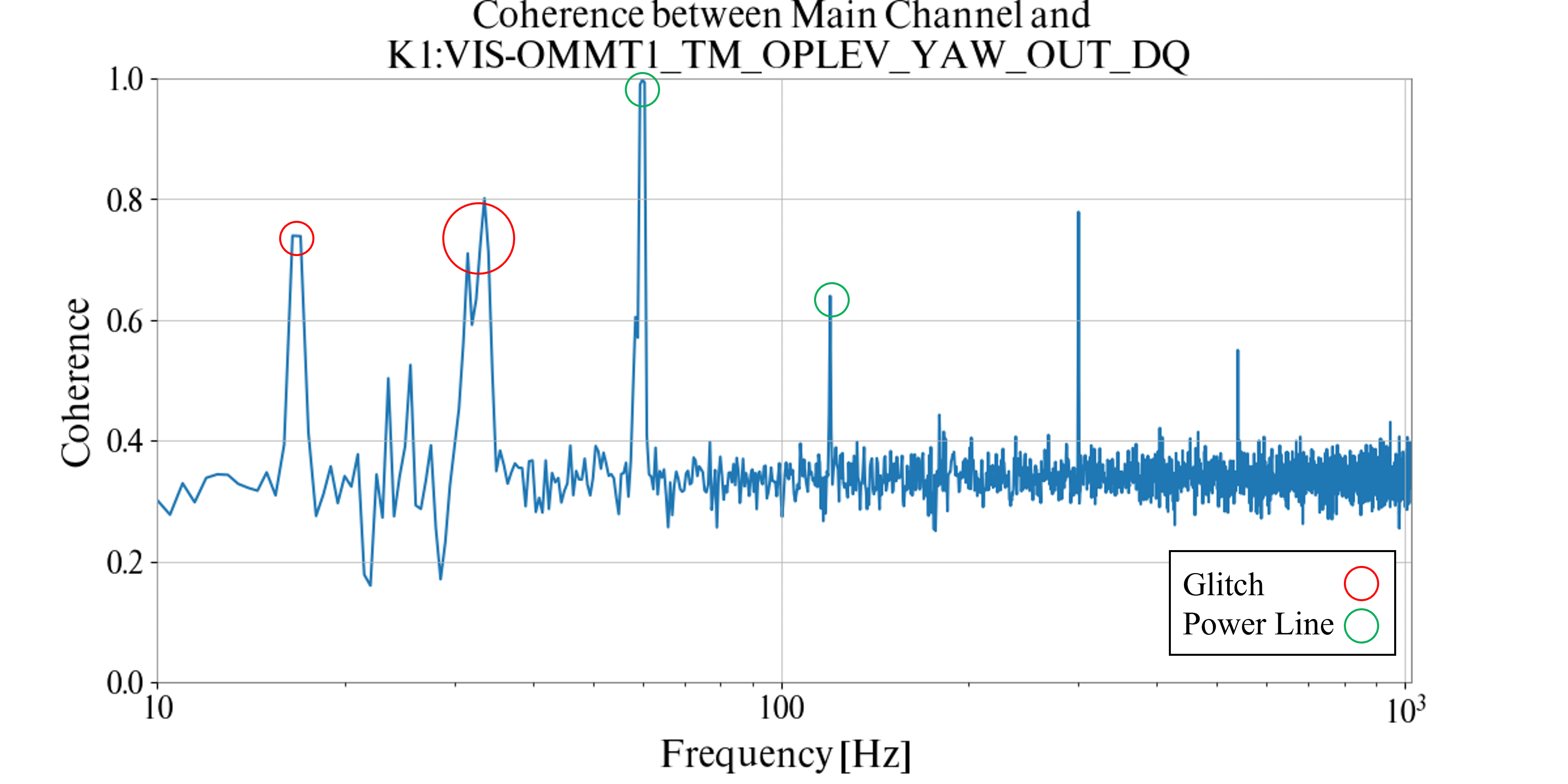}
\caption{Coherence plot between the main channel and the auxiliary channel K1:VIS-OMMT1\_TM\_OPLEV\_YAW\_OUT\_DQ.}
\label{fig:coherence_OMMT}
\end{figure}

Figure \ref{fig:coherence_OMMT} shows another average coherence plot for the auxiliary channel K1:VIS-OMMT1\_TM\_OPLEV\_YAW\_OUT\_DQ. The period (Table \ref{table3:Vetoedchannels}), of which 119 events were classified into \textit{blip} (23 events), \textbf{dot} (3 events), \textit{helix} (1 event), \textbf{line} (86 events), and \textit{scratchy} (6 events) glitch (Tables \ref{table:BlipGlitches} to \ref{table:ScratchyGlitches}: bold/italic font indicating that this auxiliary channel shows a similar/different glitch pattern to the main channel). The average coherence for this channel in Figure \ref{fig:coherence_OMMT} is calculated with the same process for the auxiliary channel as in Figure \ref{fig:coherence_ITMY}. For the coherence calculation in Figure \ref{fig:coherence_OMMT}, 19 contaminated events
(5 blip, 6 line, 1 scratchy, and 7 not classified) were excluded from the 136 vetoed events. 

Unlike Figure \ref{fig:coherence_ITMY}, Figure \ref{fig:coherence_OMMT} exhibits relatively high coherence values at 16 Hz and 32 Hz (red), while the four coherence peaks observed between 100 Hz and 300 Hz in the previous figure are absent or significantly lower. This suggests that the auxiliary channel in question is primarily sensitive to noise at 16 and 32 Hz. Furthermore, this auxiliary channel exhibits glitch patterns that differ from those typically observed in the main channel. In particular, the coherence values are relatively low in the frequency bands where blip, helix, and scratchy glitches commonly occur. This observation quantitatively supports our qualitative conclusion that the main and auxiliary channels are weakly correlated in these frequency regions. Therefore, the association between this auxiliary channel and the blip, helix, and scratchy glitches is likely to be insignificant. In particular, 16 and 32 Hz correspond to the frequencies where dot and line glitches predominantly occur in the main channel spectrograms (see Figures \ref{fig:Dotfigs} and \ref{fig:Linefigs}).

From the above two exemplary auxiliary channels, we demonstrate that the pattern similarity or difference in the spectrogram between the main channel and an auxiliary channel can be explained with more quantitative measurements, such as coherence. For example, Figure \ref{fig:coherence_ITMY} includes the auxiliary channels associated with blip, helix, and dot glitches, all of which exhibit strong visual similarity in their spectrograms. In contrast, Figure \ref{fig:coherence_OMMT} features channels related to blip, dot, helix, line, and scratch glitches, but shows strong similarity only for dot and line glitches. This distinction is also reflected in the coherence graphs: for example, in the 100–300 Hz band, where blip, helix, and scratch glitches frequently appear, the coherence values are low in Figure \ref{fig:coherence_OMMT} but higher in Figure \ref{fig:coherence_ITMY}. In contrast, at 16 and 32 Hz — where line glitches are commonly found — Figure \ref{fig:coherence_OMMT} shows stronger coherence than Figure \ref{fig:coherence_ITMY}. These observations suggest that examining the coherence between the main channel and each round winner can help identify the frequency bands associated with specific glitch types and assess whether their correlation with auxiliary channels is physically meaningful or coincidental.

In future KAGRA observing runs, coherence analysis could be combined with Hveto to enhance noise characterization, as it offers a faster and more objective alternative to spectrogram-based visual inspection. This approach may also be applicable to more recent datasets as they become available for study. The merit of using coherence is particularly evident when considering the large volume of data expected in future observing runs.

\section{Summary}\label{sec:summary}

Applying Hveto to the KAGRA O3GK data with the configuration in Table \ref{table1Configuration}, we identified 2,531 noise events, which were vetoed by 28 round winner auxiliary channels. We presented all vetoed events during the O3GK observation period in a tabular format. Tables \ref{table2:HvetoSummary} (sorted daily) and \ref{table3:Vetoedchannels} (according to the auxiliary channels in each subsystem).

For each vetoed event (or glitch), we visually investigate its pattern on the spectrogram of the main channel (i.e. strain data) and classify it into one of six glitch types (blip, dot, helix, line, scratchy, and scattered light) by using the same classification criterion as the Gravity Spy project. Of 2,531 events, only 177 glitches were not identified as one of six glitch types, but the remaining 2,354 glitches were classified into six glitch types. Table \ref{table:Glitchtypes} summarizes the number of glitches classified in each type of glitch that were found in each KAGRA subsystem. 

Because each glitch is vetoed (or identified) by a specific auxiliary channel, we also draw the glitch spectrogram for the round-winner auxiliary channel together with that for the main channel. We generated spectrogram images of both the main channel and the corresponding round-winner auxiliary channel for all glitches (i.e., all 2,531 vetoed events) as part of our analysis.
We describe individual six types of the O3GK KAGRA glitches by taking some examples. In particular, we visually examine the similarity or difference in the glitch pattern between the main channel and the auxiliary channel because it is known that a similar (different) pattern implies a strong (weak) correlation between the main channel and the auxiliary channel. We present the results of our visual inspection in bold or italic font for the round winner auxiliary channel, which corresponds to a similar or different pattern, respectively, in Tables \ref{table:BlipGlitches} to \ref{table:scattered light Glitches}. We discuss the validity of our qualitative approach based upon visual inspection by comparing with coherence, a quantitative way of measuring the correlation between the main channel and the auxiliary channel. A couple of exemplary auxiliary channels show that our results based on visual inspection are consistent with the coherence measurements. 

Although our current analysis does not directly demonstrate noise subtraction or sensitivity improvement, it provides valuable diagnostic information by identifying auxiliary channels that are statistically or morphologically correlated with glitches in the strain channel. This information supports the efforts of commissioning teams in investigating possible noise coupling paths and developing mitigation strategies. For example, as shown in~\cite{coherence_article_Derek}, identifying strongly correlated auxiliary channels is a necessary step for noise subtraction in some cases. Furthermore, by applying similar analytical methods to future observing runs, we will be able to assess whether the same correlations persist after hardware upgrades implemented after O3GK. Although comparisons with glitch characteristics observed in LIGO and Virgo may offer additional insight, we plan to incorporate such interdetector studies once KAGRA achieves comparable sensitivity and duty cycle. This will ultimately contribute to a deeper understanding of the noise environment in KAGRA and guide future sensitivity improvements.

\ack
This work was supported by MEXT, the JSPS Leading-edge Research Infrastructure Program, JSPS Grantin-Aid for Specially Promoted Research 26000005, JSPS Grant-in-Aid for Scientific Research on Innovative
Areas 2905: JP17H06358, JP17H06361 and JP17H06364, JSPS Core-to-Core Program A, Advanced Research Networks, JSPS Grants-in-Aid for Scientific Research (S) 17H06133 and 20H05639, JSPS Grant-inAid for Transformative Research Areas (A) 20A203: JP20H05854, the joint research program of the Institute
for Cosmic Ray Research, the University of Tokyo, the National Research Foundation (NRF), the Computing
Infrastructure Project of Global Science experimental Data hub Center (GSDC) at KISTI, the Korea Astronomy and Space Science Institute (KASI), the Ministry of Science and ICT (MSIT) in Korea, Academia Sinica
(AS), the AS Grid Center (ASGC) and the National Science and Technology Council (NSTC) in Taiwan under grants including the Rising Star Program and Science Vanguard Research Program, the Advanced
Technology Center (ATC) of NAOJ, and the Mechanical Engineering Center of KEK.
KJ and KK were partially supported by NRF grant (RS-2025-00516133) and Institute of Information \& communications Technology Planning \& Evaluation (IITP) grant (RS-2021-II212068, Artificial Intelligence Innovation Hub) funded by the Korea government (MSIT).  
YMKim was partially supported by NRF grant (RS-2022-NR072453) and KASI R\&D program (Project No. 2025-1-810-02) supervised by the MSIT.

\section{Appendix}

\begin{table*}[h!]
\caption{\label{table1Configuration}Omicron and Hveto Configuration.}
\centering
\resizebox{0.95\textwidth}{!}
{%
\small
\begin{tabular}{|@{}c|c|c|}
\hline
&Omicron&Hveto\\ \hline\hline
Version & 2.4.2 & 1.0.1\\ \hline
Configuration & \makecell[l]{PARAMETER TIMING 64 4 \\ PARAMETER QRANGE 4 128 \\ PARAMETER MISMATCHMAX 0.2 \\ PARAMETER SNRTHRESHOLD 6 \\ PARAMETER PSDLENGTH 128 \\ PARAMETER CLUSTERING 0.1} & 
\makecell[c]{minimum-significance = 5.0 \\ time-windows = \\ 0.01, 0.02, 0.04, 0.08, \\ 0.10, 0.20, 0.40, 0.80, 1.00 \\ snr-threshold (primary) = 8.0 }\\
\hline
\end{tabular}}
\end{table*}
In the Omicron configuration, most of the important parameters are relevant for performing the Q-transformation.
PARAMETER TIMING specifies the timing parameters for the Q-transformation. The first value (e.g., 64 in our work), which must be given as an even integer, sets the timing window size in seconds, and the second value (e.g., 4 in our work) is the overlap duration in seconds. In other words, two consecutive timing windows overlap with the second value.
PARAMETER QRANGE sets the search range of the Q value in the Q-transformation, with the two values (4 and 128) corresponding to the lower and upper bounds of the search, respectively.
PARAMETER MISMATCHMAX specifies the maximum energy mismatch between time-frequency tiles used in the Q-transformation. This parameter must be set between 0 and 1.
PARAMETER SNRTHRESHOLD sets the signal-to-noise ratio, above which the Omicron generates triggers.
PARAMETER PSDLENGTH sets the duration, in seconds, over which the power spectral density (PSD) is estimated.
PARAMETER CLUSTERING activates the clustering method with the parameter in seconds. The clustering algorithm in Omicron groups separate time intervals into predefined seconds to prevent the same event from being detected twice in two consecutive time intervals. For further details on these parameters and others, please refer to the Omicron online documentation\footnote{\url{https://virgo.docs.ligo.org/virgoapp/Omicron/index.html}}.

In the Hveto configuration,
significance evaluates the similarity of triggers between channels, and the minimum-significance value sets the criterion to measure this significance.
The time-window value sets the time interval in seconds, and we use nine different values for the time-window in this work. Hveto produces the round winners after comparing triggers between the main channel and auxiliary channels using all of these nine time-windows. Each comparison is done within the given time-window.
Hveto operates in a hierarchical manner through multiple rounds, where each round aims to statistically veto glitches by identifying one auxiliary channel most significantly associated with the main channel. The process iterates through rounds until no auxiliary channel meets the statistical threshold defined by the configuration.
The snr-threshold determines which Omicron triggers are included in the Hveto analysis; only those triggers with SNR values exceeding this threshold are used for statistical evaluation. Triggers above this value are searched for the statistical time correlation. Note that the snr-threshold in the Hveto configuration must be larger than that in the Omicron configuration. For more information on Hveto, please refer to the online documentation\footnote{\url{https://doi.org/10.5281/zenodo.3532131}}.

\bibliographystyle{ptephy}
\bibliography{2505-033-3F-KihyunJung}

\end{document}